%% file: main.tex
\documentclass{report}
\usepackage{pdflscape}  
\usepackage[T1]{fontenc}       
\usepackage[utf8]{inputenc}    

\usepackage{subcaption}        
\usepackage{graphicx}          
\usepackage{csquotes}          
\usepackage{csquotes}
\usepackage{colortbl}
\usepackage{todonotes}
\usepackage{tablefootnote}
\usepackage[export]{adjustbox}
\usepackage{wrapfig}
\usepackage[mathscr]{eucal}
\usepackage{amsmath}
\usepackage{mathtools}
\usepackage{multicol}
\usepackage{algorithm} 
\usepackage{algpseudocode}   
\usepackage{bm}
\usepackage{hyperref}
\usepackage{booktabs}
\usepackage{xparse}
\usepackage{xifthen}
\usepackage{multirow}
\usepackage{textcomp}
\usepackage{pifont}
\usepackage{soul}
\usepackage{xcolor}          
\usepackage{upgreek}
\usepackage{geometry}
\usepackage{footnote}
\makesavenoteenv{tabular}
\usepackage{lipsum}
\usepackage{amssymb}
\usepackage{comment}
\usepackage{longtable}
\usepackage{float}           
\usepackage[T1]{fontenc}
\usepackage[numbers]{natbib}
\definecolor{anti-flashwhite}{rgb}{0.95, 0.95, 0.96}
\usepackage{enumitem}

\usepackage{pdfpages}

\usepackage{titlesec}   
\usepackage{array} 

\newcolumntype{R}[1]{>{\raggedright\arraybackslash}p{#1}}
\renewcommand{\thepart}{\Roman{part}}

\titleformat{\part}[display]
  {\centering\Huge\bfseries}        
  {\partname~\thepart}              
  {1em}                             
  {}                                

\titlespacing*{\part}{0pt}{50pt}{40pt}

\newcounter{subsubsubsection}[subsubsection]

\makeatletter
\newcommand{\l@subsubsubsection}{\@dottedtocline{4}{5.5em}{3.5em}}  
\makeatother

\makeatletter
\newenvironment{breakablealgorithm}
  {%
    \begin{center}
    \refstepcounter{algorithm}
    \hrule height .8pt depth 0pt \kern 2pt%
    \renewcommand{\caption}[2][\relax]{%
      {\raggedright \textbf{\ALG@name~\thealgorithm} ##2\par}%
      \kern 2pt\hrule\kern 2pt
    }%
  }{%
    \kern 2pt\hrule%
    \end{center}
  }
\makeatother

\title{Self-Organizing Adaptive
Knowledge Evolution and Discovery:
A Dense Domain-Specific Multi-Agent,
Multi-Context Knowledge Graph Framework
for Transferable and Traceable Knowledge}
\author{
  Ryan Barron \\
  \\
  \textbf{Advisor \& Committee Chair:}\\ 
  Dr. Cynthia Matuszek\\
  \\
  \textbf{Committee Members:}\\ 
  Dr. Charles Nicholas\\
  Dr. Tim Finin\\ 
  Dr. Roberto Yus\\ 
  Dr. Boian Alexandrov\\
  Dr. Manish Bhattarai\\
  \\
  \textbf{Public Release:}\\
  LA-UR-TBD
}
\date{July 2025}

\renewcommand{\textsl}{\textit}
\newcommand{\comm}[1]{}
\newcommand\numberthis{\addtocounter{equation}{1}\tag{\theequation}}
\NewDocumentCommand{\vect}{ O{} O{} m }{\mathbf{#3}\ifthenelse{\isempty{#1}}{}{^{(#1)}}\ifthenelse{\isempty{#2}}{}{_{#2}}}
\NewDocumentCommand{\mat}{ O{} O{} m }{\mathbf{#3}\ifthenelse{\isempty{#1}}{}{^{(#1)}}\ifthenelse{\isempty{#2}}{}{_{#2}}}
\NewDocumentCommand{\ten}{ O{} O{} m }{\pmb{\mathscr{#3}}\ifthenelse{\isempty{#1}}{}{^{(#1)}}\ifthenelse{\isempty{#2}}{}{_{#2}}}
\makeatletter
\newcommand{\thickhline}{%
    \noalign {\ifnum 0=`}\fi \hrule height 1pt
    \futurelet \reserved@a \@xhline
}
\DeclareMathOperator*{\argmax}{arg\,max}

\newcommand{\mycomment}[1]{}

\def\BibTeX{{\rm B\kern-.05em{\sc i\kern-.025em b}\kern-.08em
    T\kern-.1667em\lower.7ex\hbox{E}\kern-.125emX}}
\definecolor{mygreen}{rgb}{0,0.6,0}
\definecolor{mymauve}{rgb}{0.58,0,0.82}
\definecolor{mygray}{rgb}{0.5,0.5,0.5}
\definecolor{blue}{rgb}{0.0,0.0,1.0}
\definecolor{red}{rgb}{1.0,0.0,0.0}
\usepackage{listings}
\lstdefinelanguage{cypher2}{
    sensitive=true,
    morekeywords=[1]{MATCH, RETURN, WHERE, CONTAINS},
    morekeywords=[2]{PERSON, FRIEND,Document, Keyword, Affiliation, Country},
    morestring=[b]",
    morecomment=[l]{//},
    morecomment=[s]{/*}{*/},
    morecomment=[s]{--}{\ },
}
\lstdefinestyle{cypherstyle2}{
    language=cypher2,
    basicstyle=\footnotesize\ttfamily,
    keywordstyle=\color{blue}\bfseries, 
    keywordstyle=[2]\color{red}\bfseries,   
    commentstyle=\color{mygreen},
    stringstyle=\color{mymauve},
    numberstyle=\tiny\color{mygray},
    breaklines=true,
    showstringspaces=false,
    captionpos=b
}

\usepackage{listings}
\begin{document}
\input{all_section_links}
\bibliographystyle{plain}
\renewcommand{\bibsection}{}
\addcontentsline{toc}{chapter}{{}References}
\chapter*{References}
\bibliography{references/ALL}

\newpage
\thispagestyle{empty}
\null
\newpage

\end{document}

%% file: all_section_links.tex
\input{Part1/Aknowledgements_and_Executive_Summary/0}

\input{Part0/acronyms} 
\pagenumbering{arabic}  
\part{Introduction and Background}
    \chapter{Motivation and Thesis}
    \input{Part1/Motivation_and_Thesis_Statement/1}
    \chapter{Background}
    \input{Part1/Background/2}
    \chapter{Literature Review}
    \input{Part1/Related_Work/3}

\part{Methods}
    \chapter{Data Acquisition}
    \input{Part2/Data_Aquisition/0}
    \chapter{Latent Feature Extraction \& Topic Modeling}
    \input{Part2/Latent_Feature_Extraction/1}    
    \chapter{Knowledge‐Graph \& Vector Store Retrieval Augmented Generation  }
    \input{Part2/KG_VS_Constructions/2}    
    \chapter{AI Reasoning}
    \input{Part2/AI_Reasoining/3}

\part{Applications \& Results}
    \chapter{SMART-SLIC Framework}
    \input{Part3/smart-slic/0}
    \chapter{AI-Reasoning: Missing Links}
    \input{Part3/AI_Reasoning/1}
    \chapter{Question-Answering }
    \input{Part3/QA/2}

    \chapter{Conclusion}
    \input{Part3/conclusion_future/combined_future_and_conclusion}

%% file: Part1/Aknowledgements_and_Executive_Summary/0.tex
\pagenumbering{gobble}
\renewcommand{\baselinestretch}{1}
\small \normalsize

\includepdf[pages=-,pagecommand={}]{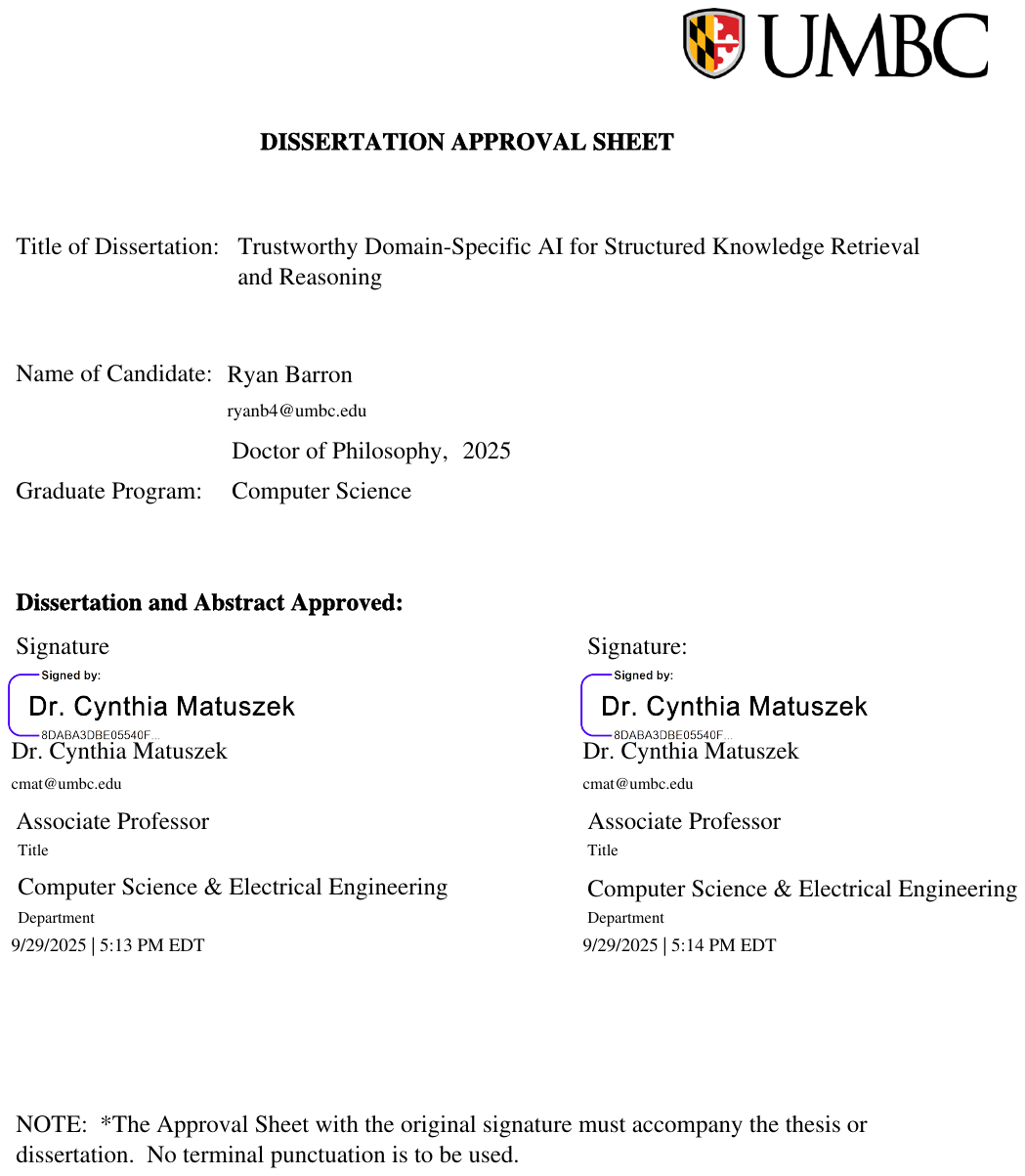}

\begin{center}
\large{{ABSTRACT}} 
\end{center}
\renewcommand{\baselinestretch}{2}
\large \normalsize

This dissertation presents a production-ready architecture that addresses the challenge of transforming vast volumes of unstructured, domain-specific text into structured and actionable knowledge representations that enable both retrieval and reasoning. Traditional pipelines for knowledge discovery remain fragmented, relying on brittle keyword extraction, ad hoc topic modeling, and disconnected reasoning modules. In response, this work proposes an integrated architecture that unifies semi-automatic corpus curation, semantic structuring, and retrieval systems into a scalable, interpretable pipeline.

The research introduces two novel techniques: Binary Bleed, an adapted binary search method that significantly reduces low-rank search complexity, particularly useful in Non-negative Matrix Factorization (NMF), and based on it, Hierarchical NMF (HNMF) with automatic selection of the number of latent features (HNMFk), a depth-adaptive topic modeling method that produces interpretable, Subject Matter Expert (SME) guided taxonomies. These representations are simultaneously instantiated as a typed, structured-based Knowledge Graph and a semantically aligned Vector Store, which includes the extracted latent features and is synchronized via an event-driven substrate.

To perform contextual information access, the research introduces a new form of  Retrieval-Augmented Generation (RAG), named Tensor-Structured RAG (T-SRAG), capable of dynamically routing queries across retrieval paths. The system improves semantic fidelity and reduces hallucinations by incorporating a contrastive loss alignment technique that maps both document and query embeddings to hierarchical topic structures.

Beyond retrieval, the framework introduces a new Artificial Intelligence (AI) reasoning by identifying and completing missing links in the knowledge graph through tensor-based link prediction. This enables inferential responses grounded in citation structure rather than generative hallucination from a Large Language Model (LLM).

The presented applications show validation across cybersecurity, law, materials science, and healthcare, demonstrating significant improvements in retrieval precision, early trend detection, as well as AI hypotheses generation and hallucination mitigation at scale.  Overall, the dissertation delivers a new deployable, modular foundation for domain-specific, trustworthy AI systems capable of both retrieving and reasoning over structured knowledge.

\thispagestyle{empty}
\newpage

\thispagestyle{empty}
\hbox{\ }
\vspace{1in}
\renewcommand{\baselinestretch}{1}
\small\normalsize
\begin{center}

\large{{Trustworthy Domain-Specific AI \\for Structured Knowledge Retrieval and Reasoning}}\\
\ \\
\ \\
\large{by} \\
\ \\
\large{Ryan C. Barron}
\ \\
\ \\
\ \\
\ \\
\normalsize
Dissertation submitted to the Faculty of the Graduate School of the \\
University of Maryland, Baltimore County in partial fulfillment \\
of the requirements for the degree of \\
Doctor of Philosophy \\
Fall 2025 
\end{center}
\vspace{4.5em}
\noindent 
\textbf{Advisor:} \\
    Dr. Cynthia Matuszek\\ \ \\ 
\noindent 
\textbf{Committee Members:} \\ 
  Dr. Charles Nicholas\\
  Dr. Tim Finin\\ 
  Dr. Roberto Yus\\ 
  Dr. Kim Rassmussen\\
  Dr. Manish Bhattarai\\

\newpage
\thispagestyle{empty}
\hbox{\ }
\vfill
\renewcommand{\baselinestretch}{1}
\small\normalsize
\vspace{-.65in}
\begin{center}
\large{\copyright \hbox{ }Copyright\\
Ryan C. Barron  
\\
2025}
\end{center}
\vfill
\newpage

\newpage
\newpage
\thispagestyle{empty}
\null
\newpage

\cleardoublepage
\pagenumbering{roman}
\setcounter{page}{2}   

\renewcommand{\baselinestretch}{2}
\small\normalsize
\hbox{\ }
 
\vspace{-.65in}

\begin{center}
\large{Dedication}
\end{center}

\begin{flushleft}
To my mother, \\
whose unwavering belief in my potential lit the path forward, \\
even when I was unsure of it myself. 

\vspace{1em}

You nurtured my love for learning long before I ever stepped into a classroom, \\
and your quiet strength as a teacher inspired my own academic journey. \\
It was through you that I learned the value of patience, of guiding others, \\
and of striving toward knowledge, not just for achievement, but for understanding.

\vspace{1em}

This work, and who I’ve become along the way, \\
are rooted in your example, your encouragement, and your love. \\
Thank you for always reaching higher and for lifting me with you.
\end{flushleft}

\newpage
\renewcommand{\baselinestretch}{2}
\small\normalsize
\hbox{\ }
 
\vspace{-.65in}

\begin{center}
\large{Acknowledgments}
\end{center} 

This dissertation has been assigned LA-UR-25-27945. Computational resources was provided by the Los Alamos National Laboratory's Institutional Computing Program supported by the U.S. Department of Energy National Nuclear Security Administration under Contract No. 89233218CNA000001.

\begin{flushleft}
I am deeply grateful to my advisor, Dr.\ Cynthia Matuszek, whose exceptional guidance, honest feedback, and steady support shaped not only this dissertation but also my growth as a researcher. Her clarity of thought, commitment to rigorous reasoning, and unwavering encouragement gave me confidence at every stage of this journey.

\medskip

I extend sincere thanks to my committee members: Dr.\ Charles Nicholas, Dr.\ Tim Finin, Dr.\ Roberto Yus, Dr.\ Kim Rasmussen, and Dr.\ Manish Bhattarai. Their thoughtful critiques, insightful questions, and generous feedback helped sharpen the scope and depth of this work.

\medskip

I am thankful to the Department of Computer Science and Electrical Engineering at the University of Maryland, Baltimore County. Their support through teaching assistantships, travel funding, and research infrastructure made this project possible. I am also especially grateful to Door to Virtue Lodge \#46 in Westminster, Maryland, for their steady presence and brotherly encouragement, which grounded me during the most demanding phases of this journey.

\medskip

Several faculty members offered mentorship that had a lasting impact on my academic development. Dr.\ Ben Johnson was a steady guide when I first arrived at the university. His encouragement, clarity, and example of instructional joy helped set the tone for my graduate path. Dr.\ Christopher Marron supported me through several teaching assistantships and served as a consistent mentor whose balance of precision and patience shaped my understanding of systems and instruction. Dr.\ Mohammad Donyaee, both as a professor and teaching mentor, provided structure, perspective, and dependable guidance during formative periods of my training. Though not one of my instructors, Dr.\ Don Engel played a meaningful role in my research development. His careful guidance in crafting and refining research papers helped elevate the quality of my work and broaden its reach into interdisciplinary spaces. Professor Tompkins also offered professional insight and encouragement, both as a teacher and a mentor.

\medskip

Other professors, instructors, and mentors who have guided and encouraged me over the years also deserve my sincere thanks. Bobby Keller, Marlene Titus, Matt Day, Robert Brown, and Janice Stencil each played a role in shaping my academic path. Whether through teaching, or mentorship their influence helped lay the foundation on which this work stands.

\medskip

This research was also shaped by the colleagues and friends with whom I shared offices, labs, and long sessions of writing and problem solving. Kasra Darvish, Ves Grantcharov, Valentine Stanev, Vladimir Valtchinov, Olga Serafimova, Padraig Higgins, Youssouf Kebe, Adam Berlier, Anjali Bhatt, Prajna Bhandary, and Nadja Bzhilyanskaya contributed sharp ideas, reliable collaboration, and a spirit of trust and friendship that sustained motivation throughout.

\medskip

I am especially thankful for the friends whose presence made this journey not only possible but also deeply meaningful. Maksim Eren offered generous support in every dimension, emotional, social, and academic, and was a constant source of encouragement and perspective. Nick Solovyev was a close collaborator, contributing thoughtful insight and a shared sense of purpose throughout the research. Colin Veison provided steady companionship and editorial feedback that helped carry this work forward. Brian Bell, Michael Geyer, Liz Ruiz Carlos, Dillon Fornaro, Nishi Patel, and Sina Sontowski offered humor, kindness, and dependable support that helped me stay grounded through challenging times. Selma Wanna contributed both academic insight and social encouragement, helping me navigate obstacles with clarity and care.

\medskip

I would like to specifically thank my undergraduate computer science advisor, Dr. Abhijit Dhutt, for encouraging me to pursue a graduate degree in computer science, which ultimately led me to complete my dissertation. 

\medskip

I am also grateful to those who supported my professional growth along the way. Jean-Marc Henriette became a mentor during the middle stages of my education, and over time, a trusted friend whose guidance and friendship shaped both my scholarly direction and personal confidence. Jeff Kite offered early opportunities to develop practical skills in a professional setting, helping me build confidence and technical fluency. Each of you contributed something essential, whether mentorship, opportunity, or belief, and I carry your influence with deep gratitude.

\medskip

I reserve special thanks for Boian Alexandrov. His insight, depth of thought, and boundless curiosity set a high standard for analytical rigor throughout this work. Boian played a central role in shaping the experimental design that underlies several key chapters. His willingness to engage with every level of the research, from conceptual foundations to technical implementation, turned challenges into opportunities. Working with him was both deeply instructive and profoundly motivating, and his influence will remain with me far beyond this dissertation.

\medskip

Each person named here made a vital contribution to this journey. Whether through guidance, encouragement, collaboration, or friendship, your presence shaped this work in meaningful ways. I am honored to have had the opportunity to complete this research in your company, and I hope these pages reflect the generosity and insight you have shared with me.
\end{flushleft}

\newpage
\tableofcontents

\cleardoublepage
  \addcontentsline{toc}{section}{\listtablename}\listoftables
\newpage
\cleardoublepage
\addcontentsline{toc}{section}{\listfigurename}\listoffigures
\newpage

%% file: Part0/acronyms.tex
\chapter*{List of Abbreviations}
\begin{longtable}{R{0.20\linewidth} R{0.75\linewidth}}
\caption{Acronyms used in this dissertation}\\
\toprule
\textbf{Acronym} & \textbf{Definition}\\
\midrule
\endfirsthead

\toprule
\textbf{Acronym} & \textbf{Definition}\\
\midrule
\endhead

\midrule
\multicolumn{2}{r}{\textit{Continued on next page}}\\
\endfoot

\bottomrule
\endlastfoot

AI & Artificial Intelligence \\
ANN & Approximate Nearest Neighbour   \\
BERT & Bidirectional Encoder Representations from Transformers \\
BUNIE & Bibliographic Utility Network Information Expansion \\
BM25 &  Best Matching 25 \\
BMF & Boolean matrix factorization \\
BNMF & Binary Non-negative Matrix Factorization \\
BNMFk &  Binary NMF with Automatic Model Determination \\
CoT & Chain of Thought   \\
CP & Connected Papers  \\
CPD & Canonical Polyadic Decomposition   \\
CSV & Comma-Separated Values \\
CV & Computer Vision \\
DOE & Department of Energy \\
DPR & Dense Passage Retrieval \\ 
GNN & Graph Neural Network \\
GPT & Generative Pre-trained Transformer \\
GloVe & Global Vectors for Word Representation \\
HDP & Hierarchical Dirichlet Processes  \\
HEAL & Hierarchical Embedding Alignment Loss \\
HITL & Human in the Loop \\ 
hLDA & hierarchical Latent Dirichlet Allocation \\
HNMF & Hierarchical Non-negative Matrix Factorization \\
HNMFk & Hierarchical NMF with Automatic Model Determination \\
HNSW & Hierarchical Navigable Small-World graph \\
ICA & Independent Component Analysis \\
IVF & Inverted File Indexing \\
KG & Knowledge Graph \\
KNN & K-Nearest Neighbors \\
LLM & Large Language Model \\
LMF & Logistic Matrix Factorization \\
LANL & Los Alamos National Laboratory \\
MF & Matrix Factorization \\
MDL & Minimum Description Length \\
MCMC & Markov Chain Monte Carlo  \\
MPI & Message Passing Interface \\
MRR & Mean Reciprocal Rank \\
nCRP & nested Chinese Restaurant Process \\
NMF & Non-negative Matrix Factorization \\
NMFk & Non-negative Matrix Factorization with Automatic Model Determination \\
NMRA & New Mexico Rules Annotated \\
PCA & Principal Component Analysis \\
PH & Pulmonary Hypertension \\
PPI & Protein–Protein Interaction   \\
PQ & Product Quantization \\ 
PR AUC & Precision–Recall – Area Under Curve \\
QA & Question Answering \\
RAG & Retrieval-Augmented Generation \\
RMSE & Root Mean Squared Error \\
RNMFk & Recommender NMF with Automatic Model Determination \\
ROC AUC & Receiver Operating Characteristic – Area Under Curve \\
SeNMFk & Semantic NMF with automatic model determination \\
SciNCL & Neighborhood Contrastive Learning for Scientific Document Representations with Citation Graph Embeddings \\
SME & Subject Matter Expert \\
SVD & Singular Value Decomposition \\
T-SRAG & Tensor-Structured Retrieval-Augmented Generation \\
TD & Tucker Decomposition \\
TF-IDF & Term Frequency–Inverse Document Frequency \\
TMD & Transition-Metal Dichalcogenide \\
t-SNE & t-Distributed Stochastic Neighbour Embedding \\
UQ & Uncertainty Quantification \\
UMAP & Uniform Manifold Approximation and Projection \\
VS & Vector Store \\
WNMF & weighted NMF \\
WNMFk &  Weighted NMF with Automatic Model Determination \\
\end{longtable}

%% file: Part1/Motivation_and_Thesis_Statement/1.tex
\section{Motivation}

Across fields like science, law, cybersecurity, and engineering, professionals are turning to AI systems to make sense of massive volumes of unstructured text. But as the scale of information grows, so does the need for systems that are not only accurate but also explainable, auditable, and aligned with domain-specific knowledge. Experts need more than just answers; they need insight that can be traced, trusted, and integrated into decision-making workflows.

While recent progress in LLMs, vector-based retrieval, and matrix factorization has been impressive, the tools often operate in silos. Keyword-based methods still let in too much irrelevant data. Flat topic models miss hierarchical structure. Dense retrieval pipelines, such as those used in RAG, frequently hallucinate because their underlying vector stores lack semantic grounding. Despite the power of GPU-accelerated computation, high-dimensional search algorithms like HNSW and instruction-tuned LLMs, these components remain loosely coupled in current implementations.

This dissertation introduces an end-to-end framework that brings these parts together into a coherent, modular pipeline. The system is designed to be transparent, scalable, and adaptable, enabling it to work across domains and be tuned and extended by domain experts themselves. By tightly integrating symbolic reasoning, embeddings, and interpretable topic hierarchies, the framework supports knowledge discovery that is both robust and explainable.

\section{Thesis Statement}

 By integrating an SME-seeded citation-expansion and embedding-pruning loop, a Binary Bleed–accelerated Hierarchical NMFk model for automatic rank discovery, and an event-synchronized dual symbolic–vector knowledge layer orchestrated by a hybrid T-SRAG agent, this dissertation delivers a scalable, interpretable, and domain-portable framework that enables both retrieval and reasoning over large, multi-domain text corpora.

\vspace{0.5em}
\noindent This claim is substantiated through:
\begin{enumerate}[leftmargin=1.4em,label=\textbf{\arabic*.}]
  \item \textbf{Corpus Curation.} A multi-stage expansion–pruning loop that combines citation-graph walks, embedding hypersphere filters, and LLM-assisted relevance voting.
  \item \textbf{Hierarchical Modelling.} The  Binary Bleed algorithm for rapid rank search and HNMFk for depth-adaptive, interpretable topic trees. 
\item \textbf{Unified Knowledge Layer.} Dual instantiation of topic–document relationships as typed triples in a knowledge graph and semantically aligned vectors in a high-performance store.
  \item \textbf{Hybrid Retrieval \& Reasoning.} A T-SRAG agent that dynamically blends graph walks, dense vector retrieval, and tensor-based link prediction for grounded answers.  
\end{enumerate}

\begin{figure}[t!]
  \centering
  \includegraphics[width=\textwidth]{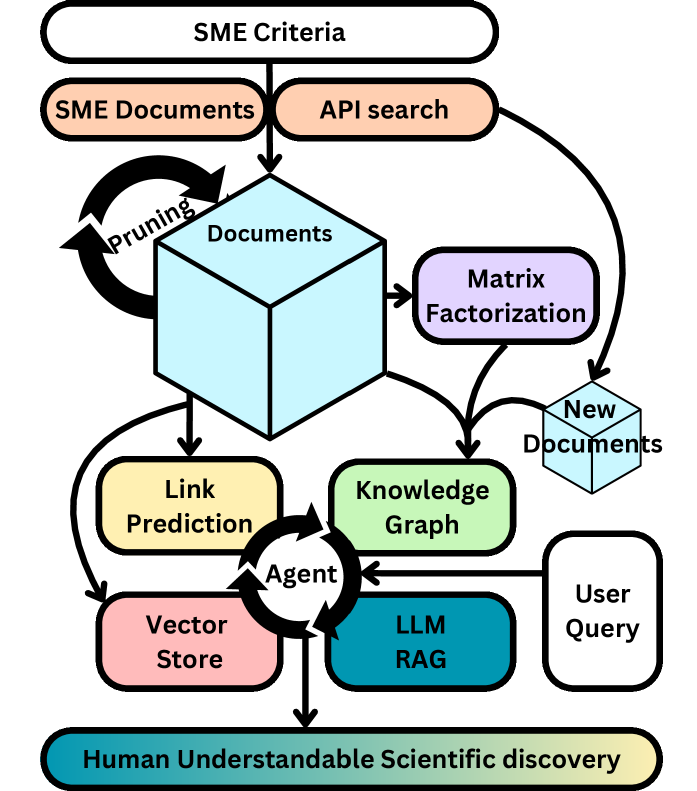}
  \caption{End-to-end pipeline: SME-seeded corpus $\rightarrow$
           Binary Bleed rank search $\rightarrow$ HNMFk topic tree
           $\rightarrow$ synchronised KG + vector store
           $\rightarrow$ hybrid RAG agent.}
  \label{fig:system_overview}
\end{figure}

\section{Design Overview}

Figure \ref{fig:system_overview} presents a high-level overview of the proposed system architecture. The pipeline begins with a curated seed corpus, selected by domain experts to ensure initial relevance and quality. From this trusted foundation, citation networks are traversed outward to collect additional candidate documents. These are subsequently filtered through a multi-stage refinement process, including keyword-based screening, embedding-based similarity scoring, and rapid expert validation via binary relevance judgments. The result is a focused and domain-specific corpus suitable for downstream analysis.

To identify the latent thematic structure within this corpus, the system employs a novel rank estimation algorithm, Binary Bleed, which efficiently determines the optimal number of topics. This method operates via a binary search strategy that iteratively narrows the candidate space, significantly reducing computational overhead compared to exhaustive approaches. Once the rank is established, the HNMFk algorithm is applied to derive a tree-structured topic hierarchy. This hierarchy captures coarse-grained topics at higher levels and increasingly fine-grained subtopics toward the leaves. Each node is automatically annotated with representative keywords to facilitate human interpretability.

The resulting document-topic associations are then dual-encoded: first, as triples embedded in a knowledge graph, and second, as dense vector representations indexed in a high-performance vector store. A synchronization mechanism ensures consistency between the symbolic representations. At the interface layer, a hybrid RAG agent acts as an intelligent query orchestrator. Given a user query, it dynamically chooses between symbolic reasoning over the knowledge graph (suitable for structured, multi-hop queries), vector-based retrieval (for semantically fuzzy queries), or a combination of both. Retrieved evidence is then passed to an LLM, which generates fluent, citation-backed responses. In summary, the system transforms a small, expert-curated set of documents into a robust, semantically organized, and interoperable knowledge system. It supports transparent, context-aware interactions through the integration of symbolic reasoning and retrieval mechanisms.

%% file: Part1/Background/2.tex
\input{Part1/Background/latent_features}

\input{Part1/Background/nmf}

\input{Part1/Background/nmfk}
\input{Part1/Background/lmf}
\input{Part1/Background/hnmfk}
\input{Part1/Background/heal}

\input{Part1/Background/binary_search}
\input{Part1/Background/kg}
\input{Part1/Background/vs}

\input{Part1/Background/link_prediction}
\input{Part1/Background/llm_agent}

%% file: Part1/Background/latent_features.tex
\section{Latent Variables and Factor Analysis}

In principle, it is challenging to directly link collected data to the underlying processes that generated it. Every dataset consists only of directly observable quantities (such as state variables), while the underlying processes are typically complex, interdependent, and manifest simultaneously across multiple observables. In most cases, it is impossible to measure even known processes because they are intertwined directly, and often there are unknown processes that remain latent, not directly observable, and need to be discovered. These hidden, unobserved, or latent variables or features are referred to as latent variables \cite{bartholomew201101}.

Latent variable extraction reduces the dimensionality of extensive collections of observables to a smaller set of latent variables, where each observable can be represented as a combination of these latent factors. This form of dimensionality reduction is central to factor analysis \citep{Harman1976}, which underpins many modern approaches in data mining and data-driven science.

\subsection{High-Dimensional Data and Tensor Representations}

Most contemporary big data sources are inherently high-dimensional and are most naturally represented as tensors rather than as two-dimensional matrices \citep{Kolda2009, Cichocki2015}. Common examples include imagery, spatiotemporal sensor networks, and large-scale text corpora. Such datasets typically encode the signatures of multiple concurrent latent processes (e.g., explosions, earthquakes, topic distributions), each imprinting distinct patterns on observable state variables (e.g., temperature, pressure, density, token counts) across various dimensions (e.g., spatial, temporal, lexical).

The identification of these latent processes requires specialized unsupervised machine learning (ML) methods capable of handling high-dimensional data. Traditional techniques such as clustering, principal component analysis (PCA) \cite{pearson_PCA}, singular value decomposition (SVD) \citep{jolliffe2002principal}, independent component analysis (ICA) \citep{Hyvarinen2001},  NMF \citep{Lee1999LearningTP}, and neural networks (NNs) have proven highly effective for matrix data. However, flattening high-dimensional tensor data into matrix form destroys essential cross-dimensional correlations. Therefore, established dimension-reduction techniques must be generalized to tensor data to preserve and exploit the inherent structure of modern datasets fully.

\subsection{Importance of Nonnegativity Constraints}

Incorporating non-negativity constraints in latent feature extraction ensures that the resulting components are physically interpretable \citep{Cichocki2016}. Such part-based representations \citep{Donoho2003} arise because, under a non-negativity constraint, reconstructions of the data can only be formed by additive combinations of the extracted features, making these features correspond to meaningful and interpretable parts of the original data. Many types of state variables, such as density, energy, spectral intensities, and population counts, are naturally nonnegative, and others can often be transformed into nonnegative forms as required.

\subsection{Example: Blind Source Separation}

Consider, for instance, a scenario in which microphones are distributed throughout a noisy room where multiple speakers are engaged in conversation. Each microphone captures a complex mixture of voices, background sounds, and noise. In this setting, the latent features correspond to the individual voices, signals that cannot be recorded independently but can be extracted through analysis of the observed mixtures. The process of extracting these latent features effectively reduces the dimensionality of the data and identifies low-dimensional subspaces \citep{Saul2003, Tenenbaum2000} that succinctly represent the entire dataset.

\subsection{Tensor Factorization: A Modern Approach}

One of the most powerful and versatile tools for latent feature extraction from high-dimensional data is tensor factorization (TF) \citep{Cichocki2016}. The goal of TF is to decompose a tensor into a set of factor matrices (one for each tensor mode) and a small core tensor. The factor matrices capture the latent features in each respective dimension of the tensor.

Two principal forms of TF are widely used:
\begin{itemize}
\item Canonical Polyadic Decomposition (CPD) \citep{Harshman1970}, also known as PARAFAC
\item Tucker Decomposition (TD) \citep{Tucker1966}, a higher-order generalization of matrix factorization
\end{itemize}

These methods extend the principles of classical factor analysis to tensor data, providing a rigorous mathematical framework for uncovering the complex latent structures embedded in modern, high-dimensional datasets.

%% file: Part1/Background/nmf.tex
\section{Non-negative Matrix Factorization (NMF)}
\label{subsection:nmf}
NMF is an unsupervised learning method based on low-rank approximation that reduces dimensionality. NMF approximates a given observation matrix $\mat{X}\in \mathbb{R}_{+}^{n \times m}$ with non-negative entries, as a product of two non-negative matrices, i.e., $\mat{X} \approx \mat{W}\mat{H}$ and $\hat{\mat{X}} = \mat{W}\mat{H}$, where $\mat{W} \in \mathbb{R}_{+}^{n\times k}$, and $\mat{H} \in \mathbb{R}_{+}^{k \times m}$, and usually $k\ll m, n$. Here, $n$ is the number of features (rows), $m$ is the number of samples (columns), and the small dimension $k$ is the low-rank of the approximation. This factorization is performed through a non-convex minimization with non-negativity constraint, utilizing the multiplicative updates algorithm \cite{Lee1999LearningTP}, and the Frobenius norm as the distance metric, with an objective function:
\begin{align*}
\label{eqn:nmf_min}
    \underset{\mat{W} \in \mathbb{R}_{+}^{n\times k}, \, \mat{H} \in \mathbb{R}_{+}^{k \times m}}{\operatorname{min}} \, ||\mat{X} - \mat{W}\mat{H}||_{F}^{2}, \numberthis 
\end{align*}
which allows NMF to be treated as a Gaussian mixture model \cite{fevotte2009nonnegative}. In Equation \ref{eqn:nmf_min}, the factors $\mat{W}$ and $\mat{H}$ are the solution of the optimization problem, the latent factors, estimated via alternative updates, with update rules in Equations \ref{eqn:NMF_updates_W} and \ref{eqn:NMF_updates_H} respectively, and the performance of the minimization is evaluated by the relative reconstruction error in Equation \ref{eqn:rel_error}:

\begin{multicols}{3}
\noindent
\begin{align}
\label{eqn:NMF_updates_W}
\mat{W} =\mat{W}\odot\frac{\mat{X}\mat{H}^T}{\mat{W} \mat{H} \mat{H}^T},
\end{align}
\begin{align}
\label{eqn:NMF_updates_H}
\mat{H} =\mat{H}\odot\frac{\mat{W}^T\mat{X}}{\mat{W}^T \mat{W} \mat{H}},
\end{align}
\begin{align}
\label{eqn:rel_error}
\text{Relative Error} =\frac{||\mat{X}-\mat{W}\mat{H}||_F^2}{||\mat{X}||_F^2}.
\end{align}
\end{multicols}

%% file: Part1/Background/nmfk.tex
\section{Non-negative Matrix Factorization with Automatic Model Determination (NMFk)}
\label{subsection:nmfk}

The NMF minimization requires prior knowledge of the latent dimensionality, $k$,  
and the number of latent features, which is often unknown. Choosing too small a value for $k$ leads to a poor approximation of the observables in $\mat{X}$ (a problem called under-fitting). At the same time, setting $k$ too large makes the extracted features hard to interpret, as they also capture noise in the data (a case of over-fitting). In essence, selecting $k$ is equivalent to estimating the number of model parameters, a well-known complex problem. 

In general, the existing partial solutions to this problem are heuristic. Among these solutions is Automatic Relevance Determination (ARD) \cite{mackay1994bayesian}, which was first modified for Principal Component Analysis \cite{bishop1999bayesian} and then for NMF \cite{morup2009tuning,tan2012automatic}.
Another approach is based on the assumed stability of the NMF solution and was proposed to identify the number of stable clusters in the observational matrix $\mat{X}$ \cite{brunet2004metagenes}. A recent model selection technique, NMFk \cite{alexandrov2020patent}, has successfully decomposed the largest dataset of human cancer genomes \cite{alexandrov2013signatures}. 
NMFk integrates classical NMF-minimization with custom clustering and Silhouette statistics \cite{ROUSSEEUW198753} and combines the accuracy of the minimization and robustness/stability of the NMF solutions when a modified bootstrap procedure (i.e., generation of a random ensemble of slightly perturbed input matrices) is applied to estimate the number of latent features. Recently, NMFk was applied to many synthetic datasets with a predetermined number of latent features, and it demonstrated its superior performance of correctly estimating $k$ compared to the other known heuristics \cite{nebgen2021neural}. The exceptional performance of the NMFk method as a model selection was also demonstrated both in practice \cite{alexandrov2020repertoire} and in a large set of synthetic cancer genomes with a predetermined number of latent features  \cite{islam2022uncovering}. In addition, it was shown that NMFk performs better than spherical k-means and other methods for topic extraction \cite{vangara2020semantic}. Therefore, NMFk is the core factorization method with automatic model selection and is adopted to WNMFk, RNMFk, and BNMFk for heuristically estimating $k$. For completeness, the pseudocode for it is in Algorithm 1, and a description of it is as follows:
\begin{algorithm}[H]
    \caption{NMFk($\mat{X}$, $k^{min}$, $k^{max}$, $M$, $Sill\_{thr}=0.8$)} 
	\begin{algorithmic}[1]
	    \Require: $\mat{X} \in \mathbb{R}_{+}^{n \times m}$ , $k^{min}$, $k^{max}$ , $r$ \label{alg:nmfk}
		\For {$k$ in $k^{min}$ to $k^{max}$} \Comment{Start and end process for NMFk}
			\For{$q$ in 1 to $M$}        \Comment{Num. of Perturbations on each k}
				\State $\ten{X}_{::q}$ = Perturb($\mat{X}$) \Comment{Resampling  $\mat{X}$ to create a random ensemble}
				\State $\ten{W}_{::kq}$,$\ten{H}_{::kq}$ = NMF($\ten{X}_{::q}$,k)
				
			\EndFor
			\State  $\ten{W}^{all}$=[$\ten{W}_{::k1}$,\ldots,$\ten{W}_{::kM}$] and $\ten{H}^{all}$=[$\ten{H}_{::k1}$,\ldots,$\ten{H}_{::kM}$]
			\State $\ten{\hat{W}}$, $\ten{\hat{H}}$ = customCluster( $\ten{W}^{all}$,$\ten{H}^{all}$)
			\State $\ten{\widetilde{W}}_{::k}$ = medians( $\ten{\hat{W}}$) 
			\State $\ten{H}^{reg}_{::k}$  = NNLS($\mat{X}$,$\widetilde{\mat{W}}_{::k}$) \Comment{Column-wise regression of $\mat{H}$ with $\widetilde{\mat{W}}$ and column of $\mat{X}$}
			\State $\vect{s}_{k}$ = clusterStability($\ten{\hat{W}}$)
			\State $\vect{err_{k}}$ = reconstructErr($\mat{X}$,$\widetilde{\mat{W}}_{::k}$ , $\mat{H}^{reg}_{::k}$ ) \Comment{Column-wise reconstruction error for L-statistics}
	\EndFor

	\State $\vect{err}^{all}$=[$\vect{err}_{k^{min}}$,\ldots,$\vect{err}_{k^{max}}$] 
	\State $k^{opt}$ = PvalueAnalysis($\vect{err}^{all}$ ,$k^{min}$,$k^{max}$,$\vect{s}_{k}$,$Sill\_{thr}$) \Comment{Predicted k value using Wilcoxon}		
    \State \textbf{return} $\ten{\widetilde{W}}_{::k^{opt}}$, $\ten{H}^{reg}_{::k^{opt}}$, $k^{opt}$
	\end{algorithmic} 

\textbf{Ensure:} $k =k^{opt}$,$\ten{\widetilde{W}}_{::k^{opt}}$ $\in \mathbb{R}_{+}^{n \times k}$ ,$\ten{H}^{reg}_{::k^{opt}}$ $\in \mathbb{R}_{+}^{k \times m}$ , $\mat{X}$ = $\ten{\widetilde{W}}_{::k^{opt}}$  $\ten{H}^{reg}_{::k^{opt}}$	
\end{algorithm}
\begin{enumerate}
    \item \textit{Resampling}: Based on the observable matrix, $\mat{X}$, NMFk creates an ensemble of $M$ random matrices, $[\ten{X}_{::q}]_{q=1,...,M}$, with means equal to the original matrix $\mat{X}$. Each one of these random matrices $\ten{X}_{::q}$ is generated by perturbing the elements of $\mat{X}$ by a small uniform noise, such that: $\ten{X}_{ijq} = \mat{X}_{ij} +\delta_{ijq}$, for each $q=1,...,M$, where $\delta_{ijq}$ is the small error that is part of random distribution, different error for each $q$.
    \item \textit{NMF minimization}: The Frobenius norm-based multiplicative updates (MU) algorithm \cite{Lee1999LearningTP} is used for different numbers of latent features, $k$, in an interval $[k^{min},k^{max}]$, for each generated $M$ random matrices.
    \item \textit{Custom clustering:} For each $k\in [k^{min},k^{max}]$, NMF minimizations of the $M$ random matrices, $[\ten{X}_{::q}]_{q=1,...,M}$, results in $M$ pairs $[\ten{W}_{::kq}; \ten{H}_{::kq}]_{q=1,...,M}$. Further, NMFk clusters the set of the $M*k$ latent features, the columns of $\ten{W}_{::kq}$. The NMFk custom clustering is similar to k-means, but it holds exactly one column for each cluster from each of the $M$ NMF solutions. This constraint is needed since each NMF minimization gives exactly one solution $\ten{W}_{::kq}$ with the same number of columns, $k$. In clustering, the cosine similarity metric measures the similarity between the columns. Here, cosine similarity determines the degree of similarity between two vectors in an inner product space.
    \item \textit{Robust $\mat{W}$ and $\mat{H}$ for each $k$:} The medians of the clusters,
    $\ten{\widetilde{W}}_{::k}$,  are the robust solution for each explored $k$. The corresponding mixing coefficients $\ten{H}^{reg}_{::k}$ are calculated by regression of $\mat{X}$ on  $\ten{\widetilde{W}}_{::k}$.
    \item \textit{Cluster stability via Silhouette statistics:} NMFk explores the stability of the obtained clusters, for each $k$, by calculating their Silhouettes \cite{ROUSSEEUW198753}. Silhouette scores quantify the cohesion and separability of the clusters. The Silhouette values range between $[-1, 1]$, where $-1$ means an unstable cluster and $+1$ means perfect stability.
    
    \item \textit{Reconstruction error:} Another metric NMFk uses is the relative reconstruction error, $R = ||\mat{X} - \ten{X}_{::k}^{rec}||/||\mat{X}||$, where $\ten{X}_{::k}^{rec} = \ten{\widetilde{W}}_{::k}*\ten{H}^{reg}_{::k}$, which measures the accuracy of the reproduction of initial data by a given solution and the number of latent features $k$. 
    \item \textit{L-statistics:} NMFk uses L-statistics \cite{vangara2021finding} to automatically estimate the number of latent features. To calculate L-statistics for each $k$, NMFk records the distributions of the column reconstruction errors, $\vect{e}_i = \Vert \mat{X}_{:j} - \ten{X}^{rec}_{:jk}\Vert/\Vert\mat{X}_{:j}\Vert$; $j=1,...,m$. L-statistics compares the distributions of column errors for different $k$ by a two-sided Wilcoxon rank-sum test to evaluate if two samples are taken from the same population \cite{Haynes2013}. 
 
    \item \textit{NMFk final solution:} The optimal number of latent features, $k^{opt}$, is the largest stable cluster count with low reconstruction error. The Wilcoxon rank-sum test computes the p-value for $k^{opt}$, ensuring subsequent column error distributions remain statistically similar, indicating noise fitting. L-statistics and a \emph{minimum} Silhouette threshold of $0.80$ guide selection, preventing overfitting. The corresponding $\ten{\widetilde{W}}_{::k^{opt}}$ and $\ten{H}^{reg}_{::k^{opt}}$ are the robust solutions for the low-rank factor matrices.

\end{enumerate}

%% file: Part1/Background/lmf.tex
\section{Logistic Matrix Factorization (LMF)}
\label{subsection:lmf}
 LMF extends the principles of matrix factorization to binary data, where the observed matrix $\mat{X} \in \{0,1\}^{n \times m}$ represents the presence or absence of interactions or a known link~\cite{Johnson2014LogisticMF}. Unlike NMF, which minimizes reconstruction error using the Frobenius norm, LMF incorporates the logistic regression model to capture the likelihood of binary observations. In the binary interaction matrix $\mat{X} \in \{0,1\}^{n \times m}$, each entry $\mat{X}_{ij}$ represents the presence ($X_{ij} = 1$) or absence ($X_{ij} = 0$) of a link between node $i$ and node $j$. However, in many real-world datasets, some interactions are unobserved or missing (i.e., the missing link). To account for this, binary mask matrix is defined $\mat{M} \in \{0,1\}^{n \times m}$, where $\mat{M}_{ij} = 1$ if $\mat{X}_{ij}$ is observed or a known link, and $\mat{M}_{ij} = 0$ for the missing links. For instance, in PPI networks, $\mat{X}_{ij} = 1$ might indicate a known interaction between proteins $i$ and $j$, while $\mat{X}_{ij} = 0$ represents the absence of an interaction. If the interaction status between proteins $i$ and $j$ has not been experimentally determined, $\mat{M}_{ij} = 0$ is used to indicate the missing entry at 
 $\mat{X}_{ij}$. In LMF, $\mat{X}$ is approximated as:
\begin{align*}
\label{eqn:lmf_eq}
\hat{\mat{X}} = \sigma(\mat{W}\mat{H} + \mat{b}_r + \mat{b}_c), \numberthis 
\end{align*}
where $\mat{W} \in \mathbb{R}^{n \times k}$ is the row latent feature matrix, $\mat{H} \in \mathbb{R}^{k \times m}$ is the column latent feature matrix, $\mat{b}_r \in \mathbb{R}^{n \times 1}$ is the row bias vector (broadcasted across columns), $\mat{b}_c \in \mathbb{R}^{1 \times m}$ is the column bias vector (broadcasted across rows), and the element-wise logistic sigmoid function:
\begin{align}
    \label{eqn:sigmoid}
    \sigma(x) = \frac{1}{1 + e^{-x}}.
\end{align}

Row and column biases are incorporated into the LMF model to account for systematic variations in the data. The row bias $\mat{b}_r \in \mathbb{R}^{n \times 1}$ captures row-specific effects, such as a node's general tendency to form links across all columns. Similarly, the column bias $\mat{b}_c \in \mathbb{R}^{1 \times m}$ models column-specific effects, such as a particular node's propensity to attract connections. These biases are added to the matrix reconstruction, allowing the model to better fit the data by accounting for individual node-specific tendencies. For example, in a PPI network, the row bias $\mat{b}_r$ could represent the inherent interaction likelihood of a specific protein across all other proteins. In contrast, the column bias $\mat{b}_c$ adjusts for the overall connectivity tendency of a target protein across all rows.
 
Each entry $\hat{\mat{X}}_{ij}$ in the reconstructed matrix represents the predicted probability of $\hat{\mat{X}}_{ij}$. The optimization objective in LMF minimizes the negative log-likelihood of the binary observations under the logistic model:
\begin{align*}
\underset{\mat{W}, \mat{H}, \mat{b}_r, \mat{b}_c}{\operatorname{minimize}} \, - \sum_{i=1}^n \sum_{j=1}^m \mat{M}_{ij} \left( \mat{X}_{ij} \log \hat{\mat{X}}_{ij} + (1 - \mat{X}_{ij}) \log (1 - \hat{\mat{X}}_{ij}) \right) + \lambda \left( ||\mat{W}||_F^2 + ||\mat{H}||_F^2 + ||\mat{b}_r||_2^2 + ||\mat{b}_c||_2^2 \right), \numberthis 
\end{align*}
where $\lambda$ is the regularization parameter to prevent overfitting. The optimization problem is solved using gradient-based methods. The gradients for each parameter are computed as follows:

\begin{align*}
\frac{\partial L}{\partial \mat{W}} = \mat{M} \odot (\hat{\mat{X}} - \mat{X}) \mat{H}^T + \lambda \mat{W}, \numberthis 
\end{align*}
\begin{align*}
\frac{\partial L}{\partial \mat{H}} = \mat{W}^T \left( \mat{M} \odot (\hat{\mat{X}} - \mat{X}) \right) + \lambda \mat{H}, \numberthis 
\end{align*}
\begin{align*}
\frac{\partial L}{\partial \mat{b}_r} = \sum_{j=1}^m \mat{M} \odot (\hat{\mat{X}} - \mat{X}) + \lambda \mat{b}_r, \numberthis 
\end{align*}
\begin{align*}
\frac{\partial L}{\partial \mat{b}_c} = \sum_{i=1}^n \mat{M} \odot (\hat{\mat{X}} - \mat{X}) + \lambda \mat{b}_c, \numberthis 
\end{align*}
where 
\begin{equation}
L = \frac{1}{2} \|\mat{M} \odot (\hat{\mat{X}} - \mat{X})\|_F^2 + \frac{\lambda}{2} \left( \|\mat{W}\|_F^2 + \|\mat{H}\|_F^2 + \|\mat{b}_r\|_F^2 + \|\mat{b}_c\|_F^2 \right). \numberthis 
\end{equation}

The parameters $\mat{W}$, $\mat{H}$, $\mat{b}_r$, and $\mat{b}_c$ are updated iteratively using gradient descent:
\begin{align*}
\mat{W} \leftarrow \mat{W} - \eta \frac{\partial L}{\partial \mat{W}}, \quad
\mat{H} \leftarrow \mat{H} - \eta \frac{\partial L}{\partial \mat{H}}, \quad
\mat{b}_r \leftarrow \mat{b}_r - \eta \frac{\partial L}{\partial \mat{b}_r}, \quad
\mat{b}_c \leftarrow \mat{b}_c - \eta \frac{\partial L}{\partial \mat{b}_c}, \numberthis 
\end{align*}
where $\eta$ is the learning rate. LMF is particularly well-suited for sparse, binary datasets, as it naturally handles probabilistic modeling of binary outcomes. Adding biases for rows and columns allows for improved flexibility in capturing systematic variations in data. 

%% file: Part1/Background/hnmfk.tex
\section{Hierarchical Non-negative Matrix Factorization with Automatic Model
Determination (HNMF{\small k})}
\label{subsection:hnmfk}
Hierarchical NMF{\small k} (HNMF{\small k}) recursively applies the model-selection variant of NMF (Section~\ref{subsection:nmfk}) to uncover a multi-resolution latent structure in $\mat{X}$.  Starting from the whole matrix, each call to NMF{\small k} produces a low-rank approximation $\mat{X}\!\approx\!\widetilde{\mat{W}}\widetilde{\mat{H}}$ and a data-driven estimate $\hat{k}$ of the number of coherent topics at the current resolution.  Samples are then assigned to topics with the maximum–loading rule (row-wise $\arg\max$ of
$\widetilde{\mat{H}}$ when clustering on $\mat{H}$, or column-wise $\arg\max$ of $\widetilde{\mat{W}}$ when clustering on $\mat{W}$).
Each topic becomes a child node that is factorized again, yielding a tree\/–\/structured decomposition $\mathcal{T}$ whose depth is controlled by (i) a user-supplied limit $d_{\max}$, (ii) a minimum-sample threshold $n_{\min}$, and (iii) the stopping condition
$\hat{k}=1$ or a single populated cluster.  The process terminates when all remaining nodes are declared leaves.  The resulting hierarchy exposes global themes near the root and increasingly specialized subtopics toward the leaves, mirroring the coarse-to-fine semantics observed in real-world corpora \cite{10.1145/3624567, Haddock2023}.   

\paragraph{Automatic model selection at every level.}
Because NMF{\small k} estimates $\hat{k}$ independently in every node, HNMF{\small k} adapts its branching factor to local data complexity: dense, heterogeneous regions of $\mat{X}$ receive more children, whereas sparse or homogeneous regions collapse early.  This avoids the under/over-fitting pitfalls of fixed-depth binary splits and has been shown to outperform both spherical $k$-means trees and heuristic hierarchical NMF baselines on text, malware, and genomic datasets \cite{10.1145/3624567,alexandrov2020patent}. \\[-0.75em]

\paragraph{Scalable HPC implementation.}
Algorithm \ref{alg:hnmfk} encapsulates the parallel implementation used in the experiments.\footnote{Source code in \cite{TELF}.} Job scheduling is handled by Message Passing Interface (MPI) \cite{mpi3standard}, allowing independent nodes of $\mathcal{T}$ to be factorized concurrently across an arbitrary number of compute nodes.  Check-pointing makes the procedure failure-resilient and
enables incremental continuation on large corpora (hundreds of millions of non-zeros).  Memory footprint is bounded because only the factors $\widetilde{\mat{W}},\widetilde{\mat{H}}$ and sample indices are stored for completed nodes; raw slices of $\mat{X}$ are regenerated on-demand via a user-supplied callback when deeper recursion is required. 

\begin{algorithm}[H]
  \caption{HNMF{\small k}$(\mat{X},K^{(0)},d_{\max},n_{\min})$}
  \label{alg:hnmfk}
  \small
  \begin{algorithmic}[1]
    \Require{$\mat{X}\in\mathbb{R}^{n\times m}_{+}$,
             candidate range $K^{(0)}$,
             depth limit $d_{\max}$,
             sample threshold $n_{\min}$.}
    \State Initialise queue $\mathcal{Q}\leftarrow\{(\texttt{Root},\mat{X},K^{(0)},0,\varnothing)\}$
    \While{$\mathcal{Q}\neq\varnothing$} \Comment{breadth-first traversal}
        \State Pop $(\texttt{name},\mat{X}_{\!c},K^{(c)},d,\mathcal{I})\leftarrow\mathcal{Q}$
        \If{$d=d_{\max}$ \textbf{or} $\min|\mat{X}_{\!c}|\le n_{\min}$}
            \State \textbf{continue} \Comment{declare leaf}
        \EndIf
        \State $(\widetilde{\mat{W}},\widetilde{\mat{H}},\hat{k})
               \leftarrow\textsc{NMF{\small k}}\bigl(\mat{X}_{\!c},K^{(c)}\bigr)$
        \If{$\hat{k}=1$}
            \State \textbf{continue} \Comment{no further separability}
        \EndIf
        \State $\text{clusters}\leftarrow
               \arg\max(\widetilde{\mat{H}})$ \Comment{or $\arg\max(\widetilde{\mat{W}})$}
        \For{$c\gets1$ \textbf{to} $\hat{k}$}
            \State $\mathcal{I}_{c}\leftarrow\{j:\text{clusters}_{j}=c\}$
            \If{$|\mathcal{I}_{c}|<n_{\min}$}
                \State \textbf{continue}
            \EndIf
            \State $K^{(c)}\leftarrow\textsc{NextRange}\!\bigl(\hat{k},|\mathcal{I}_{c}|\bigr)$
            \State $\mathcal{Q}\!\leftarrow\!
                   \mathcal{Q}\cup\bigl\{(\texttt{name}{\_}$c$,
                   \mat{X}_{:\!,\mathcal{I}_{c}},K^{(c)},d{+}1,\mathcal{I}_{c})\bigr\}$
        \EndFor
    \EndWhile
    \State \textbf{return} hierarchy $\mathcal{T}$ with factors at every node
  \end{algorithmic}
\end{algorithm}

\paragraph{Complexity.}
For each node, HNMF{\small k} inherits the $O(\hat{k}nm)$ cost of one NMF{\small k} fit (dominated by the multiplicative-update iterations) and adds $O(\hat{k}m)$ for clustering.  Parallel execution reduces the critical path to $O(\text{depth}\times\text{max-node-time})$, making HNMF{\small k} practical for matrices with millions of documents and vocabulary sizes well above $10^{5}$ on commodity clusters.

%% file: Part1/Background/heal.tex
\section{Hierarchical Embedding Alignment Loss (HEAL)}

HEAL fine-tunes a pretrained encoder so that its embedding geometry mirrors the multi-level cluster structure obtained with \textsc{HNMFk}. Documents (and synthetic queries) that share a common ancestor high in the hierarchy are pulled together, while those that diverge early are pushed apart. This is achieved by extending supervised contrastive learning with level-specific positive sets and depth-weighted penalties, injecting hierarchical semantics without altering the encoder architecture \cite{bhattarai2024heal}.

The label and positive-set definition are:
For a mini-batch of \(N\) samples \(\{(x_i,\mathbf{y}_i)\}_{i=1}^{N}\) with depth-\(L\) hierarchical labels \(\mathbf{y}_i=(y_i^{(0)},\dots,y_i^{(L-1)})\),
\[
\mathcal{P}(i,\ell)=\bigl\{\,p \mid y_p^{(\ell)} = y_i^{(\ell)},\;p\neq i\bigr\},
\]
defines the positive set for sample \(i\) at level \(\ell\).

Level-wise contrastive loss with normalised embeddings \(\mathbf{h}_i = f_\theta(x_i)/\lVert f_\theta(x_i)\rVert_2\) and temperature \(\tau\),
\[
\mathcal{L}_{i,\ell}=-
\frac{1}{|\mathcal{P}(i,\ell)|}
\sum_{p\in\mathcal{P}(i,\ell)}
\log
\frac{\exp\!\bigl(\mathbf{h}_i^\top\mathbf{h}_p/\tau\bigr)}
     {\displaystyle\sum_{a=1}^{N}\exp\!\bigl(\mathbf{h}_i^\top\mathbf{h}_a/\tau\bigr)}
\quad (\mathcal{P}(i,\ell)\neq\varnothing) \;  .
\]

Depth penalties are applied so that \(\lambda_0 > \lambda_1 > \dots > \lambda_{L-1}\), where shallower (coarser) levels receive larger weights,
\[
\lambda_\ell=\frac{2^{L-\ell-1}}{2^{L}-1},\qquad
\sum_{\ell=0}^{L-1}\lambda_\ell = 1,
\]

The overall loss is defined as:  
\[
\mathcal{L}_{\text{HEAL}}
=\frac{1}{N}
\sum_{\ell=0}^{L-1}\lambda_\ell
\sum_{i=1}^{N}\mathcal{L}_{i,\ell}.
\]

HEAL is optimised on batches containing both documents and optional synthetic queries, each inheriting the hierarchical label of its source document. Once training converges, the encoder \(f_{\theta^\ast}\) replaces the original embedding model in the retrieval index, giving hierarchy-aware nearest-neighbour search for downstream RAG.

%% file: Part1/Background/binary_search.tex
\section{Binary Search}
\label{sec:bs-bg}
Binary search is a divide-and-conquer algorithm for locating the position of a key $x$ in a sorted array $A[1\ldots n]$ in $O(\log n)$  time, improving on the linear $O(n)$ scan by systematically discarding half of the remaining search space at every step \cite{knuth1998art}.  Starting with the full interval \texttt{low}$=1$ and \texttt{high}$=n$, the algorithm iteratively computes the midpoint \texttt{mid}$=\lfloor(\texttt{low}+\texttt{high})/2\rfloor$, compares $A[\texttt{mid}]$ against $x$, and continues the search in the lower or upper sub-array accordingly.  If $A$ is immutable, the worst-case number of probes is $\lceil\log_2 n\rceil+1$, which is asymptotically optimal for any comparison-based search on ordered data \cite{bentley1975bs}.  Its cache-friendly access pattern and logarithmic complexity make binary search a building block of fundamental algorithms such as interpolation search, fractional cascading, and balanced search trees \cite{ceri2005algorithms}.  Moreover, the monotone-predicate version extends beyond arrays, enabling efficient root finding, parametric search, and optimisation when the objective is unimodal or monotone in the decision variable \cite{tarjan1983data}.  Robust implementations guard against overflow in the midpoint computation and offer variants that return insertion points for unsuccessful searches, facilitating duplicate handling and range queries \cite{bentley2000programming}.

%% file: Part1/Background/kg.tex
\section{Knowledge Graph Foundations}
\label{sec:kg-bg}

A KG is a heterogeneous, edge–labeled multi-graph $\mathcal{G}=(\mathcal{V},\mathcal{E},\mathcal{R})$ in which each edge $(h,r,t)\in\mathcal{E}$ asserts that the head entity $h\in\mathcal{V}$ and the tail entity $t\in\mathcal{V}$ participate in a semantic relation $r\in\mathcal{R}$~\cite{ehrlinger2016towards, hogan2021knowledge}.  
Nodes are typed by an ontology $\mathcal{O}$ whose schema defines class hierarchies, permitted relations, and attribute vocabularies. 
Formally, $\mathcal{O}$ is a first‑order theory that supports sub‑class (\texttt{is\_a}) and sub‑property reasoning, thereby enabling deductive closure ($\mathcal{G}\;\longmapsto\;\text{Cl}_{\mathcal{O}}(\mathcal{G})$)~\cite{hitzler2012owl}.

Downstream tasks such as entity resolution, rule mining, and knowledge graph embeddings depend on both the structural sparsity ($\lvert\mathcal{E}\rvert\ll\lvert\mathcal{V}\rvert^{2}$) and the logical regularities imposed by $\mathcal{O}$~\cite{nickel2016review}.

Construction pipelines use (I) ontology engineering (schema design, URI conventions), (ii) information extraction (token classification, dependency‑based relation detection, or LLM‑driven triple generation), and (iii) canonicalization (string normalization, synonym collapse, co‑reference clustering)~\cite{hofer2024construction}.

\section{Latent Feature Extraction via Matrix and Tensor Factorization for Knowledge Graph Construction }

The rapid growth of unstructured data in scientific, technical, and societal domains has created a need for robust frameworks to organize and extract actionable knowledge. KGs have emerged as a powerful solution, structuring entities and their relationships in a way that enables reasoning, retrieval, and analytics. However, traditional KG construction methods, particularly those based on LLMs, face substantial challenges. These include hallucination, lack of domain control, difficulty with schema alignment, and limited interpretability. To address these issues, an alternative paradigm based on matrix and tensor factorization has been proposed and successfully applied by multiple research groups.

\subsection{Benefits of Matrix and Tensor Factorization Approaches}

\subsubsection{Interpretability and Semantic Coherence}
One of the most prominent advantages of matrix factorization techniques, particularly NMF, is their ability to uncover latent features in a non-negative and additive manner. These features naturally correspond to interpretable topics or semantic clusters, which are ideal for ontology formation in KGs. Unlike embedding-based approaches, where latent dimensions are opaque, NMF topics are readily traceable to sets of keywords or phrases, enabling transparent semantic labeling.

Hierarchical extensions such as HNMFk go further by organizing the latent space into coarse-to-fine topic trees, facilitating structured and multi-scale knowledge extraction. This recursive decomposition aligns naturally with ontological hierarchies and enables the construction of multi-layered KGs, where nodes represent concepts at various levels of abstraction and edges reflect subtopic relations.

\subsubsection{Multi-Modal Knowledge Graphs}
Another key contribution of tensor/matrix factorization-based KG construction is the integration of multiple modalities of information. Observable metadata (e.g., document authors, affiliations, citations, publication venues, etc.) can be combined with latent semantic features (e.g., topics, clusters, named entities) to form richly annotated KGs. In the Smart-SLIC architecture, for instance, TF-IDF matrices are constructed from both raw text and category labels, and their joint factorization reveals topic-specific vocabularies and hidden relationships between documents.

This multimodal integration enhances the quality of the knowledge graph by grounding abstract topics in concrete data. It also facilitates domain-specific search, clustering, and missing link prediction, and improves explainability in contrast to LLM-generated graphs, which often lack such structured layering.

\subsubsection{Resistance to Hallucinations and Structural Noise}
Perhaps the most critical benefit of factorization-based KG construction is its immunity to hallucinations. Unlike LLMs, which may generate plausible-sounding but factually incorrect triples, factorization-derived topics are extracted directly from co-occurrence statistics in the input corpus. This ensures grounding in actual, true content and reduces the risk of fabricating spurious entities or relations.

In systems like HEAL, where hierarchical topics guide additional semantic allignemt in the Vector Store of the embedded documents that are in the same HNMFk leaf cluster, helps the retrieval-augmented generation. The use of factorization-derived labels significantly reduces false positive retrievals and improves factual consistency, tunning in the same time the utilized LLMs directly to the domain specific topic. Empirical evaluations have shown significant drops in hallucination severity and higher precision in question answering when KGs are built using latent topic models instead of LLM extractions \cite{bhattarai2024heal}.

This novel design also addresses additional weaknesses often found in LLM-generated KGs, including sparsity, ontological inconsistency, and ambiguity. By relying on interpretable and semantically structured decompositions, factorization-based methods ensure broader coverage of specialized topics, adherence to consistent taxonomies, and disambiguation of similarly named entities through topic context.
 

%% file: Part1/Background/vs.tex
\section{Vector Stores for Semantic Retrieval}
\label{sec:vs-bg}

A vector store (VS) maintains a set of document embeddings, each represented as a $d$‑dimensional vector $\{\mathbf{z}_{i}\}_{i=1}^{N}\subset\mathbb{R}^{d}$ associated with corresponding document payloads~\cite{karpukhin2020dense}. Given a query embedding $\mathbf{q}\in\mathbb{R}^{d}$, the VS retrieves the $k$ nearest neighbors according to a similarity metric $S(\mathbf{q},\mathbf{z}_{i})$, commonly cosine similarity or inner product~\cite{Jegou2011PQ}. For efficient querying, vector stores utilize approximate nearest-neighbor (ANN) indexing algorithms such as hierarchical navigable small-world graphs (HNSW)~\cite{MalkovYashunin2020}, inverted file indexing (IVF) combined with product quantization (PQ) ~\cite{Jegou2011PQ, Babenko2012MultiIndex}, or scalar quantization. These methods achieve sub-linear query latency and reduce computational costs while preserving high recall rates.

Typical semantic retrieval pipelines involving vector stores proceed through several distinct stages. Initially, documents undergo segmentation, frequently implemented through recursive text-splitting strategies designed to produce semantically coherent chunks adhering to token limits~\cite{lewis2020retrieval}. These segments are then transformed into embedding vectors using transformer-based language models, such as Bidirectional Encoder Representations from Transformers (BERT) or Dense Passage Retrieval (DPR) encoders~\cite{karpukhin2020dense}. The resulting embeddings are subsequently indexed via ANN structures, facilitating rapid and scalable retrieval operations~\cite{MalkovYashunin2020, Jegou2011PQ}. Lastly, hybrid scoring techniques are employed to merge lexical relevance scores (e.g., BM25 \cite{robertson2009prf}) with dense semantic embedding scores. The combined scoring function is typically expressed as $S_{\text{hyb}}=\lambda S_{\text{BM25}}+(1-\lambda)S_{\text{dense}}$, allowing systems to leverage complementary lexical and semantic strengths~\cite{Lee2023Hybrid}.

Vector stores play dual roles within contemporary retrieval architectures. First, they provide essential context for RAG, grounding large language models by supplying retrieved documents relevant to a query or prompt, thus enhancing response quality~\cite{lewis2020retrieval}. Second, vector stores serve as foundational components in knowledge-graph construction processes, where embeddings can seed graph vertices or identify semantically overlapping textual segments for establishing relational links~\cite{karpukhin2020dense, lewis2020retrieval}.

%% file: Part1/Background/link_prediction.tex
\subsection{Link Prediction in Relational Graphs}
\label{sec:lp-bg}

Let $\mathcal{A}\in\{0,1\}^{n\times n}$ denote the adjacency matrix of an undirected graph.
\textbf{Link prediction} seeks a score matrix $\widehat{\mathcal{A}}\in[0,1]^{n\times n}$ such that $\widehat{\mathcal{A}}_{uv}\approx\Pr\!\bigl[(u,v)\in\mathcal{E}\bigr]$. Approaches fall into three categories:

\begin{enumerate}[label=\alph*)]
\item  Heuristic indices (Common‑Neighbors, Adamic–Adar, Katz) exploit local walk statistics but cannot model higher‑order semantics.
\item Graph‑neural encoders learn node embeddings via message passing and decode edge existence with $\sigma(\mathbf{h}_{u}^{\top}\mathbf{h}_{v})$, achieving state‑of‑the‑art accuracy at the cost of significant GPU memory.
\item Low‑rank factorization approximates $\mathcal{A}\approx\mathbf{W}\mathbf{W}^{\!\top}$ or $\mathcal{A}\approx\mathbf{W}\mathbf{H}$, where conditioning on the Frobenius norm supplies a convex objective that scales to millions of vertices on a single node.  Boolean, weighted, and logistic variants accommodate binary, count, and implicit‑feedback graphs, respectively.
\end{enumerate}

Evaluation relies on edge‑wise metrics (ROC AUC, PR AUC) and ranking metrics (MRR, Hits@K). Confidence awareness is increasingly required; bootstrap ensembles or Bayesian posteriors furnish per‑edge uncertainty, enabling abstention policies when $\mathbb{V}\text{ar} \widehat{\mathcal{A}}_{uv}]$ exceeds a risk threshold.

These three artifacts, knowledge graphs for structured reasoning, vector stores for semantic retrieval, and uncertainty‑aware link prediction, constitute the infrastructural backbone upon which the subsequent thesis contributions are deployed.

%% file: Part1/Background/llm_agent.tex
\section{LLM-Based Agent Reasoning}
\label{subsection:llm_agents}

LLM-based agents significantly extend the capabilities of LLMs by integrating them with external environments and toolkits of callable APIs. This architecture, exemplified in systems like AutoGPT~\cite{richards2023autogpt} and OpenAI's function-calling interface~\cite{openai2023function}, enables an agent not only to interpret user inputs or environmental observations, but also to explicitly reason about which tools to invoke in order to complete complex, multi-step tasks. At the core of such an agent is a policy $\pi(c_t, m_t)$ that formulates each LLM prompt using the current context $c_t$ and retrieved memory $m_t$, an environment $\mathcal{E}$ that provides observations $o_t$ and tracks state transitions, and a toolkit $\mathcal{T}$ of callable tools or APIs for executing external actions.

At each interaction cycle, the agent first prepares the context by gathering the current observation and relevant memory, then assembles them into a structured prompt. It then uses chain-of-thought (CoT) prompting~\cite{wei2022chain} to generate intermediate reasoning steps, explicitly documenting its deliberative process. Based on this reasoning, the agent either emits a final natural language response or invokes an external tool $a_t \in \mathcal{T}$ to gather more information or perform an action. Upon receiving the outcome $o_{t+1}$ of the tool execution, the agent updates its memory with the tuple $(o_t, \text{CoT trace}, a_t, o_{t+1})$, enabling iterative refinement and long-term context awareness.

Several architectural variants of this core agent loop have been proposed to improve reasoning robustness and task performance. ReAct agents~\cite{yao2022react} interleave explicit “Thought:” and “Action:” steps within their reasoning traces, enabling dynamic tool use with fine-grained control. Reflexion agents~\cite{shinn2023reflexion} add a self-reflection step after each action to critique and adjust their strategies. Self-Refine agents~\cite{madaan2023selfrefine} perform full post-hoc reviews of their reasoning traces to iteratively correct errors before committing to final outputs. In more complex setups, multi-agent collaboration frameworks~\cite{liang2023multiagent} deploy multiple specialized agents, such as planners, coders, and critics, that exchange messages and coordinate to collectively solve complex tasks.

The performance of these LLM agents is typically evaluated along four key dimensions. Success rate measures the fraction of tasks completed correctly and reflects the agent’s overall reliability. Tool-usage efficiency measures the proportion of invoked tools that were both effective and necessary, thereby indicating cost-effectiveness. Reasoning depth refers to the average length and complexity of the chain-of-thought traces required to solve tasks, offering insight into the agent’s cognitive sophistication. Finally, latency captures the end-to-end time required per cycle, including both LLM inference and tool execution, which is critical for real-time or interactive applications. As LLM infrastructure continues to mature, researchers are exploring how to balance these trade-offs while improving memory strategies, improving response interpretability, and facilitating autonomous orchestration across environments and APIs. Together, these developments push LLM-based agents toward greater robustness, scalability, and autonomy in handling real-world workflows.

%% file: Part1/Related_Work/3.tex
\section{Refined Data Collection}
This section summarizes techniques and prior works applied in forging a highly specific dataset of research papers.

1) Topic Modeling \& Tensor Decomposition: A  common approach to topic modeling is NMF \cite{Paatero1994PositiveMF, doc_cluster_nmf},  and when applied to a document-word matrix, identifies latent patterns of the corpus. Semantic NMF with automatic model determination (SeNMFk) is an NMF extension incorporating the text's semantic structure leveraged in \cite{vangara2020semantic,vangara2021finding,erenDocEng22}. Term frequency\textendash inverse document frequency (TF\textendash IDF) and the co-occurrence/word-context matrices are often used. Although common, more advanced methods exist.

2) Document Embeddings \& Transformers:
Vector representations of a text were previously used for dimensional mapping, cross-comparisons, and similarity analysis \cite{pmlr-v37-kusnerb15}.  Standard models for learning word embeddings have been Global Vectors for Word Representation (GloVe)\cite{pennington2014glove} and Word2Vec \cite{mikolov2013efficient}. Transformers have been used for large language models (LLM) as internal states, a popular example is BERT \cite{DBLP:journals/corr/abs-1810-04805}.  Here, the SciNCL transformer generates document embeddings \cite{ostendorff2022neighborhood}, where the document mapping includes citation data.

3) Data Visualization and Tools:
Several publicly available citation networks and topic modeling tools are available to explore and analyze research papers, such as Topic Modeling Tool \cite{topic_model_tool_1},  and Stanford Topic Modeling Toolbox \cite{topic_model_tool_2}. While the tools gather topical data from inputs, they lack visualization. Alternatively, 'Connected Papers' (CP) has visuals and is a resource for discovering scientific literature~\cite{connectedPapers_gui}. While CP is applicable for co-citation and bibliographic literature exploration, this work's tool advances bibliographic utility by creating a specialized document dataset leveraging a citation network coupled with human-in-the-loop and machine learning. Another research visualization tool, designed to handle the apex of Covid-19 research production, is explained in Ref. \cite{10.1145/3395027.3419591}. The tool cleaned the text (tokenized, removed stop-words, \& punctuation \& capitalization), constructed a TF-IDF matrix, then reduced dimensions for graphing through t-distributed stochastic neighbor embedding (t-SNE).  Here,  Uniform Manifold Approximation and Projection (UMAP) \cite{mcinnes2020umap} is used to reduce 768-dimensional embeddings output by the `neighborhood contrastive learning for scientific document representations with citation graph embeddings' (SciNCL) model \cite{ostendorff2022neighborhood} to two dimensions. 

4) Human in the Loop (HITL):
User feedback has recently been adopted into several schemes, including OpenAI's ChatGPT \cite{openai2021} and Google's BARD \cite{googleBard}. In  \cite{Human_in_the_loop_Graph}, a knowledge graph is built from text-prompting a user with feedback in every response to provide an acceptable retail-item recommendation. 
Differences between the Bibliographic Utility Network Information Expansion (BUNIE) method used in this dissertation and \cite{Human_in_the_loop_Graph} exist, as the study's structure provides one recommendation, whereas BUNIE offers a dataset. Interactive modes also differ from ours in using click-and-drag selection to delete rather than prompts. Furthermore, BUNIE removes papers at the HITL phase, whereas \cite{Human_in_the_loop_Graph} requests positive and negative feedback about recommendations. 

A HITL work more similar to BUNIE in Ref. \cite{image_dataset_human_in_loop} aims to build labeled image datasets for Computer Vision (CV) applications. A user labels a few images, which are extrapolated to all in the image's cluster, and then the images are model-evaluated for reassignment. Like BUNIE, the process is iterated to convergence but differs in the direct user influence of datum retention.

\section{Factorization Innovation}
\subsection{Knowledge Graphs}
Recent studies have demonstrated the application of knowledge graphs in identifying emerging trends, uncovering hidden connections, and predicting future research directions within scientific domains \cite{manghi2021new, pingle2019relext,luan-etal-2018-multi, WANG2018112, textbook_graph, SHAO2021113764, ABUSALIH2021103076, KNOWEDU}. Integrating semantic analysis with graph-based structures allows for a more nuanced understanding of the thematic evolution in deep learning and cybersecurity \cite{opdahl2022semantic, sikos2023cybersecurity}. These advances highlight the potential of knowledge graphs to transform the exploration and exploitation of scientific literature, moving beyond traditional search mechanisms to enable a deeper, contextually rich engagement with content \cite{auer2020improving, zhu2017use}.
To achieve these results, care must be given to the specific process for constructing and analyzing knowledge graphs, as studied and explained in \cite{cyber_ontology,CHEN2019100959, k12_edu_graph, ACEKG, Knowledgegraph_theory}.

Similar to this work, \cite{Auto_web_scraping} scrapes the web for information relevant to the domain-specific knowledge. However, their domain is in medicine, and ours is in cybersecurity. 
Further, they do not use decompositions to construct the graph, but rather vocabularies. Similarly,  \cite{covid_graph} uses dates, topics, and events as entities to make the knowledge graph, but omits the use of decomposition to organize large data.
This work's decompositions are enhanced by incorporating the semantic structure of the text, allowing for the estimation of the number of topics, which enables a coherent separation of the latent topics and accurate document clustering. This method is supported by the findings in \cite{vangara2020semantic, erenDocEng22}. Building on \cite{vangara2020semantic, erenDocEng22},  this work's approach is expanded through hierarchical decomposition and incorporates semi-supervised labels of arXiv categories in joint factorization to refine this work's capabilities.

\subsection{Hierarchical Topic Modeling}

A standard method for hierarchical topic modeling is through a probabilistic approach, in which the corpus is treated as a mixture of topics that can be modeled with a probability distribution. One such method is hierarchical Latent Dirichlet Allocation (hLDA) using the nested Chinese Restaurant Process (nCRP) beforehand \cite{blei2003ncrp}. This model allows for arbitrarily large branching factors in topic hierarchies, with the nCRP aiding in inferring the appropriate depth of the topic tree. Another LDA-based approach is Hierarchical Dirichlet Processes (HDP). In this method, a set of Dirichlet processes is coupled via their base measure, which is distributed according to a Dirichlet process \cite{Teh2006HierarchicalDP}. The authors introduce two Markov Chain Monte Carlo (MCMC) sampling schemes for posterior inference under hierarchical Dirichlet process mixtures and test the approaches on multiple datasets. 

More recent approaches to hierarchical topic modeling make use of variational auto-encoders, recurrent neural networks, and transformers \cite{jin2021neural, guo2020recurrent, Viegas2020CluHTMS, Mifrah_Benlahmar_2022, grootendorst2022bertopic}. While successful, these methods face a common problem in deep learning: successful models heavily depend on the quantity and quality of the training data. HSNMFk-SPLIT is a favorable approach to hierarchical topic modeling since it is instance-based. Only the data that is being modeled is required for computation. \cite{barron2024cyber}

\section{Binary Directional Search}
This section provides a brief review of several machine learning dimensionality reduction and clustering algorithms that require hyperparameter selection for $k$. Binary Bleed uses binary search to optimize $k$, so this work reviews related binary search studies. Finally, Binary Bleed optimizes the hyperparameter $k$ over a search range, making an automatic model determination, or $k$ selection, another foundation. Other works have optimized their respective algorithms for parallel and distributed contexts in binary search and automatic model determination.

Applications of $k$ Search: Optimizing for $k$ has utility in several domains, particularly clustering, where optimal cluster count search computations can be expensive. Typically, $k$ is a user-provided parameter that must be refined through trial and error. Algorithms that require user-specified or trial-discovered $k$ include K-means Clustering \cite{MacQueen1967SomeMF}, K-medoids Clustering \cite{medioids}, K-medians Clustering \cite{jain1988algorithms}, Fuzzy C-means Clustering \cite{BEZDEK1984191}, Mini-Batch K-means \cite{mini_batch}, Spherical K-means \cite{spherical_k_means}, Elkan K-means \cite{elkan2003using}, NMF Clustering \cite{doc_cluster_nmf}, Symmetric NMF Clustering \cite{symmetric_nmf}, and RESCAL Clustering \cite{nickel2011three}.

Each algorithm can use the silhouette score to determine the ideal $k$ heuristically. In this work's experiments, detailed in the results section, Binary Bleed's performance is analyzed when operating with the silhouette score and Davies Boulding to test NMFk, RESCAL, and K-means algorithms.

Binary Search:
Several aspects of binary search are relevant to this work, including optimizing search, modifying the binary search algorithm, and parallelizing the binary search process. Noisy Binary Search \cite{Karp2007NoisyBS} minimizes the number of comparisons to find an optimal coin in a sequence of coin flips by using biased coin flips to compare elements in a sorted sequence indirectly. It identifies the position where the probability of observing heads changes from below to above a given target threshold. Flip positions are determined from previous outcomes, focusing the flips on areas with the highest uncertainty about crossing the target probability. This approach is similar to Binary Bleed, which narrows searches in the search space using a thresholding mechanism combined with an observed score. Differently, while Noisy Binary Search selects the next target based on the score of the current item, Binary Bleed is designed for methods like NMF, where silhouette scores obtained from different $k$ selections are independent of each other.
Consequently, the silhouette score cannot decide the next $k$ to visit. Instead, for the $k$ search space pruning heuristic, Binary Bleed follows the hypothesis that the silhouette score will remain low after the ideal $k$ due to the overfitting phenomenon. Once a stable $k$ is found, all the lower $k$ values are ignored, and higher or lower $k$ values require visitation until overfitting is observed.

Another work, \cite{Tyrrell1988}, sought to optimize the binary search algorithm by modifying it to check the bounds of a stack and adjust these bounds, thereby increasing or narrowing the current search space to make the search process more efficient. This was required to measure the resting position of human eyes, which changes over time. 
Similar to \cite{Tyrrell1988}, a modified binary search was described by \cite{ChadhaMM14}, which checks at both ends of a sub-array and the middle to reduce the number of iterations through many sub-arrays to find the search value. Both \cite{Tyrrell1988} and \cite{ChadhaMM14} reduce needed checks in binary search, similar to this work. However, this work precomputes the search space and iteratively prunes values. Binary Bleed does not terminate upon finding an optimal candidate. Instead, it adjusts the search space and continues exploring the parameter direction to further optimize the results.

Some works have achieved parallelization of the binary search with specific constraints. One work, \cite{efficient_parallel_binary}, partitions the search array into smaller sub-arrays, searching each in parallel, which is similar to \cite{ChadhaMM14} but operates all arrays concurrently, and so the similarities to this work are the same with the addition of parallelizing the problem. Binary Bleed may operate parallel jobs across single or multiple compute nodes, each handling a portion of the $k$ search space. Another work, Parallel Binary Search in \cite{parallel_binary} defines two arrays, $A$ and $B$, where $|A|<|B|$. Parallelization distributes $A$'s elements across resources to find the smallest element in $B$ greater than or equal to each element in $A$. While this approach to parallelization is similar to parallelizing the $k$ range in this work, the objectives differ. Binary Bleed aims to find the largest or smallest optimal value in the $k$ space rather than minimizing array indices.

Importantly, distributed and parallel binary searches are often used interchangeably but address different challenges. In a parallel search, data is small enough to fit in memory, so evaluations of different $k$ values are concurrent on other processors or nodes. In contrast, distributed search requires multiple processors or nodes to handle single $k$ evaluations for data too large for memory.
In the context of this work, comparing the search term to the current term $n$ in binary search is analogous to model computation at $k=n$. This computation at $k=n$ may exceed memory capacity for large datasets, necessitating a distributed solution where the data for a single calculation is divided among multiple resources. This uses Binary Bleed to minimize the $k$ search space with distributed NMF from this work's prior work \cite{pyDNMFk,bhattarai2023distributed}. This approach allows parallel evaluations to concurrently minimize $k$, while distributed evaluations manage the computation for each $k$ in large datasets, ensuring efficient utilization of resources.

K Search:
Several algorithms have optimized parameters or $k$, including those that utilize parallelization and distributed contexts. 
For textures,  \cite{distributed_constrained_heuristic_search} uses distributed calculations from constraints and backtracking error correction, allowing bulk calculation with minor corrections.  
While this work can distribute an NMF $k$ operation larger than memory, there is no need to backtrack in the $k$ optimization. Either an optimal $k$ is found or not, and can be done in parallel and/or distributed calculations. 

In an alternative method to approximate the number of $k$ in a dataset, \cite{distributedNearestNeighbor} uses hash functions with prototypes. This approach adaptively chooses a set of prototypes and binary codes to represent the data sample distribution, approximating neighbor relationships. The distribution in this algorithm is derived from learning the weights of the hash codes rather than selecting $k$. Therefore, while distribution is utilized in their method, this work directly employs distribution and parallelization to determine the optimal $k$ value.

Another optimization can be observed in the context of K-Nearest Neighbors (KNN) \cite{big_data_knn}, where training examples are sorted into a binary search tree based on the two points that are the furthest apart in the dataset. The furthest data points are identified at each node to construct the next partition. This method attempts to optimize the $k$ selection for KNN. However, it is peculiar to KNN rather than a general parameter search optimization method, as this work with Binary Bleed proposed.

Similar to \cite{big_data_knn}, the parallel $k$ search in \cite{sismanis2012parallel} utilizes GPU-accelerated parallel processing and specific sorting algorithms, such as bitonic sort, that are efficient under synchronous operation conditions. While this approach reduces the number of operations required in a distributed environment, it is designed explicitly for algorithms that benefit from GPU operations and truncated sorting methods. In contrast, this work aims to optimize the $k$ parameter more generally and does not rely on a specific model, sorting method, or primitive operation.

\section{Law}
This section reviews contributions across RAG domains, semantic search using vector embeddings, knowledge graph construction, NMF, and legal information systems.

\subsection{ NMF  for Pattern Discovery}
In the legal domain, NMF has proven valuable for analyzing complex textual data, such as case law and statutes, and facilitating topic discovery and clustering. For instance, \cite{Budahazy2021AnalysisOL} applied NMF to legal documents to extract latent topics and visualize relationships, demonstrating how NMF’s interpretable structure aids researchers in identifying underlying topics not readily apparent in raw text. They even applied NMF hierarchically to find fine-grained topics; however, they did not have a way to automatically approximate the number of clusters at each decomposition. More recently, \cite{Li2022GuidedSN} proposed a guided semi-supervised NMF approach for topic discovery in legal documents, using domain knowledge to steer factorization and ensure the extracted topics remain highly relevant. This semi-supervised extension bridges the gap between fully automated unsupervised techniques and expert-driven analysis.  NMF’s utility in legal contexts is particularly significant, providing an interpretable framework for analyzing large textual datasets.

\subsection{Retrieval-Augmented Generation}
RAG has emerged as a foundational approach for improving AI systems across various domains, including law. For instance, \cite{lewis2020retrieval} introduced a framework that dynamically retrieves relevant documents to ground generative outputs, achieving notable gains in accuracy. Building on this idea, \cite{guu2020realm} proposed a retrieval-augmented pretraining method that integrates external knowledge for improved downstream task performance. In contrast, \cite{izacard2021leveraging} demonstrated the effectiveness of retrieval in open-domain question answering. These advances lay the groundwork for applying RAG to the legal sector, where the method’s ability to ground LLMs in authoritative texts reduces hallucinations and increases accuracy in tasks such as case law retrieval, statutory reasoning, and judgment prediction. Notable examples include CBR-RAG, which incorporates Case-Based Reasoning to structure retrieval for legal question answering (QA)\cite{10.1007/978-3-031-63646-2_29}, and LegalBench-RAG. This benchmark suite tailors evaluation metrics to the demands of legal information synthesis \cite{pipitone2024legalbenchragbenchmarkretrievalaugmentedgeneration}. Parallel work has demonstrated RAG’s capabilities in other domains, such as malware data analysis, by combining embeddings, KGs, and NMF \cite{barron2024domain}. Other works demonstrate how LLMs can dynamically decide when and what to retrieve to enhance legal reasoning \cite{asai2023retrieve} and how multi-step legal judgment prediction can benefit from iterative retrieval and generation \cite{cui2023multistep}, further underscoring the effectiveness of combining retrieval strategies with LLMs.

\subsection{Semantic Search with Vector Embeddings}
Semantic search operates on dense vector representations to find the deeper semantic relationships in texts, going beyond keyword matching to proper context retrieval \cite{karpukhin2020dense, guu2020realm}. This search is especially valuable in the legal domain, where queries often demand conceptual rather than surface-level understanding. Domain-specific pretraining has become increasingly important, as seen in LEGAL-BERT \cite{chalkidis2020legalbert}, which outperforms general-purpose models by better capturing nuances of legal language. Benchmarks like LeCaRD \cite{ma2021lecard} show the effectiveness of dense retrievers, such as SBERT, in legal case retrieval of Chinese law and the abilities of hybrid approaches that integrate both lexical and dense retrieval methods. Hierarchical transformer architectures \cite{xiao2021hierarchical} and long-context models like Longformer \cite{beltagy2020longformer} further address the challenges of lengthy legal documents. In contrast, citation-driven approaches such as SPECTER \cite{cohan2020specter} leverage metadata to improve retrieval.
Furthermore, work on neighborhood contrastive learning for scientific document representations (SciNCL) \cite{ostendorff2022neighborhood} demonstrates how controlled nearest neighbor sampling over citation graph embeddings can provide continuous similarity. This strategy could also inform citation-based retrieval improvements in legal domains. Challenges persist in scaling, explainability, and integrating heterogeneous data sources into real-world legal workflows.

\subsection{Knowledge Graphs in Legal Research}
In the legal domain, researchers have used KGs for knowledge extraction \cite{Sovrano2020LegalKE}, constructed domain-specific graphs tailored to legal contexts \cite{Dong2021AKG}, and explored their potential for addressing trust, privacy, and transparency concerns \cite{a9b58e3cf6ea4306b01b44386cbf9c1f}. Recent studies have demonstrated the effectiveness of KGs in recommending similar legal cases \cite{dhani2024similarcasesrecommendationusing}, linking case law with statutes for improved retrieval \cite{zhou2024automaticknowledgegraphconstruction}, and using graphs to enhance legal case law search in Chinese legal systems \cite{BI2022109046}. Beyond retrieval, KGs have been incorporated into knowledge-aware machine reading systems for legal question answering \cite{long2021kalm}. These works demonstrate KGs' capacity to represent intricate legal concepts, structure regulatory frameworks, and encourage reasoning over legal data.

\subsection{Legal Systems and Case Retrieval}
Legal information systems have evolved rapidly with the advent of neural architectures and hybrid retrieval pipelines, enabling precise tasks such as precedent retrieval, statute matching, and judgment prediction \cite{chalkidis2019neural, robaldo2020neural, chalkidis2021paragraph}. Benchmarks like LeCaRD \cite{ma2021lecard} and LexGLUE \cite{chalkidis2022lexglue}, along with LEGAL-BERT-based systems \cite{chalkidis2020legalbert}, have demonstrated the capability of these neural methods to improve accuracy in analyzing legal corpora. In particular, researchers have leveraged structured reasoning with transformers and graph representations to link statutes and precedents, as evidenced in the COLIEE competition \cite{rabelo2021coliee}. Nonetheless, several limitations remain: data scarcity and jurisdictional bias continue to restrict the generalizability of such models. At the same time, resource-intensive retrievers like BERT-based cross-encoders \cite{shao2022bert} have challenges scaling to large-scale legal databases. Earlier works in juris-informatics have highlighted the potential of automating legal reasoning and document analysis \cite{ashley2017artificial}, laying the groundwork for modern approaches that fuse knowledge graphs, transformers, vector stores, and agent-oriented RAG pipelines to deliver more explainable and efficient legal research workflows.

 \section{Malware }
Recent methods for building RAG-assisted \cite{lewis2020retrieval} chatbot applications rely on unstructured text stored in vector databases for QA tasks \cite{Liu_LlamaIndex_2022}. Although the integration of KGs in AI systems is not novel \cite{kgqa_old}, increasingly, researchers are leveraging them to improve LLM reasoning while simultaneously addressing the reliability issues \cite{ VLNSurvey_Pan1, VLNSurvey_Pan2, VLNSurvey_Li}. Despite the benefits, integrating domain-specific knowledge into chatbots requires substantial effort. The prior work is reviewed for standard chatbot designs, domain knowledge integration in RAG pipelines, and the necessary steps for constructing KGs.

\subsection{KGs in RAG Pipelines}  
Building a sophisticated chatbot requires knowledge of various research fields. Prior works rarely present a fully engineered system like ours. Instead, most efforts focus on improving specific aspects of RAG pipelines, e.g., retriever design \cite{chen-etal-2017-reading, karpukhin2020dense}, query intent recognition \cite{slimproxy}, and KG reasoning \cite{ jiang-etal-2023-reasoninglm, luo2024reasoning, jin2024graphchainofthoughtaugmentinglarge, li2024kgmistral}. this work's approach resembles past methods that leverage chain-of-thought \cite{wei2023chainofthoughtpromptingelicitsreasoning} prompting on KGs \cite{jin2024graphchainofthoughtaugmentinglarge, luo2024reasoning}; in conjunction with LLM-agents to enhance reasoning capabilities \cite{yao2023react, lala2023paperqa, sanmartin2024kgragbridginggapknowledge}. In addition to incorporating these state-of-the-art techniques, this work's RAG pipeline modifies its retrieval method to use KNN with the Levenshtein metric as an entry point for context search, replacing the cosine distance.   Also, a ``highly-specific" knowledge base is constructed for targeted QA tasks.

Although expensive and time-consuming, a handful of prior works incorporate domain-knowledge into their RAG pipelines \cite{soman2023biomedical, edwards2024hybrid, kragen, Xu2024}; however, the majority either use existing KGs built broadly on medical literature \cite{soman2023biomedical, kragen}; or do not disclose any details regarding their dataset construction \cite{Xu2024}. This work's method is ``highly-specific" because it was driven by subject matter expertise, which informed its dataset curation and cleaning techniques \cite{eren2023slic, TELF}. 

\subsection{KG Development}
Developing knowledge graphs requires building a corpus, defining an ontology, and extracting the relevant entity-relation triplets from unstructured text. 

\textbf{Corpus Building.} The term ``highly-specific" and explain this work's dataset collection method. A key feature of this work's dataset collection is the use of unsupervised methods \cite{TELF} to decompose corpora into document clusters with finer specificity than the author-provided tags available on open-access websites. This differs significantly from prior approaches \cite{ABUSALIH2021103076, edwards2024hybrid, yu2022biosalgorithmicallygeneratedbiomedical}. Latent-topic information from this work's NMFk method was leveraged to filter and select the most relevant data for this work's knowledge base, as well as to prune documents based on citation information and embedding distances. This work's text cleaning pipeline is informed by subject matter experts (SME)  \cite{TELF, solovyev2023interactive}, thus going beyond standard methods by incorporating expert-derived rules for document cleaning, e.g, acronym and entity standardization. 

\textbf{KG Construction.} This work's ontology is shaped by traditional methods, i.e., relying on SME design and capturing task-specific features. Latent information from the decomposition process introduced in this work \cite{TELF} is incorporated into the knowledge graph as entities, representing an innovation over previous approaches. For entity and relation extraction, the methodology extends beyond conventional learning-based techniques \cite{ji2022} by leveraging recent advancements involving LLM-agents \cite{Liu_LlamaIndex_2022, ye2023isolationmultiagentsynergyimproving}, rather than relying on traditional LLM prompting methods \cite{wadhwa-etal-2023-revisiting, marinov2024relation, zavarella2024a, edge2024localglobalgraphrag}.  This approach yields non-sparse KGs, meaning the average out-degree of entities \cite{lv-etal-2020-dynamic, HoGRN} is high. No prior work integrates these methods into their knowledge graph construction process.

\subsection{Link Prediction}

Matrix factorization (MF) techniques have been widely used for predicting missing links in networks by leveraging latent structure and graph embedding methods to infer potential but previously unseen connections \cite{Lee1999LearningTP,chen2020robust,xu2016perturbation,peng2023negative, Johnson2014LogisticMF,nenova2013fca,miettinen2020recent}. In the context of NMF, early research addressed incomplete observations in user-rating data \cite{wnmfk_src}, introducing weighted NMF (WNMF) for collaborative filtering tasks. Further adaptations incorporated node attributes to tackle sparsity in link prediction, such as GJSNMF \cite{csse.2022.028841} and JWNMF \cite{2022_Tang_wnmfk}. In contrast, others focused on recommender systems using constraints \cite{HOSSEINZADEHAGHDAM2022116593} or federated learning \cite{eren2022fedsplit_recommender,Collaborative_Filtering}. Ensemble NMF approaches \cite{7987705} split large graphs into subproblems for better scalability, and other work introduced graph regularization or deep architectures \cite{10.5555/3305890.3305921,Yao2024EGDNMF}, often addressing extreme sparsity with additional complexity.

LMF \cite{Johnson2014LogisticMF} offers a probabilistic alternative for modeling binary or implicit feedback, with extensions that incorporate neighborhood structure \cite{LIU2019104960,10.1371/journal.pcbi.1004760} or symmetry constraints \cite{ppi_symlmf} to capture local and global relationships in network data. Beyond real-valued methods, Boolean or binary factorizations are useful in logical structures, employing 0/1 constraints to represent observed links. Other work on Boolean Matrix Factorization (BMF) investigated logical and arithmetic operations \cite{binary_factorize}. This work integrates Otsu's method \cite{4310076} and k-means clustering \cite{Jin2010} into Boolean factorization as thresholding mechanisms. Similarly, \cite{app12115377} employed Otsu's method and k-means clustering to segment brain diffusion imaging based on a reduced feature space obtained using NMF. While \cite{app12115377} applied these techniques to classify the features of interest in the images, such as white matter, gray matter, and cerebrospinal fluid, using the latent factors, this work's approach differs in that it integrates them directly into the factorization process to impose Boolean constraints on the latent factors. Rank selection remains a critical challenge in these various MF formulations, with recent attempts at cross-validation \cite{lee2022suitor}, bootstrap-based stability analyses \cite{cai2022rank}, or other regularization approaches \cite{doi:10.1137/1.9781611974348.81} that studied balancing model complexity and predictive accuracy.

Besides missing link prediction in conventional networks, MF has significantly impacted AI reasoning within KGs \cite{wang2019explainable,cao2024knowledge}, where relationships among entities are represented as triple-based structures. Symbolic reasoning methods in KGs often suffer from scalability issues and incomplete knowledge, leading to the adoption of MF-based embedding techniques that uncover latent structures for knowledge base completion \cite{nickel2016review}. Decomposing adjacency matrices corresponding to KG relations allows AI systems to predict new connections, improving coverage and inferential power. These factorizations also benefit recommendation systems by fusing user-item interactions with explicit semantic links from the KG, offering more explainable recommendations \cite{wang2019explainable}. In the life sciences, large-scale biological graphs have utilized MF to hypothesize previously unknown interactions among genes, proteins, or drugs \cite{zitnik2018polypharmacy}, highlighting the versatility of factorization-based approaches for reasoning-driven discovery.

Despite the progress of MF in missing link prediction and AI reasoning, two key challenges remain. First, Uncertainty Quantification (UQ) is critical yet largely unaddressed. Most methods yield only point estimates and cannot measure how trustworthy each prediction is, although such measures are vital in high-stakes domains like healthcare and finance \cite{LIU2019104960, NEURIPS2023_54a1495b}. Second, the problem of optimal rank selection persists, with underestimation or overestimation leading to suboptimal performance \cite{lee2022suitor,8852146}. To address these gaps, this work proposes novel Weighted, Boolean, and Recommender NMF\textit{k} factorization methods, each incorporating automated rank determination and deterministic thresholding.  These methods are further enriched through ensemble strategies and logistic components, and a bootstrap-based UQ mechanism is introduced that allows an abstention option for highly uncertain predictions. Combining thresholding, ensemble-driven rank selection, and confidence scoring aims to enhance the reliability, interpretability, and applicability of MF-based link prediction in classic networks and knowledge-graph-driven AI systems.

\subsection{Similar Systems}

\input{Part1/Related_Work/ADDED}

%% file: Part1/Related_Work/ADDED.tex
Several academic and open source systems come close to the scope of this work by combining collection pipelines, structured knowledge graphs, and vector retrieval with generation. The comparison below follows the lifecycle of a knowledge-intensive system. It considers how each system collects and filters data, how it builds and maintains a knowledge graph and a vector store, how it handles missing structure or incomplete links, and how it performs retrieval augmented generation. The discussion then explains where this work differs and how it fills the remaining gaps with a tensor-structured view of retrieval and reasoning.

A first family of systems focuses on general-purpose retrieval and generation. REALM pretrains retrieval together with masked language modeling, so lookup is active during pretraining, fine-tuning, and inference \cite{guu2020realm}. RAG joins dense retrieval with sequence-to-sequence generation over a vector index \cite{lewis2020retrieval}. DPR is a strong dual encoder retriever that supplies relevant passages \cite{karpukhin2020dense}. FiD aggregates many retrieved passages inside the decoder to strengthen answer synthesis \cite{izacard2021leveraging}. ATLAS shows that pretrained retrieval augmented models can reach strong few-shot performance on knowledge-intensive tasks \cite{izacard2023atlas}. KILT frames system-level evaluation with a shared snapshot and a common interface, which improves comparability \cite{petroni2021kilt}. These systems are useful for open domain question answering and usually use a large collection of texts specifically curated to represent broad knowledge across many subjects, and they do not prescribe expert-guided domain selection, explicit hierarchical semantics or a synchronized symbolic layer. This work of this dissertation keeps the strengths of dense retrieval and strong readers but it places domain definition and data structuring before indexing. Structure is added through an interpretable hierarchy and it synchronizes knowledge layers so that retrieval and reasoning can respect domain boundaries and semantics.

A second family increases the effectiveness of the data with its structure. KG FiD uses a knowledge graph into Fusion in the Decoder and uses graph topology with a graph neural network re-ranker for overall improvements in open domain question answering \cite{yu2022kgfid}. GraphRAG constructs a knowledge graph over a corpus, organizes it into hierarchical communities with precomputed summaries, and uses graph-guided routing together with text retrieval for both local neighborhood search and global community-level reasoning \cite{microsoft_graphrag_blog,graphrag_github_2024}. Recent hybrid approaches explicitly combine graph and vector paths so an answer can draw on both relational and textual evidence, as in HybGRAG and HybridRAG \cite{lee2025hybgrAG,hybridrag2024}. Earlier graph and text fusion, such as GRAFT Net, blends structured knowledge base signals with unstructured passages for question answering \cite{sun2018graftnet}. These systems show that structure can improve retrieval and generation quality. They do not specify how to build a domain-specific corpus or how to expose an interpretable hierarchy that can be examined and reused across domains. This work fills that gap by placing domain selection up front and further add a hierarchical topic tree with interpretable factors and by keeping a synchronized graph and vector store that shares the same semantics.

Several open source frameworks operationalize a combined graph and vector store in end to end deployments. RAGFlow integrates a conventional vector pipeline with an ingestion stage that extracts entities and relations to construct a knowledge graph and then exposes that graph at query time for multi-hop reasoning together with dense retrieval and re-ranking \cite{infiniflow_ragflow_github,ragflow_construct_kg_docs}. LightRAG builds a knowledge graph during indexing, keeps the graph aligned with the embedding store, and provides local, global, and mixed retrieval modes with an API and a web interface \cite{hku_lightrag_github}. SciPhi R2R adds an agent-driven deep research workflow on top of hybrid keyword and semantic search and automatic graph construction, which enables multi-step retrieval that pulls from both the vector index and the graph inside a production-oriented server \cite{sciphi_r2r_github,sciphi_r2r_docs}. These frameworks demonstrate how to synchronize structured and unstructured data and how to route over them. They leave open three concerns that this work addresses. They do not dictate a domain-aware collection process rooted in expert seeds and citation-guided expansion. They do not provide a transparent hierarchical representation that is used by both indexing and routing. They rarely address missing links and uncertainty in the structured layer.

Collection and corpus expansion also differentiate systems. Domain-aware pipelines often begin with subject matter expert seeds, then expand over citation networks by following document-level embeddings. Citation informed encoders such as SPECTER and SciNCL learn representations that reflect inter-paper relatedness, which supports growth that stays on topic \cite{cohan2020specter,ostendorff2022neighborhood}. Legal retrieval research also models connectivity in the citation graph, which can strengthen case retrieval as seen in work such as CaseLink \cite{tang2024caselink}. The general-purpose RAG family does not specify a collection strategy. GraphRAG and the open source frameworks above can intake a supplied corpus, and they can extract entities and relations. They do not specify a controlled citation expansion loop or a principled pruning method that keeps the resulting dataset aligned with expert intent. This work fills that gap by starting with expert seeds, expanding along citation and embedding neighborhoods, and using citation-aware encoders for control. The result is a curated domain corpus that is auditable and that is ready for structured learning.

Once a domain corpus is collected, the question is how to expose its semantics in a form that is both inspectable and useful for retrieval. Neural clustering can discover coarse themes, but it does not always produce an interpretable multilevel structure. Hierarchical nonnegative factorizations and tensor models provide transparent structure learning. A generalized hierarchical nonnegative tensor decomposition yields multi-level topic trees that can act as labels for routing and supervision \cite{vendrow2022ghntd}. Scalable NMF and RESCAL implementations such as pyDNMFk and pyDRESCALk add automatic model selection and distributed CPU and GPU execution which makes hierarchical factorization practical at large scale \cite{boureima2024pydnmfk,bhattarai2023distributed}. This work adopts that path so that topics and subtopics can be named, inspected, and versioned. It then aligns the embedding space to that hierarchy so the vector store reflects meaning at multiple levels. HEAL introduces a hierarchical embedding alignment loss that uses labels from hierarchical clustering or factorization and improves retrieval relevance and downstream classification in domain specific retrieval augmentation \cite{bhattarai2024heal}. The result is a dual data reepresentation where the graph and the vector store share the same semantics and where routing can rely on explicit structure as well as similarity.

Handling missing structure is a further gap in many systems. The research prototypes and the open source frameworks above focus on retrieval and generation over a fixed graph. They often do not include uncertainty-aware completion of missing edges or scores that enable abstention when the structure is unknown. This work addresses that gap by adding tensor and matrix factorization for link prediction in the knowledge graph. That capability supports filling in missing links with calibrated scores. It improves recall for multi-hop questions, and it relaxes strict behavior when the structured layer is sparse. The factorization stage integrates naturally with the interpretable hierarchy and with the synchronized vector store so completion is both grounded and auditable.

Prior work typically pairs a retriever with a reader. For example \cite{guu2020realm}, \cite{lewis2020retrieval}, and \cite{karpukhin2020dense}. Later variants introduce generative readers \cite{izacard2021leveraging} and scale to large collections \cite{izacard2023atlas}. Benchmarks such as KILT emphasize evidence grounding and evaluation \cite{petroni2021kilt}.

More recent systems route between text and graph retrieval. Examples include knowledge graph augmented retrievers \cite{yu2022kgfid} and GraphRAG style pipelines \cite{microsoft_graphrag_blog}. Open implementations extend this direction \cite{graphrag_github_2024}. Prior graph and text hybrids remain relevant \cite{sun2018graftnet} and newer hybrid RAG designs continue this trend \cite{hybridrag2024,lee2025hybgrAG}.

The present approach targets the same objective of selecting the best evidence per query while differing in how evidence is organized and chosen. A tensor structured view of retrieval and reasoning enables an agent to do three things. It can walk a symbolic graph. It can issue dense vector searches. It can invoke link prediction when the structure is missing. An induced topic tree supplies an interpretable routing signal across these modes.

This design treats the collection as a first-class object and unifies structure induction, completion, and retrieval. It adds and synchronizes a human-interpretable hierarchy that organizes the graph and vector store. It completes the missing structure as needed and uses tensor structured RAG to route across modes. The result is effective and transparent and is designed for domains such as legal practice, where domain tuned encoders such as Legal BERT are valuable components but not a complete structured system \cite{chalkidis2020legalbert}.

%% file: Part2/Data_Aquisition/0.tex
This chapter defines the corpus acquisition pipeline that supports every subsequent component of the dissertation. The document collection is assembled through citation-network expansion, expert filtering, and controlled vocabulary alignment, producing a broad yet pertinent input set. This corpus serves as the input to the Binary Bleed decomposition, HNMFk taxonomy induction, and Tensor-Structured RAG modules, which convert the text into provenance-aware knowledge graphs and aligned vector stores. The acquisition workflow meets two objectives: it maximizes recall so that emerging signals are retained, and it maintains the topical coherence and lexical consistency required for the low-rank matrix and tensor operations that drive the retrieval-and-reasoning pipeline.

\input{Part2/Data_Aquisition/base_bunnie}
\input{Part2/Data_Aquisition/cheetah}
\input{Part2/Data_Aquisition/super_bunnie}
\input{Part2/Data_Aquisition/acronyms_consolidations}

\section{Conclusion}
The methods described in this chapter expand a small set of expert-identified seed papers into a large-scale corpus comprising millions of documents, with standardized vocabulary, recorded provenance, and complete auditability. This fulfills the role of the chapter within the broader framework of the dissertation: to produce a clean, structured input suitable for downstream processing. The resulting corpus enables efficient decomposition of sparse term–document matrices by Binary Bleed, supports the construction of interpretable topic hierarchies through HNMFk, and provides consistent evidence streams for the event-driven infrastructure that maintains alignment between the KG and VS. With the data acquisition phase complete, the subsequent chapters shift focus to semantic structuring and hybrid inference, advancing the dissertation’s central aim of building scalable, reliable systems for knowledge retrieval and reasoning.

%% file: Part2/Data_Aquisition/base_bunnie.tex
\section{Citation–Network Expansion and Initial Pruning}
The workflow begins by seeding a compact set of publications and expanding outward along citation edges to create a broad but thematically connected corpus.  Throughout this process, human expertise and algorithmic screens cooperate to eliminate irrelevant material before it propagates downstream.

\subsection{Seeding the Core\label{sec:core_select}}
An SME selects a handful of “core’’ papers that exemplify the target topic.  Each core item is supplied as a unique identifier (e.g., DOI) and resolved via the Semantic Scholar API \cite{Kinney2023TheSS} to collect metadata such as the title, abstract, citations, and references.  
Because subsequent expansion follows citation links, the citation influence of the seed set governs the eventual corpus size: a single highly cited paper may spawn thousands of descendants, whereas several niche articles may yield a modest network.

\subsection{Traversing the Citation Graph}
Let $X$ denote the current set of papers and $X^{c}$ the set of all citations made by papers in $X$.  One network hop is defined by \[ X \gets X \cup X^{c},\] adding every cited work to the corpus.  Repeating the operation performs a breadth-first traversal of the directed citation graph.  Users may instead traverse references (i.e., reverse the edge direction) when the core comprises recent or sparsely cited work.  The hop count is user-controlled, providing a stop condition that balances recall against computational load.

\begin{figure*}[htbp]
    \centering
    \includegraphics[width=.9\textwidth]{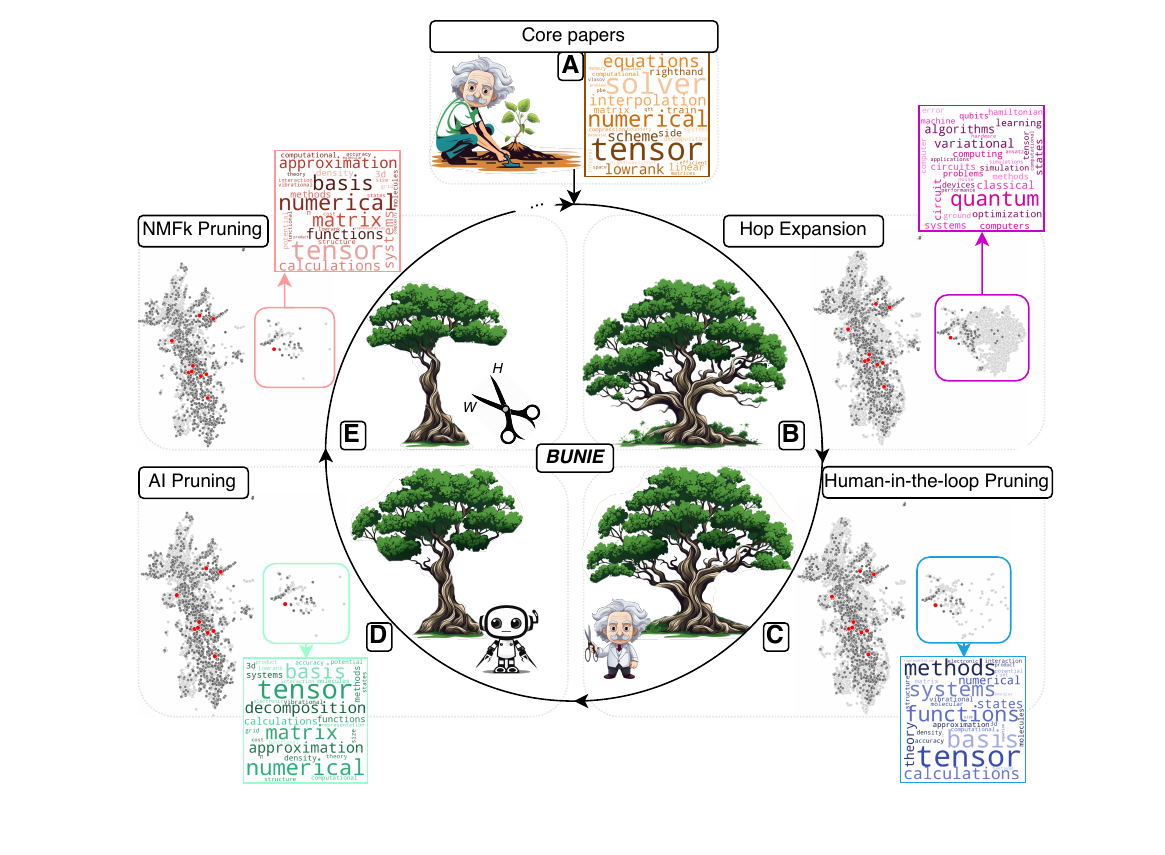}
    \caption{BUNIE's distillation pipeline.
    \textbf{Panel A}: The SME selects specific papers, shown as a bag-of-words wordcloud.  
    \textbf{Panel B}: The dataset expands through the citation network. The wordcloud of a selected subset shows how the vocabulary differs from the core.
    \textbf{Panel C}: Human-in-the-loop pruning removes data. The wordcloud begins to resemble the core. 
    \textbf{Panel D}: An embedding heuristic removes papers far from the core. The dataset becomes more compact. The wordcloud approaches parity with the original. 
    \textbf{Panel E}: NMFk topic modeling produces  H-clustering factorization to remove clusters lacking a core paper. A dense, yet substantial set of documents indicates a successful distillation. The wordcloud closely resembles the original. As indicated by the \textbf{Ellipse}, iterations can be made for refinement or the data extracted in repose for downstream analysis.}
    \label{fig:bunie_overview}
    
\end{figure*}

\subsection{Iterative Pruning Strategy}
Citation expansion allows tangential literature, through acknowledgments of foundational but off-topic publications, review articles, and so on.  To inhibit topical drift, three complementary pruning mechanisms are applied after each hop.

\subsubsection*{Human-in-the-Loop Screening}
High-dimensional embeddings of concatenated title–abstract text are computed with SciNCL \cite{ostendorff2022neighborhood} and projected to two dimensions via UMAP \cite{mcinnes2020umap}.  An interactive GUI  lets SMEs draw lassos around visible clusters, inspect a bag-of-words word-cloud, and view full metadata.  Irrelevant clusters are deselected before the next hop.

\begin{figure}[htbp]
    \centering
    \includegraphics[width=.5\linewidth]{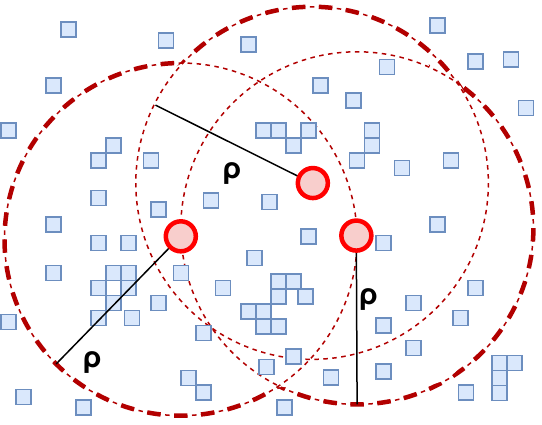}
    \caption{ 2-dimensional representation of hypersphere pruning. Core papers are red dots, citations/ references are blue squares, and hyperspheres are dotted red circles.}
    \label{fig:hypersphere_pruning}
\end{figure}

\subsubsection*{Hypersphere Pruning in Embedding Space}
To retain the geometric structure lost in the UMAP projection, pruning is repeated in the original 768-dimensional space.  Let $E=\{\boldsymbol e_1,\dots,\boldsymbol e_n\}$ be the embeddings of the $n$ core papers.  Compute all pairwise Euclidean distances, collect them in $D$, and set \[\rho \;=\; \operatorname{median}(D). \] Each core embedding becomes the centre of a hypersphere of radius $\rho$; a candidate paper is accepted if its embedding lies within at least one hypersphere (Figure~\ref{fig:hypersphere_pruning}).  The rule adapts to the intrinsic spread of the seed set and requires no manual tuning.

\subsubsection*{Topic-Model Consistency}
The remaining documents are subjected to NMFk fuzzy clustering.  Factorisation of the TF–IDF matrix $\boldsymbol X\in\mathbb R^{m\times n}_{\!\ge 0}$ yields \[
  \boldsymbol X \approx \boldsymbol W\boldsymbol H,\qquad
\] where $\boldsymbol W\in\mathbb R^{m\times k}_{\!\ge 0}$ encodes word–topic affinities and $\boldsymbol H\in\mathbb R^{k\times n}_{\!\ge 0}$ gives topic intensities per document.  Documents dominated by topics that contain no core papers are discarded.  A fixed vocabulary, established from the seed set and augmented with SME-defined term consolidations, prevents vocabulary drift and keeps computational cost stable.

\subsection{Quantifying Cohesion}
To monitor the tightening of topical focus, a compactness score \[  C \;=\; \frac{1}{n(n-1)}\sum_{i\ne j}\bigl|\cos\angle(\boldsymbol e_i,\boldsymbol e_j)\bigr| \] is computed over the cosine similarities of all document embeddings.  $C\in[0,1]$ rises as the corpus becomes more semantically coherent, providing a quantitative checkpoint before further processing.

Through citation-graph expansion coupled with iterative pruning in embedding and topic space, the procedure yields a large yet thematically faithful corpus ready for the downstream semantic filtering and LLM-assisted pruning stages described in the following sections.

%% file: Part2/Data_Aquisition/cheetah.tex
\section{Semantic Query Filtering}

Once a broad, citation-expanded corpus has been pruned, the next step is to extract sub-collections aligned with specific research intents (e.g., “laser and deflection, but not green”). This is accomplished through a transparent and auditable semantic filtering mechanism that supports Boolean logic, $n$-gram matching, and metadata-aware constraints.

\subsection{Index Construction}

For each field within the corpus, such as title, abstract, publication year, authorship, and affiliation, a corresponding inverted index is constructed. Textual fields are normalized through basic preprocessing steps, including lowercasing, punctuation removal, and whitespace normalization. Each normalized token is then mapped to the set of documents in which it appears. Non-textual fields, including categorical and numerical metadata, are indexed using similar mappings. This enables efficient lookups and rapid set-based filtering during query execution.

\subsection{Query Structure}

A query consists of an ordered sequence of clauses. Each clause can take one of the following forms:

\begin{itemize}
\item A single token, such as \texttt{laser};
\item A phrase of multiple tokens in natural word order, such as \texttt{angle deflection};
\item A structured constraint, e.g., \texttt{{"laser": $"node", "+blue" $ }}, specifying that documents must contain \texttt{laser} and \texttt{blue}, but not \texttt{node}.
\end{itemize}

Clauses are combined using intersection (default, \textsc{and}-search) or union (\textsc{or}-search). An optional linked-search mechanism allows reinforcing signals from one clause to override vetoes from another, which is particularly useful in noisy or polysemous contexts.

\subsection{$n$-Gram Matching by Sliding Window}

Multi-token phrases are resolved using a sliding window algorithm. Given a query phrase of length $w$, the algorithm iterates through all contiguous subsequences of $w$ tokens in a candidate document. A match is declared if either the tokens appear in exact order or, for unordered matches, the multisets of query and window tokens are equal. This algorithm operates in $O(mw)$ time per document, where $m$ is the number of tokens in the document and $w$ is typically small.

\subsection{Fusion of Metadata and Textual Constraints}

At evaluation time, the system constructs two candidate sets:

\begin{enumerate}
\item $S\_{\text{meta}}$, consisting of documents that satisfy all specified metadata constraints;
\item $S\_{\text{text}}$, consisting of documents matching the lexical clauses of the query, further refined by phrase validation.
\end{enumerate}

The result set is obtained by intersecting $S\_{\text{meta}}$ and $S\_{\text{text}}$, or by returning $S\_{\text{meta}}$ alone if no textual query is present. Because both sets are derived through precomputed indices, execution scales with the size of the smaller candidate set, making the process efficient and predictable.

This semantic filtering stage enables rapid and context-sensitive extraction of documents that match both lexical content and structured metadata constraints, forming the input to downstream embedding and LLM-based pruning stages.

%% file: Part2/Data_Aquisition/super_bunnie.tex
\section{Embedding and LLM-Based Corpus Pruning}

While lexical and metadata filtering effectively remove many irrelevant entries, they remain insufficient for eliminating semantically off-topic or conceptually distant documents. To refine the corpus for downstream semantic analysis, two complementary pruning techniques are employed: one based on geometric proximity in embedding space, and the other on probabilistic voting within the latent space of an  LLM. Both techniques are designed to preserve transparency by generating structured, machine-readable justification logs.

\subsection{Geometric Outlier Detection in Embedding Space}

Let $\mathcal{C}$ denote the candidate corpus and $\mathcal{R} \subset \mathcal{C}$ a trusted reference subset (e.g., documents manually verified or labeled by domain experts). A semantic encoder $f : \mathcal{C} \rightarrow \mathbb{R}^d$ maps each document to a $d$-dimensional vector space, producing embeddings $\{\boldsymbol{v}_i\}_{i \in \mathcal{C}}$.

Define the centroid of the reference set as
\[
  \boldsymbol{\mu} = \frac{1}{|\mathcal{R}|} \sum_{j \in \mathcal{R}} \boldsymbol{v}_j,
\]
and compute the average radial distance and its variance:
\[
  \bar{r} = \frac{1}{|\mathcal{R}|} \sum_{j \in \mathcal{R}} \|\boldsymbol{v}_j - \boldsymbol{\mu}\|_2,
  \qquad
  \sigma^2 = \frac{1}{|\mathcal{R}|} \sum_{j \in \mathcal{R}} \left(\|\boldsymbol{v}_j - \boldsymbol{\mu}\|_2 - \bar{r} \right)^2.
\]

A candidate document $i \in \mathcal{C}$ is retained if it satisfies the threshold:
\[
  \|\boldsymbol{v}_i - \boldsymbol{\mu}\|_2 \leq \bar{r} + k\sigma,
\]
where $k$ is a tunable parameter (default $k = 3$). This scale-invariant rule provides a nonparametric and distribution-agnostic method for identifying embedding outliers.

\subsection{Probabilistic LLM Voting}

Although the geometric filter is efficient, it cannot capture higher-level semantic cues or domain-specific discourse structures. To address this, each document excluded by the geometric criterion is evaluated by a probabilistic LLM-based voting mechanism, which determines its relevance through contextual reasoning.

\subsubsection*{Prompt Construction}

Each vote is cast using a constrained prompt composed of:

\begin{enumerate}[label=\arabic*)]
  \item A set of $m$ reference abstracts $\mathbf{x}_{1:m}$ drawn at random from $\mathcal{R}$ (typically $m = 10$);
  \item A candidate abstract $\mathbf{z}$ drawn from the rejected set $\mathcal{C} \setminus \mathcal{R}$;
  \item An instruction block enforcing output in a structured JSON format containing keys \texttt{answer} $\in \{\text{yes}, \text{no}\}$ and \texttt{reason}.
\end{enumerate}

Post-processing ensures that invalid or incomplete responses are discarded and, if necessary, resampled.

\subsubsection*{Monte Carlo Acceptance Test}

Each candidate document is submitted to $n$ independent LLM invocations at a fixed temperature $\tau$, producing a sequence of binary votes $B = \{b_1, \dots, b_n\}$, where $b_i = 1$ indicates a positive relevance decision. Define the empirical vote proportion as
\[
  \hat{p} = \frac{1}{n} \sum_{i=1}^n b_i.
\]

A candidate is accepted into the final corpus if $\hat{p} \geq p_{\mathrm{th}}$, where $p_{\mathrm{th}}$ is a predefined threshold (typically $0.6$). This ensemble-based approach enables robustness against noisy or ambiguous predictions while leveraging the deep contextual awareness of the LLM.

%% file: Part2/Data_Aquisition/acronyms_consolidations.tex
\section{Vocabulary Standardization: Acronym Resolution and Lexical Consolidation}

Even after semantic pruning, domain-specific corpora often exhibit surface-level lexical fragmentation, wherein semantically identical concepts are expressed in multiple forms (e.g., ``partial differential equation,'' ``PDE,'' ``P.D.E.''). This redundancy inflates the vocabulary, obscures term frequencies, and degrades the performance of downstream matrix- or graph-based learning methods. To address this, we apply a two-stage standardization process: (1) acronym resolution and canonicalization based on syntactic cues, and (2) lexical consolidation of morphologically similar or near-duplicate tokens based on string similarity. Both stages produce traceable substitution mappings that support downstream auditing or expert override.

\subsection{Acronym Detection and Canonicalization}

\subsubsection*{Detection Strategy}

To identify acronym–definition pairs, all $(n{+}1)$-grams ($2 \leq n \leq 8$) are scanned for patterns in which a contiguous sequence of $n$ words is immediately followed or preceded by a candidate acronym. Specifically, let $g = \langle w_1, \dots, w_{n+1} \rangle$ be a candidate phrase. The following conditions are tested:

\begin{enumerate}[label=(\alph*)]
  \item $w_{n+1}$ is an acronym whose letters match the initials of $w_1$ through $w_n$;
  \item $w_1$ is an acronym whose letters match the initials of $w_2$ through $w_{n+1}$.
\end{enumerate}

When a valid match is found, the full phrase is mapped to the corresponding acronym, and optionally replaced with a single, joined token to preserve conceptual integrity during tokenization. This replacement ensures that multi-word expressions are treated as atomic lexical units in subsequent analysis.

\subsubsection*{Standardization Output}

Across the corpus, acronym mappings are aggregated into a unified substitution table of the form $\textit{full\_form} \mapsto \textit{canonical\_token}$. This table governs all replacements applied to the corpus and may be reviewed or modified by subject-matter experts to correct overgeneralizations or contextual exceptions.

\subsection{Data-Driven Lexical Consolidation}

\subsubsection*{Motivation}

Technical corpora often contain morphological variants, typographical errors, or inconsistent stylistic conventions that fragment concept representations across tokens (e.g., \textit{tokenize/tokenisation}, \textit{normalisation/normalization}). These redundancies degrade the sparsity structure of lexical matrices and dilute the significance of term frequencies.

\subsubsection*{Partitioned Preprocessing}

Let $\mathcal{V} = \{t_1, \dots, t_{|\mathcal{V}|}\}$ denote the full vocabulary. To limit the quadratic cost of pairwise comparisons, preprocessing partitions $\mathcal{V}$ into buckets based on:

\begin{enumerate}[label=\arabic*)]
  \item \textbf{Initial character}: Tokens sharing the same first letter are grouped together, reducing the pair space to $\sum_{\alpha} |B_\alpha|^2$.
  \item \textbf{Length bands}: Within each initial character group, only tokens whose lengths differ by at most $\Delta\ell$ (typically $\Delta\ell=2$) are compared, thereby retaining plausible edit sequences.
\end{enumerate}

\subsubsection*{Similarity Scoring and Filtering}

All unordered token pairs within each group are compared using normalized Levenshtein similarity:
\[
s(t_i, t_j) = 1 - \frac{d_L(t_i, t_j)}{\max(|t_i|, |t_j|)},
\]
where $d_L$ denotes the Levenshtein edit distance. Pairs exceeding a similarity threshold $\tau$ (default $\tau = 0.90$) are flagged for consolidation. To avoid erroneous merges in ambiguous cases ($\tau - \varepsilon \leq s < \tau + \varepsilon$), suffixes such as \texttt{-ing}, \texttt{-tion}, and \texttt{-ness} are stripped before re-evaluation.

\subsubsection*{Canonical Form Selection}

For each high-similarity pair $(a, b)$, the preferred canonical form is chosen based on global corpus frequency. The less frequent token is then mapped to the selected form. The resulting collection of mappings defines a one-to-one substitution function that resolves lexical redundancy across the corpus.

\subsubsection*{Impact}

This two-stage vocabulary standardization pipeline enforces terminological coherence, reduces vocabulary dimensionality, and improves the quality of input matrices for subsequent factorization and knowledge-graph construction. The use of transparent substitution rules ensures interpretability and allows for controlled overrides in sensitive applications.

%% file: Part2/Latent_Feature_Extraction/1.tex
With the corpus established in the previous chapter, the next step is to uncover the latent structure that supports the remaining components of the system. This chapter presents the semantic structuring process, beginning with Binary Bleed, a recursive rank-selection method that identifies an efficient low-rank representation of the term–document matrix while maintaining scalability for large corpora. This selected rank initializes HNMFk, which separates domain-relevant topics across multiple levels of detail while preserving interpretability for subject-matter experts. These numerical factors are then connected to clear, human-readable labels using a language model alignment pipeline based on vector similarity. Together, these procedures transform raw term–document statistics into a structured and interpretable taxonomy, fulfilling the role of semantic structuring described in the abstract and enabling the synchronized construction of the KG and VS.

\input{Part2/Latent_Feature_Extraction/Binary_Bleed}
\input{Part2/Latent_Feature_Extraction/HNMFk}
\input{Part2/Latent_Feature_Extraction/artic_fox}

\section{Conclusion}
The procedures outlined in this chapter complete the middle layer of the dissertation’s architecture by organizing statistical patterns into a hierarchical topic structure. Binary Bleed determines a compact, data-driven rank that ensures the decomposition remains tractable at scale. Hierarchical NMFk then constructs a layered factorization that captures domain concepts at varying levels of specificity, while maintaining interpretability for subject-matter experts. These components are further supported by an automated labeling pipeline that assigns concise, readable names to each factor based on embedding similarity. Together, they form a structured representation that enables the two core capabilities defined in the abstract: accurate retrieval, addressed in the following chapter through Tensor-Structured RAG, and evidence-based reasoning, developed later through tensor-based link prediction. This chapter establishes the system’s semantic foundation, transforming unstructured term–document statistics into a stable and interpretable framework for downstream tasks.

%% file: Part2/Latent_Feature_Extraction/Binary_Bleed.tex
\section{ Binary Bleed }
The method is named Binary Bleed due to its method of pruning less optimal \( k \) values, extends traditional binary search techniques by continuing to search even after a potential optimal \( k \) is found, ensuring the maximization or minimization of the scoring function \( f: \mathcal{K} \times \mathcal{D} \rightarrow \mathbb{R} \), where \( \mathcal{K} \) is the set of candidate parameters and \( \mathcal{D} \) represents the dataset.
Binary Bleed operates recursively to identify the optimal \( k \) that maximizes or minimizes an evaluation score function \( f(k, \mathcal{D}) \). Unlike a conventional binary search that terminates upon finding a target value, Binary Bleed dynamically adjusts the search space based on an evaluation threshold. From an initial range \([k_{\text{min}}, k_{\text{max}}]\), it computes the mid-point \[
k_{\text{mid}} = \left\lfloor \frac{k_{\text{min}} + k_{\text{max}}}{2} \right\rfloor
\] and evaluates the score \[
f_{\text{mid}} = f(k_{\text{mid}}, \mathcal{D}).
\] If \( f_{\text{mid}} \) meets the threshold, the search continues in the direction of optimization. For maximization, 

the lower bound is updated to \( k_{\text{min}} = k_{\text{mid}} + 1 \) when $f(k_{\text{mid}} + 1, \mathcal{D}) > f_{\text{mid}}$, 
and conversely, the upper bound is updated to \( k_{\text{max}} = k_{\text{mid}} - 1 \) when  $f(k_{\text{mid}} - 1, \mathcal{D}) < f_{\text{mid}}$. For minimization, the process is reversed. This recursive update continues until convergence or until a predefined stopping criterion is met, allowing the algorithm to ``bleed" into higher or lower \( k \). This ensures thorough exploration and optimization of the parameter \( k \), adapting the search space dynamically to achieve the best possible evaluation score within the specified constraints.

This search does not cease upon finding a \( k \) above the threshold; instead, it continues exploring to ensure that no better solutions exist, particularly focusing on larger \( k \) values that may yield higher scores. The runtime complexity of Binary Bleed is bounded by \(\Theta(n^{\log_2{(p + 1)}} )\times (T_{\text{model}} + T_{\text{score}})\) in both the best and worst cases, as shown later in \ref{runtime}, where \( p \) is the probability of recursing twice. We apply binary search on the ordered set \(\mathcal{K}\), the search space, to find its optimal value. Mathematically, this optimization can be expressed as:
\[
k_{\text{optimal}} = \max \left\{ k \in \{1, 2, \ldots, K\} : S(f(k)) > T \right\}
\]
where \( T \) is the selection threshold, \( f(k) \) represents the model computation, and \( S(f(k)) \) is the model's scoring function.

\begin{breakablealgorithm}
\caption{Binary Bleed $k$ Search, Single Rank \& Thread}
\label{alg:binary_search_optimal_k}
\begin{algorithmic}[1]

\Require \( \mathcal{K} \) (list of \( k \) ), $data$ (dataset), $model$ (model for calculation), $scorer$ (function to quantify output), $T_{\text{select\_k}}$ (minimum score for optimal \( k \))
\Ensure Optimal $k$ value and its prediction score

\State $ranks_{\text{seen}} \gets \emptyset, k_{\text{max}} \gets \infty, k_{\text{min}} \gets -\infty$ 

\Function{BinaryBleedKSearch}{$\mathcal{K}, i_{\text{left}}, i_{\text{right}}, scorer, data, model, T_{\text{select\_k}}$}
    \If{$i_{\text{left}} \geq i_{\text{right}}$}
        \Return
    \EndIf
    \State $middle \gets i_{\text{left}} + \lfloor (i_{\text{right}} - i_{\text{left}}) / 2 \rfloor$
    \State $k_{\text{middle}} \gets \mathcal{K}[middle]$
    \If{$k_{\text{min}} < k_{\text{middle}} < k_{\text{max}}$}
        \State $score \gets scorer(model(data, k_{\text{middle}}))$
        \State Append $(k_{\text{middle}}, score)$ to $ranks_{\text{seen}}$
        \If{$score \geq T_{\text{select\_threshold}}$} 
            \State $k_{\text{min}} \gets k_{\text{middle}}$
        \EndIf
        \If{$score \leq k_{\text{stop\_threshold}}$}
            \State $k_{\text{max}} \gets k_{\text{middle}}$
        \EndIf
    \EndIf
    \If{$middle+1 \leq k_{\text{max}}$}
        \State \Call{BinaryBleedKSearch}{$\mathcal{K}, k_{\text{max}}, k_{\text{min}}, middle+1, i_{\text{right}}, ranks_{\text{seen}}, scorer, data, model, T_{\text{select\_k}}$}
    \EndIf
    \If{$middle-1 \geq k_{\text{min}}$}
        \State \Call{BinaryBleedKSearch}{$\mathcal{K}, k_{\text{max}}, k_{\text{min}}, i_{\text{left}}, middle-1, ranks_{\text{seen}}, scorer, data, model, T_{\text{select\_k}}$}
    \EndIf
\EndFunction

\end{algorithmic}
\end{breakablealgorithm}

The algorithm is presented in Algorithm~\ref{alg:binary_search_optimal_k}, showcasing its operational structure. To determine the optimal \( k \), a list of \( k \) (\( \mathcal{K} \)) is provided with the dataset, model, scorer, and score threshold. Initially, the set of visited \( k \), the maximum (\( k_{\text{max}} \)), and minimum (\( k_{\text{min}} \)) bounds are initialized (lines 1-2). The algorithm checks the base case for recursion: if the left index \( i_{left} \) is greater than or equal to the right index \( i_{right} \), the function returns, terminating the recursive search when no more \( k \) values are left to explore (lines 3-4).  The optimal \( k \) is sought by computing the middle index of $\mathcal{K}$ and its \( k \) (lines 5-6). This value is validated to ensure it is within  \( k_{\text{min}} \) and  \( k_{\text{max}} \)  (line 7). The model is evaluated at this middle \( k \)  on the dataset, and the score is added to the set of visited \( k \)  values (lines 8-9). If the score at  \( k_{\text{middle}} \) meets or exceeds the threshold, \( k_{\text{min}} \) updates to \( k_{\text{middle}} \), focusing on larger values (lines 10-12). If the score is below the stop threshold, \( k_{\text{max}} \) updates to  \( k_{\text{middle}} \), focusing on lower values (lines 13-15). The BinaryBleedKSearch function is recursively called on the narrowed \( k \) range (lines 17-19). This dynamically recursive search space adjustment allows thorough exploration of \( k \), ensuring the optimal is found based on the evaluation criteria.

Recursive Behavior Analysis:\label{runtime} Let \( T(n) \) denote the runtime of the algorithm for \( \mathcal{K} \) of size \( n \), and $T_{\text{M}}$ and $T_{\text{S}}$ are the runtimes of the model and scorer operating on data, $\mathcal{D}$.  Table \ref{tab:BBbreakdown_recursive} asymptotically itemizes the algorithm. Truncation of the search space occurs with a probability $p$ of two recursions and $1-p$ of one recursion. Therefore, the recurrence relation is:
\[T(n) = p \cdot 2T\left(\frac{n}{2}\right) + (1-p) \cdot T\left(\frac{n}{2}\right) + O(T_{\text{M}} + T_{\text{S}})\]
The model and scorer operate on $\mathcal{D}$, so their runtimes do not scale with $n$ and thus can be represented as  $O(1)$ for each subproblem. This simplifies the equation to:
\begin{center}
$T(n) = (p + 1)T\left(\frac{n}{2}\right) + O(1)$
\end{center}
Applying the  Master Theorem \cite{cormen2022introduction} with $a= p + 1$ (number of recursive calls), $b=2$ (factor by which the problem size is divided), and $f(n) = O(1)$: 
$\log_b{a} = \log_2{(p + 1)}$
 For $\log_2{(p + 1)} < 0 $ to hold, $p$ would need to be less than $0$, probabilistically impossible. Therefore, we have $c < \log_2{(p + 1)} $, and case 1 of the Master Theorem applies:
\[
T(n) = \Theta(n^{\log_2{(p + 1)}})
\]

Thus, Binary Bleed's asymptotic runtime is $\Theta(n^{\log_2{(p + 1)}})$. 

\begin{figure}[H]
  \centering
  \begin{tabular}{|c|c|c|}
    \hline
    \textbf{Step} & \textbf{Line} & \textbf{Complexity} \\
    \hline
    Initialization   & 1–2   & \(O(1)\) \\
    \hline
    Base Case        & 3–4   & \(O(1)\) \\
    \hline
    Middle/Check     & 5–7   & \(O(1)\) \\
    \hline
    Model/Score      & 8     & \(T_{\text{M}} + T_{\text{S}}\) \\
    \hline
    Update Min/Max   & 10–16 & \(O(1)\) \\
    \hline
    Bound Check      & 17,20 & \(O(1)\) \\
    \hline
    Recursion        & 18–23 & – \\
    \hline
  \end{tabular}
  \caption{Primitive operations for asymptotic analysis.}
  \label{tab:BBbreakdown_recursive}
\end{figure}

\begin{figure}[H]
  \centering
  \begin{tikzpicture}[
      scale=3,              
      transform shape,         
      level distance=4mm,
      level 1/.style={sibling distance=20mm},
      level 2/.style={sibling distance=11mm},
      level 3/.style={sibling distance=8mm},
      every node/.style={circle, draw, scale=0.5} 
  ]
    \node (a) {6}
      child {node (b) {3}
        child {node (c) {2}
          child {node (d) {1}}
          child[missing]}
        child {node (e) {5}
          child {node (f) {4}}
          child[missing]}}
      child {node (g) {9}
        child {node (h) {8}
          child {node (i) {7}}
          child[missing]}
        child {node (j) {11}
          child {node (k) {10}}
          child[missing]}};

    \foreach \x/\num in {a/0,b/1,c/2,d/3,e/4,f/5,g/6,h/7,i/8,j/9,k/10}
      \node at (\x) [circle, fill=red!30, scale=0.5, xshift=-7mm] {\num};

    \foreach \x/\num in {d/0,c/1,b/2,f/3,e/4,a/5,i/6,h/7,g/8,k/9,j/10}
      \node at (\x) [circle, fill=green!30, scale=0.5, yshift=7mm] {\num};

    \foreach \x/\num in {d/0,c/1,f/2,e/3,b/4,i/5,h/6,k/7,j/8,g/9,a/10}
      \node at (\x) [circle, fill=blue!30, scale=0.5, xshift=7mm] {\num};

  \end{tikzpicture} 
  \caption{Traversal sorts: preorder (red), inorder (green), postorder (blue).}
  \label{fig:bst_traversal}
\end{figure}
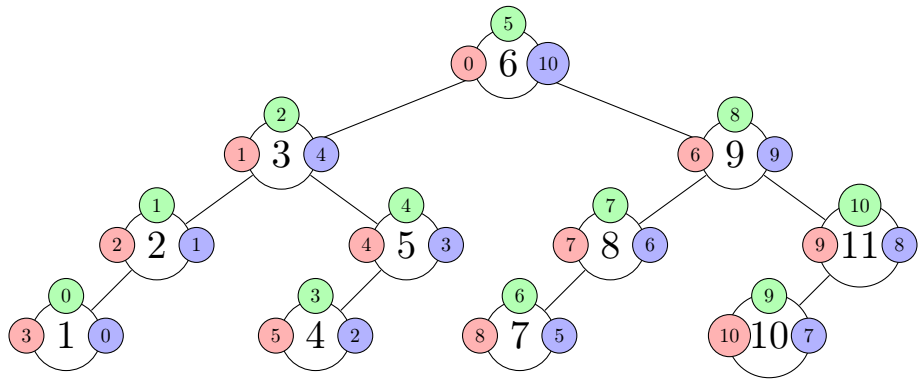

\begin{table}[H]
\centering
\caption{Chunks Pre/Post Traversal Sorts on two resources  }
\label{tab:bst_traversal}
\begin{tabular}{@{}ccc@{}}
\toprule
Order  & Operation 1                           & Operation 2 \\ \midrule

\multicolumn{1}{r}{\textbf{ T1}}  & \multicolumn{1}{c}{\cellcolor{blue!7}Traversal Order Sort} & \multicolumn{1}{c}{\cellcolor{yellow!7}Chunk Ks by Resource Count} \\
\cmidrule(lr){2-2} \cmidrule(lr){3-3}
In    &  1, 2, 3, 4, 5, 6, 7, 8, 9, 10, 11   & [1, 2, 3, 4, 5, 6] [7, 8, 9, 10, 11]   \\
\cellcolor{red!5}Pre   & \cellcolor{red!5}6, 3, 2, 1, 5, 4, 9, 8, 7, 11, 10    & [6, 3, 2, 1, 5, 4] [9, 8, 7, 11, 10]   \\
Post  & 1, 2, 4, 5, 3, 7, 8, 10, 11, 9, 6    & [1, 2, 4, 5, 3, 7] [8, 10, 11, 9, 6]   \\ 
\cmidrule(lr){2-2} \cmidrule(lr){3-3}

\multicolumn{1}{r}{\textbf{ T2}}  & \multicolumn{1}{c}{\cellcolor{blue!7}Traversal Order Sort} & \multicolumn{1}{c}{\cellcolor{green!7}Chunk Ks by Alg. \ref{alg:chunk_k}} \\
\cmidrule(lr){2-2} \cmidrule(lr){3-3}
In    & 1, 2, 3, 4, 5, 6, 7, 8, 9, 10, 11   & [1, 3, 5, 7, 9, 11] [2, 4, 6, 8, 10]   \\
\cellcolor{red!5}Pre   &\cellcolor{red!5} 6, 3, 2, 1, 5, 4, 9, 8, 7, 11, 10     & [3, 1, 5, 9, 7, 11]  [6, 2, 4, 8, 10]   \\
Post  & 1, 2, 4, 5, 3, 7, 8, 10, 11, 9, 6    & [1, 5, 3, 7, 11, 9]  [2, 4, 8, 10, 6]   \\ 
\midrule  
\multicolumn{1}{r}{\textbf{T3}} & \multicolumn{1}{c}{\cellcolor{yellow!7} Chunk Ks by Resource Count} & \multicolumn{1}{c}{\cellcolor{blue!7}Traversal  Order Sort} \\
\cmidrule(lr){2-2} \cmidrule(lr){3-3}
In    & [1, 2, 3, 4, 5, 6] [7, 8, 9, 10, 11]   & [1, 2, 3, 4, 5, 6] [7, 8, 9, 10, 11]   \\
\cellcolor{red!5}Pre   &  [1, 2, 3, 4, 5, 6] [7, 8, 9, 10, 11]     & \cellcolor{red!5}[4, 2, 1, 3, 6, 5]  [9, 8, 7, 11, 10]   \\
Post  &  [1, 2, 3, 4, 5, 6] [7, 8, 9, 10, 11]     & [1, 3, 2, 5, 6, 4] [7, 8, 10, 11, 9] \\
\cmidrule(lr){2-2} \cmidrule(lr){3-3}

\multicolumn{1}{r}{\textbf{T4}} & \multicolumn{1}{c}{\cellcolor{green!7} Chunk Ks by Alg. \ref{alg:chunk_k}} & \multicolumn{1}{c}{\cellcolor{blue!7}Traversal  Order Sort} \\
\cmidrule(lr){2-2} \cmidrule(lr){3-3}
In    & [1, 3, 5, 7, 9, 11] [2, 4, 6, 8, 10]   & [1, 3, 5, 7, 9, 11] [2, 4, 6, 8, 10]   \\
\cellcolor{red!5}Pre   &  [1, 3, 5, 7, 9, 11] [2, 4, 6, 8, 10]    & \cellcolor{red!5}[7, 3, 1, 5, 11, 9] [6, 4, 2, 10, 8]   \\
Post  &  [1, 3, 5, 7, 9, 11] [2, 4, 6, 8, 10]    & [1, 5, 3, 9, 11, 7] [2, 4, 9, 10, 6]   \\ \bottomrule
\end{tabular}
\end{table}

\begin{algorithm}[H]
\caption{Chunk $k$ values by Skip Mod Resource Count}\label{alg:chunk_k}
\begin{algorithmic}[1]
\Require \( \mathcal{K} \) (list of $k$), $num\_resources$ (resource count to split \( k \))
\Ensure List of chunks where each chunk contains integers assigned to a resource
\Function{ChunkKs}{$K, num\_resources$}
    \State $K\_chunks \gets []$
    \For{$i = 0$ \textbf{to} $num\_resources - 1$}
        \State $K\_chunks[i] \gets$ empty list
    \EndFor
    \For{$k = 0$ \textbf{to} $Ks - 1$}
        \State $resource\_id \gets k \mod num\_resources$
        \State Append $k$ to $K\_chunks[resource\_id]$
    \EndFor
    \State \textbf{return} $K\_chunks$
\EndFunction
\end{algorithmic}
\end{algorithm}

\subsection{ Binary Bleed Multi-threaded, Multi-rank}
A parallel implementation of Binary Bleed can be obtained by extending the serial implementation depicted in Algorithm \ref{alg:binary_search_optimal_k} with the following changes: Make $k_{min}$, $k_{max}$ and the list of visited $k$ $rank_{seen}$ global using a distributed cache such as reddis, and $i_{left}$ and $i_{right}$ are now computed by each MPI rank or thread as $i_{left}=k_{min}+idx \times (k_{max}-k_{min})/size$ and $i_{rigth}=min(i_{left}+size, k_{max})$ where $idx$ is the thread index or MPI rank, and $size$ is the total number of threads or MPI ranks. The resulting parallel implementation will work well, but will not be effective when the list of $k$ is sparse.

A more robust algorithm can be obtained by replacing recursions in Algorithm \ref{alg:binary_search_optimal_k} by a $k$-sort, illustrated in Figure \ref{fig:bst_traversal}, where $k$ values are sorted using in-order, pre-order, or post-order binary tree traversal, indicated by the color on three sides of each rank's $k$ value.

Further, \( \mathcal{K} \) must be chunked into available resources using Algorithm \ref{alg:chunk_k}.  By appropriately sorting and chunking the $k$ values, the Binary Bleed algorithm can efficiently operate in a multi-threaded and multi-rank environment, enhancing its scalability and performance.

 Logistics of Chunking \( \mathcal{K} \) and Traversal Sort : To illustrate the optimal function sequence, assume $k=[1-10]$, with two operating resources. The four splits of the data are:

\begin{enumerate}
  \item Sort, then data chunk by resources (Table \ref{tab:bst_traversal}'s \textbf{T1})
  \item Sort, then data chunk by Algorithm \ref{alg:chunk_k} (  Table \ref{tab:bst_traversal}'s \textbf{T2})
  \item Data chunk by resources, then sort (Table \ref{tab:bst_traversal}'s \textbf{T3})
  \item Data chunk by Algorithm \ref{alg:chunk_k}, then sort (Table \ref{tab:bst_traversal}'s \textbf{T4})
\end{enumerate}

Table  \ref{tab:bst_traversal} distributes $k$ on two resources, showing that in-order traversal monotonically increases, leading to inadequate ordering for early termination of $k$.
Truncation removes unvisited smaller values, which is impossible for in-order traversal. Furthermore, \textbf{T1} demonstrates the insufficient distribution of \( k \) values across resources. For example, the second resource may find an optimal result in truncation for all \( k \) in the first resource, leading to one resource being idle while the other operates.
The remaining splits present \textbf{T3} as the least optimal, as the values are similarly partitioned to \textbf{T1}, resulting in higher optimal resources. \textbf{T2} and \textbf{T4} are distinguished by order, so operation logistics are a lower priority than the use of Algorithm \ref{alg:chunk_k} as a load-balanced partition. Table \ref{tab:bst_traversal} shows that traversal selections are ambiguous. However, Section \ref{sec:C_results} indicates pre-order traversals visit fewer \( k \) values. Algorithm \ref{alg:threading} coordinates with Algorithm \ref{alg:chunk_k} to distribute and process $k$ values across multiple resources efficiently.

\begin{breakablealgorithm}
\caption{Multiple Ranks and Threads Binary Bleed}
\label{alg:threading}
\begin{algorithmic}[1]
\Require $Ks$ (sorted $k$ values), $N$ (ranks in network), $t$ (threads per rank)
\State Initialize all ranks with unique identifiers and communication between ranks

\Procedure{InitializeRankKs}{$Ks, N, t, data$}

    \State $K_{\text{chunks}} \gets$ \Call{ChunkKs}{$Ks, N$}
 
    \For{each rank $n$ in the network}
        \State Initialize $Ks_{\text{bst}} \gets []$ 
        
        \State \Call{TraversalSort}{$K_{\text{chunks}}[n], length(Ks), \&Ks_{\text{bst}}$}
        \State \Call{StartThreads}{$data, Ks_{\text{bst}}, t$}  
     
    \EndFor
\EndProcedure
\Procedure{StartThreads}{ $data, Ks, num_{\text{threads}} $}  
    \State $ Ks\_size \gets |Ks|$, $mutex \gets \text{CreateMutex()}$ 
    
    \For{$i \gets 0 \text{ to } Ks\_size$ } 
        \State \textbf{start thread} \Call{BinaryBleedMulti}{ 
        
        $Ks_{\text{bst}}[i \% num_{\text{threads}}],  data, \&k_{\text{optimal}}, mutex$
        }
    \EndFor
    \State \textbf{wait for all threads to finish, then} \textbf{process} $k_{\text{optimal}}$
\EndProcedure
\Procedure{BroadcastK}{$k_{\text{optimal}}, sender$}
    \For{each rank $n$ in the network}
        \If{$n \neq sender$}  Send $k_{\text{optimal}}$ to rank $n$
        \EndIf
    \EndFor
\EndProcedure
\Procedure{ReceiveKCheck}{ $k_{\text{receiver}}, receiver$}
    \If{message from network}  Receive $k_{\text{sender}}, sender$
    \Else \textbf{ return} $ k_{\text{receiver}}$
    \EndIf
    
    \If{$ k_{\text{receiver}} < k_{\text{sender}} $}
       $ k_{\text{receiver}} \gets k_{\text{sender}}$
    \Else \textbf{ Send} $k_{\text{receiver}}$ to rank $sender$
    \EndIf
\EndProcedure
\end{algorithmic}
\end{breakablealgorithm}

First, in Algorithm \ref{alg:threading}, the  $k$ list, number of ranks, and threads per rank are needed. Ranks are initialized with IDs and communication setup (lines 1-2). Algorithm  \ref{alg:chunk_k} chunks the  $k$ (line 3). Each rank initializes a binary search tree (BST) (lines 4-5) of sorted  $k$ and starts threads on the data (lines 7-9). The thread initializes variables for $|\mathcal{K}|$ and a mutex (lines 10-11). It iterates over the threads to start Binary Bleed on each $k$ value, waits for completion, and processes the optimal value (lines 12-16). The broadcast function iterates the ranks, sending the optimal  $k$ to others (lines 17-22). The function for receiving communication is defined (line 23). On receiving a message, the rank checks if the received  $k$ exceeds the current optimal, updating or returning the higher optimal as needed (lines 24-30). The second change needed for parallel Binary Bleed is the communication of pruned $k$ values to other resources. Algorithm \ref{alg:threading}, in conjunction with Algorithm \ref{alg:binary_search_multi}, contains the set of calls to complete the communication of pruning in parallel and distributed Binary Bleed.

\begin{breakablealgorithm}
\caption{Binary Bleed $k$ Search, Multi-rank \& Thread }\label{alg:binary_search_multi}
\label{alg:threading2}
\begin{algorithmic}[1]
\Require $k$ (value to compute), $data$ (dataset ), $model$ (calculator), $scorer$ (quantifies output), $T_{\text{select\_k}}$ (minimum $k$ score), $rank\_id$ (id of rank)
\Ensure Optimal $k$ value and its prediction score
\State $k_{\text{optimal}} \gets  null$  

\Function{BinaryBleedMulti}{
$k, k_{\text{optimal}}, data, mutex, T_{\text{select\_k}}$}
    \State Initialize $model, scorer$, $report \gets 0$  
    \State $\text{Lock}(mutex)$ \textbf{and}  $k_{\text{received}} \gets k_{\text{optimal}}$  \textbf{then}   $\text{Unlock}(mutex)$

    \State $k_{\text{received}} \gets$ \Call{ReceiveKCheck}{    
    $k_{\text{receiver}}, rank\_id$ }
    
    \State $\text{Lock}(mutex)$
        \If{$k_{\text{received}} \neq k_{\text{optimal}}$}
            \If{$\neg k_{\text{optimal}}$ \textbf{or} $k_{\text{received}} > k_{\text{optimal}}$} $k_{\text{optimal}} \gets k_{\text{received}}$
            \Else  \textbf{ } $report \gets 1$
            \EndIf
        \EndIf
    \State $\text{Unlock}(mutex)$

    \If{$report$} \Call{BroadcastK}{$k_{\text{optimal}}, rank\_id$} \textbf{and} $report \gets 0$ 
    \Else
        \If{$k_{\text{optimal}} > $k$ $ } \textbf{return} \EndIf
    \EndIf
    \State $score \gets scorer(model(data, k))$
    \If{$score \geq k_{\text{select\_threshold}}$}  $\text{Lock}(mutex)$
         \If{$\neg k_{\text{optimal}}$ \textbf{or} $k > k_{\text{optimal}}$} $k_{\text{optimal}} \gets k$ \textbf{and}  $report \gets 1$
        \EndIf
        \State $\text{Unlock}(mutex)$
        \If{$report$} \Call{BroadcastK}{$k_{\text{optimal}}, rank\_id$}
        \EndIf
    \EndIf
\EndFunction
\end{algorithmic}
\end{breakablealgorithm}

Algorithm \ref{alg:binary_search_multi} requires the \( k \) range, data, model, scorer, selection threshold, and rank ID. Lines (1-3) initialize the optimal \( k \), define the Binary Bleed multi-function, and initialize the model, scorer, and report flag. Line (4) locks the mutex, sets the received \( k \) as optimal, and unlocks the mutex. Line (5) checks network messages and sets the received \( k \). Lines (6-12) lock the mutex, compare optimal and received \( k \), update the optimal if different, and possibly set the report flag before unlocking the mutex. Lines (13-17) check if a report is needed for other resources and reset the report flag if necessary. Line (18) operates the model, data, and scorer on \( k \). Lines (19-26) lock the mutex if the score exceeds the threshold, update the optimal \( k \) if larger, and broadcast if the optimal was updated. In multi-threading, the rank propagates the optimal \( k \) from the finding thread to others, pruning smaller \( k \) values. In multi-rank, any rank propagates the optimal \( k \) to other ranks. In both cases, threads report to ranks, which then report to other ranks and threads. In an HPC system, resources are processes or compute ranks. In multi-threading, the optimal  \( k \) a thread finds is propagated by the controlling rank to all threads, which prune smaller  \( k \) values. In multi-rank, any rank finding the optimal  \( k \) shares it with all ranks. Threads report to ranks, then communicate the optimal  \( k \) to other ranks and threads.

\begin{figure}[H]
\centering
\begin{tikzpicture}[
    scale=2.5,
    transform shape,
    level distance=3.5mm,
    level 1/.style={sibling distance=20mm},
    level 2/.style={sibling distance=10mm},
    level 3/.style={sibling distance=7mm},
    every node/.style={circle, draw, scale=0.6}
]
\node (a) {6}
  child {node (b) {3}
    child {node (c) {2}
      child {node (d) {1}}
      child[missing]}
    child {node (e) {5}
      child {node (f) {4}}
      child[missing]}}
  child {node (g) {9}
    child {node (h) {8}
      child {node (i) {7}}
      child[missing]}
    child {node (j) {11}
      child {node (k) {10}}
      child[missing]}};

\foreach \x/\num in {i/0, f/1, d/2, k/3}
  \node at (\x) [circle, fill=red!30, scale=0.6, yshift=-6mm] {\num};
\foreach \x/\num in {h/0, e/1, c/2, j/3}
  \node at (\x) [circle, fill=green!30, scale=0.6, yshift=-6mm] {\num};
\foreach \x/\num in {a/0, b/1, g/2}
  \node at (\x) [circle, fill=blue!30, scale=0.6, yshift=-6mm] {\num};
\end{tikzpicture}
\caption{Vanilla: Part 1}
\label{fig:vanilla_part1}
\end{figure}
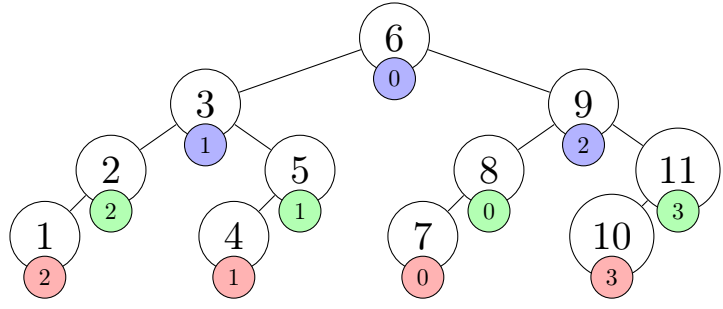

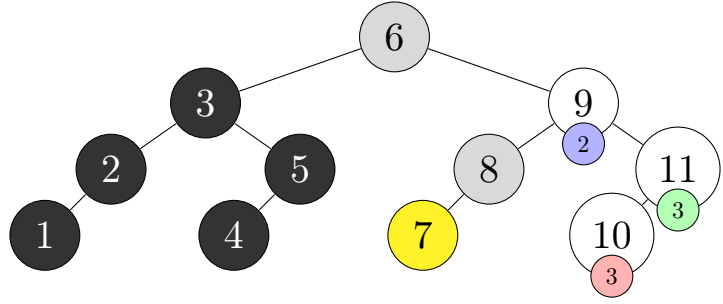
\begin{figure}[H]
\centering
\begin{tikzpicture}[
    scale=2.5,
    transform shape,
    level distance=3.5mm,
    level 1/.style={sibling distance=20mm},
    level 2/.style={sibling distance=10mm},
    level 3/.style={sibling distance=7mm},
    every node/.style={circle, draw, scale=0.6}
]
\node [fill=gray!30] (a) {6}
  child {node[fill=black!80, text=white] (b) {3}
    child {node[fill=black!80, text=white] (c) {2}
      child {node[fill=black!80, text=white] (d) {1}}
      child[missing]}
    child {node[fill=black!80, text=white] (e) {5}
      child {node[fill=black!80, text=white] (f) {4}}
      child[missing]}}
  child {node (g) {9}
    child {node[fill=gray!30] (h) {8}
      child {node[fill=yellow!90] (i) {7}}
      child[missing]}
    child {node (j) {11}
      child {node (k) {10}}
      child[missing]}};

\foreach \x/\num in {k/3}
  \node at (\x) [circle, fill=red!30, scale=0.6, yshift=-6mm] {\num};
\foreach \x/\num in {j/3}
  \node at (\x) [circle, fill=green!30, scale=0.6, yshift=-6mm] {\num};
\foreach \x/\num in {g/2}
  \node at (\x) [circle, fill=blue!30, scale=0.6, yshift=-6mm] {\num};
\end{tikzpicture}
\caption{Vanilla: Part 2}
\label{fig:vanilla_part2}
\end{figure}

\begin{figure}[htb]
    \centering
    \includegraphics[width=0.8\columnwidth]{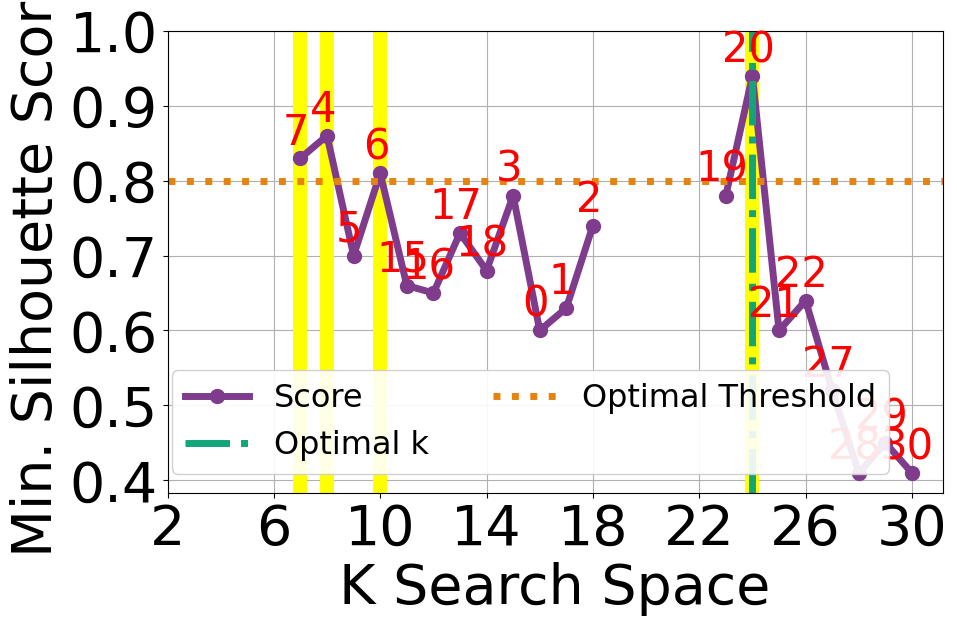}
    \caption{Binary Bleed Vanilla.}
    \label{fig:multi}
\end{figure}

Operation Dynamics: Figure \ref{fig:vanilla_part1} Pre-order sorts \( k =  \{1,2,\cdots,11\}\) on three resources after Algorithm \ref{alg:chunk_k} (\textbf{T4} \ref{tab:bst_traversal}). $\forall k$ are the possible optimal.  Resource allocation (r,g,b) is by visit order.  In part 2, Figure \ref{fig:vanilla_part2}, the score of $k=7$ is greater than the threshold, so it is optimal,  and the scores of $k=\{6,8\}$ are less than the threshold.  $k=\{1,2,\cdots,5\}$ are pruned for being less than the optimal. The upper $k$ range, $k=\{9,10,11\}$ continues. Figure \ref{fig:multi} schedules \( k \) four resources chunk values, then pre-order sorted. 
The graph shows the optimal selection threshold as a dashed gold line.  Yellow vertical bars indicate truncation is reported in other resources, where a \( k \) score passed the threshold. The gold threshold is crossed at four values: \( k = \{7, 8, 10, 24\} \). When the first \( k = 8 \) passes the threshold, all \( k < 8 \) are pruned. However, the implementation shown does not prune \( k \) values after the model begins execution; therefore, \( k = 7 \) is complete. The \( k \) values [7, 10] have no remaining \( k \) values to prune. Due to pre-order sorting, \( K = 24 \) is run before \( k =  \{18,19,\cdots,22\}\). Since \( k = 24 \) is above the selection threshold, those lower priority \( k \) values are pruned. The upper \( k \) range continues without crossing the selection threshold, so \( K = 24 \) remains optimal.

\subsection{ Binary Bleed Early Stop}
\label{Early_stopping}
Specific domains allow for an additional heuristic. Rather than pruning the lower \( k \) values, setting a bound on the other extreme of the data will allow for truncation of the upper values. This heuristic is based on domain knowledge, where if any scores cross the stop threshold, they will never rise to cross the selection threshold, and subsequent \( k \) values can be ignored. Mathematically, it can be represented as:
\[
k_{\text{optimal}} = \max \left\{ k \in \{1, 2, \ldots, K\} : \forall i \leq k, S(k_i) > U \right\}
\]
where \( S(k_i) \) is the score of the \( i^{th} \) \( k \), and \( U \) is the stop bound.

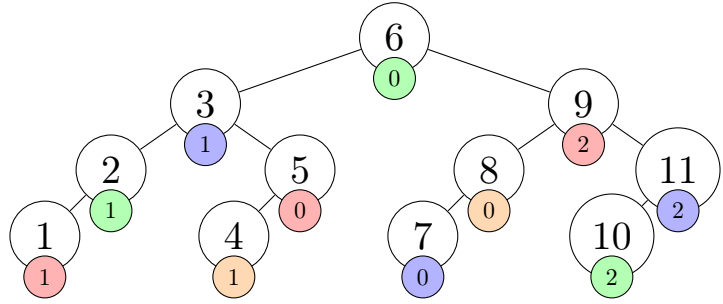
\begin{figure}[H]
\centering
\begin{tikzpicture}[
    scale=2.5,
    transform shape,
    level distance=3.5mm, 
    level 1/.style={sibling distance=20mm},
    level 2/.style={sibling distance=10mm}, 
    level 3/.style={sibling distance=7mm}, 
    every node/.style={circle, draw, scale=0.6}
]
\node (a) {6}
  child {node (b) {3}
    child {node (c) {2}
      child {node (d) {1}}
      child[missing]}
    child {node (e) {5}
      child {node (f) {4}}
      child[missing]}}
  child {node (g) {9}
    child {node (h) {8}
      child {node (i) {7}}
      child[missing]}
    child {node (j) {11}
      child {node (k) {10}}
      child[missing]}};

\foreach \x/\num in {e/0, d/1, g/2}
  \node at (\x) [circle, fill=red!30, scale=0.6, yshift=-6mm] {\num};
\foreach \x/\num in {a/0, c/1, k/2}
  \node at (\x) [circle, fill=green!30, scale=0.6, yshift=-6mm] {\num};
\foreach \x/\num in {i/0, b/1, j/2}
  \node at (\x) [circle, fill=blue!30, scale=0.6, yshift=-6mm] {\num};
\foreach \x/\num in {h/0, f/1}
  \node at (\x) [circle, fill=orange!30, scale=0.6, yshift=-6mm] {\num};
\end{tikzpicture}
\caption{Early Stop: Part 1}
\label{fig:earlystop_part1}
\end{figure}

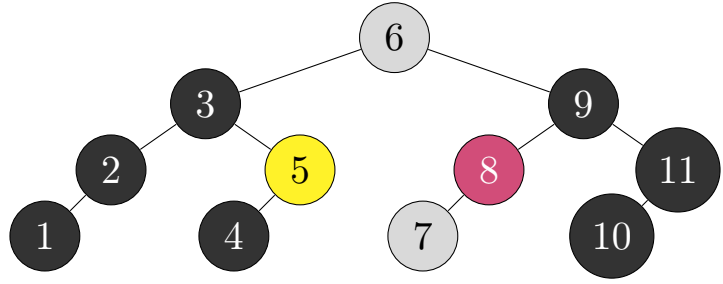
\begin{figure}[H]
\centering
\begin{tikzpicture}[
    scale=2.5,
    transform shape,
    level distance=3.5mm, 
    level 1/.style={sibling distance=20mm},
    level 2/.style={sibling distance=10mm}, 
    level 3/.style={sibling distance=7mm}, 
    every node/.style={circle, draw, scale=0.6}
]
\node [fill=gray!30] (a) {6}
  child {node[fill=black!80, text=white] (b) {3}
    child {node[fill=black!80, text=white] (c) {2}
      child {node[fill=black!80, text=white] (d) {1}}
      child[missing]}
    child {node[fill=yellow!90] (e) {5}
      child {node[fill=black!80, text=white] (f) {4}}
      child[missing]}}
  child {node[fill=black!80, text=white] (g) {9}
    child {node[fill=purple!70, text=white] (h) {8}
      child {node[fill=gray!30] (i) {7}}
      child[missing]}
    child {node[fill=black!80, text=white] (j) {11}
      child {node[fill=black!80, text=white] (k) {10}}
      child[missing]}};
\end{tikzpicture}
\caption{Early Stop: Part 2}
\label{fig:earlystop_part2}
\end{figure}

 Figure \ref{fig:earlystop_part1} Pre-order sorts $k=[1-11]$ after Algorithm \ref{alg:chunk_k} (\textbf{T4} \ref{tab:bst_traversal}), on four resources. In part 2, Figure \ref{fig:earlystop_part2}, $k=5$ exceeds the selection threshold, so the optimal is set and $k=[1-4]$ (dark nodes) are pruned.  $k=8$ crosses the stop threshold, so $k=[9-11]$ are pruned. Therefore, the optimal remains 5.

\subsection{ Additional Considerations }
In multi-resource computing, an optimal $k$ may be larger than an executing \( k \). For long computations, checks can be pushed into the model to terminate such $k$ early.
Consider the scoring distribution, like silhouettes, which can speed up the process. True optimal values are found faster when scores above the threshold resemble a square wave. Mathematically:

\[
S(k_i) = \frac{\text{sgn}(k_0 - k_i) + 1}{2}
\]

where \( S(k_i) \) is the evaluation score at the \( i^{th} \) index of \( k \) values, \( k_0 \) is the optimal \( k \), and \(\text{sgn}\) is the signum function, which evaluates to \( +1 \) for \( k_i < k_0 \) and \( -1 \) for \( k_i \geq k_0 \).
In the worst case, a Laplacian score distribution will peak at the optimal \( k \) while other \( k \) scores will be below the score selection threshold. Lower \( k \) values may be pruned if the peak is visited before lower values. Otherwise, all values will be visited in the order of the sort. Despite the score distribution, Binary Bleed will not visit more \( k \) values than a linear search.

%% file: Part2/Latent_Feature_Extraction/HNMFk.tex
\subsection{HNMF{\small k} implementation details}
\label{subsec:methods-hnmfk}
Building on the theoretical foundation in Section~\ref{subsection:hnmfk}, we deploy HNMF{\small k} as the workhorse of the topic–concept decomposition layer.  The input is the TF-IDF-weighted document–term matrix $\mat{X}\!\in\!\mathbb{R}^{n\times m}_{+}$ whose vocabulary has been pre‐filtered for corpus‐specific stop-words and rare terms (occurrence $<\!5$).  An adaptive candidate range $K^{(0)}=\{2,3,\dots,32\}$ is selected via the rank–heuristic $K^{(0)}=\!\{2,\dots,\lceil\sqrt{\min(n,m)}\rceil\}$ and refined on‐the‐fly by the \textsc{NextRange} policy in Algorithm~\ref{alg:hnmfk}.  Each NMF{\small k} fit uses \emph{$ \beta $‐divergence} ($\beta=1$) minimisation with $200$ multiplicative‐update iterations, followed by $10$ HALS2 refinement sweeps to accelerate convergence \cite{cichocki2009nmf}.  The rank‐selection criterion is the minimum description length (MDL) of the residual spectrum, evaluated over a held‐out validation fold to avoid optimistic bias \cite{squires2019mdl}.  All factors are re‐normalised to unit $\ell_1$ length before the maximum–loading assignment to stabilise clustering against scale drift.

Distributed execution: HNMF{\small k} is embedded in an \texttt{mpi4py}–based breadth‐first scheduler that maps each active node of the hierarchy to a distinct Slurm task.  The checkpointed factor pairs $(\widetilde{\mat{W}},\widetilde{\mat{H}})$ and index sets $\mathcal{I}_{c}$, so that failed tasks can be restarted without re‐factoring ancestors.  For corpora with $>10^{6}$ documents, we slice $\mat{X}$ row-wise into vocabulary shards of at most $10^{5}$ terms, factorize each shard independently, and merge the shard-specific hierarchies through HEAL matching \cite{bhattarai2024heal}.

Post-processing and export:
Every non-leaf node is enriched with (i) the top‐$20$ loading terms in $\widetilde{\mat{W}}_{c}$, (ii) a centroid embedding obtained by averaging sentence-BERT vectors of member documents, and (iii) a list of citation identifiers propagated from the raw corpus.  These artefacts seed the graph‐reasoning layer (Section \ref{subsec:graph-reasoning}) and the dense vector store.  The complete hierarchy is serialised in a compact JSON–LD dialect (6–8× smaller than plain CSV) and pushed to the \texttt{RabbitMQ} message bus, enabling asynchronous downstream consumers without blocking the factorisation frontier.

%% file: Part2/Latent_Feature_Extraction/artic_fox.tex
\section{Automatic Topic Labelling with Embedding–Aligned LLM Prompts}
\label{subsec:topic_labelling}

Hierarchical factorisation yields a set of coherent but unnamed clusters.  To replace ad-hoc manual naming, we introduce a two–stage label–selection pipeline that unites distributional semantics with  LLM generation.  The key idea is to sample candidate phrases from an LLM conditioned on the cluster’s most salient terms and then rank those phrases by vector proximity to the cluster centroid in the same embedding space used for downstream retrieval.

\paragraph{Notation.}
Let $\mathcal{D}=\{d_1,\dots,d_N\}$ be a corpus segmented into $K$ latent topics by HNMF$k$, and let $f:\mathcal{D}\!\cup\!\mathcal{L}\to\mathbb{R}^{d}$ denote a sentence-level encoder (SciNCL or SPECTER) that maps both document texts and candidate label phrases $\ell\in\mathcal{L}$ into the same $d$-dimensional space.

\subsubsection*{Step~1: Cluster Centroid Estimation}
For each topic $c$, compute the centroid \[   \boldsymbol\mu_c \;=\;   \frac{1}{|\mathcal{D}_c|}\sum_{d\in\mathcal{D}_c} f(d), \] or alternatively a medoid when \texttt{center\_metric} is set to \texttt{"closest"}. The index of the document closest to $\boldsymbol\mu_c$ is cached for explanatory displays.

\subsubsection*{Step 2: Candidate Generation via LLM Prompting}
Given the $m$ highest-probability tokens $\langle w_1,\dots,w_m\rangle$ from the factor-loading column $\boldsymbol W_{\cdot,c}$, we build a few-shot prompt \[ \text{\small\texttt{Create a concise label for the following words (order matters):}} \;   w_1,\dots,w_m. \]

Optional criteria: minimum/maximum word count, must-contain or must-not-contain lists, and auxiliary metadata (e.g., \ venue, year span) may be embedded in the prompt. The framework supports both local Ollama models (\texttt{llama3.2:3b-instruct-fp16}) and remote OpenAI models; here we use the former for reproducibility. We draw $n_\text{trial}=10$ independent samples at temperature $\tau=0.4$ and discard any sample that violates the criteria (Algorithm 1).

\begin{algorithm}[H]
\caption{Sampling valid candidate labels}\label{alg:label_sampling}
\begin{algorithmic}[1]
\Function{GenerateValidLabels}{$\textit{LLM},\;p,\;\kappa,\;n$}
  \State $\mathcal L\gets\varnothing$
  \While{$|\mathcal L|<n$}
      \State $\ell\gets \textit{LLM.invoke}(p)$
      \If{$\textsc{Validate}(\ell,\kappa)$} $\mathcal L\gets\mathcal L\cup\{\ell\}$ \EndIf
  \EndWhile
  \State\Return $\mathcal L$
\EndFunction
\end{algorithmic}
\end{algorithm}

\subsubsection*{Step 3: Embedding-Consistent Selection}
Each candidate $\ell$ is embedded by $f$ on CPU or GPU, depending on \texttt{embedds\_use\_gpu}.  The cosine (or user-supplied) distance to the centroid is computed, and the \emph{closest} label is selected: \[   \hat\ell_c = \arg\min_{\ell\in\mathcal L_c}       \; \mathrm{dist}\bigl(f(\ell),\boldsymbol\mu_c\bigr). \]

This embedding-aligned LLM labelling scheme closes the loop between distributional topic discovery and human-readable semantics, providing interpretable, high-fidelity labels without manual curation.

%% file: Part2/KG_VS_Constructions/2.tex
The previous chapter identified the latent structure of the corpus; this chapter builds on that foundation by constructing a knowledge representation that supports querying, versioning, and inference. It begins by defining a typed, schema-aware knowledge graph in which edges reflect the topic structures produced by the earlier decomposition methods. After examining the limitations of language model–generated graphs, including issues such as hallucination, semantic drift, and schema violations, the focus shifts to a matrix and tensor factorization approach that grounds all nodes and relations in observed co-occurrence statistics. In parallel, a topic-partitioned vector store is constructed using aligned embeddings that preserve the same hierarchical semantics. Together, these two representations form a consistent foundation for downstream retrieval and reasoning tasks.

\section{Unified Data-Infrastructure for Retrieval-Augmented Generation}
\label{sec:methods}
A scalable retrieval-augmented generation (RAG) system emerges from coupled layers that may be re-used, with minimal parameter changes, across domains as diverse as biomedical literature, legal corpora, or cyber-security reports.

\begin{enumerate}[label=\textbf{\arabic*.},leftmargin=1.6em]
  \item \textbf{Hierarchical topic discovery.}  HNMF$k$ extracts a multi-level latent taxonomy whose outputs seed the subsequent layers.
  \item \textbf{Knowledge-graph (KG) construction.}  A labelled–property graph unifies entities, provenance, and ontological types.
  \item \textbf{Vector-store (VS) construction.}  Topic-sharded indices provide sub-second dense retrieval.
  \item \textbf{Hybrid retrieval orchestration.}  An agent loop routes queries between symbolic and sub-symbolic stores.
\end{enumerate}

\subsection{Schema Design for a Versioned Knowledge Graph}
\label{sec:schemas}

The KG is formalised as
\[
\mathcal G=(\mathcal V,\mathcal E,\mathcal L,\mathcal P),
\]
where $\mathcal V$ is the node set, $\mathcal E\subseteq\mathcal
V\times\mathcal V$ the directed edge set, $\mathcal L$ the labelling
function, and $\mathcal P$ a key–value map (e.g.\ \texttt{doi},
\texttt{version}, \texttt{confidence}).  Table~\ref{tab:kg_schema}
summarises the principal node labels and edge types.

\begin{table}[ht]
  \centering
  \small
  \renewcommand{\arraystretch}{1.05}
  \begin{tabular}{@{}ll@{}}
    \toprule
    \multicolumn{2}{@{}l}{\textbf{Node labels}} \\ \midrule
    \texttt{Document}    & \{doi,title,year,version\} \\
    \texttt{Author}      & \{author\_id,name\}  \\
    \texttt{Affiliation} & \{aff\_id,name,country\} \\
    \texttt{Venue}       & \{venue\_id,name\} \\
    \texttt{Topic}       & \{topic\_id,level,coherence\} \\
    \texttt{Keyword}     & \{term\} \\ \midrule
    \multicolumn{2}{@{}l}{\textbf{Edge types}} \\ \midrule
    \texttt{WRITTEN\_BY}      & Document $\to$ Author \\
    \texttt{AFFILIATED\_WITH} & Author   $\to$ Affiliation \\
    \texttt{PUBLISHED\_IN}    & Document $\to$ Venue \\
    \texttt{CITES}            & Document $\to$ Document \\
    \texttt{MENTIONS}         & Document $\to$ Keyword \\
    \texttt{IN\_TOPIC\,[w]}   & Document $\to$ Topic (weight $w$) \\
    \bottomrule
  \end{tabular}
  \caption{Core KG schema with provenance and monotone
           \texttt{version} properties.}
  \label{tab:kg_schema}
\end{table}

\paragraph{Entity extraction \& resolution.}
Named-entity recognition (spaCy) yields raw mentions $E=\operatorname{NER}(T)$ from the corpus $T=\{t_1,\dots,t_n\}$.  Ambiguity is resolved by a hybrid distance \[ \operatorname{dist}(e_i,e_j)=\alpha\, operatorname{SimHash} +(1-\alpha)\,\operatorname{Lev},\qquad 0\le\alpha\le1, \] merging pairs below a threshold $\tau$.  Canonicalised entities are then batch-inserted via \texttt{UNWIND}, each edge recording a \emph{provenance–confidence} pair $(v,c)$ to guarantee auditability.

\subsection{Data Sources}
\label{sec:data_sources}

Our pipeline ingests heterogeneous artefacts:

\begin{itemize}[nosep,leftmargin=1.25em]
  \item \textbf{Structured metadata} from SCOPUS and Semantic Scholar APIs, including citations and affiliation hierarchies.
  \item \textbf{Born-digital full texts} (PDF or XML) harvested through publisher APIs compliant with Crossref’s \textit{TDM} licences.
  \item \textbf{Grey literature} (e.g.\ policy briefs) fetched via focused web-crawls and subsequently filtered by a domain-specific classifier.
\end{itemize}

All sources are normalised into a canonical \texttt{Document} record whose fields feed both KG nodes and VS payloads.

\subsection{Full-Text Acquisition and Segmentation}
\label{sec:full_texts}

Full texts are parsed with GROBID; boiler-plate sections (references, appendices) are removed by a rule-based filter.  Each document’s token sequence $\langle w_1,\dots,w_L\rangle$ is split into windows $S_i$ of $B=384$ tokens with an overlap of $s=48$ tokens, \[ S_i=\langle w_{iB+1},\dots,w_{(i+1)B}\rangle,\quad  S_{i+1}\leftarrow S_{i+1}\cup\text{tail}(S_i). \] The pair $(doc\_id,par\_id)$ serves as a globally unique key shared by the KG and VS, thereby enabling $O(1)$ cross-store dereferencing.

\subsection{Special Graph Constructions}
\label{sec:special_graphs}
Beyond the base KG, two derived graphs support downstream tasks:
\begin{description}[style=unboxed,leftmargin=1.25em]
  \item[Co-authorship graph.]  An undirected, weighted projection of the \texttt{WRITTEN\_BY} relation; edge weights are TF-normalised collaboration counts.
  \item[Citation-topic graph.]  A bipartite graph linking \texttt{Topic} nodes to cited \texttt{Document} nodes, capturing topic-specific influence patterns.
\end{description}

\subsection{Vector Store Construction}
\label{sec:vs}

Each text segment $S_i$ is embedded once with a domain–tuned transformer encoder,
yielding a single vector representation $\mathbf v_{S_i} \in \mathbb R^d$.
After \mbox{HNMF$k$} assigns a dominant topic $T_k$ to every segment,
all vectors belonging to $T_k$ are written directly into a dedicated
Milvus collection that uses an HNSW index.  
Partitioning by topic keeps each query inside a semantically coherent subset and
reduces latency; index parameters are selected empirically to balance
recall against throughput.

New vectors are inserted straight into the persistent Milvus layer—
no intermediate caching tier is required.  
Vector-store identifiers are aligned with those in the knowledge graph so that
\[
  \mathrm{KGVersion}(d,p) \;=\; \mathrm{VSVersion}(d,p)
\]
for every document $d$ and paragraph $p$, ensuring consistent cross-store look-ups.

\subsection{Knowledge Graph Population}
\label{sec:kg}

The topics produced by \mbox{HNMF$k$} are materialised as 
\texttt{:Topic} nodes annotated with their hierarchy depth and coherence score.
For each topic, the most salient keywords are instantiated as
\texttt{:Keyword} nodes and connected to the topic by \texttt{MENTIONS}
edges weighted by keyword importance.
A document $d$ links to its predominant topic via an
\texttt{IN\_TOPIC} edge whenever the corresponding entry $H_{dk}$ in the loading
matrix exceeds a threshold chosen to balance coverage and precision.

All nodes and edges carry \texttt{provenance}, \texttt{confidence},
and \texttt{version} properties, enabling complete audit trails and
time-travel queries over historical graph states.

\subsection{Hybrid Retrieval and Agent Orchestration}
\label{sec:hybrid}

\paragraph{Query routing.}
An instruction-tuned classifier predicts an execution plan \[ Q(q)=\{\text{VS}:\ell,\; \text{KG}:m\},\] where $\ell$ is the number of vector hops and $m$ the number of graph traversals, guided by lexical cues and topic priors.

\paragraph{Agent loop.}
Adopting the ReAct paradigm, the agent alternates reasoning $R_t$ and action $A_t$, \[ (R_t,A_t)=\operatorname{Agent}(q,\text{history}_{<t}), \] until its self-assessed faithfulness $s\ge0.7$.  Available tools are \[\text{VS-Search}(q,k),\; \text{VS-ReRank}(q,\mathcal C),\; \text{KG-Traverse}(\psi),\] enabling dense retrieval, cross-encoder re-ranking, and symbolic Cypher execution.  The final answer concatenates ranked passages, structured graph facts, and citations.

\subsection{Contribution}
\label{sec:contribution}
By unifying topic-aware KG edges, topic-sharded VS indices, and an agent-mediated hybrid retrieval loop, we deliver a reproducible blueprint for low-latency, high-accuracy RAG systems.  The design supports strict versioning, full provenance, and plug-and-play domain adaptation, thereby lowering the barrier to production-grade knowledge-centred language models.

\section{Conclusion}
By integrating a typed knowledge graph, provenance-aware metadata, and a topic-partitioned vector store within a unified, versioned pipeline, this chapter completes the central layer of the dissertation’s architecture. The knowledge graph provides a structured and interpretable representation grounded in co-occurrence statistics, while the vector store enables efficient semantic retrieval through aligned embeddings. Shared identifiers across both components ensure consistency and traceability, allowing for reliable interaction between symbolic and embedding-based views of the corpus. This alignment sets the stage for the next phase of the system, the Tensor-Structured Retrieval-Augmented Generation module, and provides the structured basis for tensor-based link prediction. In summary, this chapter translates abstract statistical patterns into a coherent, extensible knowledge representation, supporting the dissertation’s broader objective of building AI systems that retrieve and reason over domain-specific information with transparency and scale.

%% file: Part2/AI_Reasoining/3.tex
Having established a consistent alignment between the symbolic graph and the vector store in the previous chapter, this chapter addresses the gaps that remain in the knowledge representation. It introduces a family of tensor factorization models, WNMFk, RNMFk, and BNMFk, that frame missing links as low-rank completion problems. Each model selects its rank automatically, estimates its confidence, and combines complementary perspectives such as Boolean, real-valued, and probabilistic to produce calibrated link predictions across large-scale graphs. The focus remains on achieving high link prediction accuracy and quantifying uncertainty, while more complex forms of graph reasoning are intentionally deferred until the graph has been made denser and more complete through this stage of inference.

\input{Part2/AI_Reasoining/link_prediction}

\input{Part2/AI_Reasoining/graph_method_reasoning}

\section{Conclusion}
The techniques presented in this chapter transform the knowledge graph from a static structure into an evidence-augmented representation, where newly inferred links are accompanied by explicit confidence estimates. These enriched connections improve recall in downstream retrieval tasks, as developed in the next chapter, while providing a carefully expanded foundation for subsequent reasoning. By focusing on accurate link completion without introducing premature complexity, this chapter fulfills the dissertation’s objective of identifying and completing missing links in a way that preserves interpretability and supports reliable inference.

%% file: Part2/AI_Reasoining/link_prediction.tex
\section{Link Prediction}
\label{sec:F_methods}
This section discusses the methodologies and techniques underlying the approaches. Sections \ref{subsection:wnmfk}, \ref{subsection:rnmfk}, and \ref{subsection:bnmfk} present the WNMFk, RNMFk, and BNMFk methods, respectively. The ensemble approaches are described with these methods by integrating LMF in Section \ref{subsection:lmf_ensemble}. In Section \ref{subsection:boolean_perturbations}, the Boolean perturbation technique is introduced, and in Section \ref{subsection:boolean_clustering}, the Boolean clustering scheme employed in the methods when performing decomposition under Boolean settings is examined. Additionally, Section \ref{subsection:boolean_thresholding} introduces the adoption of Otsu thresholding and k-means clustering for thresholding latent factors in Boolean decomposition, along with a description of the coordinate descent-based approach. Note that the Boolean operations discussed in Sections \ref{subsection:boolean_perturbations}, \ref{subsection:boolean_clustering}, and \ref{subsection:boolean_thresholding} are integrated into BNMFk and are utilized to enable its functionality in Boolean settings. In the experiments, these operations are also applied to WNMFk and RNMFk to evaluate their performance under Boolean conditions. Conversely, the Boolean settings in BNMFk can be turned off to assess the method's performance under non-Boolean optimization, providing a comprehensive analysis of its adaptability across different data environments. Finally, the integration of the UQ framework into the system is presented in Section \ref{subsection:uq}. For ease of reference, a summary of the notations used throughout the paper is provided in Table \ref{table:notations}.

\begin{table*}[t!]
\small
\caption{Summary of the notation styles used in the paper.}
\label{table:notations}
\begin{tabular}{c|l||c|l}
\hline
\textbf{Notation}  & \textbf{Description}  & \textbf{Notation}  & \textbf{Description}  \\ \hline
$x$                & Scalar                 & $\vect{x}_i$       & $i$th element in the vector\\
$\vect{x}$         & Vector                 & $\mat{X}_{ij}$     & Entry at row $i$, column $j$\\
$\mat{X}$          & Matrix                 & $\mat{X}_{i:}$     & $i$th row\\
$\ten{X}$          & Tensor                 & $\mat{X}_{:j}$     & $j$th column\\
$\ten{X}_{::i}$    & $i$th slice (3rd dim)  & $\ten{X}_{:::i}$   & $i$th slice (4th dim)\\
$\ten{X}^{\textit{name}}$  & Superscript identifier & $*$ & Dot product\\
$\odot$           & Element-wise product    & $\otimes_B$ & Boolean matrix multiplication\\
\hline
\end{tabular}
\end{table*}

\subsection{Weighted Non-Negative Matrix Factorization with Automatic Model Determination (WNMFk)}
\label{subsection:wnmfk}

While NMF, as summarized in Section \ref{subsection:nmf}, is an effective method for extracting meaningful latent features, it minimizes the distance between $\mat{X}$ and the approximation $\mat{W}\mat{H}$ as defined in Equation \ref{eqn:nmf_min}. This minimization includes zero entries in $\mat{X}$, causing the approximation $\mat{W}\mat{H}$ to be forced toward zero for those entries. However, in link prediction tasks, the network nodes interact with only a small subset of other nodes, resulting in a highly sparse matrix $\mat{X}$. Standard NMF minimization fails because the zeros in $\mat{X}$, representing missing values, are precisely the links to predict rather than those to minimize. Instead, the optimization must consider only the known entries in $\mat{X}$. Estimating the missing values in this way, as recommendations or predictions of missing links, constitutes a matrix completion problem \cite{xu2012alternating}.

WNMFk extends the NMF framework by incorporating element-wise weights during factorization. This modification enables WNMFk to account for varying confidence levels in the observed data, providing robustness in scenarios where some matrix elements are less reliable than others. Additionally, WNMFk extends standard WNMF \cite{wnmfk_src} by integrating the NMFk framework for automatic rank determination, addressing the challenge of selecting the optimal number of latent features, $k$. 

Given a matrix $\mat{X} \in \mathbb{R}_{+}^{n \times m}$ and an associated weight matrix $\mat{M} \in \{0,1\}^{n \times m}$, where $\mat{M}_{ij} > 0$ indicates that the corresponding entry $\mat{X}_{ij}$ is observed or a known link and $\mat{M}_{ij} = 0$ indicates that the entry is unobserved or a missing link, WNMFk seeks to approximate $\mat{X}$ as the product of two non-negative matrices such that $\mat{X} \approx \mat{W}\mat{H}$, where $\mat{W} \in \mathbb{R}_{+}^{n \times k}$ and $\mat{H} \in \mathbb{R}_{+}^{k \times m}$. The final prediction matrix is obtained as:
\begin{align*}
\label{eqn:wnmfk_eq}
\hat{\mat{X}} = \mat{W}\mat{H}. \numberthis 
\end{align*}
The weight matrix $\mat{M}$ prioritizes reconstructing observed entries while ignoring the missing ones. When the mask $\mat{M}$ is binary, with values $0$ and $1$, instead of continuous weights, WNMFk naturally performs missing link prediction by focusing solely on reconstructing the observed entries ($\mat{M}_{ij} = 1$) while ignoring the unobserved entries ($\mat{M}_{ij} = 0$). The objective function for WNMF is defined as:
\begin{align*}
\underset{\mat{W}, \mat{H} \geq 0}{\operatorname{minimize}} \, ||\mat{M} \odot (\mat{X} - \mat{W}\mat{H})||_F^2 + \lambda(||\mat{W}||_F^2 + ||\mat{H}||_F^2), \numberthis 
\end{align*}
where $\lambda$ is a regularization parameter to prevent overfitting. The optimization problem is solved iteratively using multiplicative update rules for $\mat{W}$ and $\mat{H}$. At each iteration, the residual matrix $\mat{R}$ is computed as $\mat{R} = \mat{X} - \mat{W}\mat{H}$. For each latent factor $k$, the updates for $\mat{W}$ and $\mat{H}$ are as follows, and $\mat{W}$ and $\mat{H}$ are constrained to remain non-negative:
\begin{align*}
W_{:, k} \leftarrow \frac{(\mat{M} \odot \mat{R}) \mat{H}_{k, :}^\top}{\mat{M}\mat{H}_{k, :}^\top \mat{H}_{k, :} + \lambda}, \numberthis 
\end{align*}
\begin{align*}
H_{k, :} \leftarrow \frac{(\mat{M} \odot \mat{R})^\top W_{:, k}}{\mat{M}^\top W_{:, k} W_{:, k} + \lambda}. \numberthis 
\end{align*}

To automatically determine the optimal rank $k$, WNMFk leverages the NMFk framework, as described in Section \ref{subsection:nmfk}. More specifically, line 4 in Algorithm 1, where $\ten{W}_{::kq}$,$\ten{H}_{::kq}$ = NMF($\ten{X}_{::q}$,k), is replaced by the factorization procedure as described in this section. WNMFk identifies the rank that balances reconstruction accuracy and stability across perturbations by applying the perturbation-based ensemble approach and clustering stability metrics. Furthermore, with WNMFk, the mask matrix $\mat{M}$ can contain non-Boolean entries, allowing continuous values to be used as weights during approximation. For instance, if specific entries in the observed matrix $\mat{X}$ are less reliable or should contribute less to the optimization process, their corresponding weights in $\mat{M}$, such as $\mat{M}_{ij}$, can be assigned lower values. For example, in a recommendation system, if a user's rating for a specific item is suspected to be influenced by noise (e.g., a randomly assigned rating rather than a genuine preference), the corresponding entry in $\mat{M}$ could be assigned a lower weight to reduce its influence on the factorization. This flexibility makes WNMFk particularly useful in scenarios with varying confidence levels across data points.

\subsection{Recommender NMF with Automatic Model Determination and Biases (RNMFk)}
\label{subsection:rnmfk}

RNMFk incorporates biases to account for systematic user and item tendencies. Like WNMFk, RNMFk integrates the NMFk framework to automatically determine the optimal rank $k$, addressing the challenge of model selection. While WNMFk focuses on reconstructing observed entries in the data matrix using a weight mask, RNMFk introduces additional bias terms to improve performance in scenarios where user- and item-specific effects must be explicitly modeled. The mask $\mat{M}$ contains boolean values (0s or 1s) to signify the missing and known entries. The biases help capture variations such as individual user preferences or global item popularity, which cannot be represented solely by the latent factors $\mat{W}$ and $\mat{H}$. In this work, the Collaborative NMF algorithm presented in the \textit{Surprise} package \cite{Hug2020} is extended and modified, and the code is adopted to leverage vector multiplication in the Python package T-ELF, and integrated with the NMFk framework.

Given a data matrix $\mat{X} \in \mathbb{R}_{+}^{n \times m}$, where $n$ represents users or number of rows and $m$ represents items or number of columns, RNMFk aims to approximate $\mat{X}$ by factoring it into:
\begin{align*}
\label{eqn:rnmfk_eq}
\hat{\mat{X}}_{i,j} = \mat{W}_{i,:}\mat{H}_{:,j} + b_{W_i} + b_{H_j} + \mu, \numberthis 
\end{align*}
where: $\mat{W} \in \mathbb{R}_{+}^{n \times k}$ is the user latent factor matrix, $\mat{H} \in \mathbb{R}_{+}^{k \times m}$ is the item latent factor matrix, $b_{W_i}$ and $b_{H_j}$ are the biases for user $i$ and item $j$, respectively, $\mu$ is the global average rating, representing the group bias. This formulation extends the standard NMF objective by explicitly modeling biases, which account for systematic user- or item-level variations. The optimization objective in RNMFk is given by:
\begin{align*}
\underset{\mat{W}, \mat{H} \geq 0, \, b_W, \, b_H}{\operatorname{minimize}} \, ||\mat{X}_{i,j} - (\mat{W}_{i,:}\mat{H}_{:,j} + b_{W_i} + b_{H_j} + \mu)||_F^2 + \alpha||\mat{W}||_F^2 + \beta||\mat{H}||_F^2 + \gamma||b_W||_F^2 + \delta||b_H||_F^2, \numberthis 
\end{align*}
where $\alpha$, $\beta$, $\gamma$, and $\delta$ are regularization parameters for $\mat{W}$, $\mat{H}$, $b_W$, and $b_H$, respectively. The minimization is performed only over the observed entries in $\mat{X}$, making RNMFk suitable for matrix completion tasks. The parameters $\mat{W}$, $\mat{H}$, $b_W$, and $b_H$ are estimated iteratively using the following updates:
\begin{multicols}{2}
\noindent
\begin{align}
\label{eqn:bias_W_updates}
b_W \leftarrow b_W + \eta_W \sum_{j=1}^m (\mat{err}_{:,j} - \gamma b_W),
\end{align}
\begin{align}
\label{eqn:bias_H_updates}
b_H \leftarrow b_H + \eta_H \sum_{i=1}^n (\mat{err}_{i,:} - \delta b_H),
\end{align}
\end{multicols}

\begin{multicols}{2}
\noindent
\begin{align}
\label{eqn:masked_W_updates}
\mat{W} \leftarrow \mat{W} \odot \frac{\mat{X}\mat{H}^\top}{\hat{\mat{X}}\mat{H}^\top + \alpha \mat{W}},
\end{align}
\begin{align}
\label{eqn:masked_H_updates}
\mat{H} \leftarrow \mat{H} \odot \frac{\mat{W}^\top\mat{X}}{\mat{W}^\top\hat{\mat{X}} + \beta \mat{H}},
\end{align}
\end{multicols}
where $\hat{\mat{X}}_{i,j} = \mat{W}_{i,:}\mat{H}_{:,j} + b_{W_i} + b_{H_j} + \mu$ is the predicted matrix, and $\mat{err}_{i,j} = \mat{X}_{i,j} - \hat{\mat{X}}_{i,j}$. To integrate automatic model selection to RNMFk, the factorization procedure of NMFk is replaced, line 4 in Algorithm 1, where $\ten{W}_{::kq}$,$\ten{H}_{::kq}$ = NMF($\ten{X}_{::q}$,k), with the factorization procedure described in this section.

\subsection{Boolean Non-Negative Matrix Factorization with Automatic Model Determination (BNMFk)}
\label{subsection:bnmfk}

 BNMFk extends the NMF framework to Boolean settings, where the data matrix $\mat{X} \in \{0,1\}^{n \times m}$ consists of binary values and integrates NMFk for automatic rank determination. Unlike WNMFk and RNMFk, which use weighted and biased approaches, BNMFk applies constraints to maintain the Boolean structure in both the data and the latent factor matrices during factorization.
Given a binary matrix $\mat{X} \in \{0,1\}^{n \times m}$, BNMFk approximates $\mat{X}$ as $\mat{X} \approx \mat{W}\otimes_B\mat{H}$, where $\mat{W} \in \{0,1\}^{n \times k}$ is the Boolean row latent factor matrix, and $\mat{H} \in \{0,1\}^{k \times m}$ is the Boolean column latent factor matrix. The final prediction matrix was also obtained using the Equation \ref{eqn:wnmfk_eq}. BNMFk aims to minimize the reconstruction error while maintaining the binary constraints on $\mat{W}$ and $\mat{H}$. Unlike traditional NMF, the factorization involves thresholding operations to ensure Boolean values. Boolean-specific clustering and thresholding operations are introduced in Sections \ref{subsection:boolean_perturbations}, \ref{subsection:boolean_clustering}, and \ref{subsection:boolean_thresholding}. In the experiments, these Boolean settings for BNMFk are turned off and turned on for WNMFk and RNMFk. The optimization problem for BNMFk is defined as:

\begin{align*}
\underset{\mat{W}, \mat{H} \geq 0}{\operatorname{minimize}} \, ||\mat{X} - (\mat{W} \otimes_B) \mat{H})||_F^2 + \alpha ||\mat{W}||_F^2 + \beta ||\mat{H}||_F^2, \numberthis 
\end{align*}
where $\alpha$ and $\beta$ are regularization parameters that penalize large values in $\mat{W}$ and $\mat{H}$. The Boolean constraints are enforced through adaptive thresholding during updates. BNMFk employs multiplicative updates with adaptive thresholding to ensure the Boolean structure of the latent factor matrices:
\begin{align}
\label{eqn:masked_W_updates_boolean}
\mat{W} \leftarrow \text{threshold} \left(\mat{W}\odot\frac{\mat{X} \mat{H}^\top}{\mat{W} \mat{H}^\top \mat{H} + \alpha}, \tau_\text{low}, \tau_\text{high} \right),
\end{align}
\begin{align}
\label{eqn:masked_H_updates_boolean}
\mat{H} \leftarrow \text{threshold} \left(\mat{H}\odot\frac{\mat{W}^\top \mat{X}}{\mat{W}^\top \mat{W} \mat{H} + \beta}, \tau_\text{low}, \tau_\text{high} \right),
\end{align}
where the \texttt{threshold} function, described in Section \ref{subsection:boolean_thresholding}, for  restricting $\mat{W}$ and $\mat{H}$ to Boolean values. Here, $\tau_\text{low}$ and $\tau_\text{high}$ represent the lower and upper thresholds used in the thresholding operation for binarizing the matrices $\mat{W}$ and $\mat{H}$. BNMFk integrates the NMFk framework to determine the optimal rank $k$ automatically. Like WNMFk and RNMFk, the NMF procedure in NMFK is replaced, line 4 in Algorithm 1, to integrate an automatic model determination system.  

\subsection{LMF Extensions for Ensemble BNMFk, RNMFk, and WNMFk}
\label{subsection:lmf_ensemble}

An ensemble approach for BNMFk, RNMFk, and WNMFk is introduced by integrating LMF from Section \ref{subsection:lmf} into these methods ($\text{WNMFk}_{\text{\textbf{lmf}}}$, $\text{BNMFk}_{\text{\textbf{lmf}}}$, and $\text{RNMFk}_{\text{\textbf{lmf}}}$). This extension leverages the predicted rank from the automatic model determination process in plain BNMFk, RNMFk, and WNMFk, solves for LMF with the same rank, and combines the reconstructed matrix $\hat{\mat{X}}$ from the original decomposition with the biases learned from LMF. The combined output is then passed through a sigmoid function to produce probabilistic predictions.

The ensemble approach works as follows:
\begin{enumerate}
    \item $\mat{X}$ is first factorized with one of the models introduced above, BNMFk, RNMFk, or WNMFk, to obtain the predicted rank $k$ and $\hat{\mat{X}}$ (Equation \ref{eqn:wnmfk_eq} for BNMFk and WNMFk, and Equation \ref{eqn:rnmfk_eq} for RNMFk) from this decomposition.
    \item The predicted rank $k$ from the automatic model determination step is used to initialize LMF, where the same $\mat{X}$ is decomposed at the predicted rank $k$ using LMF.
    \item LMF solves the optimization problem to learn row and column biases ($b_{\text{r}}$ and $b_{\text{c}}$ from Equation \ref{eqn:lmf_eq}).
    \item Then the final prediction matrix $\tilde{\mat{X}}_{\text{final}}$ is calculated as: 
    \begin{align*}
    \tilde{\mat{X}}_{\text{final}} = \sigma\left(\hat{\mat{X}} + b_{\text{row}} + b_{\text{col}}\right). \numberthis 
    \end{align*}
\end{enumerate}

This ensemble approach aims to combine the strengths of the original matrix factorization method, BNMFk, RNMFk, or WNMFk, with LMF, effectively combining the structural insights and adding the probabilistic modeling capabilities of LMF.

\subsection{Boolean Perturbations for Boolean Factorization}
\label{subsection:boolean_perturbations}

In the context of Boolean matrix factorization, Boolean perturbations are applied to generate an ensemble of slightly modified versions of the binary input matrix $\mat{X}$. These perturbations introduce controlled noise by flipping randomly selected entries in $\mat{X}$, ensuring diversity in the perturbed matrices while maintaining the Boolean structure. The perturbation process can be described as follows:
\begin{itemize}
    \item \textbf{Positive Noise (Additive):} A fraction of entries with value $0$ in $\mat{X}$ are randomly flipped to $1$.
    \item \textbf{Negative Noise (Subtractive):} A fraction of entries with value $1$ in $\mat{X}$ are randomly flipped to $0$.
\end{itemize}

The perturbed matrix $\mat{Y}$ is generated by applying these noise components $\mat{Y} = \text{boolean}(\mat{X}, \epsilon)$ where $\epsilon = [\epsilon_{\text{pos}}, \epsilon_{\text{neg}}]$ represents the proportion of positive and negative noise to be added. For each noise type:
\begin{itemize}
    \item $\epsilon_{\text{pos}}$: Proportion of $0$s in $\mat{X}$ flipped to $1$s.
    \item $\epsilon_{\text{neg}}$: Proportion of $1$s in $\mat{X}$ flipped to $0$s.
\end{itemize}
Line three on Algorithm 1 of NMFk is replaced, where $\ten{X}_{::q}$ = Perturb($\mat{X}$), with Boolean perturbation when applying Boolean factorization.

\subsection{Boolean Clustering for Boolean Factorization}
\label{subsection:boolean_clustering}

In the context of Boolean factorization, the clustering procedure used in the standard NMFk framework, specifically line 7 of Algorithm 1 where $\ten{\hat{W}}$, $\ten{\hat{H}}$ = customCluster( $\ten{W}^{all}$,$\ten{H}^{all}$), is replaced with a Boolean clustering approach described in this section. This adjustment ensures that the clustering step respects the Boolean nature of the data, making it more suitable for binary datasets. Boolean clustering operates on the latent factor matrices $\mat{W}$ generated from multiple perturbed instances of the input matrix $\mat{X}$. Let $\ten{W}_{\text{all}} \in \{0,1\}^{n \times k \times M}$ represent a three-dimensional tensor containing the perturbed $\mat{W}$ matrices, where: $n$ is the number of rows in $\mat{W}$, $k$ is the number of latent features, $M$ is the number of perturbations, and $\mat{W}$ and $\mat{H}$ are Boolean following the thresholding techniques that will be introduced in Section \ref{subsection:boolean_thresholding}. Boolean clustering aims to iteratively align and compute the Boolean centroids for the perturbed $\mat{W}$ matrices using a distance metric tailored for Boolean data, such as Hamming distance \cite{hamming}. The Boolean clustering algorithm alternates between two main steps: distance calculation and centroid computation, repeated until convergence or a maximum number of iterations is reached.

\paragraph{1. Distance Calculation:}
For each perturbed $\mat{W}$, the algorithm computes the distance between the current centroids and the columns of $\mat{W}$ using a Boolean distance metric, such as Hamming distance. The columns of each perturbed $\mat{W}$ are then reordered to minimize the total distance to the centroids.

\paragraph{2. Centroid Computation:}
The centroids are updated based on the reordered $\mat{W}$ matrices. Boolean centroids are computed by aggregating the binary values of each column across all perturbations, ensuring that the centroids reflect the majority consensus of the binary data.

Let $\ten{W}_{\text{all}}$ represent the tensor of perturbed $\mat{W}$ matrices and $\mat{C}$ represent the centroids. Then the steps are:
\begin{enumerate}
    \item \textbf{Initialization:} The centroids are initialized using the first perturbed $\mat{W}$ matrix in $\ten{W}_{\text{all}}$.
    \item \textbf{Distance Calculation:} For each perturbed $\mat{W}$, compute the distance between the centroids and the columns of $\mat{W}$ and reorder the columns to minimize the distance.
    \item \textbf{Centroid Update:} Compute the Boolean centroids by aggregating the reordered columns of $\mat{W}$ across all perturbations.
    \item \textbf{Convergence Check:} If the column ordering stabilizes across all iterations, terminate the algorithm; otherwise, repeat the process.
\end{enumerate}

Boolean clustering replaces the standard custom clustering step in NMFk, enabling BNMFk (or WNMFk and RNMFk when Boolean settings are used) to effectively analyze Boolean datasets while maintaining consistency with the underlying data structure.

\subsection{Boolean Latent Factor Thresholding}
\label{subsection:boolean_thresholding}

Boolean latent factor thresholding ensures that the latent factor matrices $\mat{W}$ and $\mat{H}$ retain their Boolean structure while minimizing the reconstruction error. This section presents three thresholding techniques--Otsu's method, k-means clustering, and coordinate descent--each of which determines thresholds to binarize \(\mat{W}\) and \(\mat{H}\) while preserving a close approximation to the observed matrix \(\mat{X}\). Binarization converts continuous or multi-valued data into binary values (0s and 1s) based on a predefined threshold.

\subsubsection{Otsu's Method}
Otsu's method \cite{4310076} determines the optimal threshold for binarizing the latent factor matrices $\mat{W}$ and $\mat{H}$ by maximizing the between-class variance in their values. Specifically, for each column $\mat{W}_{:,i}$ or row $\mat{H}_{i,:}$, Otsu's method computes a threshold $t^*$ that maximizes the separability between binary classes, ensuring that $\mat{W}, \mat{H} \in \{0, 1\}$. The threshold $t^*$ is defined as:
\begin{align*}
t^* = \argmax_t \left[ \pi_0(t) \pi_1(t) \left( \mu_0(t) - \mu_1(t) \right)^2 \right], \numberthis 
\end{align*}
where $\pi_0(t)$ and $\pi_1(t)$ are the probabilities (normalized counts) of values in $\mat{W}_{:, i}$ or $\mat{H}_{i,:}$ below and above the threshold $t$, respectively, $\mu_0(t)$ and $\mu_1(t)$ are the means of the values below and above the threshold $t$, respectively. Otsu's method finds the optimal threshold by exhaustively evaluating all possible threshold values and selecting the one that maximizes the between-class variance $\sigma_B^2(t)$, defined below. For a given threshold $t$, the method partitions the data into two classes: values below $t$ and values above $t$. It then calculates the probabilities $\pi_0(t)$ and $\pi_1(t)$ and the means $\mu_0(t)$ and $\mu_1(t)$ of the two classes. The threshold that maximizes the separability of these two classes, as measured by $\sigma_B^2(t)$, is chosen as the optimal threshold. This ensures that the binary partitioning captures the most significant difference between the two groups. For each component $i$ of $\mat{W}$ or $\mat{H}$:
\begin{itemize}
    \item Compute the histogram of the values in $\mat{W}_{:,i}$ or $\mat{H}_{i,:}$.
    \item Calculate $\pi_0(t)$, $\pi_1(t)$, $\mu_0(t)$, and $\mu_1(t)$ for each potential threshold $t$.
    \item Select $t^*$ that maximizes the between-class variance:
    \begin{align*}
    \sigma_B^2(t) = \pi_0(t) \pi_1(t) \left( \mu_0(t) - \mu_1(t) \right)^2.
    \end{align*}
\end{itemize}

After determining $t^*$, binarization is applied as follows:
\begin{align*}
\mat{W}_{ij} = 
\begin{cases} 
1, & \text{if } \mat{W}_{ij} \geq t^* \\
0, & \text{if } \mat{W}_{ij} < t^*
\end{cases}
\quad ,
\mat{H}_{ij} = 
\begin{cases} 
1, & \text{if } \mat{H}_{ij} \geq t^* \\
0, & \text{if } \mat{H}_{ij} < t^*. \numberthis 
\end{cases}
\end{align*}

This approach ensures that the thresholds for $\mat{W}$ and $\mat{H}$ effectively partition the data into binary groups, maximizing the separability between classes and preserving the Boolean structure of the latent factors.

\subsubsection{K-Means Clustering}
K-means clustering \cite{pedregosa2011scikit} thresholds each component of $\mat{W}$ and $\mat{H}$ by clustering their values into two groups corresponding to 0 and 1. For a vector $\vect{z}$ (e.g., $\mat{W}_{:,i}$ or $\mat{H}_{i,:}$), k-means clustering identifies two cluster centers $c_0$ and $c_1$. The threshold is then defined as:
\begin{align*}
t^* = \frac{c_0 + c_1}{2}. \numberthis 
\end{align*}
The binarization is performed by assigning:
\begin{align*}
z_j = 
\begin{cases} 
1, & \text{if } z_j \geq t^* \\
0, & \text{if } z_j < t^*. \numberthis 
\end{cases}
\end{align*}

\subsubsection{Coordinate Descent Thresholding (search)}
Coordinate descent thresholding iteratively adjusts the thresholds for $\mat{W}$ and $\mat{H}$ to minimize the reconstruction error $\mat{X} \approx \mat{W} \odot \mat{H}$. For each component $i$, the reconstruction error is computed as:
\begin{align*}
\text{Error} = ||\mat{X} - (\mat{W}_{:,i} \mat{H}_{i,:})||_F^2. \numberthis 
\end{align*}
The thresholds $t_W[i]$ and $t_H[i]$ for $\mat{W}_{:,i}$ and $\mat{H}_{i,:}$ are optimized iteratively:
\begin{itemize}
    \item Fix $\mat{H}$ and optimize $\mat{W}_{:,i}$ by selecting $t_W[i]$ to minimize the reconstruction error.
    \item Fix $\mat{W}$ and optimize $\mat{H}_{i,:}$ by selecting $t_H[i]$ to minimize the reconstruction error.
\end{itemize}
The process repeats until the thresholds converge or the maximum number of iterations is reached. The final binarization is applied as:
\begin{align*}
\mat{W}_{:,i} = 
\begin{cases} 
1, & \text{if } \mat{W}_{:,i} \geq t_W[i] \\
0, & \text{if } \mat{W}_{:,i} < t_W[i], \numberthis 
\end{cases}
\end{align*}
and similarly for $\mat{H}_{i,:}$. In the experiments, \textit{search} is used as a term for this thresholding technique.

\subsection{Uncertainty Quantification (UQ)}
\label{subsection:uq}

UQ is a critical component of robust predictive modeling, providing insights into the reliability of model predictions. In the context of NMFk and its variants WNMFk, RNMFk, and BNMFk presented in this paper, UQ evaluates the stability of the reconstructed matrix $\hat{\mat{X}}$ (Equation \ref{eqn:wnmfk_eq} for BNMFk and WNMFk, and Equation \ref{eqn:rnmfk_eq} for RNMFk) across perturbations, using data augmentation to define the confidence, enabling the identification of confident and uncertain predictions. The UQ framework leverages the idea that truly confident predictions are stable under perturbation or data augmentation \cite{10028760, Bahat2018ConfidenceFI}, which hypothesizes that stable predictions across multiple perturbations are more likely to be accurate. During the NMFk process, perturbations of the input matrix $\mat{X}$ generate an ensemble of latent factor matrices $\ten{W}_{\text{all}}$ and $\ten{H}_{\text{all}}$, corresponding to different realizations of $\mat{W}$ and $\mat{H}$. 

Under non-Boolean settings, perturbations are generated by uniformly sampling from $\mat{X}$ to create modified versions that are a controlled distance away from the original matrix $\mat{X}$. This is achieved by scaling the entries of $\mat{X}$ with random noise drawn uniformly from the range $[1 - \epsilon, 1 + \epsilon]$, where $\epsilon$ controls the magnitude of the perturbation. Then, the perturbed matrix $\mat{Y}$ is defined as:
\begin{align*}
\mat{Y} = \mat{X} \odot \left(1 - \epsilon + 2\epsilon \cdot \text{rand}(\text{shape}(\mat{X}))\right), \numberthis 
\end{align*}
where: $\text{rand}(\text{shape}(\mat{X}))$ generates random values uniformly distributed in $[0, 1]$, and $\epsilon$ is the perturbation parameter. In the experiments, $\epsilon=0.015$. Here, the $\epsilon=0.015$ hyperparameter is selected as it is shown to give a stable region for $k$ selection \cite{nebgen2021neural}. The perturbation method described in Section \ref{subsection:boolean_perturbations} is used for Boolean settings. This approach modifies the Boolean structure of $\mat{X}$ by flipping selected entries (from $0$ to $1$ or $1$ to $0$) based on the specified noise proportions $\epsilon_{\text{pos}}$ and $\epsilon_{\text{neg}}$ ($(\epsilon_{\text{pos}}$, $\epsilon_{\text{neg}})=(0.015, 0.015)$ in the experiments). For each perturbation $p$, the reconstructed matrix $\hat{\mat{X}}^{(p)}$ is computed as:
\begin{align*}
\hat{\mat{X}}^{(p)} = \mat{W}^{(p)} \mat{H}^{(p)},\\
\hat{\ten{X}}_{::p} = \ten{W}_{::p}\ten{H}_{::p}, \numberthis 
\end{align*}
where $\mat{W}^{(p)} \in \mathbb{R}_{+}^{n \times k}$ and $\mat{H}^{(p)} \in \mathbb{R}_{+}^{k \times m}$ are the latent factors for perturbation $p$. For $P$ perturbations, there are $\ten{W}^{all}=[\ten{W}_{::1},\ldots, \ten{W}_{::p} , \ldots,\ten{W}_{::P}]$ and $\ten{H}^{all}=[\ten{H}_{::1},\ldots, \ten{H}_{::p} , \ldots,\ten{H}_{::P}]$, where $\ten{W}^{all}$ and $\ten{H}^{all}$ are three dimensional tensors of size $n \times k \times P$ and $k \times m \times P$. Across $P$ perturbations, the ensemble of reconstructed matrices $\{\hat{\mat{X}}^{(p)}\}_{p=1}^P$ reflects the variability in the model's predictions, where $\hat{\ten{X}}^{all}=[\hat{\ten{X}}_{::1},\ldots, \hat{\ten{X}}_{::p}, \ldots,\hat{\ten{X}}_{::P}]$, and where $\hat{\ten{X}}^{all}$ is a three dimensional tensor size of $n \times m \times P$. To quantify uncertainty, the standard deviation of the reconstructed values for each entry $(i,j)$ in $\hat{\mat{X}}$ is computed:
\begin{align*}
\label{eq:UQ_matrix}
\mat{U}_{ij} = \sqrt{\frac{1}{P} \sum_{p=1}^P \left( \hat{\mat{X}}_{ij}^{(p)} - \bar{\hat{\mat{X}}}_{ij} \right)^2}, \numberthis 
\end{align*}
where: $\hat{\mat{X}}_{ij}^{(p)}$ is the $(i,j)$ entry of $\hat{\mat{X}}^{(p)}$, $\bar{\hat{\mat{X}}}_{ij} = \frac{1}{P} \sum_{p=1}^P \hat{\mat{X}}_{ij}^{(p)}$ is the mean of the reconstructed values at $(i,j)$ across all $P$ perturbations. The resulting uncertainty matrix $\mat{U}$ captures the variability of predictions at each entry of $\hat{\mat{X}}$. Low uncertainty values or standard deviation indicate consistent predictions across perturbations, suggesting high confidence in the model's output at those entries. Conversely, high uncertainty values signal variability and reduced confidence in predictions. Based on the hypothesis that truly confident predictions are stable under error \cite{10028760}, predictions with low uncertainty are more likely to be accurate and reliable.

\subsection{Evaluation}
Given $y$ total samples, we define the training set size as a proportion $\text{\textit{train}}_{\text{size}} \in [0.1, 0.2, \dots, 0.9]$. The number of samples in the training set is $y_{\text{train}} = y \times \text{\textit{train}}_{\text{size}}$. The test set consists of the remaining samples, with size $y_{\text{test}} = y \times (1 - \text{\textit{train}}_{\text{size}})$. For missing link prediction, we define the observation matrix $\mathbf{X}$, where known links are represented by nonzero values ($\mathbf{X}_{ij} \neq 0$), and known negative links (i.e., known absence of a connection) are represented by $\mathbf{X}_{ij} = 0$. The locations of known links are determined by the mask matrix $\mathbf{M}$, where $\mathbf{M}_{ij} = 1$ indicates an observed link in the training set, and $\mathbf{M}_{ij} = 0$ represents a missing link. To construct the training and test sets, we define the index sets:
\begin{enumerate}
\item Let $\mathcal{I}_{\text{pos}}$ be the set of indices where $\mathbf{X}_{ij} \neq 0$ (known positive links).
\item Let $\mathcal{I}_{\text{neg}}$ be the set of indices where $\mathbf{X}_{ij} = 0$ (known negative links).
\item The test set index set, $\mathcal{I}_{\text{test}}$, is chosen such that the number of indices satisfies $|\mathcal{I}_{\text{test}}| = y_{\text{test}}$. 
\end{enumerate}

The test set is constructed by stratified sampling from both known positive and known negative links:
\begin{enumerate}
    \item \textbf{Positive known links:} A subset of $\mathcal{I}_{\text{pos}}$ is randomly selected and placed in $\mathcal{I}_{\text{test}}$.
    \item \textbf{Negative known links:} A subset of $\mathcal{I}_{\text{neg}}$ is randomly selected and placed in $\mathcal{I}_{\text{test}}$.
    \item \textbf{Training set}: The remaining indices belong to the training index set $\mathcal{I}_{\text{train}} = (\mathcal{I}_{\text{pos}} \cup \mathcal{I}_{\text{neg}}) \setminus \mathcal{I}_{\text{test}}$, where $\setminus$ represents the set difference.
\end{enumerate}
The mask matrix $\mathbf{M}$ is defined such that test set locations are masked out:
\begin{equation}
\mathbf{M}_{ij} =
\begin{cases}
0, & \text{if } (i,j) \in \mathcal{I}_{\text{test}} \\
1, & \text{otherwise}.
\end{cases}
\end{equation}

The training matrix is then:

\begin{equation}
\mathbf{X}^{\text{train}} = \mathbf{X} \odot \mathbf{M},
\end{equation}

where $\odot$ represents the element-wise (Hadamard) product. The test matrix consists of the missing edges:

\begin{equation}
\mathbf{X}^{\text{test}} = \mathbf{X} \odot (1 - \mathbf{M}).
\end{equation}

A prediction matrix $\hat{\mathbf{X}}$ is computed using a model trained on $\mathbf{X}^{\text{train}}$. The performance is evaluated by comparing the predicted values at test locations:

\begin{equation}
\hat{\mathbf{X}}_{\text{test}} = \hat{\mathbf{X}} \odot (1 - \mathbf{M}),
\end{equation}

where we extract only the entries corresponding to the test set to measure prediction performance.

On the Swimmer dataset, we keep the test-set size fixed at 10\% and do not test for data sparsity, as the computational time required for this dataset is significantly longer due to the larger matrix. We set the test set size for the PPI dataset to 20\%, the same as in \cite{ppi_symlmf}. The missing links are randomly sampled 10 times to ensure robust evaluation. All results are reported with an average over the cross-validations with coverage intervals (CIs).

\subsection{Metrics}
\label{subsec:metrics}
Performance metrics include rank $k$ predictions, Root Mean Squared Error (RMSE), abstained sample fraction, RMSE on non-abstained samples, and Pearson Correlation between UQ entries and reconstruction error. We also report ROC AUC and PR AUC, including UQ-based extensions, where UQ values from $\mat{U}$ at test set points $\mathcal{I}_{\text{test}}$ serve as weights, simulating confidence-based predictions.

\subsubsection*{Rank $k$ Predictions}
We assess the accuracy of the model's automatic determination of the true rank $k$. Correct rank prediction is critical for capturing the underlying structure of the data. The results use a Violin plot \cite{Hintze01051998} to show how the rank predictions are distributed.

\subsubsection*{Root Mean Squared Error (RMSE)}
RMSE measures the reconstruction error, quantifying how closely the reconstructed matrix $\hat{\mat{X}}$ approximates the observed matrix $\mat{X}$ at the test set points $\mathcal{I}_{\text{test}}$:
\begin{align*}
\text{RMSE} = \sqrt{\frac{1}{|\mathcal{I}_{\text{test}}|} \sum_{(i,j) \in \mathcal{I}_{\text{test}}} \left( \mat{X}_{ij} - \hat{\mat{X}}_{ij} \right)^2}. \numberthis 
\end{align*}
A lower RMSE indicates better reconstruction accuracy. We use RMSE to report the performance of predicting the missing links.

\subsubsection*{Fraction of Abstained or Rejected Samples}

This metric evaluates the proportion of predictions the model abstains from due to high uncertainty, focusing on confident predictions while disregarding less certain ones. The fraction of abstained samples is calculated as follows:
\begin{align*}
f_\text{abstain} = \frac{|\mathcal{I}_\text{abstain}|}{|\mathcal{I}_{\text{test}}|}, \numberthis
\end{align*}
where $f_\text{abstain}$ is the fraction of abstained predictions, $|\mathcal{I}_\text{abstain}|$ is the number of abstained predictions, and $|\mathcal{I}_{\text{test}}|$ is the total number of test set samples. The coverage rate can be calculated with $1-f_\text{abstain}$, referring to the fraction of non-abstained samples. The threshold for rejecting predictions in missing links is based on the uncertainty values in $\mathbf{U}$. For a given uncertainty matrix $\mathbf{U}$, the threshold for abstaining (reject-option) $\tau$ is defined as:
\begin{align*}
\tau = \frac{1}{|\mathcal{I}_{\text{train}}|} \sum_{(i,j) \in \mathcal{I}_{\text{train}}} \mathbf{U}_{ij}, \numberthis
\end{align*}
where $\mathcal{I}_{\text{train}}$ is the set of all training-set indices (known links), and $\mathbf{U}_{ij}$ is the uncertainty value at location $(i,j)$. Predictions at test-set indices $(i,j) \in \mathcal{I}_{\text{test}}$ are abstained if:
\begin{align*}
\mathbf{U}_{ij} > \tau, \numberthis
\end{align*}
where $\mathcal{I}_{\text{test}}$ represents the set of test-set indices. This ensures that the model avoids making predictions in locations with higher uncertainty than the average certainty observed in the training set (known links).

\subsubsection*{RMSE on Non-Rejected or Non-Abstained Samples}
This metric calculates the RMSE only on the predictions that are not abstained:
\begin{align*}
\text{RMSE}_{\text{Non-Abstained}} = \sqrt{\frac{1}{|\mathcal{I}_{\text{test}}'|} \sum_{(i,j) \in \mathcal{I}_{\text{test}}'} \left( \mathbf{X}_{ij} - \hat{\mathbf{X}}_{ij} \right)^2}, \numberthis
\end{align*}
where $\mathcal{I}_{\text{test}}'$ is the subset of $\mathcal{I}_{\text{test}}$ containing non-abstained predictions. This highlights the accuracy of the most confident predictions.

\subsubsection*{ROC, AUC, and PR AUC}
Receiver Operating Characteristic (ROC) AUC and Precision-Recall (PR) AUC are standard metrics for evaluating classification performance. ROC AUC measures the ability to distinguish between positive and negative classes, while PR AUC evaluates precision and recall trade-offs:
\begin{align*}
\text{ROC AUC} = \int_0^1 \text{TPR}(FPR) \, d(\text{FPR}), \quad \text{PR AUC} = \int_0^1 \text{Precision}(Recall) \, d(Recall). \numberthis
\end{align*}
The results include their UQ-based extensions incorporating UQ values $\mathbf{U}$ as weights, simulating a scenario where confidence is assigned to predictions. The weights are calculated based on the UQ values and normalized to account for overall uncertainty. Specifically, the weight for each prediction is defined as:
\begin{align*}
w_{ij} = \frac{\mathbf{U}_{ij}}{1 + \text{median}(\mathbf{U}_{kl} \text{ for } (k,l) \in \mathcal{I}_{\text{train}})}. \numberthis
\end{align*}
The normalized weight is given by:
\begin{align*}
w_{ij}^{\text{norm}} = \frac{1}{1 + w_{ij}}. \numberthis
\end{align*}

Using these normalized weights, the UQ-based ROC AUC and PR AUC metrics are computed as:
\begin{align*}
\text{UQ - ROC AUC} = \frac{\sum_{(i,j) \in \mathcal{I}_{\text{test}}} w_{ij}^{\text{norm}} \cdot \text{TPR}(\text{FPR})}{\sum_{(i,j) \in \mathcal{I}_{\text{test}}} w_{ij}^{\text{norm}}}, \numberthis
\end{align*}
\begin{align*}
\text{UQ - PR AUC} = \frac{\sum_{(i,j) \in \mathcal{I}_{\text{test}}} w_{ij}^{\text{norm}} \cdot \text{Precision}(\text{Recall})}{\sum_{(i,j) \in \mathcal{I}_{\text{test}}} w_{ij}^{\text{norm}}}. \numberthis
\end{align*}

We incorporate these weights into the ROC AUC and PR AUC calculations by utilizing the \texttt{sample\_weight} parameter provided in Scikit-learn \cite{pedregosa2011scikit}. This allows us to account for the uncertainty-based weights during the evaluation, ensuring that higher-confidence predictions contribute more significantly to the metrics. In contrast, less confident predictions have a reduced impact, reflecting the effect of uncertainty on classification performance.

%% file: Part2/AI_Reasoining/graph_method_reasoning.tex


\section{Graph reasoning over latent knowledge graphs}
\label{subsec:graph-reasoning}
While the vector store supports rapid semantic matching, many queries require multi-hop logical inference that is opaque to pure embedding
similarity.  We therefore integrate a  graph reasoning module  that operates on the heterogeneous knowledge graph
$\mathcal{G}=(\mathcal{V},\mathcal{E},\mathcal{R})$ produced by the extraction pipeline .  Each
document cluster from HNMF{\small k} contributes a topic  node, connected via edges to its constituent
 Document nodes and via  has\_keyword  edges to the topics recorded above.

\paragraph{Symbolic path enumeration.}
Given an entity-centric query $q$, we identify a start set $S\subset\mathcal{V}$ using deterministic matchers (regular
expressions for statute citations, Neo4j full-text for free text).   Paths of length $\le\!L{=}3$ are enumerated in Cypher with
edge-type constraints that privilege explanatory relations (\texttt{cites}, \texttt{is\_case\_of}, \texttt{has\_method}) over
associative ones (\texttt{co\_occur}).  Each path n $p=v_{0}\!\xrightarrow{r_{1}}\!v_{1}\!\dots v_{\ell}$ receives an initial score
$\sigma_{\mathrm{sym}}(p)=\prod_{i=1}^{\ell}w(r_{i})$ where $w(r)$ is a learnt edge reliability weight.

\paragraph{Neuro-symbolic re-ranking.}
The candidate subgraph is further re-ranked by an R-GCN encoder \cite{schlichtkrull2018rgcn} that projects every node to a
$d$-dimensional vector, optimised with a margin-ranking loss against ground-truth relevance triples obtained from expert annotations in the
cybersecurity and legal domains.  The final path score is a convex combination $\sigma(p)=\lambda\,\sigma_{\mathrm{sym}}(p)+(1-\lambda)\,
\frac{1}{\ell}\sum_{i=0}^{\ell}\! \mathrm{cos}\bigl(\mathbf{h}_{q},\mathbf{h}_{v_{i}}\bigr)$, where $\mathbf{h}_{q}$ is the query embedding from LegalBERT and $\lambda=0.4$ tuned on a validation set.  Paths above a threshold $\tau=0.15$ supply supporting documents whose passages are forwarded
to the RAG composer.

\paragraph{Incremental materialisation.}
To keep latency bounded, the reasoning engine maintains an \texttt{openCypher} materialised view cache of the $10^{5}$ most
frequently traversed patterns, refreshed nightly, and invalidated when the upstream KG ingest reports a structural delta $>\!0.5\%$.  This
policy yields a $2\times$ reduction in average query time on the jurisprudence workload (median $210$ ms vs.\ $440$ ms) without sacrificing recall.

\paragraph{Alignment with the vector store.}
Before retrieval, the top-ranked subgraph is converted into dense prompts by concatenating relation labels and node names, encoded with the same encoder that built the VS.  De-duplication against the $k$‐NN neighbours ensures that symbolic and dense channels contribute
disjoint evidence, realising the hybrid RAG architecture.

%% file: Part3/smart-slic/0.tex
The previous chapters established the architectural foundation of the system. This chapter turns to its practical significance by demonstrating how the full pipeline operates on real-world data. It traces the workflow from Binary Bleed parameter selection through HNMFk taxonomy construction and the synchronization of the knowledge graph and vector store. Rather than emphasizing numerical benchmarks, the focus is on qualitative outcomes. Binary Bleed reduces the search space, making large-scale factorization computationally routine. HNMFk generates topic hierarchies that are interpretable by subject-matter experts. The combined knowledge graph and vector store remain consistent even when applied to diverse corpora, including scientific articles, legal texts, and cybersecurity reports. Together, these demonstrations show the transition from a conceptual architecture to a functioning system, advancing the dissertation's central claim that a unified, interpretable pipeline for domain-specific knowledge discovery can replace fragmented and brittle workflows.

\input{Part3/smart-slic/binarybleed}
\input{Part3/smart-slic/cyber-security}

\input{Part3/smart-slic/Legal_KG_RAG}

\input{Part3/smart-slic/Malware}

\section{Conclusion}
Viewed in its entirety, the set of case studies presented in this chapter affirms the dissertation’s core contributions: scalability, interpretability, and applicability across distinct domains. The same methods that support large-scale factorization on high-performance computing resources also enable usable systems for legal practitioners and materials scientists, without relying on opaque or ad hoc techniques. By emphasizing detailed evidence over isolated metrics, the discussion highlights the broader value of the approach, a unified architecture capable of ingesting, structuring, and presenting knowledge in a form that users can understand and trust. With this empirical foundation established, the dissertation now turns to the final component, showing how Tensor-Structured RAG builds on the same graph and vector infrastructure to support reliable, context-aware question answering. This completes the transition from document retrieval to reasoned response that defines the overarching goal of the system.

%% file: Part3/smart-slic/binarybleed.tex
\section{Binary Bleed}
\label{sec:C_results}
We gauge the efficacy of unsupervised learning methods in single-node and distributed settings. We evaluate  NMF and K-means in the former, then NMF and RESCAL in the latter.
\begin{figure}[H]
    \centering
    \begin{minipage}[b]{0.48\columnwidth}
        \centering
        \includegraphics[width=\columnwidth]{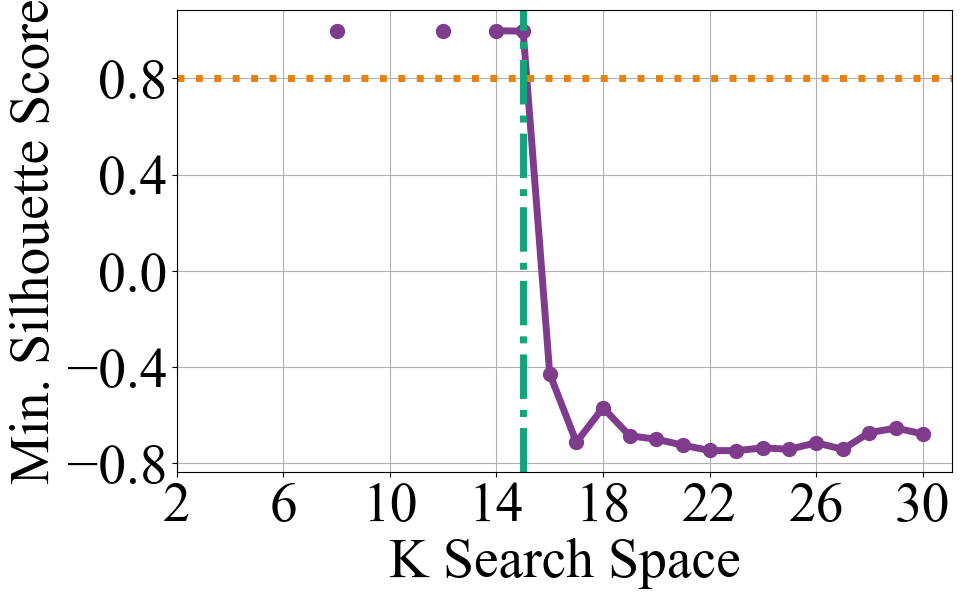}
    \end{minipage}
    \hfill
    \begin{minipage}[b]{0.48\columnwidth}
        \centering
        \includegraphics[width=\textwidth]{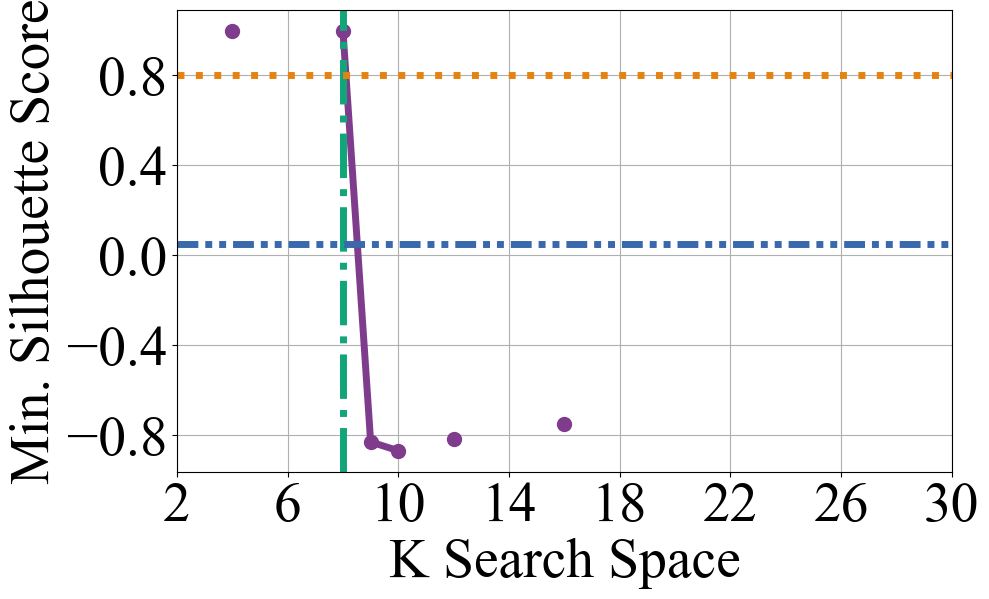}
    \end{minipage}
    \vskip\baselineskip
    \begin{minipage}[b]{0.48\columnwidth}
        \centering
        \includegraphics[width=\columnwidth]{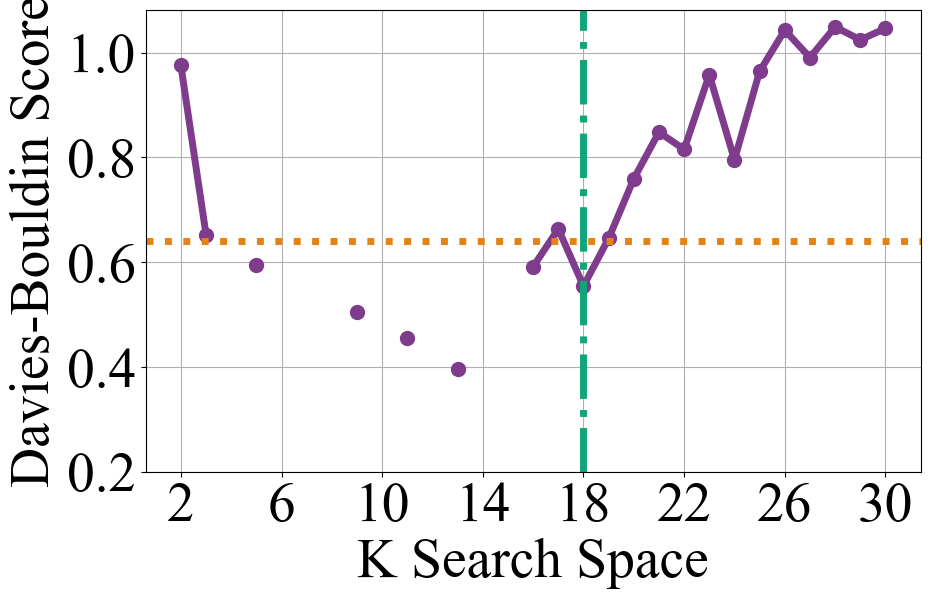}
    \end{minipage}
    \hfill
    \begin{minipage}[b]{0.48\columnwidth}
        \centering
        \includegraphics[width=\columnwidth]{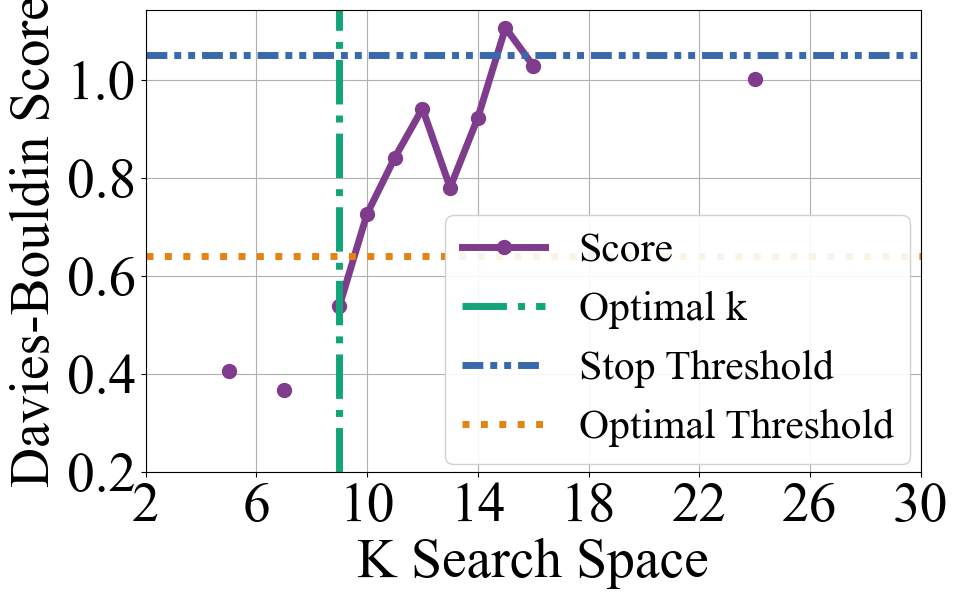}
    \end{minipage}
    \caption{NMFk (top row) and K-means (bottom row), Vanilla (Left) and Early stopping (Right).  $\forall k_{optimal}=k_{true}$.}
    \label{fig:2x2grid}
\end{figure}

\subsection{Single-node Setting}
\begin{figure}[ht]
    \centering
    \begin{minipage}[b]{0.49\columnwidth}
         \begin{figure}[H]
        \centering
         \includegraphics[width=\columnwidth]{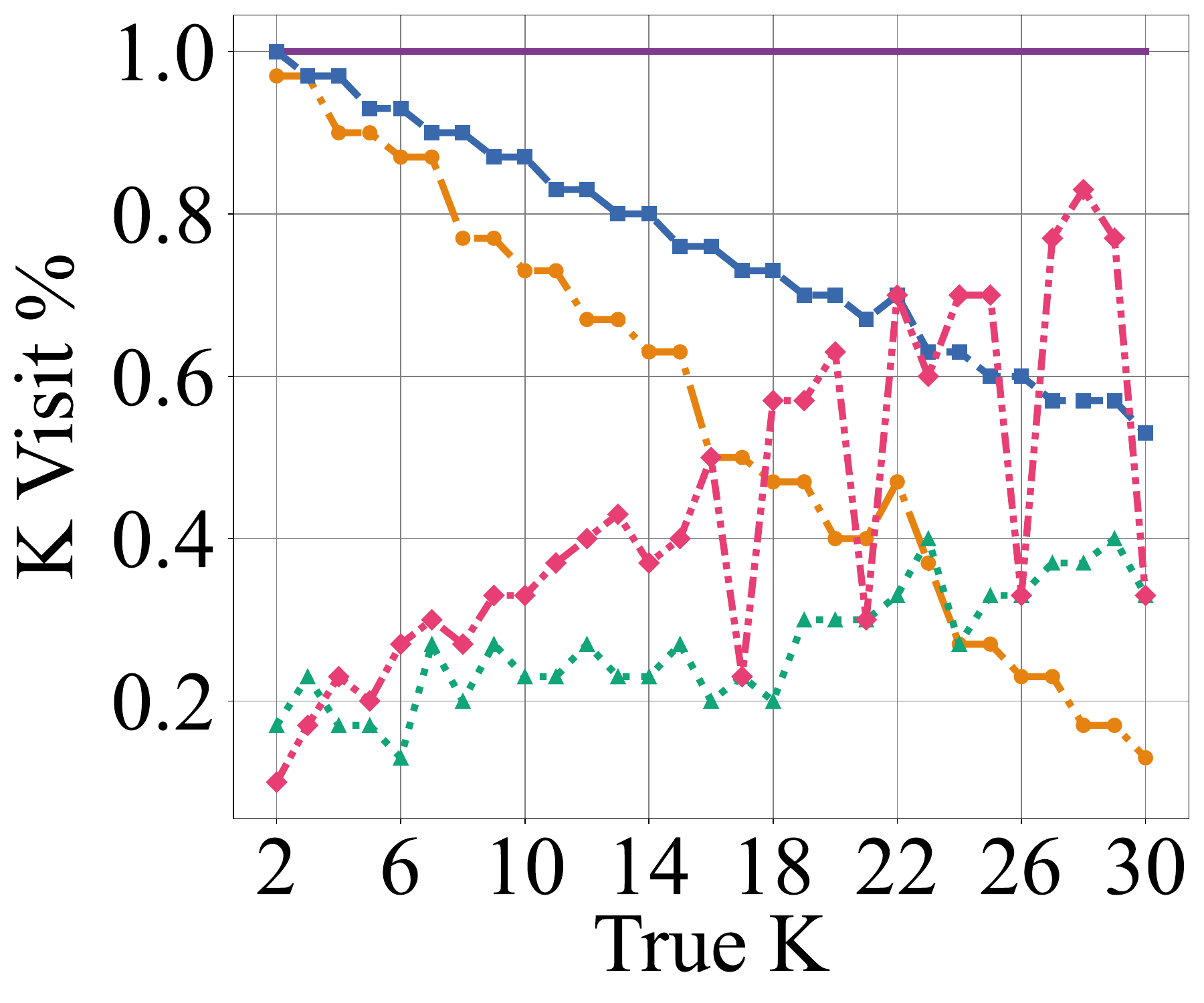}
        
            \label{fig:preorder_results}
        \end{figure}
    \end{minipage}
\hfill
\centering
    \begin{minipage}[b]{0.49\columnwidth}
          \begin{figure}[H]
            \centering
             \includegraphics[width=\columnwidth]{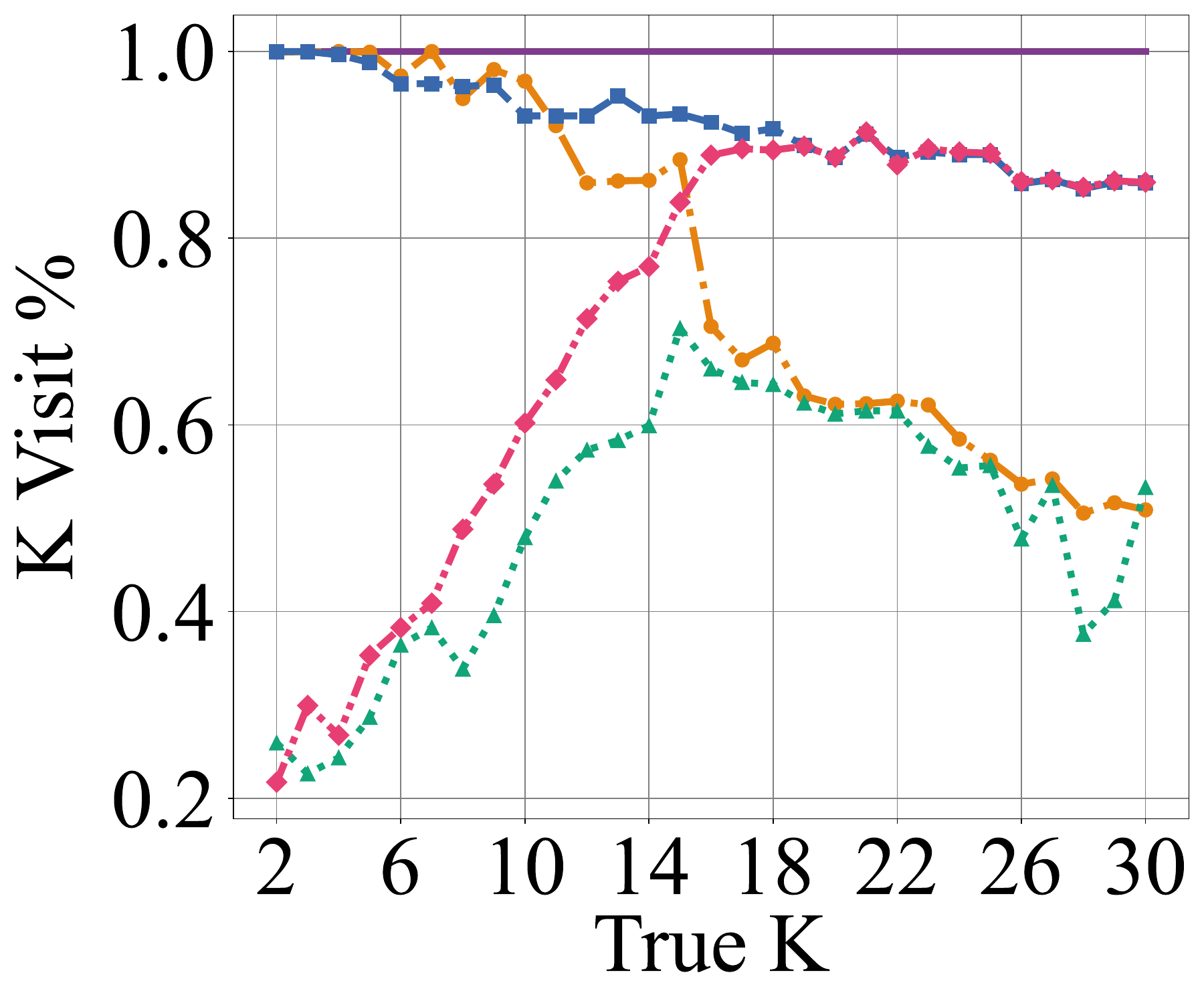}
            
                \label{fig:kmeans_ressults}
            \end{figure}
    \end{minipage}
\hfill
\vspace{0pt} 
\centering
    \begin{minipage}[b]{0.97\columnwidth}
          \begin{figure}[H]
            \centering
             \includegraphics[width=\columnwidth]{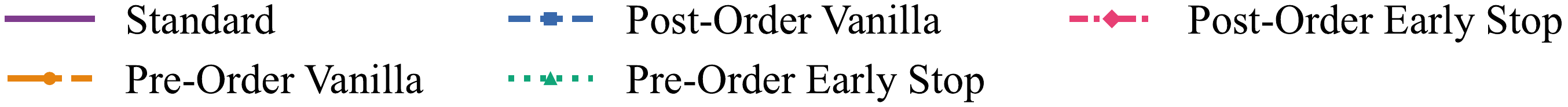}
            
                \label{fig:axis_result}
            \end{figure}
    \end{minipage}
    
\caption{ Standard NMFk (left) and K-means (right), Vanilla, and Early Stop over Pre-order, and Post-order traversal sorting.}
\label{fig:combined_results_kmean_nmfk}
\end{figure}

\textbf{NMFk}: Data was generated using a synthetic data generator with random Gaussian features for a predetermined $k$, where $k_{true}=\{2,3,\cdots,30\}$.  The shape of 30 matrices was sized 1000 by 1100, resulting in 1.1 million non-negative entries. The $k_{true}$ predetermined the number of clusters in the data, typically showing a drastic drop-off in the scoring metric for subsequent $k$ values. The evaluation criterion was the silhouette scores of the proposed clusters for the visited $k$. For the NMFk Binary Bleed trials, three separate instances were operated over the same data: Standard NMFk, Binary Bleed Vanilla, and Binary Bleed Early Stopping. Each instance was evaluated on all 30 matrices with $k$ search space $\mathcal{K}=\{2,3,\cdots,30\}$. The results corresponding to NMFk Vanilla and NMFk Early Stopping for $k = 15$ and $k = 8$, respectively, are Figure~\ref{fig:2x2grid}'s top row. It can be seen that Binary Bleed prunes multiple $k$ values, whereas the standard method must visit all $\mathcal{K}$.
In the overview, Figure \ref{fig:combined_results_kmean_nmfk}, orange and blue lines show the Binary Bleed Vanilla algorithm and the downward trend of $k$ visits relative to the standard, where Pre-order finds the optimal $k$ in fewer overall visits. Similarly, Early Stop in pink and green have lower overall visits than Vanilla, with Pre-order benefiting more despite a slightly increasing trend for both Early Stop lines over $k_{true}$. Interestingly, the post-order Early Stop is as fast as the Pre-order and visits more $k$ than the Binary Bleed Vanilla, which can be attributed to the number of compute resources paired with which values are $k_{true}$ and the ordering of $\mathcal{K}$.  Overall, the algorithms visit the following percentages of $\mathcal{K}$: Pre-order Vanilla: 56\%, Post-order Vanilla: 76\%, Pre-order Early Stop: 27\%, Post-order Early Stop: 44\%, while Standard NMFk visits 100\% of $\mathcal{K}$.  This shows pre-ordering $\mathcal{K}$ with Early Stop executes fastest.

\textbf{K-means}: The data was generated by simulating Gaussian-distributed clusters with a standard deviation of .5 and the true $k$ cluster count. Overlaid random noise introduces variability and ensures robustness. $k_{true}=\{2,3,\cdots,30\}$ and  $\mathcal{K}=\{2,3,\cdots,30\}$. Given the stochastic scoring, we cluster fifty times for each $k_{true}$ on every method-ordering pair (Vanilla, K-means Standard, Early Stop) in pre- and Post-order configurations. Davies-Bouldin scoring is used to evaluate the cluster quality and to determine the correct $k$. The Root Mean Square Error (RMSE) scores of correct $k$ were as follows: Post-Order Early Stop: 1.08, Pre-order Early Stop: 2.11, Post-Order Vanilla: 1.08, Pre-order Vanilla: 1.72, and baseline Standard k-means: 1.32. These results indicate the accuracy in identifying $k_{true}$, with all Binary Bleed iterations having RMSEs of 0.79 or less, close to the standard RMSE of 1.32.
 
 K-means Vanilla and K-means Early Stopping results for $k = 18$ and $k = 9$, respectively, are shown in Figure~\ref{fig:2x2grid}. In the overview, Figure \ref{fig:combined_results_kmean_nmfk}, Early Stop, pink and green, dominate the speed-up in lower $k$, but the dominating factor transitions to Pre-order sorting after $k=15$. Average $k$ visit percentages are-- Pre-order Vanilla: 77\%, Post-order Vanilla: 92\%, Pre-order Early Stop: 50\%, Post-order Early Stop: 71\%. As reported, the percent $k$ visited shows the reduced number of visits $k$ needed to complete the optimization process.

\subsection{Multi-node Setting}
 We demonstrate a reduction in $k$ visits for topic modeling of over 2 million scientific abstracts from arXiv with NMFk from \cite{10527237}. We used the LANL Chicoma super-computer cluster on the  GPU partition and allocated ten nodes, with four NVIDIA A100s per node.  NMFk with Binary Bleed Early Stop and standard run on  $\mathcal{K}=\{2,3,\cdots,100\}$. Early stop visited 60\% of the total compared to Standard NMFk. Both agreed on the $k_{optimal}=71$ for the vocabulary size 10,280.
 
\subsection{Distributed Setting}
To showcase the efficacy of the approach on the largest dataset, we utilized the distributed NMF framework, pyDNMFk, results, and the distributed RESCAL framework, pyDRESCAL, as referenced in \cite{pyDNMFk} and \cite{bhattarai2023distributed}, respectively. Large datasets need substantial computational resources to factorize. For instance, without Binary Bleed, pyDNMFk required 2 hours with 52,000 cores to estimate $k$ using standard NMF for a 50TB dataset, averaging 17.14 minutes per $k$ for $\mathcal{K}=\{2,3,\cdots,8\}$. Similarly, pyDRESCALk required 3 hours with 4,096 cores to factorize 11.5TB of synthetic data using standard RESCAL, averaging 18 minutes per $k$ for $\mathcal{K}=\{2,3,\cdots,11\}$. Binary Bleed Vanilla and Early Stop results on this distributed data were identical, so only the former's results were considered, given that the stop thresholds were crossed on the last $k$. For both RESCAL and NMFk, the selected $k$ matched the standard.

\textbf{RESCAL}: Binary Bleed was applied to the silhouette and relative error metrics with pre- and Post-order traversal. 

\begin{figure}[htb]
    \centering
    \includegraphics[width=0.8\columnwidth]{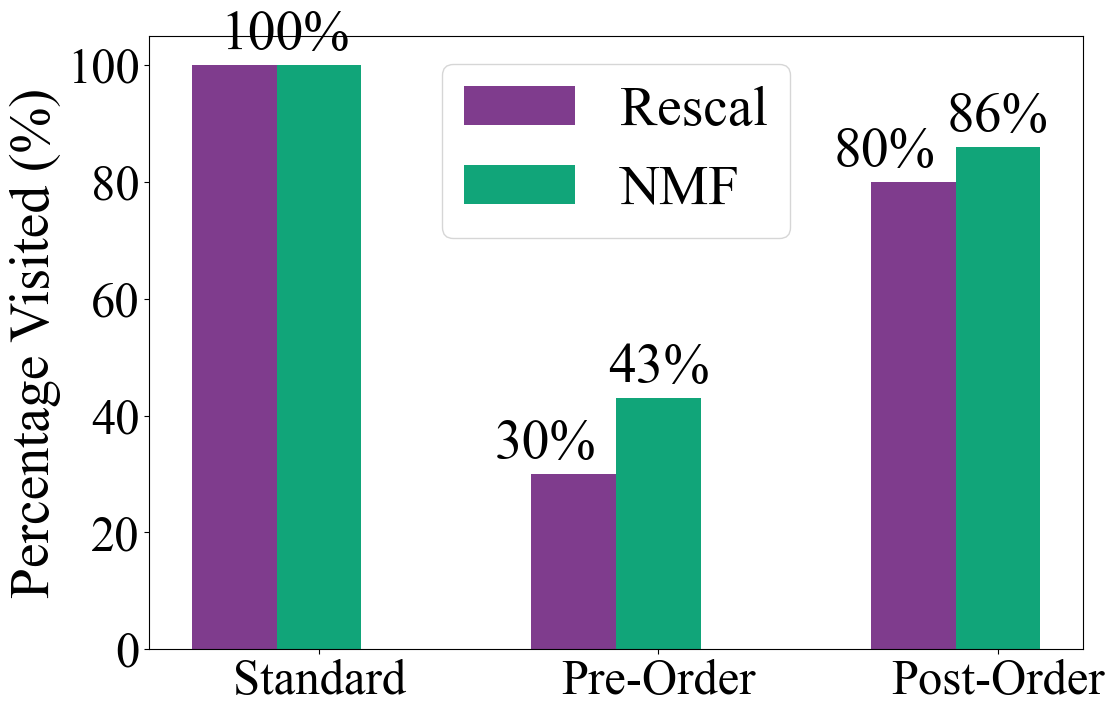}
    \caption{ Binary Bleed reduction on distributed Rescal (Purple), and distributed NMF (Green).}
    \label{fig:distributed_results}
\end{figure}
 For Pre-order traversal, 30\% of the total $k$ values were visited, resulting in an average runtime of 54 minutes, compared to the 180 minutes required for the standard RESCAL. In contrast, Post-Order traversal visited 80\% of the total $k$ values, with an average runtime of 144 minutes, as in Figure~\ref{fig:distributed_results}. 

\textbf{NMF:} Similarly, the results from \cite{pyDNMFk} were analyzed for $\mathcal{K}=\{2,3,\cdots,8\}$. Using Pre-order and post-order traversal, we evaluated the Binary Bleed Vanilla algorithm for the silhouette score. For Pre-Order traversal, 43\% of the total $k$ values were visited, resulting in an average runtime of 51.43 minutes, compared to the 120 minutes required for the standard NMF. Post-order traversal visited 86\% of the total $k$ values, with an average runtime of 102.86 minutes, as shown in Figure~\ref{fig:distributed_results}. These results further emphasize the efficiency of the proposed approach in reducing computational time while ensuring effective factorization.

%% file: Part3/smart-slic/cyber-security.tex
\section{Cyber-Security KG via HNMFk}
\begin{figure*}[h]
\centering
\includegraphics[width=.95\linewidth]{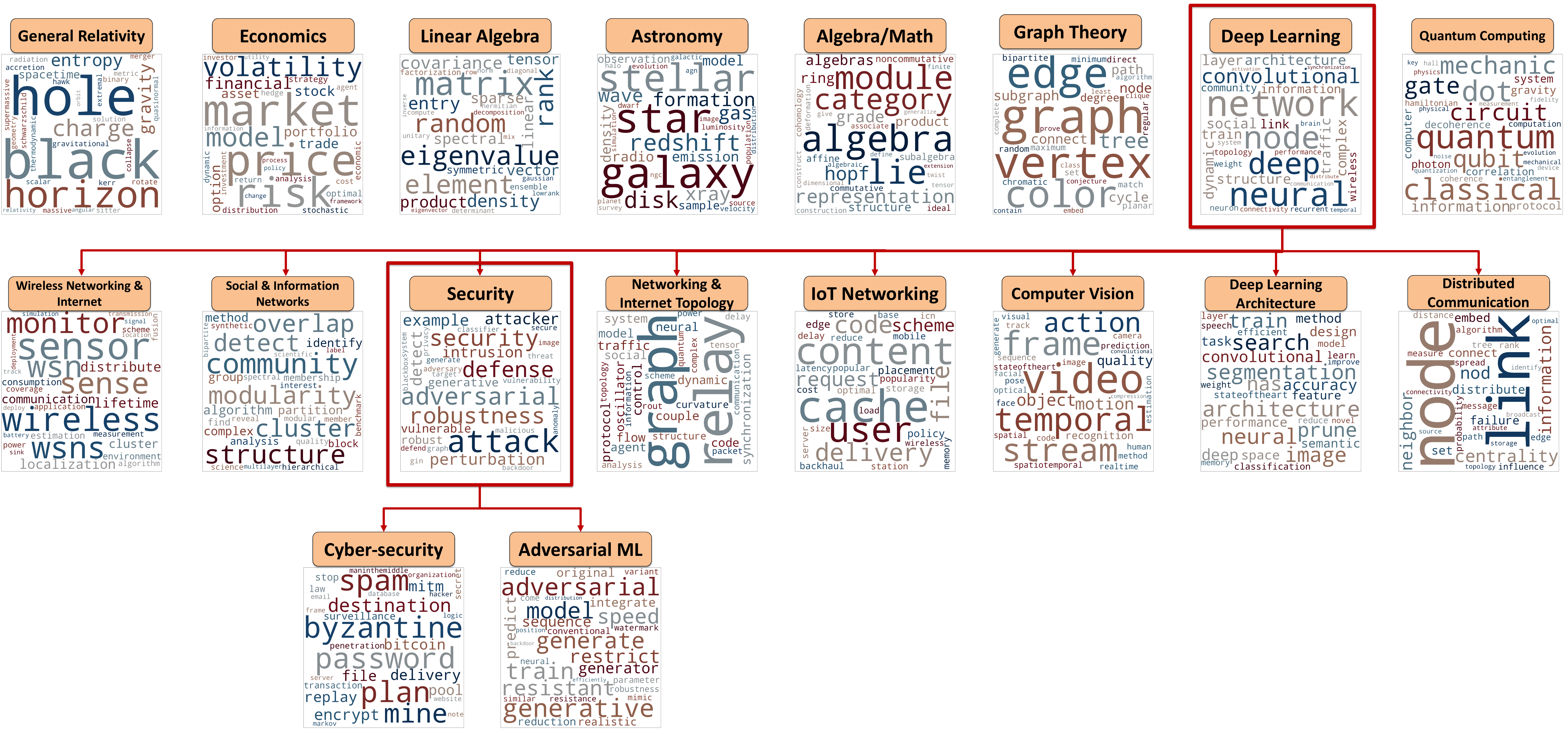}
\captionsetup{width=0.95\textwidth}
\caption{Word cloud of selected topics and their interpreted labels. The method is hierarchically applied three times, selecting a topic to extract its subtopics at each level. Selected topics from each level are shown in each row of the figure. \label{fig:topics}}
\vspace{-1em}
\end{figure*}

\begin{figure}[H]
\centering
\includegraphics[width=.7\columnwidth]{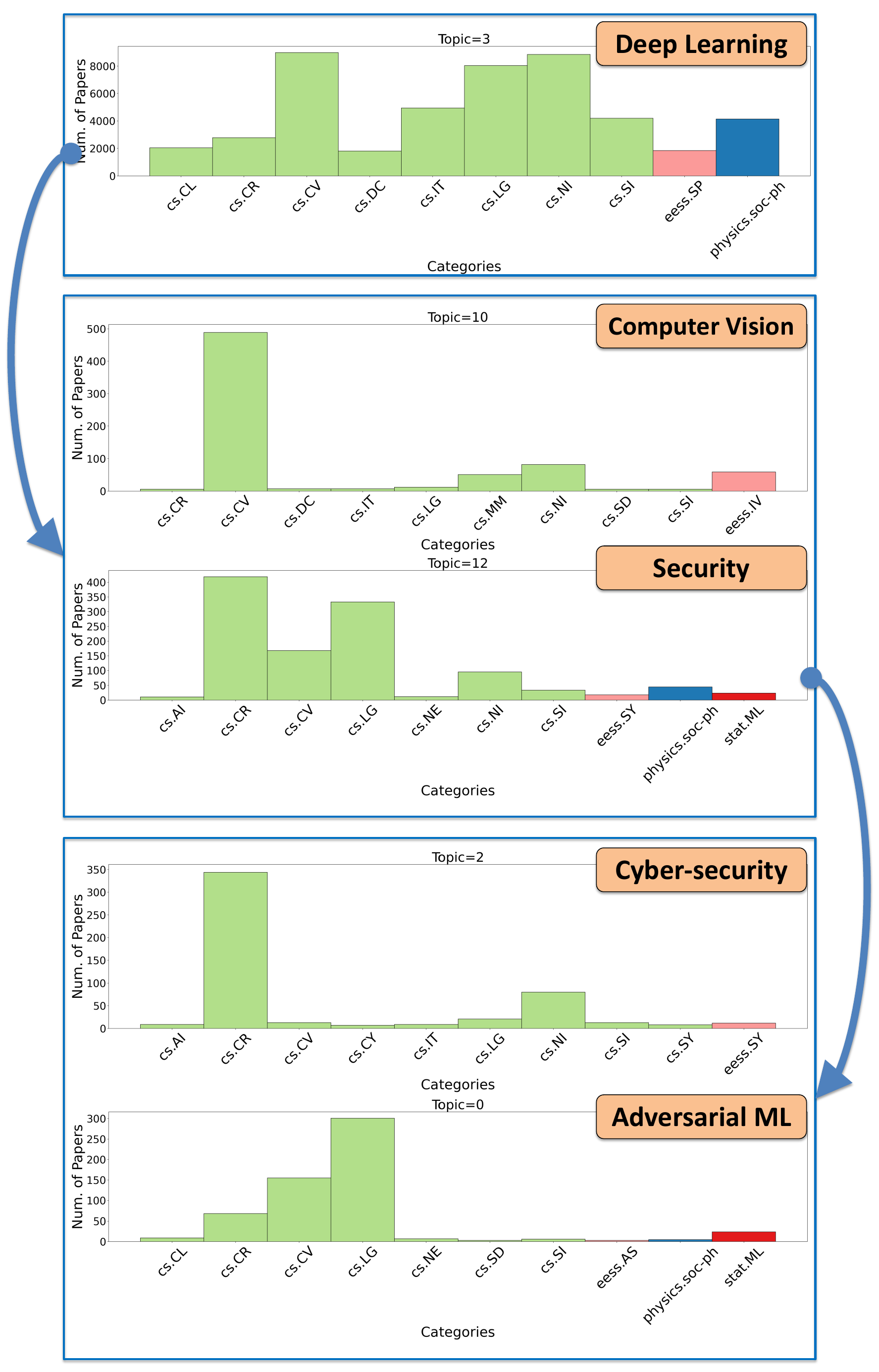}

\caption{Distribution of the top ten document categories, based on the arXiv author assignments. Each topic corresponds to the word clouds from Figure \ref{fig:topics}. Selected topics from each level are shown in each row of the figure. \label{fig:categories}}
\vspace{-1em}
\end{figure}

Results are shown in Figure \ref{fig:topics} with the word clouds (most prominent words corresponding to each column of $\mat{W}$), and with the distribution of categories in each topic and sub-topic in Figure \ref{fig:categories} (the top ten categories are shown for the given topic). While word clouds highlight the interpretability and quality of the extracted topics, the categories' distribution shows the documents' specificity, and the sub-categories are also examined. The method was hierarchically applied three times, each level selecting one topic to expand further and identify its subtopics. Since the initial $\mat{X}$ is large,  chunk the documents into 20 separate matrices and apply semantic joint factorization to determine the number of latent topics automatically. Individually, the 20 matrices revealed between 30 and 80 latent topics. After joint factorization, 24 total super-topics were identified in 2 million+ documents. 

The topic related to Deep Learning was selected. The first row of Figure \ref{fig:topics} shows the word cloud for this topic, and the first row of Figure \ref{fig:categories} displays the distribution of categories in this topic. The categories are mainly represented in diverse sub-fields of Computer Science, ranging from Cyber (\textit{cs.CR}) to Computer Vision (\textit{cs.CV}). At the second level, the sub-topic of Deep Learning produced 29 total subtopics, which include Computer Vision and Security. The sub-topic that can be interpreted as Computer Vision based on the tokens in the word cloud is now narrowed down to the \textit{cs.CV}. On the other hand, the Security sub-topic is mainly distributed between the categories \textit{cs.CR} (Cyber), \textit{cs.CV} (Computer Vision), and \textit{cs.LG} (Language); therefore, the Security sub-topic was expanded to the third level, further separating it into three sub-sub-topics, two of which can be interpreted as Cyber-security and Adversarial ML based on the word clouds. Similarly, at the third level, the Cyber-security topic category distribution is narrowed to \textit{cs.CR}. The Adversarial ML topic contains documents from three main categories of \textit{cs.CR}, \textit{cs.CV}, and \textit{cs.LG}, which is reasonable based on this field and how research focuses on adversarial computer vision and language models.

In the domain-specific leaf decomposition, 1383 vocabulary words were used to discover three latent topics: the counts of the entities in the graph. These three topics contain their own unique, non-overlapping set of documents. Specifically, topic 0 had 597 documents, topic 1 had 114 documents, and topic 2 had 565 documents. These topics comprised 1383 vocabulary words, but the probability of the words occurring in each topic changes according to the documents. The knowledge graph contained NER entities such as 245 organizations, seven events, 64 persons, 3 locations, 679 products, and 22 geopolitical entities. Overall, the graph produced 3758 node entities and 9428 edge relationships. Queries could be asked of the knowledge graph, which must first be translated from natural language into Cypher. For example, 'Which papers mention MNIST?' returns all of the documents with MNIST as an NER, a count of 50 documents. Similarly, 'Which documents in the security decomposition are in the 'cs.SE' (software engineering) arXiv category' finds seven documents, which are further linked to 2 organizations, an event, topics 0 and 2, a product, and seven additional categories.

%% file: Part3/smart-slic/Legal_KG_RAG.tex
\section{Law}
This section presents the resulting legal texts and their hierarchical decomposition, offering a detailed breakdown of the sections and cases within each document type. The following results illustrate the scope and depth of the collected data, providing a foundation for further exploration of trends and patterns across the legal corpus.
\begin{figure}[ht]
    \vspace{-.9em}
    \centering
    \includegraphics[width=.9\columnwidth]{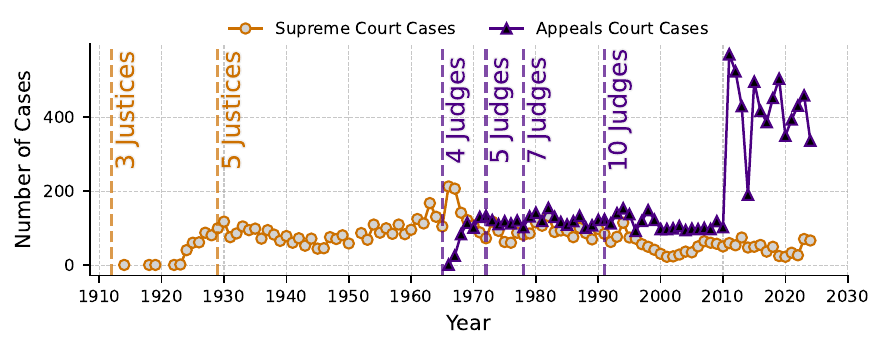}
    \vspace{-1.5em}
    \caption{New Mexico Supreme/Appeals case counts per year.}
    \vspace{-1.5em}
    \label{fig:court_cases}
\end{figure}

\subsection{Dataset}
After collecting and structuring the data, the legal documents were organized into 265 Constitutional provisions, 28,251 Statutory sections, and Case Law, comprising 10,072 Court of Appeals cases and 5,727 Supreme Court cases. Figure~\ref{fig:court_cases} shows the trends in these Supreme Court and Court of Appeals cases over the years, as available from Justia \cite{justia}, which also includes the expansions and creations of the courts themselves.

\begin{figure}[htbp]
    \centering
    \includegraphics[width=0.45\linewidth]{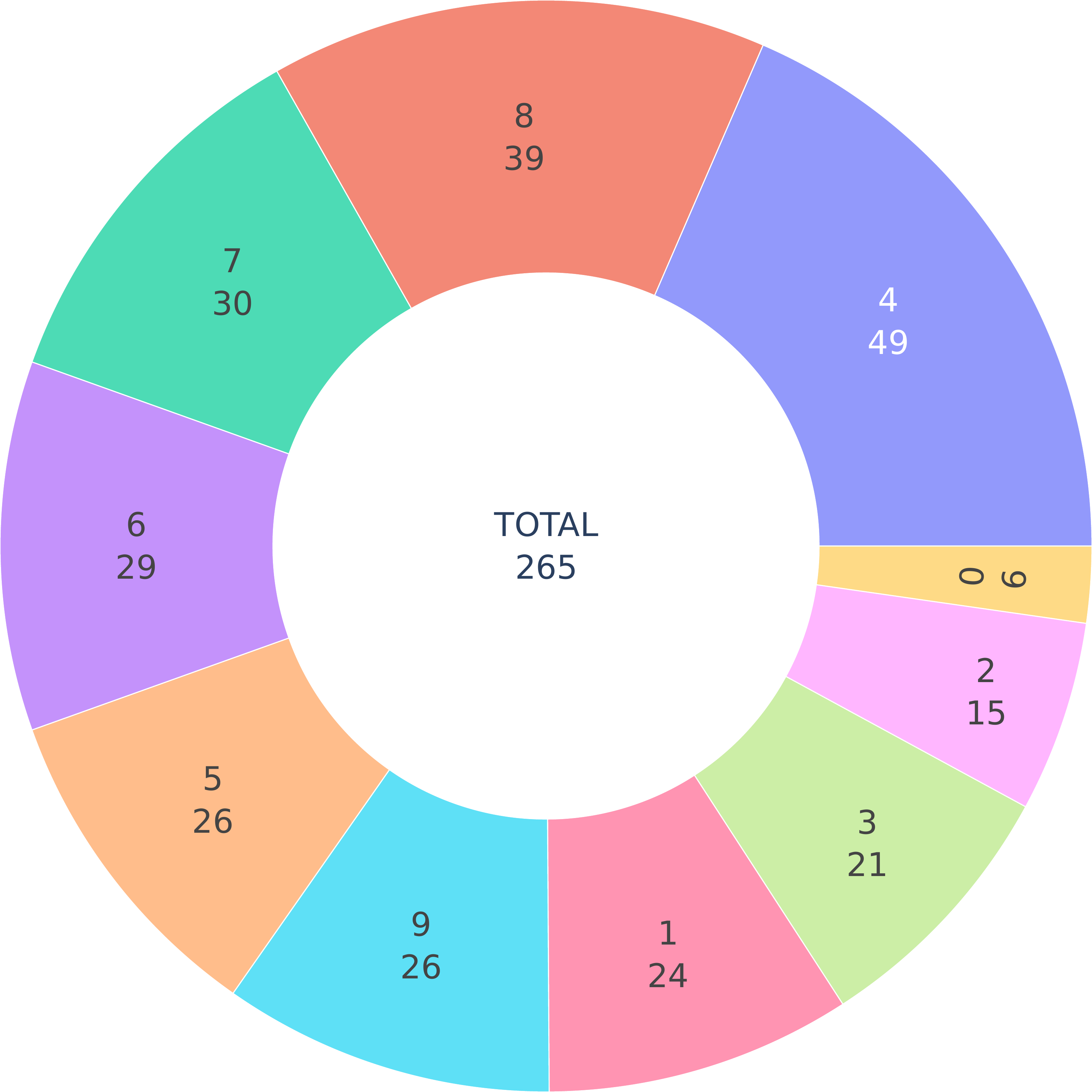}
    \caption{10 latent topics from \textbf{Constitutional Provisions}.}
    \label{fig:decomp_const}
\end{figure}

\begin{figure}[htbp]
    \centering
    \includegraphics[width=0.45\linewidth]{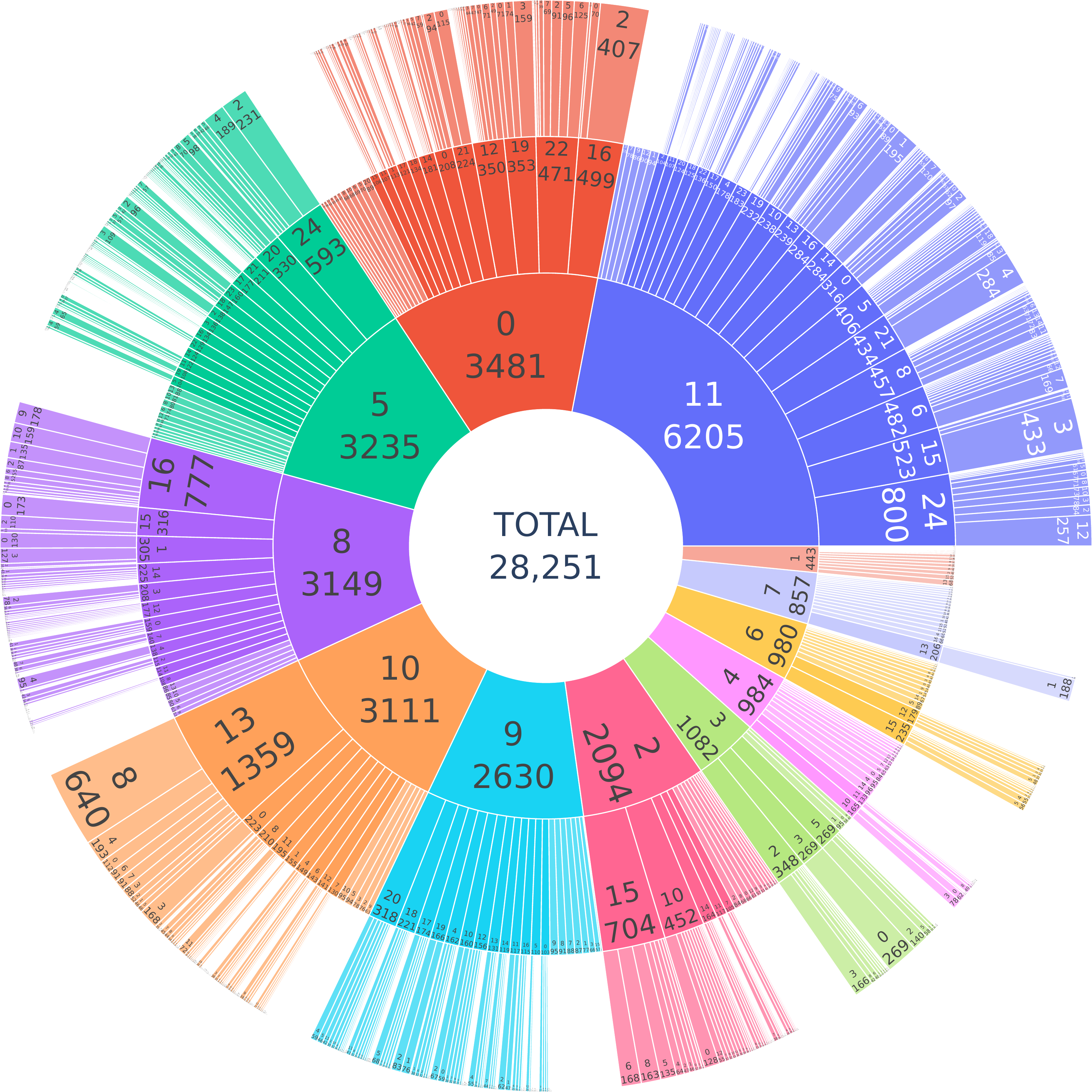}
    \caption{985 latent topics from \textbf{Statutory Sections}.}
    \label{fig:decomp_stat}
\end{figure}

\begin{figure}[htbp]
    \centering
    \includegraphics[width=0.45\linewidth]{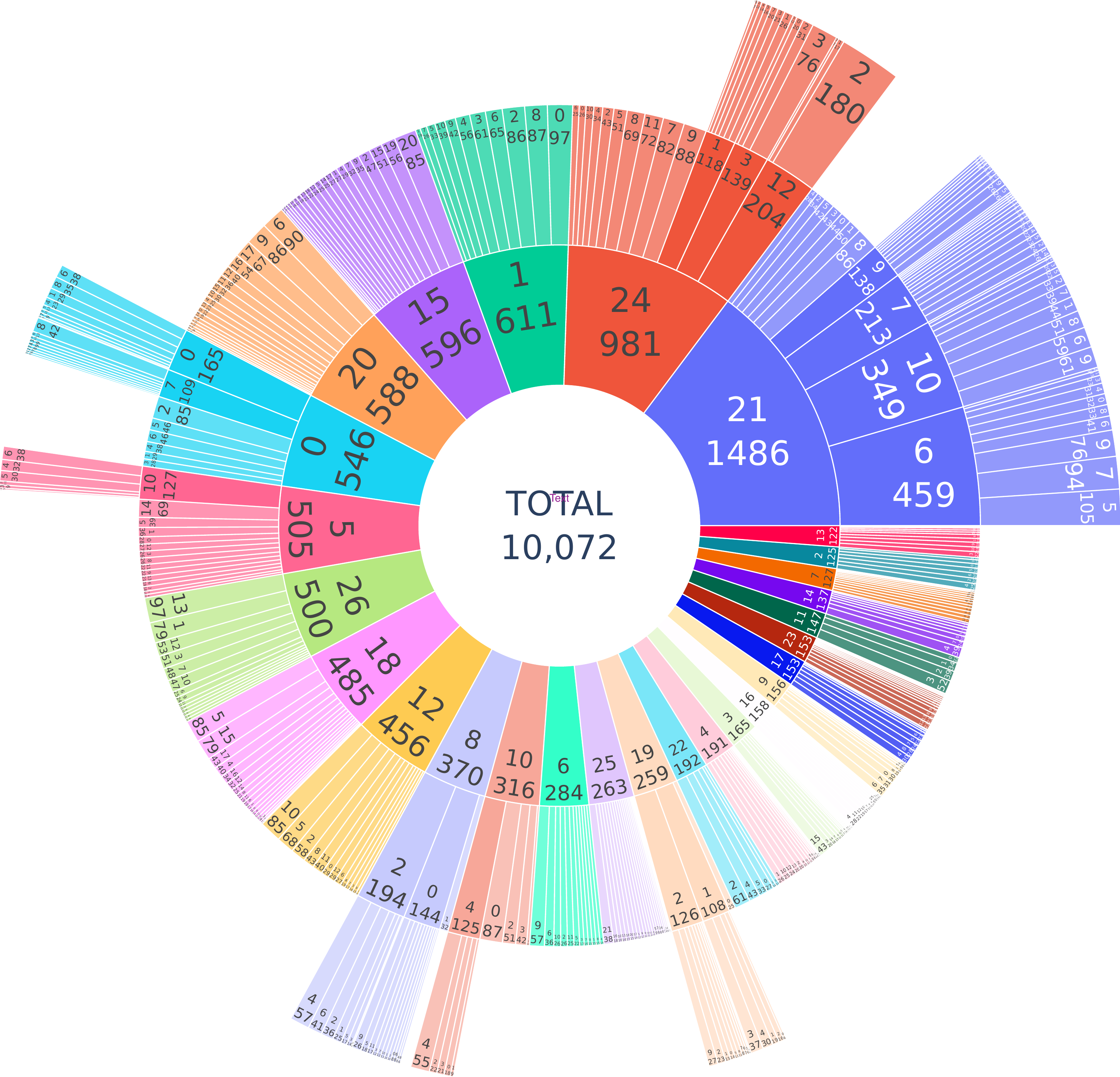}
    \caption{420 latent topics from \textbf{Court of Appeals Cases}.}
    \label{fig:decomp_app}
\end{figure}

\begin{figure}[htbp]
    \centering
    \includegraphics[width=0.45\linewidth]{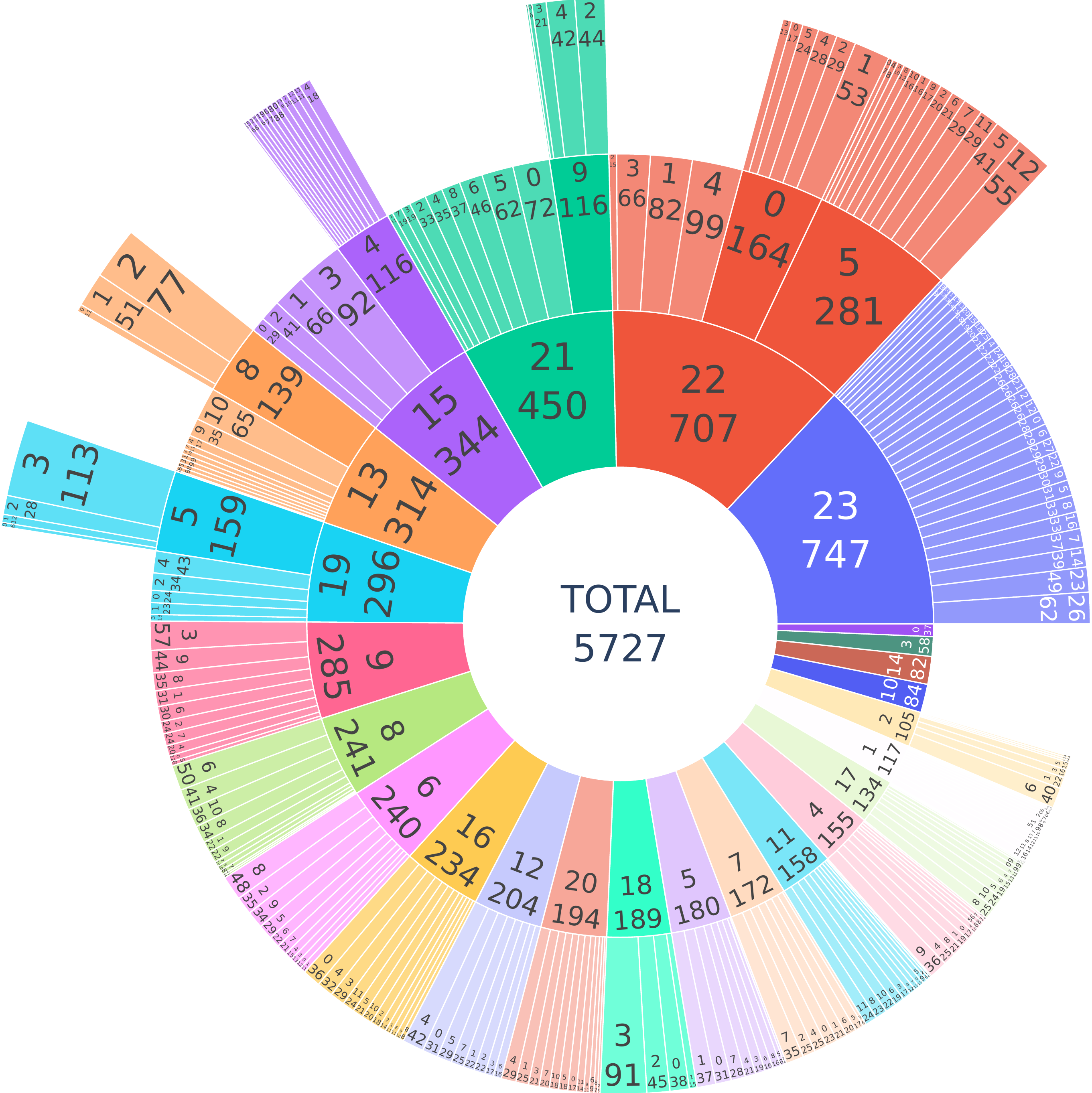}
    \caption{248 latent leaf topics from \textbf{Supreme Court Cases}.}
    \label{fig:decomp_sup}
\end{figure}

\subsection{Decomposition}
The four component data types were decomposed hierarchically with NMFk. The vocabulary for building the TF-IDF matrix was collected using specific parameters for each part. For the Constitution vocabulary, the minimum token document frequency (DF) was set to 5 documents, with a maximum token DF of 80\% of the corpus, resulting in a final size of 416 tokens. For the Statutes vocabulary, the minimum token DF was set to 30 documents, with a maximum token DF of 70\% of the corpus, yielding 7,508 tokens. The Court of Appeals vocabulary used a minimum token DF of 50 documents (cases) and a maximum token DF of 70\% of the corpus, resulting in 10,189 tokens. Last, the Supreme vocabulary employed a minimum token DF of 50 papers and a maximum token DF of 70\% of the corpus, with a final size of 8,425 tokens.
The maximum depth was set to 2, of which only the constitution sections did not reach due to the limited number of sections. The limiting factor of further decompositions was 100 documents, so if the preceding cluster had $\geq 100$, the cluster would decompose or otherwise stop. As seen in Figure \ref{fig:decomp_const}, the largest H-cluster is from cluster 4 in dark blue with 49 sections, with the fewest documents in yellow with six sections in H-cluster 0. The other three decompositions can be examined in Figures~\ref{fig:decomp_const}, \ref{fig:decomp_stat}, \ref{fig:decomp_app}, and \ref{fig:decomp_sup}, where the court cases, there are 10 leaf H-clusters in the constitution decomposition, 985 leaf H-clusters in the statutes decomposition,  420 leaf H-clusters in the Court of Appeals cases, and 248 leaf H-clusters in the Supreme Court cases. From the methods in \cite{topic_tag}, each H-cluster throughout the decomposition hierarchies has LLM-generated labels for ease of reference and quick insight. Labels for the first decomposition depth for the constitution can be observed in Table \ref{tab:Constitution},  the statutes in Table \ref{tab:statutes}, the Supreme Court in Table \ref{tab:supreme}, and the appeals in Table \ref{tab:appeals}. The depth-0 H-clustering corresponds to the first ring radial from the center totals in the several Figures~\ref{fig:decomp_const}, \ref{fig:decomp_stat}, \ref{fig:decomp_app}, and \ref{fig:decomp_sup}.

\definecolor{darkgreen}{rgb}{0.0, 0.6 , 0.0}
\definecolor{lightgreen}{rgb}{0.0, 0.9, 0.0}
\definecolor{goldyellow}{rgb}{1.0, 0.84, 0.0}
\definecolor{darkorange}{rgb}{1.0, 0.55, 0.0}
\definecolor{lightblue}{rgb}{0.0, 0.9, 1.0}
\definecolor{darkpurple}{rgb}{0.4, 0.0, 0.4}

\begin{longtable}{@{}p{0.02\linewidth}p{0.89\linewidth}@{}}
\caption{NM Constitutional Depth-0 H-Clusters} \label{tab:Constitution} \\
\toprule
\textbf{\#} & \textbf{Constitution Label} \\
\midrule
\endfirsthead
\toprule
\textbf{\#} & \textbf{Constitution Label} \\
\midrule
\endhead
\midrule
\multicolumn{2}{r}{\textit{Continued on next page}} \\
\endfoot
\bottomrule
\endlastfoot
0 & Irrigation and Water Resource Management Principles \\
1 & Regulation of Private Sector Influence on Public Schools and Education Services \\
2 & Branches of Government Structure and Functionality \\
3 & Education Funding for New Mexico's Educational Institutions \\
4 & Legislative Proceedings and Lawmaking Activities Enacted During Sessions \\
5 & Territorial Transition: Constitutional Ratification and Statehood Provisions \\
6 & Governance, Land, and Taxation Framework \\
7 & Municipal and County Financial Obligations and Liabilities Management \\
8 & Judicial Power and Jurisdictional Frameworks \\
9 & Public Service Election Governance Structure and Processes \\
\end{longtable}

\begin{longtable}{@{}p{0.02\linewidth}p{0.89\linewidth}@{}}
\caption{NM Statutory Depth-0 H-Clusters} \label{tab:statutes} \\
\toprule
\textbf{\#} & \textbf{Statute Label} \\
\midrule
\endfirsthead
\toprule
\textbf{\#} & \textbf{Statute Label} \\
\midrule
\endhead
\midrule
\multicolumn{2}{r}{\textit{Continued on next page}} \\
\endfoot
\bottomrule
\endlastfoot
0  & Municipal Court Civil Cases Involving Children's Rights \\
1  & Public Education Infrastructure Management Systems \\
2  & Criminal Codes, Local Governance Boundaries, Licensing Rules \\
3  & Comprehensive Emergency Health and Human Services Response Framework \\
4  & Taxation and Revenue Collection Oversight \\
5  & Military Decorations, Licensing Procedures, Governance Boards \\
6  & Regulatory Insurance Contract Law and Policy Analysis \\
7  & Public Obligations Financing and Project Bonds Issuance \\
8  & State Government Investment Grants for Education and Economic Development \\
9  & Irrigation and Water Rights Regulations \\
10 & Military Honors, Discrimination Penalties, and Trust Authority \\
11 & Corporation Governance Framework and Regulatory Compliance \\
\end{longtable}

\begin{longtable}{@{}p{0.02\linewidth}p{0.89\linewidth}@{}}
\caption{NM Appeals Court Case Law Depth-0 H-Clusters} \label{tab:appeals} \\
\toprule
\textbf{\#} & \textbf{Appeals Court Label} \\
\midrule
\endfirsthead
\toprule
\textbf{\#} & \textbf{Appeals Court Label} \\
\midrule
\endhead
\midrule
\multicolumn{2}{r}{\textit{Continued on next page}} \\
\endfoot
\bottomrule
\endlastfoot
0  & Real Property Rights and Interests \\
1  & Civil Liability and Injury Issues Arising from Healthcare Services \\
2  & Parental Rights and Custody Proceedings Involving Disputed Parental Fitness \\
3  & Motor Vehicle Insurance and Liability Claims Processing Procedures \\
4  & Mortgage Foreclosure and Secured Lending Frameworks \\
5  & Arraignments, Tribal Jurisdiction, Divorce, Bond Conditions, Motor Licensing \\ 
6  & Courtroom Advocacy and Representation Strategies \\
7  & Petitioner's Right to Parental Decision Making \\
8  & Fourth Amendment Protections Against Unlawful Searches and Seizures \\
9  & Taxation of Gross Receipts and Sales Transactions in a Business Context \\
10 & Key Legal Concepts Related to Sexual Offenses \\
11 & Child Protective Services Laws and Regulations \\
12 & Work-Related Injury and Disability Compensation Process \\
13 & Healthcare Contract Dispute Resolution Processes \\
14 & Marital Property and Financial Disposition \\
15 & Legal Proceedings Involving Jury Determination of Guilt \\
16 & Probation Supervision and Monitoring Oversight Process \\
17 & Juvenile Protection, Miranda Rights, Evidence Collection \\
18 & Firearms, Substances, DWI, and Sentencing \\
19 & Sentencing Enhancements for Serious Repeat Offenders \\
20 & Traffic Stop Under Suspicion with Mandatory Blood Alcohol Testing \\
21 & Criminal Proceedings Trial Litigation Documentation and Record-Keeping Procedures \\
22 & Workers' Rights and Insurance Benefits \\
23 & Children's Welfare and Family Reunification Efforts \\
24 & Public Municipal Legal Frameworks and Governance Structure \\
25 & Business Disputes, Wrongful Injury, Taxation Appeals, Workers’ Compensation \\ 
26 & Administrative Disputes, Native American Legal Memorandums \\
\end{longtable}

\begin{longtable}{@{}p{0.02\linewidth}p{0.89\linewidth}@{}}
\caption{NM Supreme Court Case Law Depth-0 H-Clusters} \label{tab:supreme} \\
\toprule
\textbf{\#} & \textbf{Supreme Court Label} \\
\midrule
\endfirsthead
\toprule
\textbf{\#} & \textbf{Supreme Court Label} \\
\midrule
\endhead
\midrule
\multicolumn{2}{r}{\textit{Continued on next page}} \\
\endfoot
\bottomrule
\endlastfoot
0  & Arbitration of Contract Disputes and Judicial Decision-Making in Motor Vehicle Cases \\
1  & Revenue and Taxation Frameworks in Governance and Administration \\
2  & Mineral Rights Leases \\
3  & Native American Self-Governance and Tribal Jurisdictional Frameworks \\
4  & Municipal Zoning Ordinances and Regulations of Local Governance Areas \\
5  & Damages Award for Wrongful Conduct Against Business Partners \\
6  & Jurisdictional Appeals, Juvenile Sentencing, Felony Enhancements, Parole Terms \\
7  & Post-Divorce Asset Distribution and Management Strategies \\
8  & Constitutional Challenges to Public Education Governance \\
9  & Secured Financial Instruments and Property Rights \\
10 & Water Resource Allocation and Management \\
11 & Electric Utility Rate Regulation Oversight Authority \\
12 & Law Enforcement Procedures and Rights Protection under Fourth Amendment Protections \\
13 & State Legislative Review Process Decisions \\
14 & Denial of Petition for Habeas Corpus \\
15 & Workers' Compensation Process for Work-Related Injuries and Disabilities \\
16 & Real Estate Boundary Disputes and Conveyance Matters \\
17 & Parental Rights and Legal Guardianship Proceedings \\
18 & Appellant's Argument Against the Ruling of a Trial Court \\
19 & Document Terms and Conditions Regarding Business Transactions \\
20 & Mutual Insurance Policies for Vehicle and Individual Coverage \\
21 & Accident resulting from driver error on public roadways leads to legal consequences \\
22 & Court proceedings involving criminal trials and testimonial evidence \\
23 & Civil Litigation Proceedings and Trials \\
\end{longtable}

\subsection{Knowledge Graph}
The four data parts, 265 constitutional provisions, 28,251 Statute sections, 5,727  Supreme Court cases, and 10,072 Court of Appeals cases, were inserted into the neo4j \cite{neo4j2023} knowledge graph. 
The graph's number of nodes and edges can be seen in Table \ref{tab:neo4j_counts}, where edge counts are where the triplet's tail originates with the row's node. The legal citations were collected by iterating the text of each case or section into chat-gpt-3.5-turbo with the following prompt: "You are an expert legal document analyzer. Your job is to find all references to the Constitution, Case Law, or Statutes in the text." The result was that the LLM acted like a named entity extractor, such that any citations in the text were pulled out in an enumerated list. The citations mainly included the cases, statutes, and constitution of New Mexico, but also had references to the United States Constitution, New Mexico Administrative Code, and New Mexico Rules Annotated (NMRA). The NMRA had many references to Uniform Jury instructions and the rules of criminal and civil proceedings.

\begin{figure}[H]
    \centering
    \includegraphics[width=\columnwidth, trim=15 0 15 5, clip]{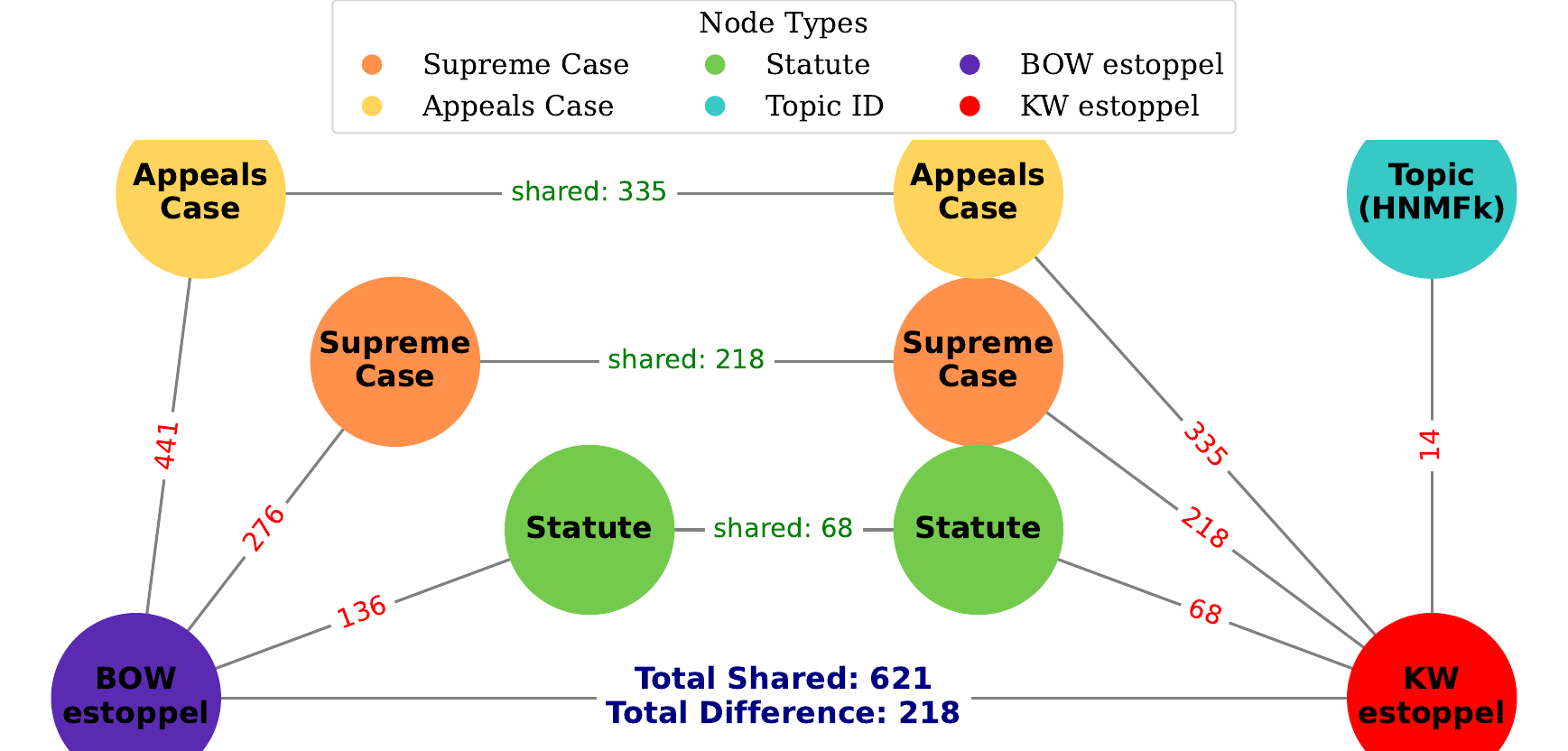}
    \caption{ Examination of `Estoppel' relating to being a keyword in topics, vs bag of words vocabulary. }
    \label{fig:estoppel}
\end{figure}

In figure \ref{fig:estoppel}, the NMFk topic keyword and a bag-of-words vocabulary were both queried for `estoppel.' The NMFk keyword is red, the topics show in a light blue node, and the BOW node is purple. Three node types occur for this keyword: the statutes are green nodes, the Court of Appeals cases are yellow, and the Supreme Court cases are orange. The constitution neither clustered over the term nor mentions it, which is not represented in Figure \ref{fig:estoppel}. There is a difference in the data that have topics associated with estoppel vs all of the documents that mention 'estoppel'. Still, not all documents mentioning `estoppel' were clustered with the word, which means that other terms and concepts from those terms had more importance for the documents on the left side of the image than `estoppel'. There are 14 topics, 441 Court of Appeals cases,  276 Supreme Court Cases, and 136 Statutes in Figure \ref{fig:estoppel}. 

\begin{table}[t!]
\centering
\caption{Neo4j Node and Edge Overview}
    \vspace{-1em}
\label{tab:neo4j_counts}
\definecolor{headerblue}{HTML}{CFE2F3} 
\definecolor{lightgray}{gray}{0.92}     
\begin{tabular}{@{}lrrr@{}}
  \toprule
  \textbf{Node Type}      & \textbf{Nodes}   & \textbf{Out Edges}   & \textbf{Legal Cites} \\
  \midrule
  NMFk Topics             & 2,469            & 92,634               & --                   \\
  NMFk Keyword            & 11,076            & 8,281,843                   & --                   \\
  BOW Vocabulary          & 132,423          & --                   & --                   \\
  Constitution     & 265              & 9,067                & 41                   \\
  Statute          & 28,251           & 1,930,707            & 81,353               \\
  Supreme Court Case      & 5,727            & 2,437,161            & 76,478               \\
   Court of Appeals Case       & 10,072           & 4,176,288            & 131,230              \\
  \midrule
  \textbf{Total Unique}   & \textbf{190,283} & \textbf{16,927,700}  & \textbf{289,102}     \\
  \bottomrule
\end{tabular}
\end{table}

These topics have `estoppel' in their top keywords, whereas if every topic that contained `estoppel' in BOW words were called, there would be 328 topics. Of the 14 topics, one was connected to the Statutes, \scriptsize ``Collection and recovery of liabilities made to board members with errors and omissions''. 
\normalsize Three of the 14 topics were connected to Supreme Court cases: 
\scriptsize ``Employment Rights and Property Disputes in New Mexico Municipal Affairs'', ``Public Corporation Property Taxation Matters and Disputes with Licensing Authorities'', ``Drilling and Gas Agreement Terms Regarding Oil Wells''. 
\normalsize Finally, the remain 10 of the 14 topics with `estoppel' in its top words were connected to cases from the Court of Appeals: \scriptsize "Litigation outcomes and jurisdictional limitations", "Administrative License Revocation Proceedings by Division Officers", "Legal Proceedings and Litigation Issues in a Medical Context", "Corporate Governance and Financial Management Matters", "Dispute Resolution Process for Agricultural Property Transactions", "Motor Vehicle Administrative License Actions", "Insanity Defense Expert Witness Testimony", "Malpractice claim within time constraints", "Employer Liability for Federal Disability Claims Against Administration Agencies", "Criminal offenses and doctrine involve multiple types of larceny charges", "Collection and recovery of liabilities made to board members with errors and omissions", "Public Corporation Property Taxation Matters and Disputes with Licensing Authorities", "Employment Rights and Property Disputes in New Mexico Municipal Affairs", "Drilling and Gas Agreement Terms Regarding Oil Wells".
\normalsize
\vspace{-1em}

\input{Part3/smart-slic/law_LLM_responses/revised_result_subsection}

%% file: Part3/smart-slic/law_LLM_responses/revised_result_subsection.tex
\subsection{Domain-Specific Evaluation}
Questions were formulated to query information about legal concepts from different LLM channels. In one experiment, five questions from each data source, namely, the constitutional provisions, the statutes, the Court of Appeals opinions, and the Supreme Court cases, were followed by five additional quantity-based questions. 

To assess our domain-specific RAG system's accuracy in retrieving quantitative data without hallucinations from our KG and to verify the system's integrity, we generated the first group of 25 questions using ChatGPT-3.5. These questions were then verified through the SME lawyer. This method reflects established practices where LLMs create evaluation questions. Additionally, our evaluation targeted the system's ability to answer highly specific queries assessable through non-expert term searches, highlighting the limitations of general models in addressing such inquiries.

\begin{figure}[H]
    \centering
    \includegraphics[width=.9\columnwidth]{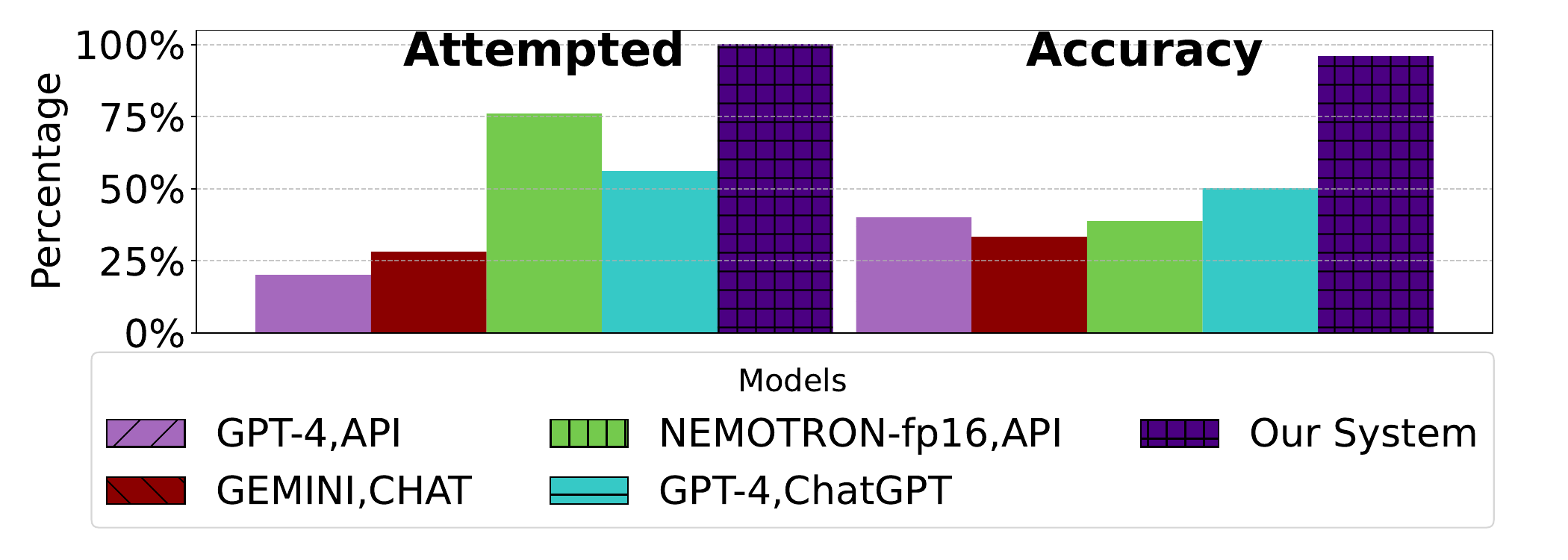}
    \caption{25 questions queried across five different LLM channels. ``Attempts'' indicate the percentage of responses that tried to answer the question, and ``accuracy'' represents the percentage of correct attempts.}
    \label{fig:attempted_accuracy_comparison}
\end{figure}

The first 25 questions were used to generate the attempted accuracy results shown in Figure~\ref{fig:attempted_accuracy_comparison}. In another experiment, a total of 60 questions were formulated by a legal subject matter expert. 

The models evaluated differ slightly between the two experiments. For the 25-question study (see Figure~\ref{fig:attempted_accuracy_comparison}), the systems compared were:
OpenAI's GPT-4 accessed via the API, Google's Gemini via web chat, Nvidia's nemotron:70B-Instruct, OpenAI's GPT-4 accessed via a web chat, and our system.

Answering legal questions becomes increasingly challenging with larger datasets. For example, while statutes have a natural hierarchical organization that enables LLMs to train on and summarize them internally, case law consists of unstructured, lengthy texts that are more difficult to process. Although constitutional questions were more manageable, many models mentioned only articles (rather than specific sections) when finer detail was requested, citing frequent changes in how sections are enumerated.

Our evaluation procedure marks an attempt as zero if a model states it cannot answer the question, and as one for any non-empty attempt. Accuracy is then measured on a scale from 0 to 3: 
\begin{itemize}
    \item 3 is awarded for an entirely correct or nearly correct answer (allowing for some uncertainty in absolute quantities),
    \item 2 for a response that is primarily correct but with minor misunderstandings,
    \item 1 when some truth is present despite significant error,
    \item and 0 for completely wrong or hallucinated responses.
\end{itemize}
Occasionally, models were given points for close numeric estimates even if those were based on flawed ``database lookups.'' In contrast, our system provides traceable reasoning by referencing a KG and by analyzing the decomposed hierarchical structure of the documents.

\begin{figure}[H]
    \centering
    \includegraphics[width=\columnwidth]{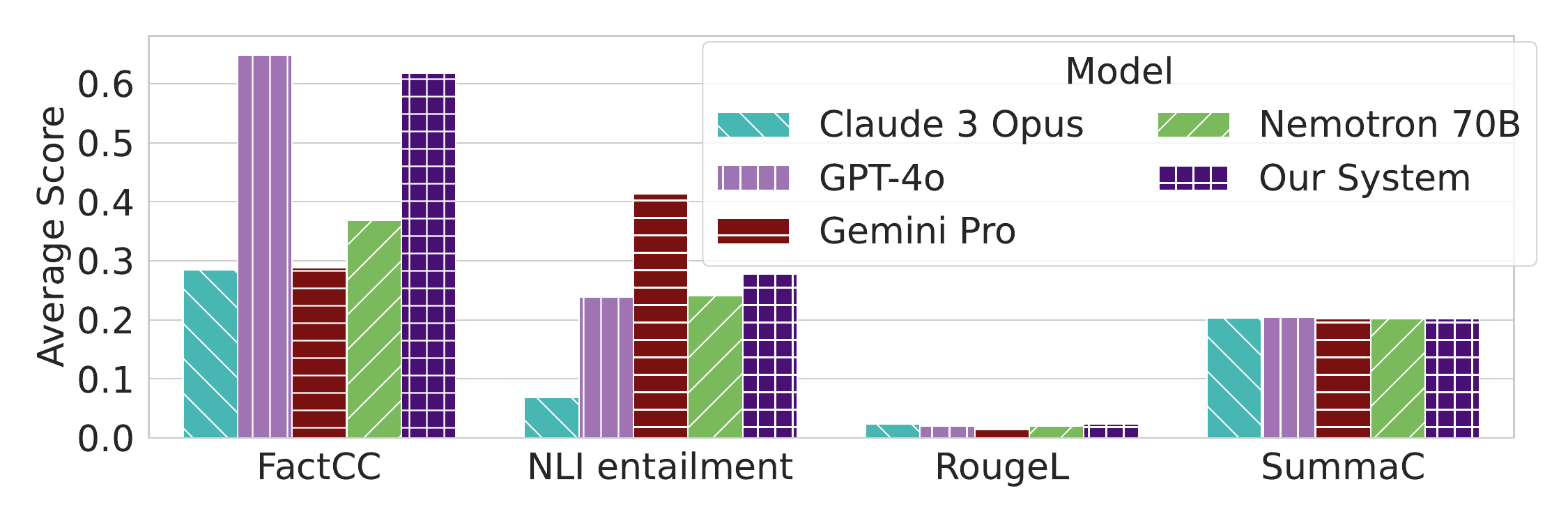}
    \caption{Comparison of metrics (ROUGEL, NLI entailment, SummaC, and FactCC) across models: Nemotron--70B-Instruct, Claude 3 Opus, GPT-4o, Gemini Pro, and our system.}
    \label{fig:model_performance}
\end{figure}

%% file: Part3/smart-slic/Malware.tex
\section{Malware Analysis Research}

This section discusses the identification of optimal clusters for tensor decomposition, vectorization of the dataset, construction of KG, and compares the system with GPT-4-instruct \cite{openai_api}  as the operating model of \textbf{SMART-SLIC} to answer research questions. The same model was used to answer without RAG as well. The findings highlight the accuracy and reliability of the \textbf{SMART-SLIC}'s RAG.

\begin{longtable}{|p{0.02\linewidth}|>{\centering\arraybackslash}p{0.56\linewidth}|>{\centering\arraybackslash}p{0.10\linewidth}|>{\centering\arraybackslash}p{0.08\linewidth}|}
\caption{Labels for Topic Clusters} \label{tab:topic_clusters} \\
\hline
\# & Label & \# Docs. & Percent \\
\hline
\endfirsthead
\hline
\# & Label & \# Docs. & Percent \\
\hline
\endhead
\hline
\multicolumn{4}{r}{\textit{Continued on next page}} \\
\endfoot
\hline
\endlastfoot

\cellcolor{green!10}0  &\cellcolor{green!10} Malware Behavioral Analysis           &\cellcolor{green!10}158 &\cellcolor{green!10}1.80\\
1                     & Cybersecurity Challenges                                   & 305                    & 3.47\\
\cellcolor{green!10}2 &\cellcolor{green!10} Cybersecurity Research                 &\cellcolor{green!10}114 &\cellcolor{green!10}1.30\\
3                     & Botnet Detection Techniques                                & 142                    & 1.62\\
\cellcolor{green!10}4 &\cellcolor{green!10} Malware Feature Selection And Extraction &\cellcolor{green!10}353 &\cellcolor{green!10}4.02\\
5                     & Network Intrusion Detection                                & 134                    & 1.52\\
\cellcolor{green!10}6 &\cellcolor{green!10} Evaluation of Malware Classifiers      &\cellcolor{green!10}301 &\cellcolor{green!10}3.42\\
7                     & Malicious Code Analysis                                    & 827                    & 9.41\\
\cellcolor{green!10}8 &\cellcolor{green!10} Artificial Intelligence for Malware    &\cellcolor{green!10}888 &\cellcolor{green!10}10.10\\
9                     & Nonnegative Matrix Decomposition                           & 520                    & 5.92\\
\cellcolor{green!10}10&\cellcolor{green!10} Security Threat Mitigation             &\cellcolor{green!10}180 &\cellcolor{green!10}2.05\\
11                    & Deep Learning for Malware                                  & 113                    & 1.29\\
\cellcolor{green!10}12&\cellcolor{green!10} Machine Learning Techniques            &\cellcolor{green!10}275 &\cellcolor{green!10}3.13\\
13                    & Education Technology                                       & 447                    & 5.09\\
\cellcolor{green!10}14&\cellcolor{green!10} Unsupervised Anomaly Detection         &\cellcolor{green!10}372 &\cellcolor{green!10}4.23\\
15                    & Ransomware Prevention                                      & 147                    & 1.67\\
\cellcolor{green!10}16&\cellcolor{green!10} Temporal Graph Forecast                &\cellcolor{green!10}307 &\cellcolor{green!10}3.49\\
17                    & Mobile Malware Detection                                   & 230                    & 2.62\\
\cellcolor{green!10}18&\cellcolor{green!10} Adversarial Defense Strategy           &\cellcolor{green!10}358 &\cellcolor{green!10}4.07\\
19                    & IoT Security                                               & 238                    & 2.71\\
\cellcolor{green!10}20&\cellcolor{green!10} Privacy Protection Challenge           &\cellcolor{green!10}628 &\cellcolor{green!10}7.14\\
21                    & Sparse Tensor Decomposition                                & 212                    & 2.41\\
\cellcolor{green!10}22&\cellcolor{green!10} Backdoor Detection                     &\cellcolor{green!10}350 &\cellcolor{green!10}3.98\\
23                    & Neural Network Architecture                                & 581                    & 6.61\\
\cellcolor{green!10}24&\cellcolor{green!10} Malware Analysis Techniques            &\cellcolor{green!10}610 &\cellcolor{green!10}6.94\\
\end{longtable}

\subsection{ Dataset}
The SME initially selected 30 documents specializing in large-scale malware analysis and anomaly detection with tensor decomposition fields as the core documents to construct the data. These documents were expanded along the citation/reference network two times. The final dataset included 8,790 scientific publications. From the cleaned corpus, the tensor object was generated.

\subsection{ Extraction of Latent Features}
After setting up the tensor, the most coherent grouping is determined by iterating through a range of \( k = \{1, 2, 3, \ldots, 45\} \) clusters to decompose. The analysis determined that 25 topic clusters represented the optimal division across all evaluated \( k \) values.  The decomposition was executed using \textbf{T-ELF} on high-performance computing resources, specifically two AMD EPYC 9454 48-Core Processors. This setup provided 192 logical CPUs, enabling us to complete the entire decomposition process in approximately 2 hours. After the decomposition, post-processing refined and defined clusters for the topics listed in Table \ref{tab:topic_clusters}.

\subsection{ Vector Store}
The 8,790 documents were vectorized and ingested into the Milvus vector store. When questions are posed to the framework, they are also vectorized using this model. Of the total documents, 22\% had full-texts available, vectorized into the Milvus. Each document and full-text had a DOI, with the full-texts also including paragraph identifiers.

\begin{figure}[h]
    \centering
    \includegraphics[width=.9\columnwidth]{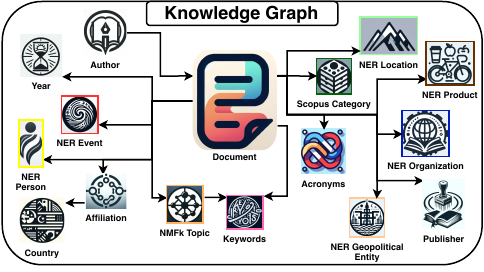}
     \vspace{-1.3em}
    \caption{  The KG schema. 
    Images generated with DALL·E \cite{dalle_tensor_decomp_arxiv_images}. }
    \label{fig:KG_SCHEMA}
\end{figure}

\subsection{ Knowledge Graph}
\begin{figure}[H]
    \centering
    \includegraphics[width=\columnwidth]{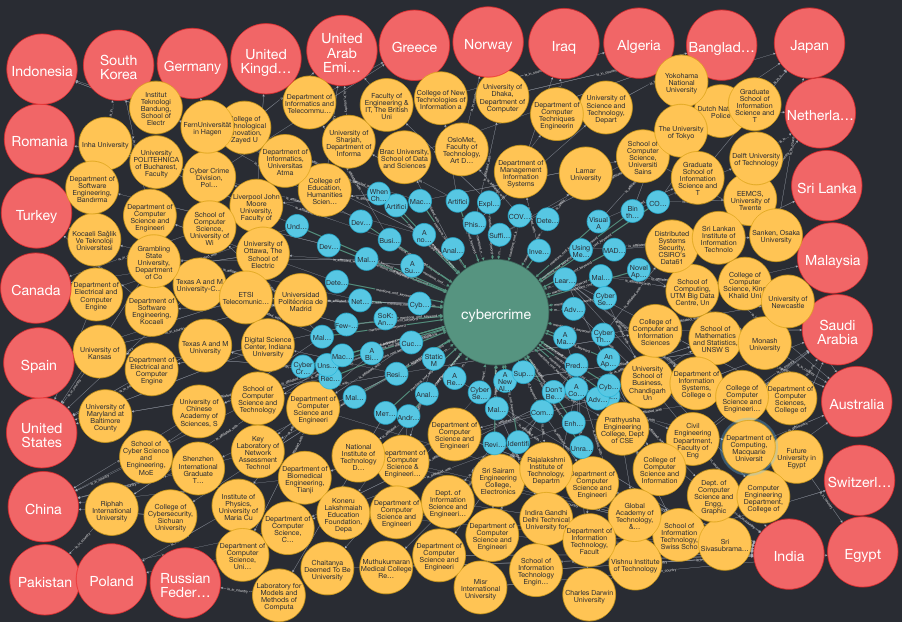}
    \vspace{-2em}
    \caption{  Keyword 'cybercrime' graph search. A single keyword (green) and linked documents (light blue) are returned. The documents also link affiliated institutions (yellow) and the country of the institutions (red). }
    \label{fig:KG_data}
\end{figure}

From the 25 clusters output from \textbf{T-ELF}, the data was formatted into 1,457,534 triplets. Once injected into the knowledge graph, there were 321,122 nodes and 1,136,412 edge relationships.  The nodes injected into the graph are represented in Figure \ref{fig:KG_SCHEMA}, where they are organized into 16 base categories, referred to as labels, that define the foundational classes for the injection process. Once the graph was built, it was directly queried for information as Structured Query Language (SQL) is directly queryable outside of an application. In Figure \ref{fig:KG_data}, the knowledge graph for the SME keyword related to cybercrime is queried. The query is structured as:
 
\begin{center}
\begin{lstlisting}[style=cypherstyle2, xleftmargin=.05\textwidth]
MATCH  (k:Keyword)-[r1]-(d:Document)-[r2]
-(aff:Affiliation)-[r3]-(c:Country )
WHERE k.term CONTAINS 'cybercrime'
RETURN k,r1,d,r2,aff,r3,c
\end{lstlisting}
\end{center}

Several relationships were navigated to retrieve the country nodes from a keyword. First, from the keyword to documents, then from documents to affiliations, and finally from the affiliations to the countries. In the cypher query, these links are denoted as an $r$ with an integer following, where $r$ is the relationship identifier.  The syntax is ()-[]-()-[]-()-[]-(), where brackets are relations and parentheses are nodes. In the first part of the ``where" clause, the keyword label is further tailored to the keyword node, such that it must contain ``cybercrime." Overall, this can answer the question, Which countries have published papers that mention cybercrime?"  The question's retrieved nodes in Figure \ref{fig:KG_data} have 29 countries in red, 99 affiliated institutions in yellow, and 65 published documents in blue.

%% file: Part3/AI_Reasoning/1.tex
This chapter shifts the emphasis from demonstrating functionality to evaluating reliability.  
Using a spectrum of graphs from controlled synthetic constructs (Dog, Swimmer, Gaussian) through biologically rich protein–protein-interaction (PPI) networks to a newly curated corpus of TMDs for superconductivity link prediction, we probe how the WNMFk, RNMFk, and BNMFk families behave under varying graph size, sparsity, and noise.  
Three themes consistently emerge.  First, Binary Bleed continues to identify a compact rank that preserves signal while containing computational cost.  Second, Boolean and real-valued decompositions reveal complementary latent factors that remain interpretable across domains.  Third, ensemble-based uncertainty estimates stay well-calibrated, enabling the system to defer judgment when evidence is thin. This is an ability confirmed not just on synthetic data and PPIs, but also on the TMD link-prediction task, where masked superconducting links are rediscovered with near-perfect top-3 precision.  Together, these findings strengthen the dissertation’s claim that a unified, principled architecture generalizes across domains without manual hyperparameter tuning.

The results are evaluated on the synthetic datasets (Dog, Swimmer, and Gaussian Data) by randomly sampling the missing links 10 times for cross-validation to ensure statistical significance. For the Boolean datasets, Dog Data and Swimmer Data, missing links are randomly sampled by stratifying on $0$s and $1$s. This enables us to test the methods' predictive capabilities on negative and positive interactions or known links. Additionally, we test the methods against data sparsity on the Dog and Gaussian datasets by systematically increasing the test-set size from 10\% to 90\% in increments of 10\%. 

\subsection{System Configuration}
The experiments were conducted on an HPC cluster named Dracarys at the Los Alamos National Laboratory (LANL).  Dracarys uses the AMD EPYC 9454 48-core processor at a clock speed of 3.81GHz. There are 192 virtual processors and a total physical RAM of 1.97 TeraBytes (TBs). The system also comprises 8 NVIDIA H100 GPUs with VRAM memory of 82 GigaBytes (GBs) each.

This section presents method performances, starting with synthetic datasets under Boolean and non-Boolean settings. We evaluate rank prediction, link prediction, performance under increasing sparsity, and UQ utility in controlled settings with pre-determined ranks. Next, we showcase results on PPI networks, demonstrating real-world effectiveness. 

\input{Part3/AI_Reasoning/ppi}
\input{Part3/AI_Reasoning/dog}
\input{Part3/AI_Reasoning/swimmer}
\input{Part3/AI_Reasoning/gaussian}
\input{Part3/AI_Reasoning/materials}

\section{Conclusion}
Across synthetic stress tests, two biological PPI graphs, and a domain-specific TMD corpus, the experiments reveal a stable pattern: factorization methods that respect data structure remain accurate, interpretable, and uncertainty-aware even as workload and sparsity shift.  Binary Bleed supplies information-efficient ranks, Boolean and real matrices expose mutually supportive latent factors, and calibrated confidence scores allow the system to withhold predictions when evidence is weak.  These outcomes confirm that dependable retrieval and reasoning start with transparent, well-behaved building blocks.  With robustness established, the next chapter turns to T-SRAG, showing how the same confidence-aware graph underpins responsive, low-hallucination question answering, completing the journey from raw text to trustworthy, domain-specific insight.

%% file: Part3/AI_Reasoning/ppi.tex
\section{PPI Datasets}
\begin{table*}[t!]
\caption{The statistics of the five PPI datasets include the number of proteins and positive and negative interaction pairs of each PPI network when represented as a matrix.}
\label{table:ppi_dataset}
\resizebox{\textwidth}{!}{
\begin{tabular}{l|c|c|c|c|c}
\hline
\rowcolor[HTML]{D3D3D3} 
\textbf{Properties/Dataset} & \textbf{H.sapiens-extended} & \textbf{Brain} & \textbf{Disease of Metabolism} & \textbf{Liver} & \textbf{Neurodegenerative Disease} \\ \hline
Number of proteins          & 14,407                       & 11,167         & 1,036                           & 10,627          & 820                                \\
Number of positive pairs    & 157,950                      & 225,200        & 5,131                           & 218,239         & 5,881                               \\
Number of negative pairs    & 157,300                      & 223,130        & 5,123                           & 215,984         & 5,879                               \\ 
\hline  
\end{tabular}
}
\end{table*}

This paper uses five PPI network datasets from \cite{ppi_symlmf}. Specifically, the PPI networks include Brain, Disease of Metabolism, \textit{H. sapiens}-extended, Liver, and Neurodegenerative Disease. The statistics of these five PPI networks are summarized in Table \ref{table:ppi_dataset}. Pre-processing of these datasets involves removing conflicting or inconsistent interactions. For example, cases where protein $i$ is shown to interact positively with protein $j$, but the dataset also includes a second entry indicating a negative interaction between the same proteins, are excluded. Additionally,  proteins that interact with fewer than five other proteins are removed, serving as a pruning approach to ensure the inclusion of proteins with sufficient interactions to allow meaningful predictions \cite{Collaborative_Filtering}.

The experiments benchmark against the symNMF method results reported in \cite{ppi_symlmf}. At the same time, it is unclear whether \cite{ppi_symlmf} applied similar pre-processing considerations, notably removing conflicting protein interactions. Therefore, the analysis also includes a comparison with LMF (symLMF without the symmetry constraint) on the pre-processed datasets used in this paper.

\begin{table}[t]
\centering
\tiny
\setlength\tabcolsep{3.5pt}      
\renewcommand{\arraystretch}{1.12} 

\begin{tabular}{lccccccccc}
\toprule
\textbf{Dataset / Metric} &
\rotatebox{60}{BNMFk} &
\rotatebox{60}{RNMFk} &
\rotatebox{60}{WNMFk} &
\rotatebox{60}{$\text{BNMFk}_{\text{lmf}}$} &
\rotatebox{60}{$\text{RNMFk}_{\text{lmf}}$} &
\rotatebox{60}{$\text{WNMFk}_{\text{lmf}}$} &
\rotatebox{60}{LMF} &
\rotatebox{60}{symLMF} \\
\midrule
\rowcolor{gray!10}\multicolumn{9}{l}{\textbf{H.sapiens‑extended}}\\
ROC AUC      & 0.755 ± 0.022 & 0.941 ± 0.001 & 0.951 ± 0.003 & \textbf{0.959 ± 0.002} & 0.955 ± 0.001 & 0.955 ± 0.001 & 0.949 ± 0.001 & 0.944 ± 0.001 \\
PR  AUC      & 0.884 ± 0.009 & 0.957 ± 0.001 & 0.964 ± 0.003 & \textbf{0.969 ± 0.002} & 0.965 ± 0.001 & 0.965 ± 0.001 & 0.934 ± 0.001 & 0.955 ± 0.001 \\
UQ – ROC AUC & 0.762 ± 0.027 & 0.942 ± 0.001 & 0.954 ± 0.003 & 0.962 ± 0.004 & 0.956 ± 0.001 & 0.957 ± 0.001 & – & – \\
UQ – PR AUC  & 0.880 ± 0.010 & 0.958 ± 0.001 & 0.966 ± 0.002 & 0.969 ± 0.004 & 0.966 ± 0.001 & 0.967 ± 0.001 & – & – \\
\rowcolor{gray!10}\multicolumn{9}{l}{\textbf{Brain}}\\
ROC AUC      & 0.721 ± 0.011 & 0.943 ± 0.001 & 0.941 ± 0.001 & \textbf{0.955 ± 0.001} & 0.954 ± 0.001 & 0.953 ± 0.001 & 0.931 ± 0.001 & 0.952 ± 0.001 \\
PR  AUC      & 0.849 ± 0.004 & 0.953 ± 0.001 & 0.948 ± 0.001 & \textbf{0.962 ± 0.001} & 0.960 ± 0.001 & 0.959 ± 0.002 & 0.921 ± 0.001 & 0.957 ± 0.001 \\
UQ – ROC AUC & 0.727 ± 0.010 & 0.944 ± 0.001 & 0.945 ± 0.002 & 0.960 ± 0.001 & 0.955 ± 0.001 & 0.956 ± 0.002 & – & – \\
UQ – PR AUC  & 0.840 ± 0.004 & 0.953 ± 0.001 & 0.952 ± 0.002 & 0.960 ± 0.002 & 0.961 ± 0.001 & 0.962 ± 0.002 & – & – \\
\rowcolor{gray!10}\multicolumn{9}{l}{\textbf{Disease of Metabolism}}\\
ROC AUC      & 0.771 ± 0.025 & 0.958 ± 0.006 & 0.959 ± 0.003 & \textbf{0.969 ± 0.007} & 0.966 ± 0.008 & 0.966 ± 0.007 & 0.968 ± 0.002 & 0.911 ± 0.006 \\
PR  AUC      & 0.861 ± 0.014 & 0.958 ± 0.007 & 0.957 ± 0.003 & \textbf{0.968 ± 0.008} & 0.963 ± 0.009 & 0.963 ± 0.008 & 0.943 ± 0.003 & 0.926 ± 0.005 \\
UQ – ROC AUC & 0.785 ± 0.024 & 0.961 ± 0.007 & 0.962 ± 0.003 & 0.974 ± 0.006 & 0.968 ± 0.008 & 0.967 ± 0.007 & – & – \\
UQ – PR AUC  & 0.863 ± 0.017 & 0.961 ± 0.009 & 0.957 ± 0.003 & 0.969 ± 0.006 & 0.965 ± 0.009 & 0.962 ± 0.008 & – & – \\
\rowcolor{gray!10}\multicolumn{9}{l}{\textbf{Liver}}\\
ROC AUC      & 0.712 ± 0.007 & 0.944 ± 0.001 & 0.941 ± 0.001 & \textbf{0.956 ± 0.001} & 0.954 ± 0.001 & 0.953 ± 0.001 & 0.933 ± 0.001 & 0.952 ± 0.0004 \\
PR  AUC      & 0.844 ± 0.004 & 0.954 ± 0.001 & 0.948 ± 0.001 & \textbf{0.962 ± 0.001} & 0.961 ± 0.001 & 0.960 ± 0.001 & 0.925 ± 0.001 & 0.958 ± 0.0004 \\
UQ – ROC AUC & 0.719 ± 0.009 & 0.945 ± 0.001 & 0.945 ± 0.002 & 0.961 ± 0.002 & 0.955 ± 0.001 & 0.957 ± 0.002 & – & – \\
UQ – PR AUC  & 0.833 ± 0.003 & 0.955 ± 0.001 & 0.951 ± 0.002 & 0.960 ± 0.002 & 0.962 ± 0.001 & 0.963 ± 0.001 & – & – \\
\rowcolor{gray!10}\multicolumn{9}{l}{\textbf{Neurodegenerative Disease}}\\
ROC AUC      & 0.765 ± 0.035 & 0.968 ± 0.006 & 0.957 ± 0.005 & \textbf{0.976 ± 0.003} & 0.973 ± 0.002 & 0.974 ± 0.002 & 0.974 ± 0.002 & 0.941 ± 0.005 \\
PR  AUC      & 0.873 ± 0.015 & 0.973 ± 0.006 & 0.961 ± 0.005 & \textbf{0.981 ± 0.003} & 0.978 ± 0.002 & 0.979 ± 0.002 & 0.963 ± 0.002 & 0.952 ± 0.004 \\
UQ – ROC AUC & 0.769 ± 0.029 & 0.970 ± 0.006 & 0.959 ± 0.005 & 0.979 ± 0.003 & 0.975 ± 0.002 & 0.975 ± 0.001 & – & – \\
UQ – PR AUC  & 0.859 ± 0.016 & 0.975 ± 0.006 & 0.962 ± 0.006 & 0.978 ± 0.003 & 0.979 ± 0.002 & 0.979 ± 0.002 & – & – \\
\bottomrule
\end{tabular}
\caption{Performance comparison of BNMFk, RNMFk, WNMFk, and their LMF‑based extensions across five protein–protein interaction datasets.  Bold values indicate the best classical score per block; dashes indicate metrics not reported by uncertainty‑quantified variants.}
\label{tab:metrics_LMF}
\end{table}

Next, benchmarks are presented for the methods against LMF and symLMF \cite{ppi_symlmf}. Table \ref{tab:metrics_LMF} presents the benchmark results on the PPI networks: \textit{H. sapiens}-extended, Brain, Disease of Metabolism, Liver, and Neurodegenerative Disease. For BNMFk, we employ k-means clustering for Boolean thresholding, as described in Section \ref{subsection:boolean_thresholding}. At the same time, RNMFk and WNMFk are evaluated without Boolean settings, following their original implementations as described in Sections \ref{subsection:wnmfk} and \ref{subsection:rnmfk}, respectively. Additionally, we include the LMF-extended results for $\text{WNMFk}_{\text{\textbf{lmf}}}$, $\text{BNMFk}_{\text{\textbf{lmf}}}$, and $\text{RNMFk}_{\text{\textbf{lmf}}}$, as described in Section \ref{subsection:lmf_ensemble}. Performance metrics include ROC AUC, PR AUC, and UQ-based scores, detailed in Section \ref{subsec:metrics}. 

Table \ref{tab:metrics_LMF} shows BNMFk has the lowest performance, while RNMFk and WNMFk consistently match or outperform LMF and symLMF across datasets. For example, on H.sapiens-extended, RNMFk achieves an ROC AUC of 0.941, compared to 0.949 for LMF and 0.944 for symLMF, with a higher PR AUC of 0.957 (vs. 0.934 for LMF and 0.955 for symLMF). Ensemble models yield higher performance, with $\text{BNMFk}_{\text{\textbf{lmf}}}$ achieving the highest scores across datasets, showing that combining models with LMF improves performance over standalone versions. On the Brain PPI dataset, BNMFk attains a PR AUC of 0.849, RNMFk 0.953, and WNMFk 0.948, while their LMF extensions reach 0.962, 0.960, and 0.959, respectively. These scores surpass LMF alone, indicating complementary model strengths under the ensemble framework. UQ-based weighting slightly improves scores in some cases, as shown by underlined ROC AUC and PR AUC values, though the gains are small for these datasets. These results on five real-world PPI networks validate the efficacy of the methods beyond synthetic datasets. The following section summarizes the technical capabilities and software development considerations behind the Python library.

%% file: Part3/AI_Reasoning/dog.tex
\section{Synthetic Data}

\begin{figure}[ht]
\centering
\includegraphics[width=0.6\textwidth]{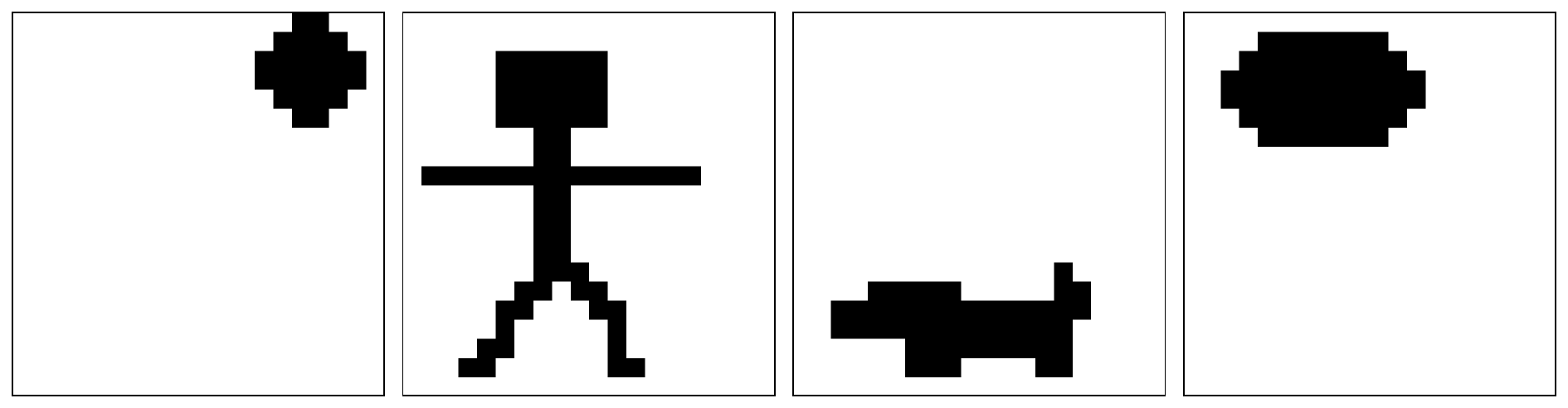} %
\caption{\textbf{Dog Dataset -} Four binary images are used as Boolean latent features to generate the synthetic data of shape $400 \times 16$.  \label{fig:dog_data}}
\end{figure}

The first dataset used in the experiments, referred to as the Dog dataset \cite{binary_factorize}, is constructed by mixing four binary images. This dataset includes four distinct images--sun, person, dog, and cloud, each with dimensions $20 \times 20$ pixels. The Dog dataset is illustrated in Figure \ref{fig:dog_data}. To generate the dataset, the columns of each image are stacked along the first axis. Then they create 16 unique samples by combining these stacked representations using Boolean addition, considering all possible combinations of the four images. This results in a matrix $\mat{X} \in \{0, 1\}^{400 \times 16}$. Since the dataset comprises four distinct images, the true Boolean rank of the Dog dataset is $k = 4$.

\subsection{Dog Data}
\label{subsec:dog_results}
\begin{figure}[t!]
\centering
\includegraphics[width=1.0\textwidth]{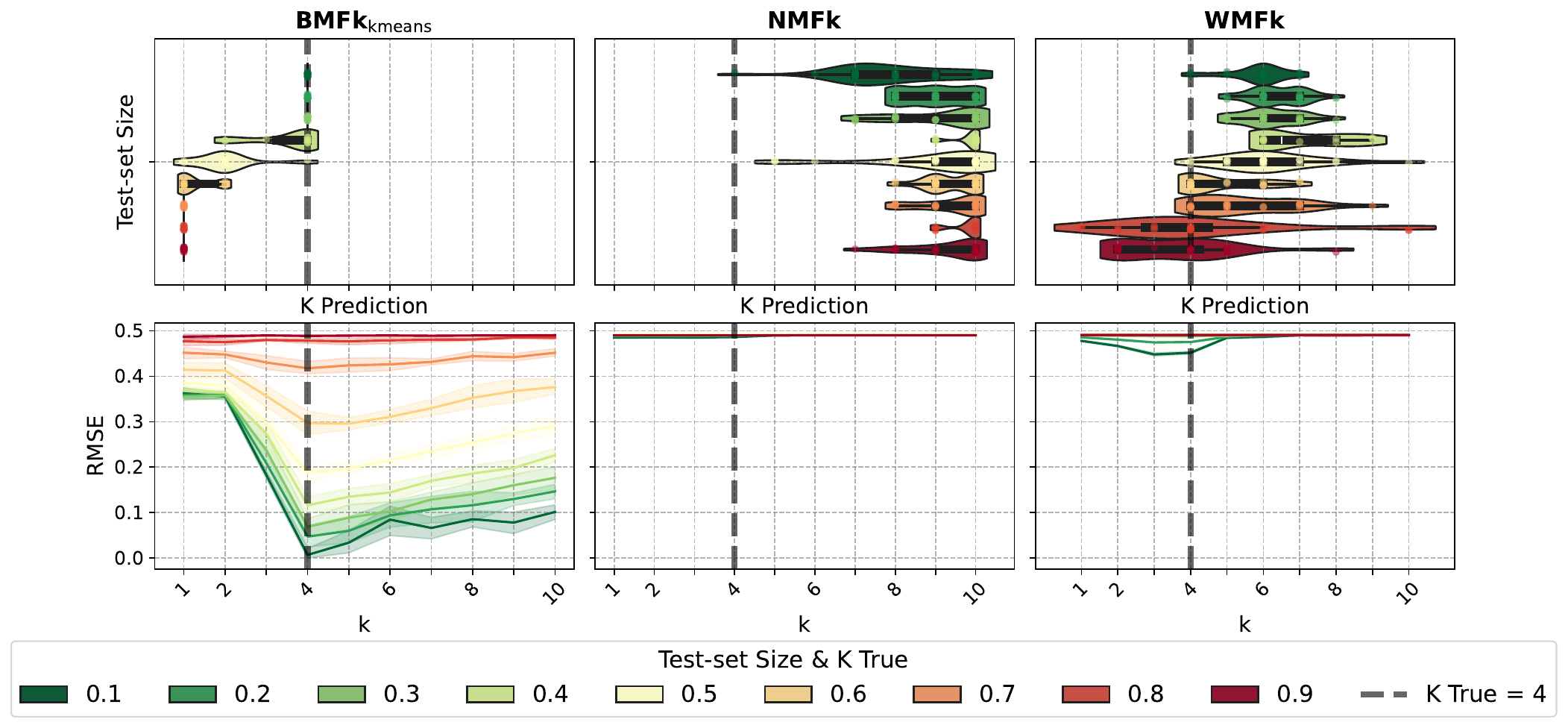} %
\caption{Results for Dog Data across methods including $\text{BNMFk}_{\text{\textbf{kmeans}}}$, NMFk, and WNMFk. Boolean thresholding is not used for NMFk and WNMFk. The first row presents violin plots visualizing the rank $k$ predictions at different test-set size levels. The second row displays RMSE scores for the test set, demonstrating the missing link prediction performance. The results are reported for each rank $k$ on the x-axis, with the dark/dashed vertical line across columns being the true rank $k=4$.
 \label{fig:dog_data_results}}
\end{figure}

\begin{figure}[t!]
\centering
\includegraphics[width=1.0\textwidth]{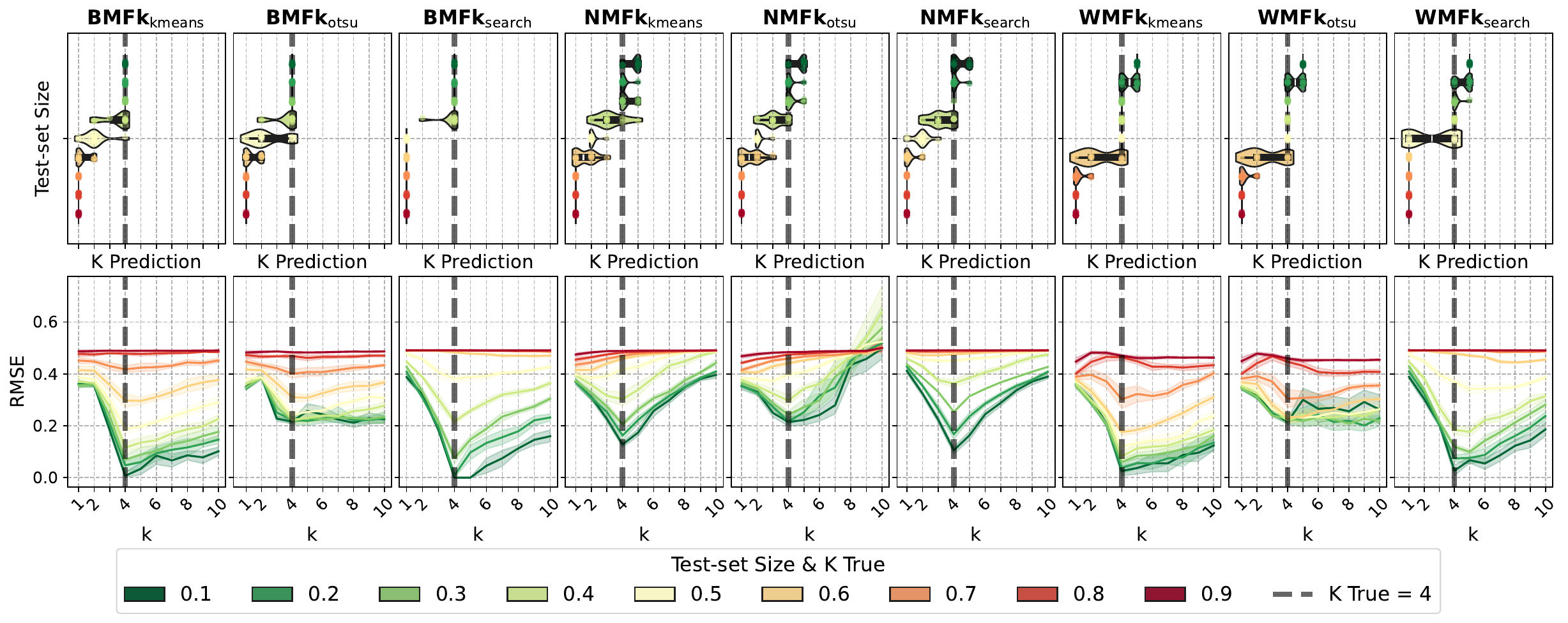} %
\caption{Results for Dog Data across methods, including BNMFk, NMFk, and WNMFk, evaluated under different Boolean thresholding techniques. The Boolean thresholding techniques are denoted with the subscripts of \textbf{kmeans}, \textbf{otsu}, and \textbf{search} (coordinate descent). The first row presents violin plots visualizing the rank $k$ predictions at different test-set sizes. The second row displays RMSE scores for the test set, demonstrating the missing link prediction performance. The results are reported for each rank $k$ on the x-axis, with the dark and dashed vertical line across each column representing the true rank $k=4$.
 \label{fig:dog_data_results_booleans}}
\end{figure}

\begin{figure}[t!]
\centering
\includegraphics[width=1.0\textwidth]{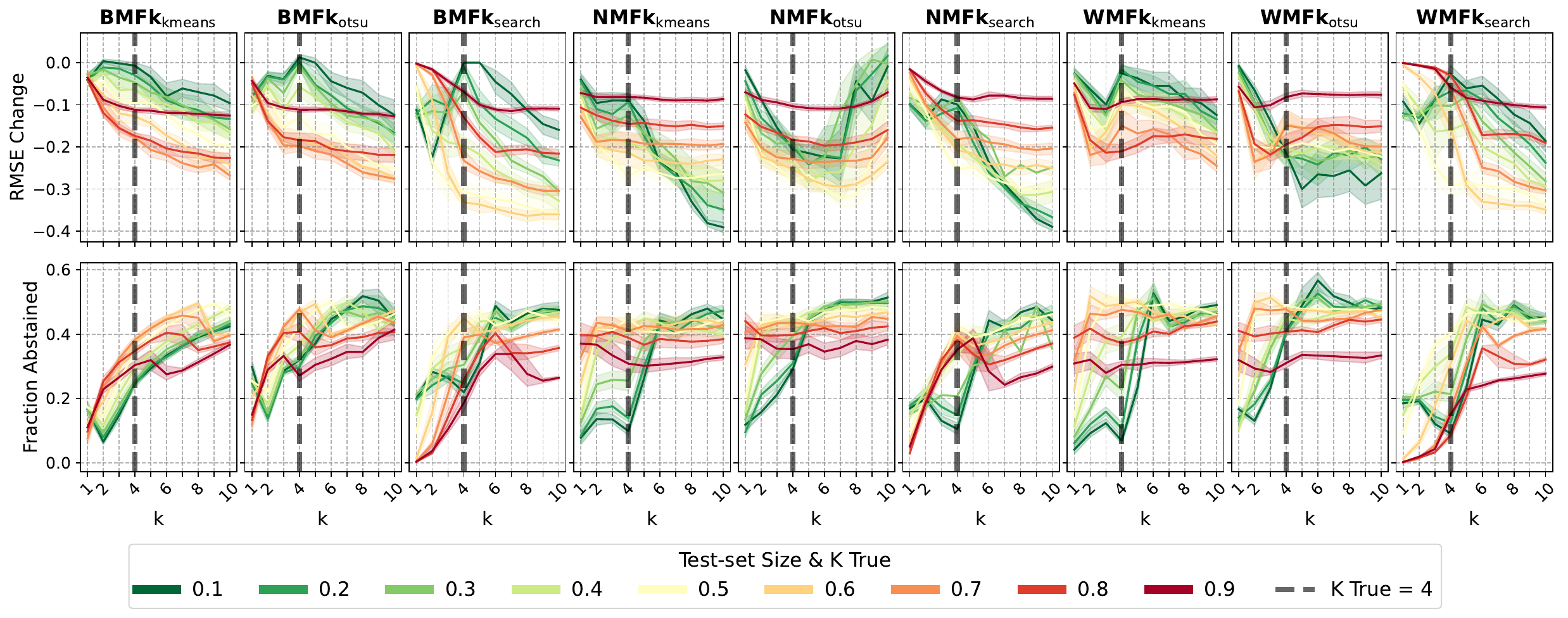} %
\caption{Results for Dog Data across methods, including BNMFk, NMFk, and WNMFk, evaluated under different Boolean thresholding techniques. The Boolean thresholding techniques are denoted with the subscripts of \textbf{kmeans}, \textbf{otsu}, and \textbf{search} (coordinate descent). The first row presents the change in RMSE after making predictions on the non-abstained samples, where a negative change indicates improved performance (lower RMSE) for the given fraction of abstained samples. The second row illustrates the fraction of abstained predictions, representing the proportion of cases where the model opted not to make a prediction due to uncertainty. The results are reported for each rank $k$ on the x-axis, with the dark and dashed vertical line across each column indicating the true rank $k=4$.
 \label{fig:dog_data_results_booleans_reject_option}}
\end{figure}

The results on the Dog Data are presented in Figures \ref{fig:dog_data_results}, \ref{fig:dog_data_results_booleans}, and \ref{fig:dog_data_results_booleans_reject_option}, where each column displays the performance of the methods under different settings. 

Figure \ref{fig:dog_data_results} compares $\text{BNMFk}_{\text{\textbf{kmeans}}}$, NMFk, and WNMFk, where the k-means subscript denotes Boolean thresholding via k-means clustering (Section \ref{subsection:boolean_thresholding}). In the first row, $\text{BNMFk}_{\text{\textbf{kmeans}}}$ predicts the true rank $k=4$ more consistently than NMFk and WNMFk, as expected for a Boolean decomposition method on Boolean Dog Data. However, for test-set sizes above 0.4, $\text{BNMFk}_{\text{\textbf{kmeans}}}$ shifts toward $k=1$, while NMFk and WNMFk frequently predict ranks above $k=4$, capturing non-Boolean structures. Both $\text{BNMFk}_{\text{\textbf{kmeans}}}$ and WNMFk predict lower ranks as sparsity increases. The second row shows RMSE scores, where lower values indicate better link prediction. $\text{BNMFk}_{\text{\textbf{kmeans}}}$ achieves the lowest RMSE at smaller test sets, with RMSE increasing as test-set size grows, limiting pattern learning. RMSE is minimized near $k=4$, emphasizing the importance of correct rank selection.

Figure \ref{fig:dog_data_results_booleans} expands on these results, comparing Boolean thresholding techniques for BNMFk, NMFk, and WNMFk (Section \ref{subsection:boolean_thresholding}). Each column represents a method using a different thresholding approach, denoted by subscripts \textbf{kmeans}, \textbf{otsu}, and \textbf{search}, with \textbf{search} referring to coordinate descent-based thresholding. For instance, $\text{BNMFk}_{\text{\textbf{otsu}}}$ applies Otsu thresholding. The first row presents rank prediction violin plots, while the second row reports RMSE scores. Compared to Figure \ref{fig:dog_data_results}, Boolean thresholding helps NMFk and WNMFk predict $k=4$ more frequently, though they still tend to overestimate ranks at smaller test sizes, whereas BNMFk predicts $k=4$ or lower. RMSE results confirm Boolean settings improve link prediction, especially for $\text{WNMFk}_{\text{\textbf{kmeans}}}$ and $\text{WNMFk}_{\text{\textbf{search}}}$. BNMFk remains the best for rank prediction and lower RMSE, except with Otsu thresholding, which yields higher RMSE values. Across all methods, a trend of improved RMSE scores at or around the true rank can be observed in this figure, further highlighting the importance of rank selection.

Figure \ref{fig:dog_data_results_booleans_reject_option} highlights UQ's role in improving predictions. The first row shows the RMSE change after applying UQ, where negative values indicate improved performance (reduced RMSE value) by abstaining from uncertain predictions. The second row presents the fraction of abstained predictions at each test-set size, with higher fractions indicating lower coverage. UQ effectively filters uncertain predictions, lowering RMSE in several cases compared to Figure \ref{fig:dog_data_results_booleans} (without UQ). Methods like $\text{BNMFk}_{\text{\textbf{search}}}$ and $\text{WNMFk}_{\text{\textbf{kmeans}}}$ show notable RMSE reduction. Abstention rates generally increase with test-set size, reflecting greater uncertainty in sparse data, until around a test-set size of 0.8. Beyond this point, the training set becomes too small, resulting in no meaningful abstentions. However, at the true rank $k=4$, abstention rates decline for several methods, which, combined with observations in Figure \ref{fig:dog_data_results_booleans} showing improved performance at true rank, indicate higher confidence in predictions at the correct rank.

\begin{figure}[t!]
\centering
\includegraphics[width=1.0\textwidth]{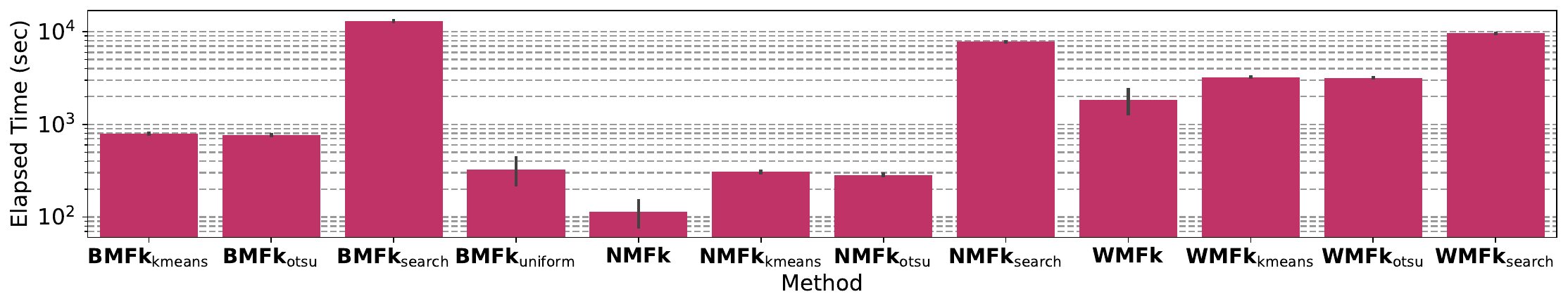} %
\caption{Computation times for each method are shown in the log scale on the Dog Data. The results indicate that methods leveraging coordinate descent (search) for Boolean thresholding require significantly longer computation times than other approaches.
\label{fig:dog_data_compute_time}}
\end{figure}

Figure \ref{fig:dog_data_compute_time} shows the computational time for each method in seconds. While Figure \ref{fig:dog_data_results} demonstrated that coordinate descent-based Boolean thresholding yields better results, its improvement comes at significantly higher computational time. As shown, methods using the search setting have substantially longer run times. Given this, we exclude the search method from subsequent experiments.

%% file: Part3/AI_Reasoning/swimmer.tex
\subsection{Swimmer Data}
\label{subsec:swimmer_results} 
\begin{figure}[ht]
\centering
\includegraphics[width=1.0\textwidth]{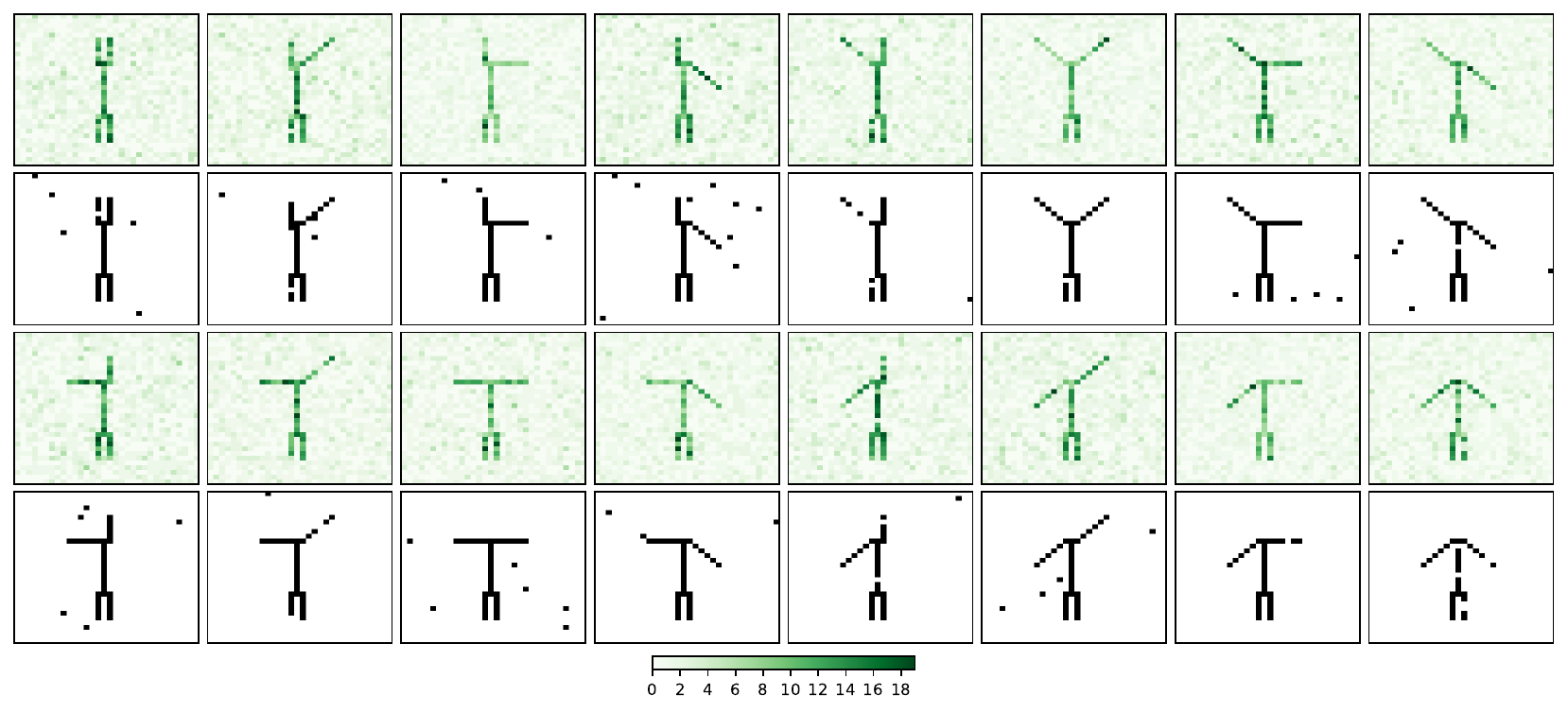} %
\caption{\textbf{Swimmer Dataset -} Dataset of 16 swimmer images. The first and third rows are the images with real-valued intensities ranging from 0 to 19. The second and fourth rows display the Boolean versions obtained after applying Otsu thresholding. For analysis, the Boolean versions of the dataset are represented as a matrix of size $1024 \times 256$.  \label{fig:swimmer_data}}
\end{figure}

The second dataset used in the experiments is the Swimmer dataset \cite{probabilistic_binary}, which contains 16 synthetic images of a "stick figure" swimming. Each image has dimensions of $32 \times 32$ pixels. Examples from this dataset are displayed in Figure \ref{fig:swimmer_data}. As shown in the first and third rows of Figure \ref{fig:swimmer_data}, the images contain "splashes" appearing as high-intensity values away from the swimmer or locations with low values of the swimmer figure, representing that location being under the water. Otsu thresholding is applied to convert the dataset into a Boolean format, with the resulting Boolean images displayed in rows 2 and 4 of Figure \ref{fig:swimmer_data}. The locations of the "splashes" and the places where the swimmer is under the water are visible in the Boolean version of the dataset. To construct the Boolean dataset,  each Boolean image is stacked along the first axis and combined using Boolean addition into 256 different samples, resulting in a matrix $\mat{X} \in \{0, 1\}^{1024 \times 256}$. The true rank of the dataset is $k = 16$, corresponding to the 16 distinct swimmer figures.

\begin{figure}[ht]
\centering
\includegraphics[width=1.0\textwidth]{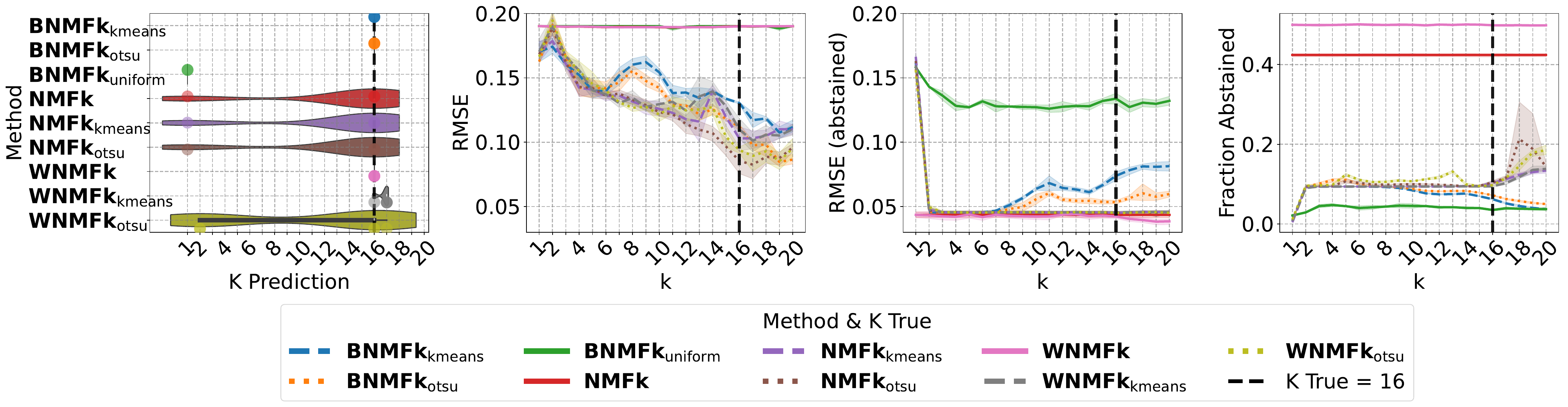} %
\caption{The performance of each method on the Swimmer Data is shown. The first column features a violin plot displaying the distribution of $k$ predictions for each method, with the y-axis representing the methods and the x-axis representing the predicted $k$ values. The second through final columns present results at each $k$ decomposition value, with the x-axis corresponding to the rank $k$. The dark dashed vertical lines indicate this dataset's true rank of $k=16$. The Boolean thresholding techniques are denoted with the subscripts of \textbf{kmeans}, \textbf{otsu}, and \textbf{search} (coordinate descent), while the \textbf{uniform} subscript refers to running BNMFk without Boolean thresholding on the latent factors. NMFk and WNMFk, without subscripts, are the results of not using Boolean techniques with these methods.
 \label{fig:swimmer_data_results}}
\end{figure}

The Swimmer dataset is analyzed with results in Figure \ref{fig:swimmer_data_results}. The test-set size is fixed at 0.1 (10\%) without sparsity analysis. Figure \ref{fig:swimmer_data_results} shows $k$-predictions (first column), RMSE scores (second), RMSE with abstention (third), and the fraction of abstained samples (fourth). The true rank $k=16$ is marked by a dark dashed vertical line, with the x-axis representing $k$ values. Each method is color-coded, while line styles distinguish Boolean settings in columns 2-4. Figure \ref{fig:swimmer_data_results} shows that BNMFk with Boolean settings consistently identified $k=16$, as did WNMFk without Boolean settings. WNMFk with k-means thresholding predicted both $k=16$ and $k=17$, while $\text{BNMFk}_{\text{\textbf{uniform}}}$ (without Boolean settings) always predicted $k=1$. Other methods produced $k$ values between $k=1$ and $k=17$. Some non-Boolean methods predicted $k=16$ because the Swimmer dataset's non-negative and Boolean ranks are equal.

In the second column of Figure \ref{fig:swimmer_data_results}, $\text{NMFk}_{\text{\textbf{kmeans}}}$, $\text{WNMFk}_{\text{\textbf{otsu}}}$, $\text{WNMFk}_{\text{\textbf{kmeans}}}$, and $\text{BNMFk}_{\text{\textbf{kmeans}}}$ improve as $k$ nears 16, after which RMSE increases. Notably, $\text{BNMFk}_{\text{\textbf{kmeans}}}$ continues improving until $k=18$ before RMSE rises, mirroring Dog Data trends. Non-Boolean decompositions maintain higher RMSE (lower performance) across all $k$, confirming Boolean methods perform better on Boolean data. The third column shows UQ significantly improves RMSE, even for non-Boolean methods, though less so for $\text{BNMFk}_{\text{\textbf{uniform}}}$. The RMSE remains stable across ranks as models abstain from uncertain predictions. However, as a trade-off, non-Boolean methods like WNMFk and NMFk exhibit a higher reduction in coverage, shown by a greater fraction of abstained samples in the last column.

%% file: Part3/AI_Reasoning/gaussian.tex
The third synthetic dataset is randomly sampled matrices $\mat{X}$ from a normal distribution, leveraging the framework introduced in \cite{nebgen2021neural}. This dataset is referred to as the Gaussian Dataset. The generated matrices $\mat{X} \in \mathbb{R}_{+}^{50 \times m}$, where $m \in \{50, 100, 150, 200, 250, 300, 350, 400\}$ (8 distinct shapes), correspond to an increasing number of features. For each matrix size, the true rank is set as $k \in \{2, 3, 4, 5, 6\}$ (5 distinct true rank values). Each matrix size and rank combination is randomly sampled ten times with different random seeds, resulting in $8 \times 5 \times 10 = 400$ matrices. While the two Boolean datasets mentioned above and one non-Boolean synthetic dataset are used to assess the performance of the methods under controlled conditions, larger real-world datasets are used to evaluate further and validate the methods in practical scenarios.

\subsection{Gaussian Data}
\label{subsec:ben_results}
\begin{figure}[t!]
\centering
\includegraphics[width=0.9\textwidth]{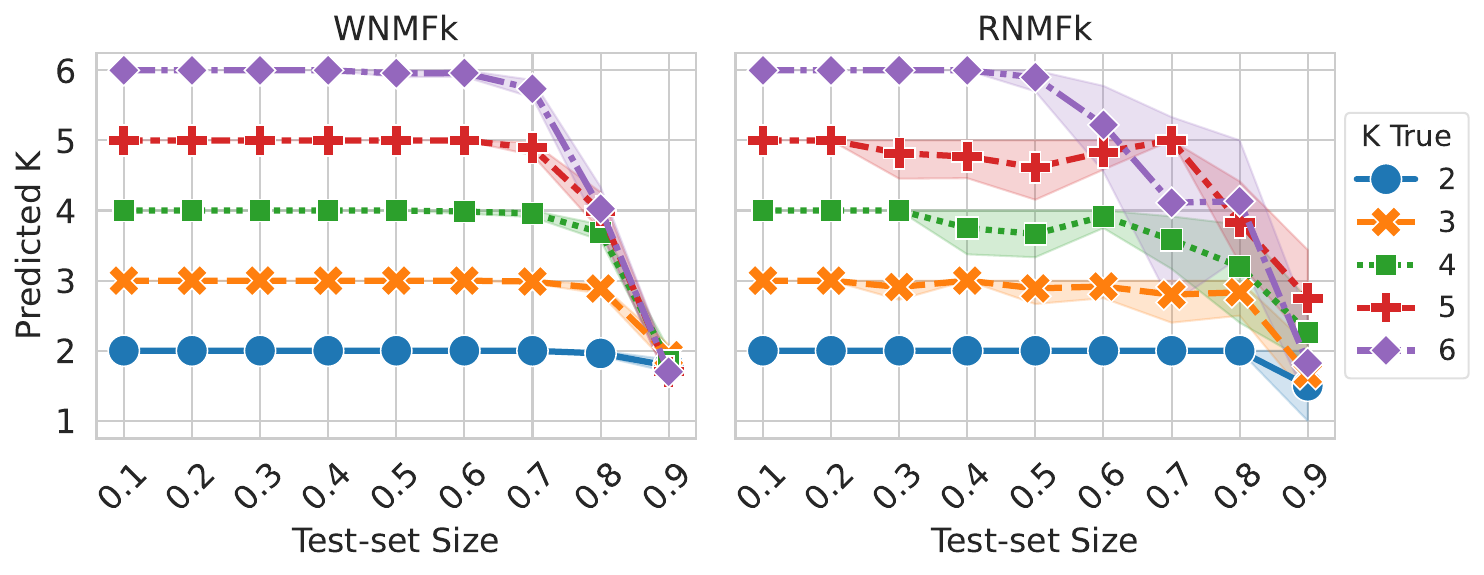} %
\caption{Results for $k$ predictions on Gaussian Data are shown for WNMFk and RNMFk methods across different data sparsity levels, represented by the increasing test-set size on the x-axis. The y-axis indicates the predicted $k$ values. The results demonstrate that $k$ predictions remain accurate for both methods until high levels of sparsity are reached, while WNMFk results in more accurate $k$ predictions. Line color and style differentiate between different true $k$ values. For instance, a solid line with circle markers represents decompositions where the matrix rank was $k=2$.
 \label{fig:ben_k_predictions}}
\end{figure}

The final synthetic dataset, Gaussian Data, is analyzed in Figure \ref{fig:ben_k_predictions} for WNMFk and RNMFk, assessing their rank prediction at different test-set sizes. No Boolean settings are used. Color and line style differentiate true ranks; for example, an orange line with $x$ markers represents $k=3$. WNMFk predicts the correct rank until a test-set size of 0.7, after which predictions skew toward $k=1$. RNMFk follows a similar trend but loses rank prediction accuracy earlier at higher true ranks. For instance, at $k=6$ (purple line with diamond markers), lower rank predictions appear after a test-set size of 0.5.

\begin{figure}[t!]
\centering
\includegraphics[width=1.0\textwidth]{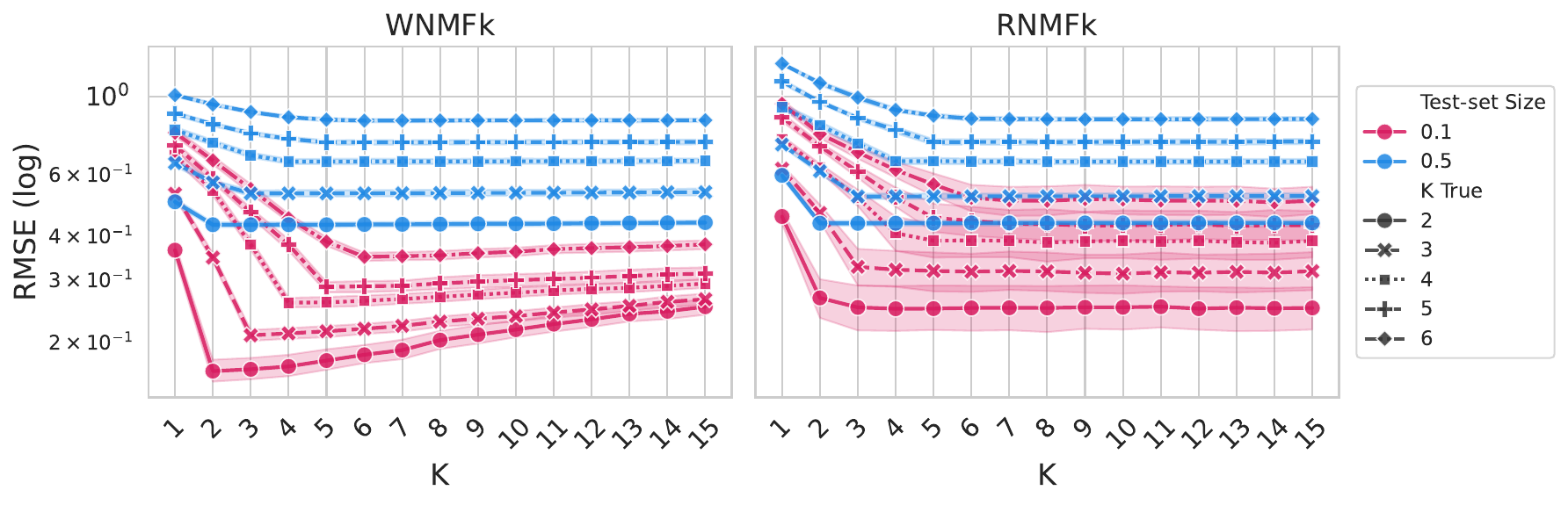} %
\caption{Link prediction performance on test-set sizes of 10\% (red lines) and 50\% (blue lines) is shown for WNMFk and RNMFk, with RMSE scores plotted on a log scale. The x-axis represents the $k$ values, displaying results at different $k$ decompositions. Line markers are used to differentiate between the true $k$ values. For example, solid red lines with circle markers represent matrices with a true rank of $k=2$ and a test-set size of 10\%.
\label{fig:ben_k_rmse}}
\end{figure}

We examine link prediction performance using RMSE on a log scale for test-set sizes 0.1 and 0.5 in Figure \ref{fig:ben_k_rmse}, where blue lines represent 0.5 and red lines 0.1. Shapes differentiate true ranks; for example, square markers indicate $k=4$, with blue squares for $k=4$ at 0.5. For WNMFk, RMSE increases beyond the true rank, reflecting reduced performance, with the best results at the correct rank. This trend holds at higher sparsity (0.5 test-set size) but is less pronounced due to overall performance decline. RNMFk follows a similar pattern but shows more stable link prediction, as indicated by flatter RMSE lines, suggesting more consistent performance across ranks.

\begin{figure}[t!]
\centering
\includegraphics[width=1.0\textwidth]{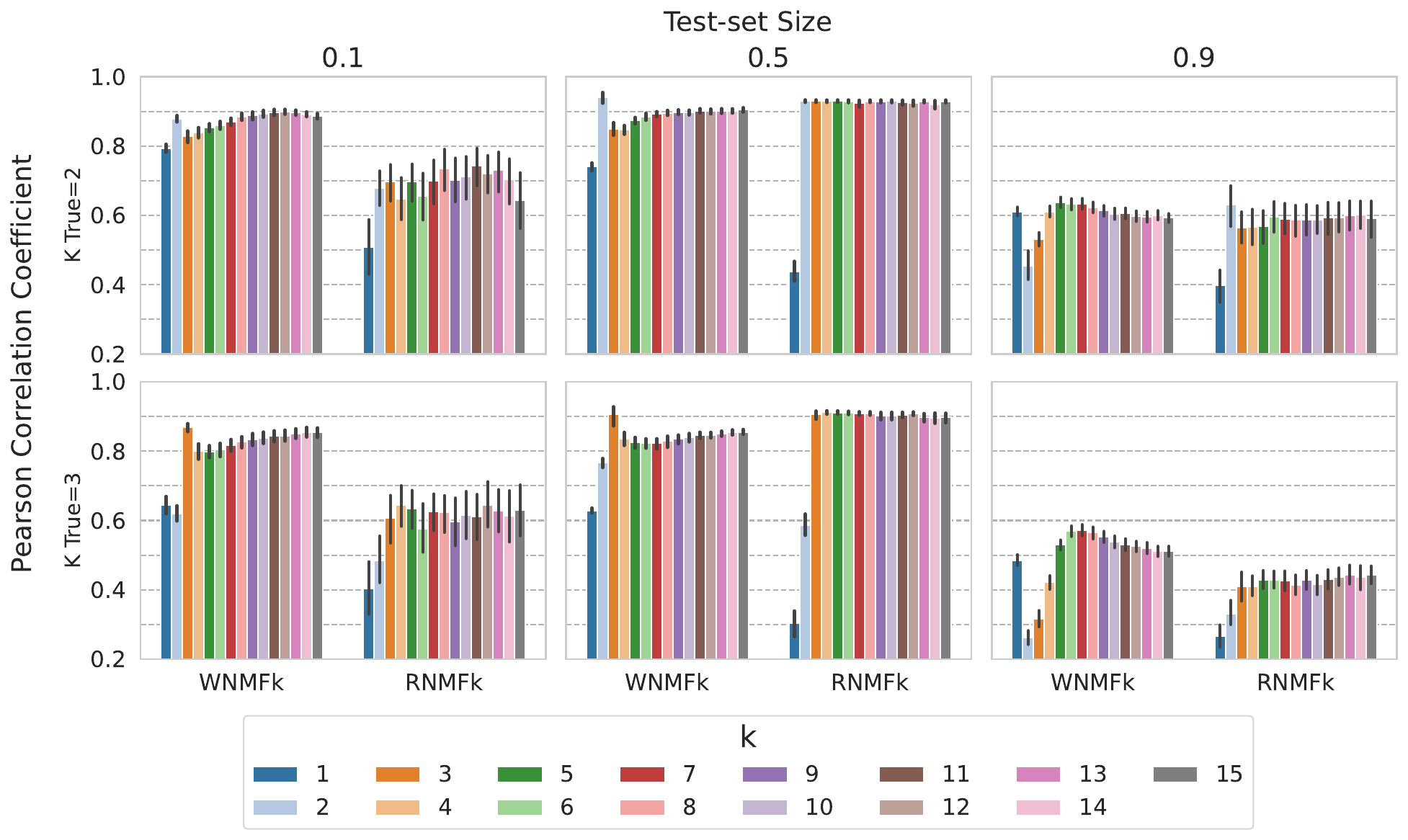} %
\caption{Pearson Correlation Coefficients with CI between UQ and errors at the test set are reported across different sparsity levels or test-set sizes that include 0.1, 0.5, and 0.9 (columns) and true matrix ranks $k$ including $k \in [2,3]$ (rows). The shared x-axis differentiates between the methods WNMFk and RNMFk using hues, while the shared y-axis represents the Pearson correlation values. Bars with different colors correspond to results at various $k$ decompositions. 
\label{fig:ben_uq_correlation}}
\end{figure}

For Gaussian Data, we analyze the correlation between UQ and errors in Figure \ref{fig:ben_uq_correlation} to assess whether models assign higher uncertainty to higher-error points. Figure \ref{fig:ben_uq_correlation} shows Pearson Correlation Coefficients and CI between UQ and relative reconstruction errors (Equation \ref{eqn:rel_error}). Columns represent test-set sizes, rows correspond to true ranks $k=2$ and $k=3$, and the y-axis denotes correlation coefficients. Hue differentiates WNMFk and RNMFk, with color bars indicating decomposition results for each $k$.

Figure \ref{fig:ben_uq_correlation} shows that correlations between certainty and error remain high for RNMFk and WNMFk, confirming that UQ effectively assigns higher uncertainty to high-error points. Correlation is lower before the true rank and peaks at the correct rank, especially for WNMFk and smaller test-set sizes, indicating reliable certainty estimates. In the second row ($k=3$), the orange bars for WNMFk are notably higher, highlighting improved performance at this rank. Another key trend in Figure \ref{fig:ben_uq_correlation} is the reduction in correlation between certainty and error at the largest test-set size of 0.9. The hypothesis for this behavior is that at higher sparsity levels, where the training set is significantly reduced, the uncertainty estimates may saturate at consistently high values across all test points, thereby reducing variability. We have also observed a relevant and similar result in Figure \ref{fig:dog_data_results_booleans_reject_option}, where the fraction of abstained samples reduced at higher test set sizes.  When uncertainty does not vary meaningfully, its correlation with actual errors naturally diminishes. This saturation reflects a breakdown in the model's ability to provide meaningful uncertainty estimates, signaling that the model is operating in a regime of extreme sparsity. Under such conditions, there are significantly more missing links to predict while relying on minimal information from the training set. As a result, the model's uncertainty quantification methods become less reliable.

\begin{figure}[t!]
\centering
\includegraphics[width=0.7\textwidth]{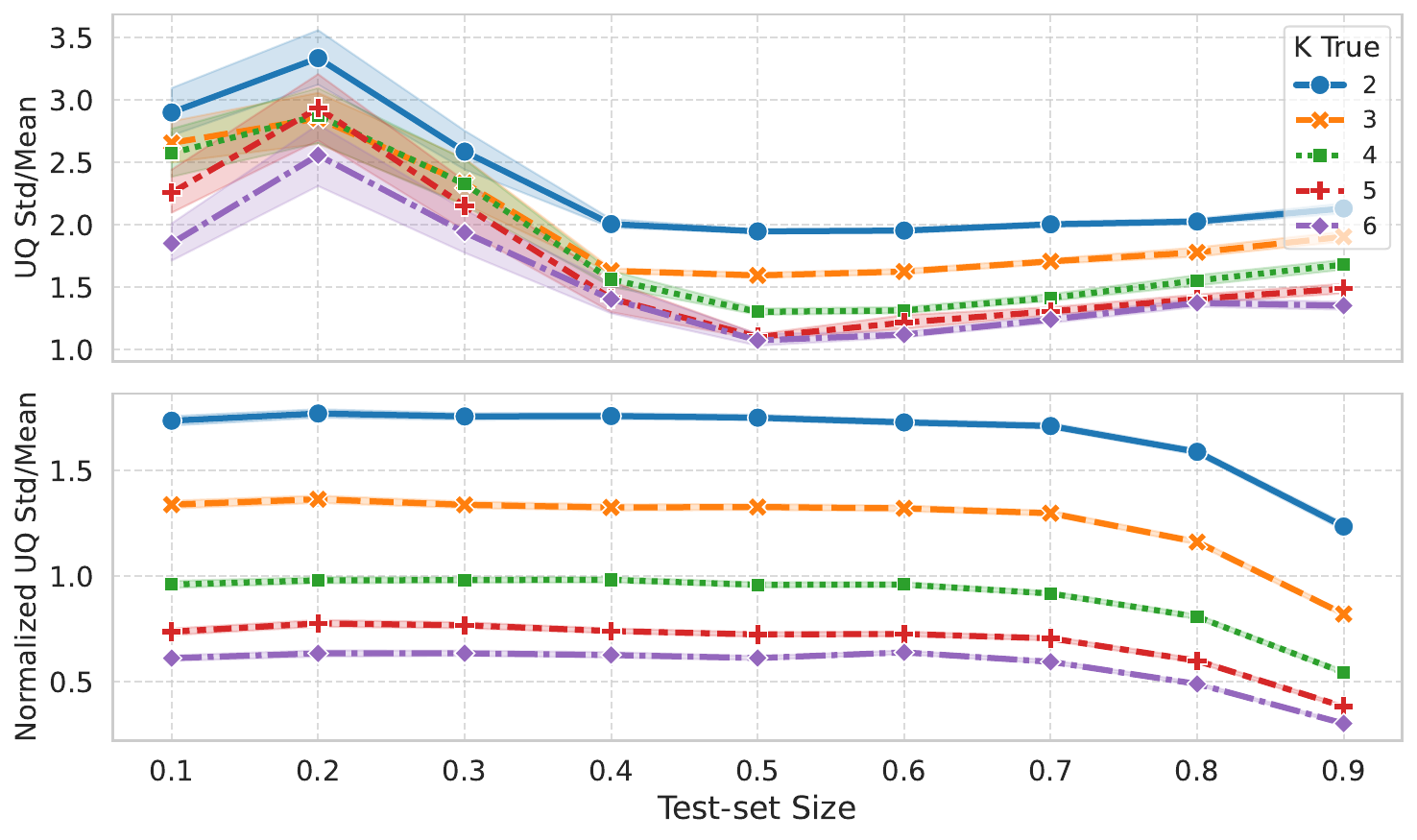} %
\caption{\textbf{Relative-Std-to-Mean Ratio of Uncertainty plot} provides a quantitative perspective on the behavior of uncertainty estimates across varying sparsity levels (or test set size) for each $k$ true. The top figure shows the standard deviation (std) at test points divided by the overall average of certainty (UQ). The bottom figure investigates certainty saturation by normalizing the std-to-mean uncertainty ratio at the top plot with the correlation between the UQ and errors at test points. \label{fig:uq_std_to_mean}}
\end{figure}

The "Relative-Std-to-Mean Ratio of Uncertainty" plot, shown in the top row of Figure \ref{fig:uq_std_to_mean}, provides a quantitative perspective on the behavior of uncertainty estimates across varying levels of sparsity (or test-set size). This metric is calculated for the test locations based on the uncertainty matrix $\mathbf{U}$ and is defined as:
\begin{align*} 
\text{SMR} = \frac{\sigma_{\mathbf{U}}}{\mu_{\mathbf{U}}}, \numberthis 
\end{align*}
where $\text{SMR}$ is the standard deviation (std)-to-mean ratio, $\sigma_{\mathbf{U}}$ is the standard deviation of UQ values at test locations, and $\mu_{\mathbf{U}}$ is the mean of UQ values at test locations, calculated as:
\begin{align*} 
\sigma_{\mathbf{U}} = \sqrt{\frac{1}{|\mathcal{I}_{\text{test}}|} \sum{(i,j) \in \mathcal{I}_{\text{test}}} (\mathbf{U}{ij} - \bar{\mathbf{U}})^2}, \quad \mu_{\mathbf{U}} = \frac{1}{|\mathcal{I}_{\text{test}}|} \sum{(i,j) \in \mathcal{I}_{\text{test}}} \mathbf{U}{ij}, \numberthis 
\end{align*}
where $\mathcal{I}_{\text{test}}$ is the set of test locations, and $\bar{\mathbf{U}}$ is the mean uncertainty value at the test locations.
This metric tracks the relationship between the variability of uncertainty (measured as the standard deviation of uncertainty) and the average uncertainty (mean of UQ) at the test points. By normalizing variability through its mean, the std-to-mean ratio captures whether the relative dispersion in uncertainty estimates decreases significantly as sparsity increases. In the top row of Figure \ref{fig:uq_std_to_mean}, the std-to-mean ratio drops until around a 0.5 test-set size and stabilizes. Meanwhile, the correlation between errors and UQ remains high until a test-set size of 0.7 or 0.8. The drop in the std-to-mean ratio suggests that the variability of uncertainty at test locations is becoming more consistent relative to its mean. However, it does not necessarily indicate saturation (since the correlation remains high). The normalized std-to-mean ratio plot, shown in the second row of Figure \ref{fig:uq_std_to_mean}, refines this measure by incorporating the correlation between UQ and errors. This metric is defined as:
\begin{align*} 
\text{NSMR} = \frac{\text{SMR} \cdot r}{\max(r)}, \numberthis 
\end{align*}
where $r$ is the vector of correlation values between UQ and reconstruction error. This normalization adjusts the standard deviation-to-mean ratio (SMR) by weighting it with the correlation between UQ and error, scaled by the maximum correlation value observed. This scaling emphasizes variability in uncertainty estimates that predict error rather than those that fluctuate without meaningful correlation to errors. By incorporating the correlation term, this metric provides a refined measure of how uncertainty dispersion (the std-to-mean ratio) relates to actual predictive reliability. Their significance diminishes if uncertainty estimates are highly variable but poorly correlated with errors. Conversely, when variability in uncertainty remains well-correlated with errors, it suggests that the uncertainty estimates are informative.

In the second row of Figure \ref{fig:uq_std_to_mean}, higher values, observed until around a test-set size of 0.7, indicate that the uncertainty estimates remain variable and meaningfully related to errors. This suggests that UQ values are informative about prediction confidence. The observed decline beyond a test-set size of 0.8, where both the std-to-mean ratio in the first row of Figure \ref{fig:uq_std_to_mean} and the correlation in Figure \ref{fig:ben_uq_correlation} are low, suggests that uncertainty estimates are saturating. This saturation implies that the model's certainty values no longer distinguish between correct and incorrect predictions, reducing their usefulness at higher sparsity levels.

%% file: Part3/AI_Reasoning/materials.tex
\section{Materials }
The experiments test three hypotheses: 
(i) an \textsc{HNMFk}$\rightarrow$\textsc{BNMFk}$+$\textsc{LMF} pipeline can rediscover masked superconducting links; 
(ii) the posterior scores of the ensemble cleanly separate positive from negative edges; and 
(iii) the same model ranks unseen compounds so that known superconductors appear above chemically similar non-superconductors.  
These objectives guide the design choices detailed below.

\subsection{Corpus construction and pre-processing}
We build on a domain-specific corpus of 46,862 peer-reviewed articles that mention at least one of the 73 TMDs extracted by the BUNIE pipeline.  
Titles and abstracts are lower–cased, tokenised, stop-words removed, and converted to a $V{=}48,712$-term TF–IDF matrix with \verb|min_df = 3| and \verb|max_df = 0.98|.  Documents with fewer than 20 non-zero terms are discarded, yielding a final $46,715 \times 48,712$ matrix with $0.12\%$ density.  

\subsection{Topic hierarchy (\textsc{HNMFk})}
The TF–IDF matrix is factorised recursively with \textsc{HNMFk}.  Stable rank selection operates in the range $k\in[2,6]$ at every node, and the recursion stops automatically when any leaf contains $<150$ documents.  
Bootstrap stability (200 resamples) identified an optimal depth $L=3$, producing $k_1{=}3$, $k_2{=}5$, and $k_3{=}18$ clusters on successive levels.  The $|\mathcal{T}|=815$ leaf topics constitute the row index of the subsequent Materials-Property matrix.

\subsection{Link-prediction Model}
The ensemble first fits \textsc{BNMFk} with automatic rank selection to $M$, then freezes that rank and refines the reconstruction with Logistic Matrix Factorisation (LMF).  The two reconstructions are combined as \[ \hat{M} \;=\; \sigma\!\bigl(W H + b_r + b_c\bigr), \] where $\sigma$ is the logistic link.  Hyper-parameters ($\lambda{=}0.1$, \verb|lr=0.01|, \verb|epochs=500|) were tuned on a held-out validation set.
 
\subsection{Materials Property Matrix}
\label{sec:data_materials}
The TMDs corpus comprises 46,862 peer-reviewed articles.  
TMDs, with the general formula $MX_2$ ($M$ = transition metal, $X$ = chalcogen), are foundational to 2D materials research, known for their exceptional electronic, optical, and mechanical properties \cite{gupta2025two}. Despite only 72 known compounds, the combinatorial space of possible TMDs is vast and remains largely underexplored.

For hierarchical modeling, we convert titles + abstracts of the selected by BUNIE scientific texts to TF–IDF matrix and factorize it with HNMFk. A depth of \textbf{three} levels (\(L=3\)) proved optimal by bootstrap stability analysis, yielding coarse “super-topics” at the root and progressively finer subtopics at lower depths.

\begin{figure}[ht]
    \vspace{-.9em}
    \centering
    \includegraphics[width=.8\columnwidth]{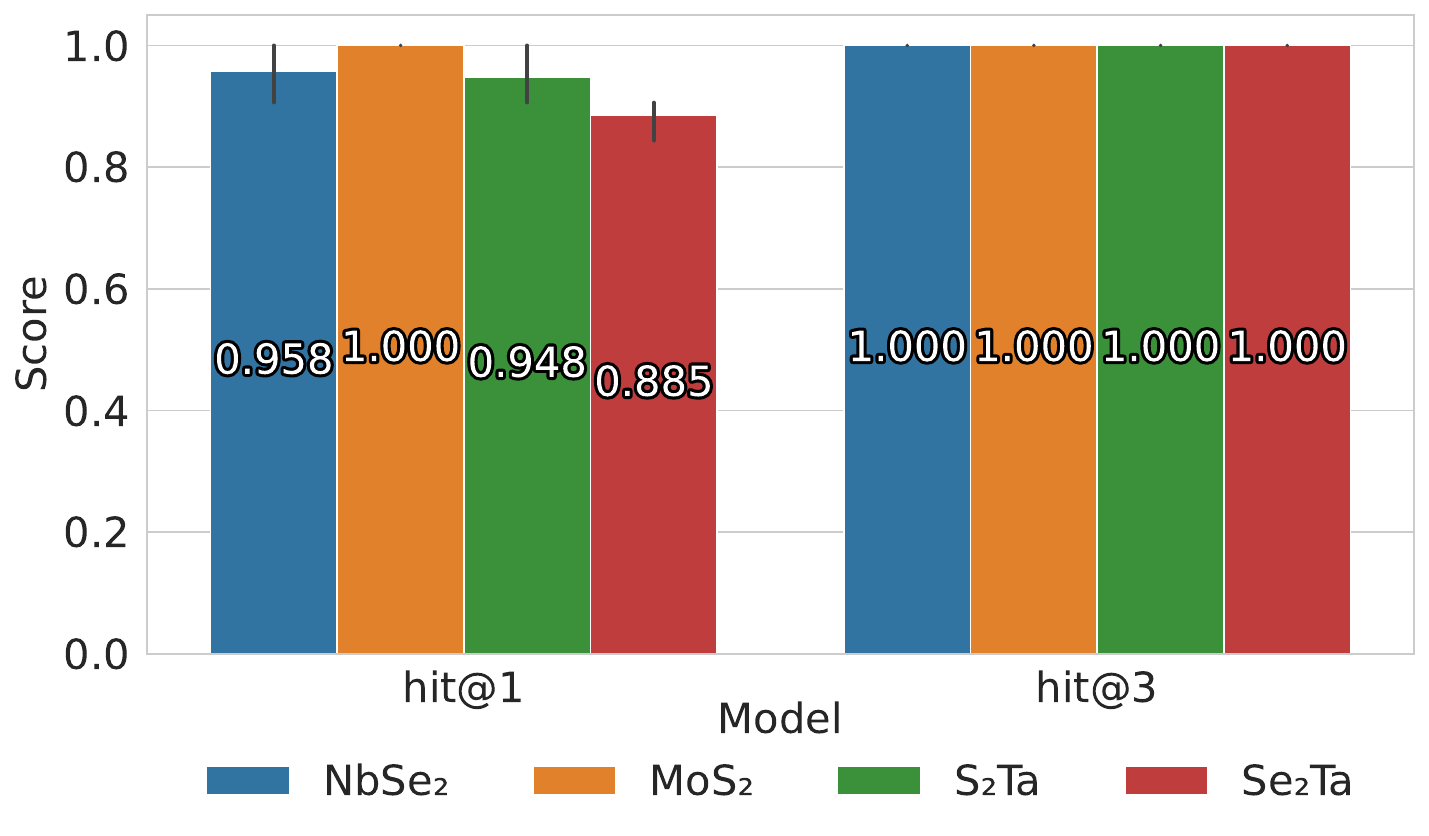}
    \vspace{-1.5em}
    \caption{Mean hit@1 and hit@3 scores (bars, annotated) of the
\textsc{BNMF}$+$\textsc{LMF} ensemble after masking every known superconducting link and an equal number of random zeros for each of the four benchmark compounds }
    \vspace{-1em}
    \label{fig:hit_scores}
\end{figure}

\subsection{Top \(k\) retrieval of hidden links}
\label{sec:topk_retrieval}
To evaluate whether our \(\textsc{BNMF}_k + \textsc{LMF}\) ensemble can \emph{rediscover} true but withheld superconducting links, we perform a leave-out experiment on four benchmark transition-metal dichalcogenides: NbSe\(_2\), MoS\(_2\), S\(_2\)Ta, and Se\(_2\)Ta.

\begin{enumerate}[label=(\arabic*)]
  \item \textbf{Masking.} For each compound, remove all verified superconducting links from the compound–superconductor cluster matrix, and randomly sample an equal number of non-superconducting (zero) entries as negatives.
  \item \textbf{Training and scoring.} Train the ensemble on the masked matrix and compute predicted scores \(s_{ij}\) for each masked entry \((i,j)\).
  \item \textbf{Metric.} Define
    \[
      \mathrm{hit@}k
      \;=\;
      \frac{1}{\lvert \mathcal P\rvert}
      \sum_{(i,j)\in\mathcal P}
      \mathbf{1}\bigl\{\mathrm{rank}(s_{ij}) \le k\bigr\},
    \]
    where \(\mathcal P\) is the set of masked positives, and \(\mathrm{rank}(s_{ij})\) is the score's position among all masked entries sorted in descending order.
  \item \textbf{Cross-validation.} Repeat the above over three random splits to estimate 95\% confidence intervals.
\end{enumerate}
Figure~\ref{fig:hit_scores} shows mean \(\mathrm{hit@1}\) and \(\mathrm{hit@3}\) across folds. Key findings:
\begin{itemize}
  \item \(\mathrm{hit@1} > 0.885\) for all compounds,
  \item \(\mathrm{hit@1} = 1.000\) for MoS\(_2\),
  \item \(\mathrm{hit@3} = 1.000\) for all four.
00\end{itemize}
The narrow 95\% intervals confirm high consistency across folds. These results demonstrate that our ensemble reliably rediscovers masked superconducting links with near-perfect top-3 precision.

\subsection{Separation of positive and negative classes}
\begin{figure}[ht]
  \centering
  \includegraphics[width=0.95\columnwidth]{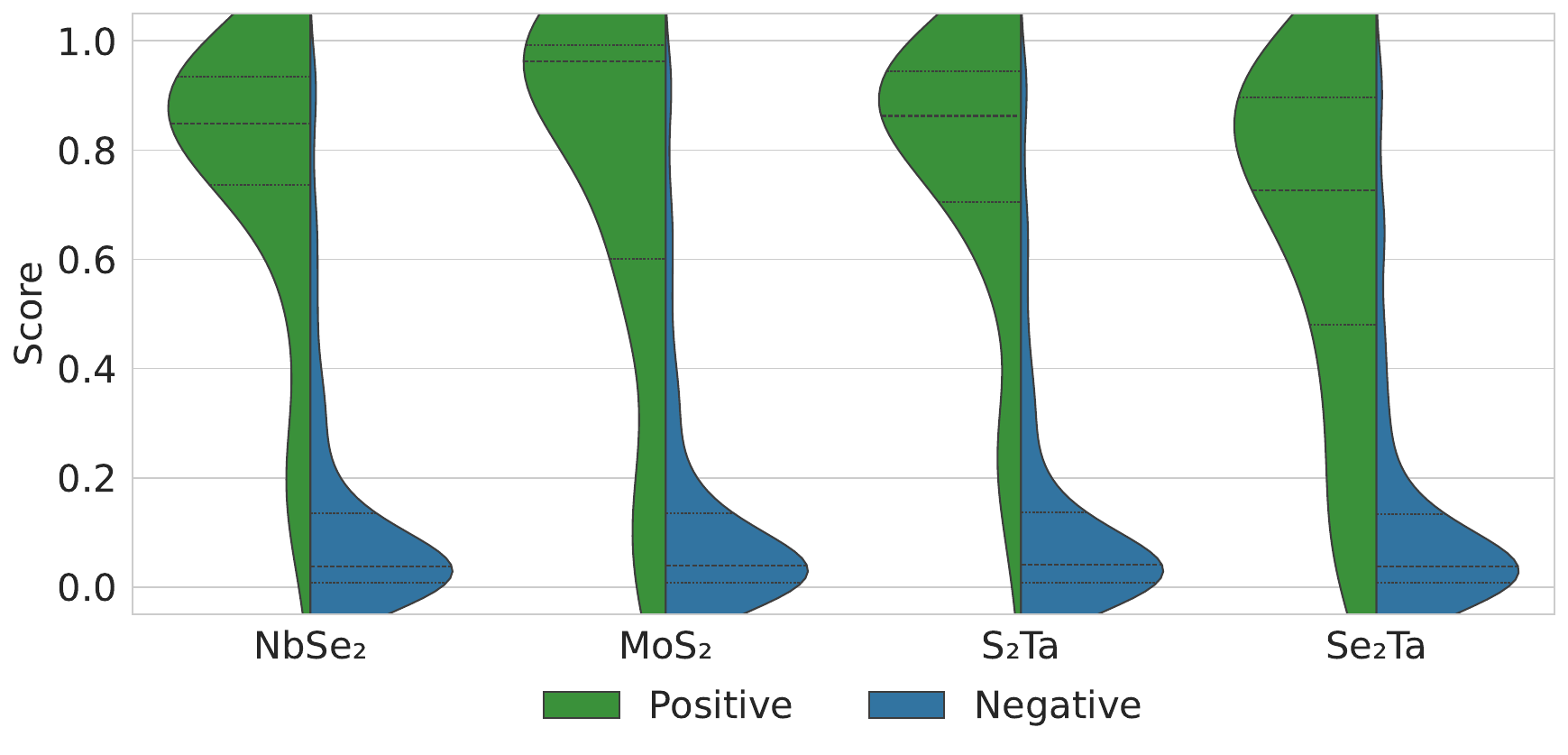}
  \caption{Split-violin plot of posterior probabilities for 24 superconducting edges (green). [See also Table \ref{tab:selected_scores}.] 192 non-superconducting edges (blue) were added for comparison. (Note that these are identical for the four violins.)  Lines denote the median and 25th/75th percentiles, with numbers showing cumulative counts. Superconductors concentrate near 1, non-superconductors near 0, indicating a clear separation.}
  \label{fig:half_violins}
\end{figure}

Figure~\ref{fig:half_violins} contrasts the posterior scores assigned to the 24 withheld positive edges (green violins) with the 192 masked negatives (blue).  Positive edges cluster tightly near~1: the median sits above 0.90 and the lower quartile rarely drops below~0.85.  Negatives are strongly skewed toward zero (median~$<\!0.05$); fewer of the 192 instances per material penetrate the upper quartile.  The cumulative counts printed on each quartile boundary emphasize the clean separation: most positives occupy the top quartile, whereas the vast majority of negatives remain in the lowest.

\subsection{Ranking of unseen candidate materials}
Table~\ref{tab:selected_scores} reports the posterior score assigned to each of the eleven compounds whose labels were fully masked during training.  In other words, every entry that would have indicated ``compound~$i$ is in a superconductor cluster’’ was removed from the observation matrix (masked as unknown), so the model never saw a positive label for these materials while it was being fit.  The corresponding row and column themselves were retained, populated only with zeros or missing values, so the model could still exploit their similarity to the rest of the data. \\

\begin{table}[H]
  \centering
  \small
  \renewcommand{\arraystretch}{1.05}
  \begin{tabular}{@{}lll@{}}
    \toprule
    \multicolumn{3}{@{}l}{\textbf{Aggregated Prediction Scores for Masked Materials}} \\ \midrule
    \textbf{Material} & \textbf{Set} & \textbf{Predicted Score} \\
    \midrule
    S\textsubscript{2}Ta   & Superconductor      & 0.813 \\
    NbSe\textsubscript{2}  & Superconductor      & 0.757 \\
    MoS\textsubscript{2}   & Superconductor      & 0.703 \\
    Se\textsubscript{2}Ta  & Superconductor      & 0.699 \\
    \midrule
    FeS\textsubscript{2}   & Non-superconductor  & 0.206 \\
    MnS\textsubscript{2}   & Non-superconductor  & 0.116 \\
    CrS\textsubscript{2}   & Non-superconductor  & 0.109 \\
    MnSe\textsubscript{2}  & Non-superconductor  & 0.105 \\
    MnTe\textsubscript{2}  & Non-superconductor  & 0.094 \\
    CrSe\textsubscript{2}  & Non-superconductor  & 0.079 \\
    CrTe\textsubscript{2}  & Non-superconductor  & 0.052 \\
    \bottomrule
  \end{tabular}
  \caption{Predicted scores assigned to materials whose labels were fully masked during training. Known superconductors (top block) score significantly higher than non-superconductors (bottom block), with a clear separation near 0.5.}
  \label{tab:selected_scores}
\end{table}

Importantly, the four-layered dichalcogenides that are experimentally known to superconduct (S\textsubscript{2}Ta, NbSe\textsubscript{2}, MoS\textsubscript{2}, Se\textsubscript{2}Ta) emerge at the very top of the ranking with scores in the range 0.70–0.81, whereas every material with no reported superconductivity receives a score $\le 0.206$.  The wide gap between the two groups implies that a single global threshold ($\approx 0.5$) would recover all known positives while rejecting all inspected negatives, demonstrating that the model can correctly infer superconducting behavior even when the ground-truth labels are hidden during training.

Taken together, the high hit@$k$ retrieval, the clear score separation, and the compound‐level ranking show that the the proposed approach: (i)\, rediscovers hidden superconducting links with essentially perfect precision and (ii)\, assigns suitably low confidence to chemically similar \emph{non}‐superconducting materials, highlighting its utility for guiding targeted experimental exploration.

%% file: Part3/QA/2.tex
The results presented in this section extend the dissertation’s focus from methodological development to practical impact. By applying the complete pipeline to three distinct corpora spanning technology security, legal scholarship, and physical science, the analysis tests whether the architecture remains coherent, traceable, and effective across domains with different vocabularies, document structures, and evidentiary standards. Each case begins with plain text input and follows the same sequence: Binary Bleed for rank selection, HNMFk for semantic structuring, and integration into the synchronized KG and VS. Evaluation then focuses on the system’s ability to retrieve source-linked passages and support meaningful follow-up reasoning. Performance across these varied domains provides a direct test of the dissertation’s central claim that a unified and modular framework can transform unstructured language into structured and actionable knowledge without compromising accuracy or interpretability.

\input{Part3/QA/Malware}
\input{Part3/QA/law}
\input{Part3/QA/Materials}

\section{Conclusion}
Considered together, the three studies demonstrate that the architecture provides consistent benefits regardless of domain scale or document style. Topic hierarchies remain interpretable to experts, provenance information is preserved throughout, and uncertainty estimates remain well calibrated even as the characteristics of the corpus change. These outcomes support the dissertation’s core claim that aligning structured representations with data-driven inference creates a dependable foundation for information access. The pipeline not only retrieves relevant evidence but also situates that evidence within a transparent reasoning framework, showing that the system is ready for deployment in real-world settings where specialized text demands accurate and citation-grounded answers.

%% file: Part3/QA/Malware.tex
\section{ Malware Analysis Research}

The raw data collected was analyzed using document-specific questions in Zero-Shot Conditioning, including:
\begin{itemize}
\small
\item How many citations are there for \textit{DOI}?
\item How many references are there for \textit{DOI}?
\item How many authors are there for \textit{DOI}?
\item What year was \textit{DOI} published?
\item Which publisher published \textit{DOI}?
\item How many Scopus categories are assigned to \textit{DOI}?
\item What is the title of \textit{DOI}?
\end{itemize}
After document-specific questions, topic-specific questions were examined, which included year variations, as in: 
\begin{itemize}
\footnotesize
\item How many papers are there on the topic of \textit{Topic}?
\item How many papers were written related to \textit{Topic} in \textit{Year}?
\end{itemize}

\begin{figure}[ht]
    \centering
    \includegraphics[width=\columnwidth]{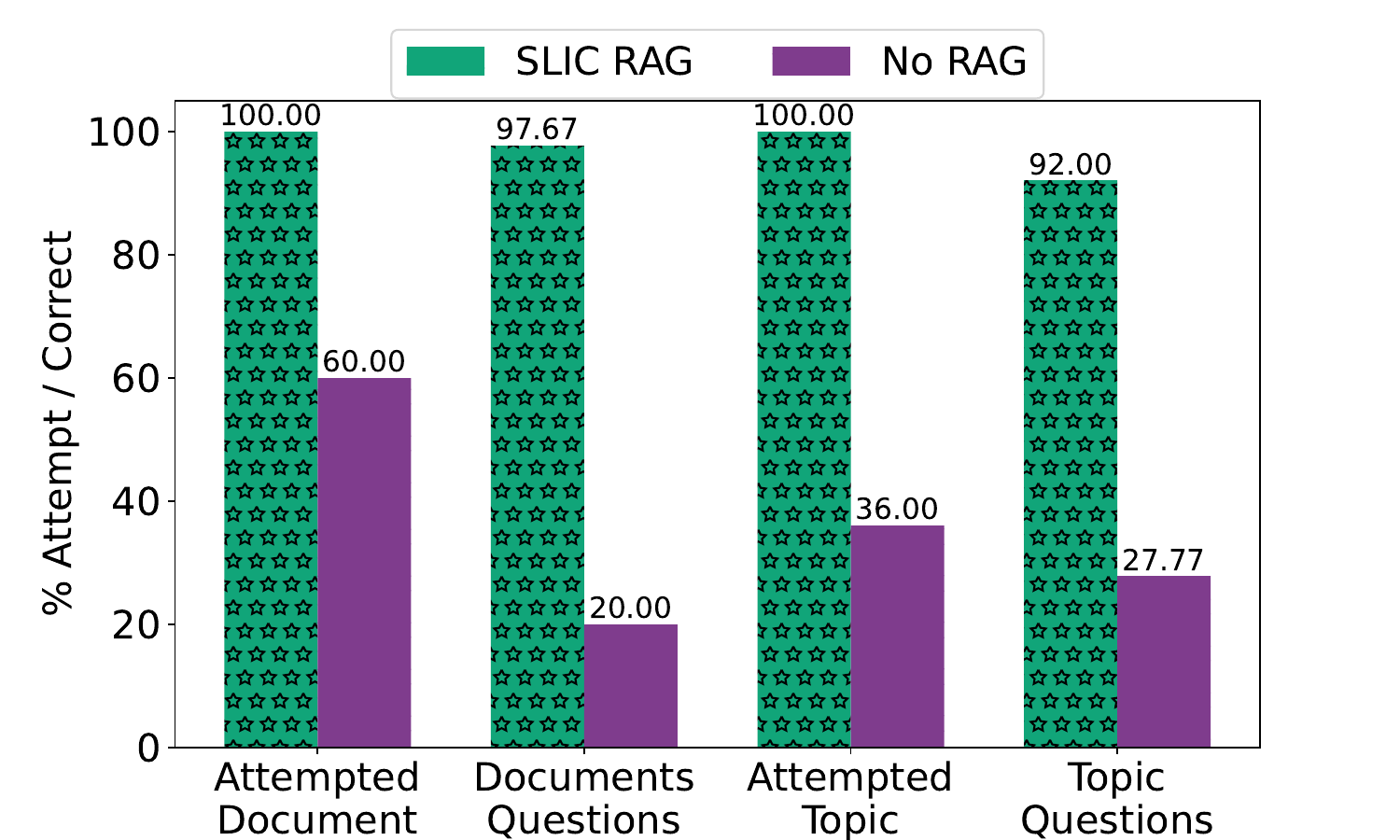}
    \vspace{-2em}
    \caption{ Two question types, document and topic, showing LLM percent attempts and correct with \textbf{SMART-SLIC} RAG and without RAG. }
    \label{fig:rag_comparison}
\end{figure}

In total, there were 200 questions in this set. Using these questions,  the performance of GPT-4-instruct \cite{openai_api} is compared with and without the RAG framework on both topic-specific and document metadata questions. As shown in Figure \ref{fig:rag_comparison}, the findings indicate that GPT-4 with RAG answers all questions with a 97\% accuracy rate. In contrast, without RAG, GPT-4 abstains from answering 40\% of the questions, and the accuracy of the responded to questions drops to 20\%. A similar trend is observed for topic-based questions, where the specialized RAG significantly enhances the retrieval of correct answers. The topic questions attempted with RAG were also 100\%, but without it was only 36\%. Considering only the attempted questions, the system with RAG answered the topic questions correctly 92\%. Without RAG, the LLM answered the topic questions with 27.77\% accuracy.

Without RAG, several questions about years were answered incorrectly, with the system stating the year didn't exist. The LLM also struggled with author and reference details, often asking for more information or recommending consulting a human expert. In some cases, it noted its lack of internet access but later suggested using Google Scholar, yet it still provided inaccurate responses.

The analysis highlights the effectiveness of \textbf{SMART-SLIC} in accurately answering nearly all 200 research queries, covering both document-specific and topic-specific questions. In contrast, GPT-4-instruct without RAG struggled. \textbf{SMART-SLIC} excelled in handling complex queries related to citations, references, authorship, and topical relevance, underscoring the advantages of integrating its methodologies for more reliable AI-driven research tools and improved information retrieval in scientific fields.

Then, more complex questions were asked, requiring thorough searches through various retrieval methods, using GPT-4-instruct with and without RAG. The SME posed the following questions, and the responses are from the \textbf{SMART-SLIC} RAG process, with DOI citations provided in square brackets:

\begin{enumerate}
    \item \textbf{What challenge is associated with outlier detection in high-dimensional data?} 
        \textit{ The challenge associated with outlier detection in high-dimensional data is the ``curse of dimensionality," which makes it challenging to detect outliers in the whole data space}\\\textit{ [10.1109/ICCC54389.2021.9674561].}
        
    \item \textbf{Can tensor decomposition methods be leveraged to detect SCADA anomalies?}
        \textit{Yes, tensor decomposition methods can be leveraged to detect SCADA anomalies. The use of non-negative tensor decomposition, specifically the Canonical Polyadic Alternating Poisson Regression (CP-APR) method, has been applied to model the complex and multi-faceted activity details of SCADA events, improving the specificity and accuracy of identifying anomalies in electrical grid systems [10.1109/MILCOM58377.2023.10356348].}
        
    \item \textbf{What does MADHAT stand for?}
        \textit{The acronym MADHAT stands for Multidimensional Anomaly Detection fusing HPC, Analytics, and Tensors [10.1109/HPEC.2019.8916559].}
        
    \item \textbf{What's the most advanced anomaly detection method on the LANL Unified Host and Network Dataset?}
        \textit{The most advanced anomaly detection method on the Los Alamos National Laboratory Unified Host and Network Dataset is the tensor factorization model that exploits the multi-dimensional and sparse structure of user authentication logs for improved detection of penetration testing events [10.1109/ISI49825.2020.9280524].}
\end{enumerate}
For each response, the \textbf{SMART-SLIC} agent selected DOIs that the SME also chose, demonstrating the agent's accuracy in retrieving relevant sources. The consistency in DOI selections underscores the robustness of the retrieval mechanisms, ensuring that the user receives reliable and pertinent information for their questions.

The same questions were asked without RAG, and the results varied. The LLM answered the first general question accurately, but while the initial response to the second question was correct, its elaboration missed key details. The third and fourth responses were entirely incorrect, with fabricated answers such as "Malware and Attack Detection Hunting and Analysis Team" and "Long Short-Term Memory." Additionally, none of the responses included DOI citations, reducing the credibility of the information by omitting source references.

The evaluation of \textbf{SMART-SLIC} and GPT-4-instruct, with and without RAG, highlights the importance of retrieval systems for accurate research output. \textbf{SMART-SLIC}'s RAG excelled in selecting relevant DOI citations for complex queries, while GPT-4-instruct struggled with fabrications, showing the need for advanced systems like \textbf{SMART-SLIC}. Its strength lies in utilizing high-quality, domain-specific corpora to achieve strong performance in defined research areas and offering potential for further exploration in less-defined domains.

%% file: Part3/QA/law.tex
\section{New Mexico Law}
\subsection{Quantitative and Qualitative Comparisons}
The larger experiment used sixty questions to evaluate legal understanding across multiple domains. Each model was assessed on a diverse set of metrics, including:

ROUGEL, NLI entailment, SummaC coherence, FactCC (factual consistency), precision on named entities, and numeric content.
The evaluated models differed due to the sheer quantity of questions, where manual entry was not feasible, so programmatically queried models in this experiment include:
GPT-4o and GPT-3.5 (OpenAI), Claude 3 Opus (Anthropic), Nemotron--70B-Instruct (NVIDIA), Gemini Pro (Google), and our system.
 Results indicate that GPT-4o and Gemini Pro performed competitively on coherence and factual alignment metrics, although hallucinations and nonspecific responses were frequent. Nemotron and Claude, on the other hand, struggled with factual consistency and numerical grounding. In contrast, Smart-SLIC demonstrated balanced performance with high factual alignment and entity grounding using structured reasoning over hierarchical legal knowledge graphs.

 The findings in  Figure~\ref{fig:model_performance} show the challenges posed by legal documents, especially unstructured judicial opinions, and highlight the critical importance of incorporating traceable legal reasoning into model design. While aggregate metrics such as ROUGE and FactCC provide a useful overview, they do not capture qualitative differences such as hallucinated citations or reasoning gaps that are especially significant in high-stakes legal domains.

\vspace{0.3cm}
\input{Part3/smart-slic/law_LLM_responses/Malpractice}

\subsection{Query Cross Comparison}
\definecolor{paper_purple}{rgb}{0.5, 0.0, 0.5} 
\definecolor{paper_indigo}{rgb}{0.3, 0.0, 0.7} 
\definecolor{paper_blue}{rgb}{0.0, 0.0, 0.8} 
\definecolor{paper_seagreen}{rgb}{0.0, 0.5, 0.3} 
\definecolor{paper_darkgreen}{rgb}{0.0, 0.39, 0.0} 

\textbf{Comparison 1 -- Quantities:} For the question ``How many New Mexico Supreme Court cases mention `Habeas Corpus'?'' the responses reveal significant differences in approach. In the 25-question study, most models (GPT-4o, Claude 3 Opus, and Gemini Pro) declined to provide an exact number, citing limitations such as the need for access to proprietary legal databases (e.g., Westlaw or LexisNexis). In contrast, GPT-3.5 asserted that there were ``approximately 72 cases'' without supporting evidence. Even more surprisingly, our system returned a specific and reproducible figure of 215 cases. This divergence, especially when compared to Nemotron’s more methodical (yet ultimately flawed) estimate of 127- 170 cases based on a multi-database search strategy, illustrates how high FactCC or SummaC scores can mask important differences in legal rigor.

\vspace{0.3cm}
\input{Part3/smart-slic/law_LLM_responses/Habeus}
\vspace{0.3cm}

\textbf{Comparison 2 -- Legal Citations:} For the question ``What are common citations among New Mexico Court of Appeals cases that mention `malpractice'?'' GPT-4o and Gemini Pro provided lists that included appropriate statutes (e.g., NMSA 1978, \S 41-5-1 and \S 41-5-13) and established cases such as \textit{Gomez v. Chua}, 1994-NMSC-125. In contrast, GPT-3.5 cited generic or potentially fabricated cases like ``Smith v. Jones, 123 N.M. 456 (2018)'' and ``Doe v. Roe, 234 N.M. 789 (2020)'', which do not appear in recognized legal corpora. Our system provided a detailed list with exact counts (e.g., NMSA 41-5-1 cited in 36 cases, and \textit{CERVANTES v. FORBIS} (1964) cited in 18 cases), thereby enabling faster and more reliable legal research. Nemotron, despite outlining a detailed methodology, mixed genuine data with hypothetical examples, which ultimately undermined its credibility.

\begin{figure}[H]
    \centering
    \includegraphics[width=.9\columnwidth]{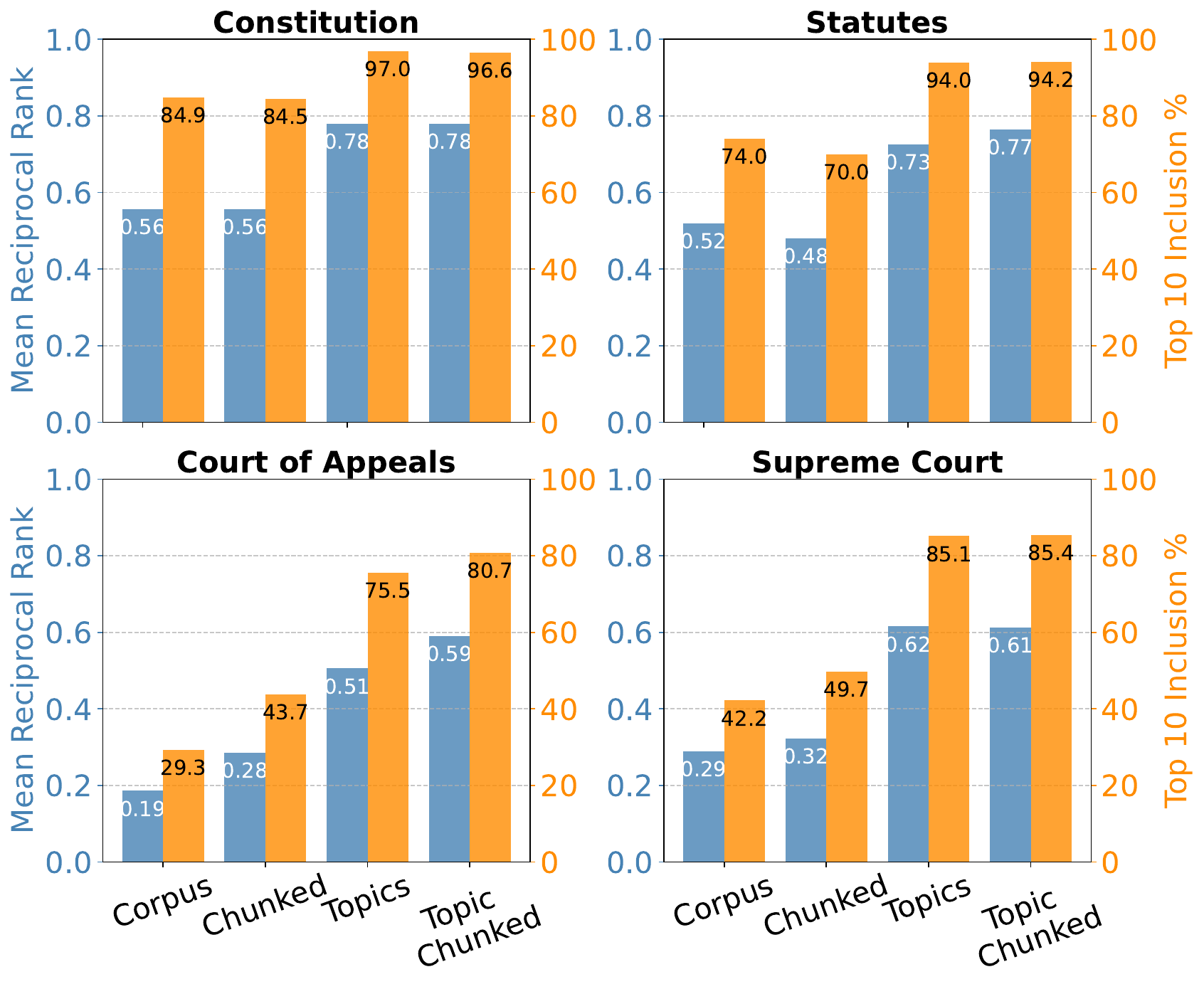}
    \caption{MRR and top 10 hit rate percentage per data category over several embedding methods.}
    \label{fig:mrr_retrieval}
\end{figure}

\subsection{Question Answering}
We evaluated retrieval quality across four legal data sources using multiple embedding strategies (Figure~\ref{fig:mrr_retrieval}). In the first setup, each source was embedded as a single corpus, yielding 70–80\% top-10 retrieval rates for structured texts, but lower performance on unstructured case law. Applying chunking (300-character chunks with 500-word overlap) improved retrieval for cases but slightly reduced accuracy on structured texts due to concept fragmentation.

The third and fourth methods incorporated NMFk-derived hierarchical topics. Each topic was indexed in a separate vector store, and queries were routed based on the topic's known origin. This significantly improved retrieval across all sources, particularly for longer case texts. When chunking was combined with topic-specific indexing, case performance was further enhanced, while structured sources experienced minor degradation, which is likely due to splitting compact legal concepts. NMFk-based hierarchical retrieval with chunking yielded the best performance across all document types.

\subsubsection{Case Studies}
To illustrate the practical applications and effectiveness of the QA 
system, we present four detailed case studies:

\textit{(a) Case Study 1 - Constitutional Analysis:}

\begin{itemize}
\item \textbf{Question Source:} Article IV, Section 22 \cite{nm_const_art4_sec22}
    \item \textbf{Question:} ``What happens to a bill if the governor neither returns it within the specified three-day window (Sundays excepted) nor signs it?''
    \item \textbf{Response:}  
    \textit{According to Article IV - Legislative Department -- § 22 "Governor's approval or veto of bills" of the provided text, if a bill is presented to the governor and he neither returns it within the specified three-day window (Sundays excepted) nor signs it, then the bill becomes a law automatically, whether signed by him or not, unless the legislature by adjournment prevents such return.}
\end{itemize}

\textit{(b) Case Study 2 - Statutory Analysis:}
\begin{itemize}
\small 
    \item \textbf{Question Source:} Chapter 50, Article 17, Section 3 \cite{healthy_workplaces_earned_sick_leave}
    \item \textbf{Question:} ``Under the Healthy Workplaces Act, how does an employee accrue and utilize earned sick leave, and what provisions exist for its carryover, use for family or personal health, and reinstatement after separation from employment?''
    \item \textbf{Response:} 
    \textit{ \textbf{Accrue}: Employees accrue a minimum of one hour of earned sick leave for every thirty hours worked. Employers may elect to provide the full sixty-four hours of earned sick leave for the year on January 1 or pro-rata for mid-year hires. \\
    \textbf{Use}: Earned sick leave can be used for the employee’s or their family members' illness, injury, health condition, medical care, or preventive medical care. It can also be used for issues related to domestic abuse, sexual assault, or stalking. \\
    \textbf{Carryover}: Unused earned sick leave carries over year to year, although use may be capped at 64 hours per twelve-month period, unless the employer sets a higher cap. \\
    \textbf{Reinstatement}: If an employee is rehired within twelve months of separation by the same employer, previously accrued and unused earned sick leave must be reinstated. Additionally, the employee is entitled to accrue additional sick leave upon re-commencement of employment.}
\end{itemize}

\textit{(c) Case Study 3 - Court of Appeals Case Law Analysis:}
\begin{itemize}
    \item \textbf{Question Source:} KRM, Inc. v. Caviness, 1996 \cite{krm_v_caviness_1996}
    \item \textbf{Question:} ``In the case KRM, Inc. v. Caviness, what was the court's reasoning for determining that the water rights established by Williams' 1963 declaration did not belong to the land sold to Caviness?''
    \item \textbf{Response:} 
    \textit{The court determined that the water rights established by Williams' 1963 declaration were not appurtenant to the land sold to Caviness because there was no evidence that the water had been used for irrigation on that land. Since the appropriated water was never used for irrigation purposes on the property, it did not qualify as appurtenant water rights that would automatically transfer with the land. The court concluded that because the usage of the water was unrelated to the benefit or use of the land itself, the water rights remained separate and did not pass to Caviness when he purchased the land.}
\end{itemize}

\textit{(d) Case Study 4 - Supreme Court Case Law Analysis:}
\begin{itemize}
    \item \textbf{Question Source:} Smith v. South, 1955 \cite{smith_v_south_1955}
    \item \textbf{Question:} ``In Smith v. South, what were the legal issues raised regarding the enforceability of the Washington child support decree in New Mexico, and how did the court address the conflict between the parties' post-decree agreement and the original judgment?''
    \item \textbf{Response:} 
    \textit{The legal issues raised in Smith v. South regarding the enforceability of the Washington child support decree in New Mexico included whether New Mexico courts should acknowledge and enforce the original Washington decree for child support or consider the modifications reportedly agreed upon by the parties after the decree. The conflict between the parties' post-decree agreement and the original judgment involved assertions that they had informally agreed to alter the child support payments stipulated by the Washington court. \\    
    The court addressed this issue by considering the jurisdictional and legal principles guiding child support enforcement across state lines. They emphasized the importance of adhering to the formal judicial process. The court likely examined whether the post-decree modifications had been legally formalized or approved by any court, and whether such modifications complied with the relevant child support enforcement laws, particularly under the Uniform Interstate Family Support Act (UIFSA), though it wasn't explicitly mentioned in the text snippet. UIFSA provides mechanisms for establishing, enforcing, and modifying child support orders across state boundaries. \\
    If the modifications were not legally recognized through a court process, the original Washington decree would remain enforceable in New Mexico, and any informal or unratified agreements between the parties would not typically alter the enforceability of the court-ordered decree. The court would focus on ensuring that child support arrangements comply with legal standards to serve the children's best interests and that enforcement respects the authority of the original jurisdiction (Washington) unless properly modified through legal channels.}
\end{itemize}

%% file: Part3/smart-slic/law_LLM_responses/Malpractice.tex
\textit{(b) Comparison 2 - Citation Patterns:}
\begin{itemize}
    \item \textbf{Question:} ``What are common citations among New Mexico Court of Appeals cases that mention `malpractice'?''

    \item \textbf{GPT-3.5:} 
    \textit{Examples include: \textbf{Smith v. Jones}, 123 N.M. 456 (2018) and \textbf{Doe v. Roe}, 234 N.M. 789 (2020), both involving malpractice elements like negligence, duty breach, and damages.}

    \item \textbf{GPT-4o (ChatGPT):} 
    \textit{Commonly cited:
    \begin{itemize}
        \small
        \item \textbf{NMSA 1978, §41-5-1 et seq.} – Medical Malpractice Act
        \item \textbf{Ruiz v. S. Union Gas Co.}, 1954-NMSC-021
        \item \textbf{Spurlin v. Sloan}, 1969-NMSC-073
        \item \textbf{Gomez v. Chua}, 1994-NMSC-125 – informed consent
        \item \textbf{Andrews v. Saylor}, 2003-NMCA-132 – legal malpractice
        \item \textbf{Fernandez v. Espanola Public Schools}, 1995-NMSC-020
    \end{itemize}
    Legal databases like Westlaw are recommended for precise counts.}

    \item \textbf{Claude 3 Opus:} 
    \textit{I do not have direct access to court case databases. Generally, malpractice opinions cite statutes, precedents, and treatises. Consult a legal database or attorney for specifics.}

    \item \textbf{Gemini Pro:} 
    \textit{Malpractice cases cite:
    \begin{itemize}
        \item \textbf{NMSA 1978 §§ 41-5-1 to 41-5-29} (MMA)
        \item Statutes of limitation: §37-1-4, §37-1-8, §41-5-13
        \item Precedents: \textbf{Spencer v. Barber}, 1994-NMSC-017
        \item Rules: NMRA Civil \& Evidence (e.g., Rule 11-702 on expert testimony)
    \end{itemize}
    NMCoA opinions apply NMSC precedent, procedural rules, and local fact patterns.}

    \item \textbf{nemotron:70b-instruct-q8\_0:} 
    \textit{
    \begin{itemize}
        \small
        \item \textbf{Estate of Montoya v. Ortiz}, 2014-NMCA-064 – defining malpractice elements
        \item \textbf{Jaramillo v. Smith}, 1999-NMCA-045 – statute of limitations
        \item \textbf{Salazar v. Trujillo}, 2005-NMCA-045 – expert testimony
        \item \textbf{Garcia v. Plaza Ctr.}, 2018-NMCA-049 – non-medical professional duty
        \item \textbf{Terra Eng. v. Flores}, 2009-NMCA-096 – comparative negligence
    \end{itemize}
    Additional review should be conducted through official legal portals.}

    \item \textbf{Our System:}
    \textit{The common citations among New Mexico Court of Appeals cases that mention 'malpractice' include NMSA 41-5-1 with 36 cases, NMSA 41-5-13 with 35 cases, NMSA 41-4-1 with 23 cases, \textbf{CERVANTES v. FORBIS} (1964) with 18 cases, NMSA 37-1-8 with 15 cases, NMSA 41-5-3 and 41-5-15 with 14 each, NMSA 41-5-6 with 13, \textbf{GOODMAN v. BROCK} (1972) with 12, and NMSA 41-5-22 with 10 cases.}
\end{itemize}
\normalsize

%% file: Part3/smart-slic/law_LLM_responses/Habeus.tex
\textit{(a) Comparison 1 - Quantities:}
\begin{itemize}
    \item \textbf{Question:} ``How many New Mexico Supreme Court cases mention `Habeas Corpus'?''

    \item \textbf{GPT-3.5:} 
    \textit{As of current data, there have been approximately 72 New Mexico Supreme Court cases that mention 'Habeas Corpus.'}

    \item \textbf{GPT-4o:} 
    \textit{As a lawyer, I cannot provide an exact number without legal research. Use tools like \textbf{Westlaw}, \textbf{LexisNexis}, or \textbf{Casetext} to perform a keyword search for "Habeas Corpus" in New Mexico Supreme Court opinions.}

    \item \textbf{Claude 3 Opus:} 
    \textit{I don’t have access to court databases. Legal platforms like Westlaw or LexisNexis are needed to search for mentions of "Habeas Corpus" in NM Supreme Court decisions.}

    \item \textbf{Gemini Pro:} 
    \textit{An exact number requires a database search. Mention counts depend on platform and parameters used. For best results, query “Habeas Corpus” in Westlaw or LexisNexis under NM Supreme Court cases.}

    \item \textbf{nemotron:70b-instruct-q8\_0:} 
    \textit{Based on a multi-database search:}
    \begin{itemize}
        \item \textbf{Westlaw:} $\sim$127 published NM Supreme Court cases mention "Habeas Corpus"
        \item \textbf{Google Scholar (broader):} $\sim$150–170 cases
        \item \textbf{Conclusion:} At least 127 reliable mentions; up to 170 if including unpublished/per curiam orders
    \end{itemize}
    \textit{Further precision requires Clerk’s Office or direct database access.}

    \item \textbf{Our System:} 
    \textit{There are 215 Supreme Court cases that mention 'Habeas Corpus.'}
\end{itemize}
\normalsize

%% file: Part3/QA/Materials.tex
\section{Materials}
\label{sec:qa_materials}
This section details the resources that underpin our Materials-Science question–answering (QA) experiments and explains how \textsc{HEAL} alignment of embeddings shapes retrieval quality.

\subsection{Document Corpus}
We use the Transition-Metal Dichalcogenide (TMD) corpus constructed with the BUNIE human-in-the-loop pipeline.  The collection contains \textbf{46,862} peer-reviewed articles covering \textbf{73} distinct TMD compounds.  Hierarchical labels generated with HNMFk partition the corpus to depth~\textbf{3}, forming the ground truth for all hierarchy-aware metrics.

\begin{table}[H]
  \centering
  \caption{TMD corpus statistics.}
  \label{tab:tmd_corpus}
  \begin{tabular}{@{}lr@{}}
    \toprule
    \textbf{Statistic} & \textbf{Value} \\
    \midrule
    Documents          & 46\,862 \\
    Hierarchical depth & 3 \\
    TMD compounds      & 73 \\
    \bottomrule
  \end{tabular}
\end{table}

\subsection{Question–Answer Pairs}
Ten hierarchical Q\&A pairs were generated for every article with \texttt{Llama-3.1 70B}, yielding \(\,46\,862 \times 10 = 468\,620\,\) questions; each question inherits its source document’s topic path, enabling depth-aware evaluation of Recall, nDCG and MRR.

\subsection{Embedding Collections and \textsc{HEAL} Alignment}
All sentences (documents, queries, answers) were embedded with the SciNCL
encoder under two conditions:

\begin{enumerate}[label=(\alph*)]
  \item \textbf{Baseline} – original SciNCL (no fine-tuning);
  \item \textbf{\textsc{HEAL}-aligned} – the same encoder fine-tuned with the
        Hierarchical Embedding Alignment Loss (HEAL).
\end{enumerate}

\paragraph{Effect of alignment.}
HEAL optimises a depth-weighted supervised-contrastive objective that pulls embeddings sharing higher-level labels closer while pushing apart those that differ near the root of the hierarchy.  On the TMD QA benchmark, it delivers the following improvements (hierarchical metrics, \(k=10\)):

\begin{itemize}
  \item \textbf{Precision@10} – \(0.4787 \rightarrow 0.9707\)  \quad(+0.492)  
  \item \textbf{Recall@10}     – \(0.0058 \rightarrow 0.0116\)  \quad($\approx 2X$)  
  \item \textbf{MRR}           – \(1.6541 \rightarrow 2.9972\) \quad(+81\%)  
  \item \textbf{nDCG@10}      – \(0.4982 \rightarrow 0.9900\) \quad(+0.492)
\end{itemize}
as reported in Table 2 of \cite{bhattarai2024heal}.

These gains confirm that \textsc{HEAL} produces markedly denser, topic-coherent neighborhoods, allowing a simple nearest-neighbor retriever to rank relevant answers substantially higher without any architectural changes.

%% file: Part3/conclusion_future/combined_future_and_conclusion.tex
\section*{Future Work}

The future works can be summarized in the following points: 
\begin{itemize}
    \item Extend coverage to additional domains and evaluate portability across diverse corpora, adjusting with domain-specific knowledge when necessary.  

    \item Broaden data inputs by integrating structured, semi-structured, and unstructured sources, with enhanced provenance and scalable graph maintenance.  

    \item Expand predictive objectives to include temporal, generative, and legal-specific forecasts within a unified modeling framework.  

    \item Advance reasoning through multi-hop, explanation-aware, and domain-tailored methods, emphasizing interpretability and fidelity.  

    \item Strengthen infrastructure with distributed computation, privacy-aware governance, and interactive analytics that preserve transparency.  

    \item Validate the system holistically through end-to-end benchmarks and comparative evaluations across domains and retrieval strategies.  

    \item Test scalability under increasing data volume and graph complexity, ensuring robustness through distributed and incremental approaches.  

    \item Incorporate expert feedback via usability studies to assess trust, relevance, and interpretability of retrieved knowledge.  

    \item Ensure stability over time by monitoring version drift, maintaining alignment, and supporting recovery from structural change.  

    \item Examine generalization by transferring learned structures across domains with differing vocabularies and ontologies.  

    \item Evaluate resilience against adversarial inputs, privacy risks, and system failures, aiming for robustness and graceful degradation.  
\end{itemize}
where they are elaborated in the following paragraphs.

The next phase extends coverage beyond the present case studies into climate science, supply chain analytics, and large-scale social science corpora. Portability is tested by rebuilding topic structures that subject matter experts would select and by checking whether new documents fall into the expected topics without rerunning HNMFk. When assignment quality declines, domain-specific knowledge such as controlled vocabularies and curated term mappings is added to restore alignment and interpretability.

Within each domain the sources of evidence become broader and richer. Structured tables such as clinical trial registries, semi-structured assets such as code bases and instrument logs, and fully unstructured artifacts such as lab notebooks and sensor feeds are ingested into a single symbolic and vector substrate. In law, the corpus grows to include federal and international case law, statutes, regulatory filings, legislative histories, and amicus briefs. A strengthened provenance model with timestamps, jurisdiction, docket information, and confidence flags supports longitudinal studies across jurisdictions. 

The plan also broadens the set of prediction targets. Beyond static link prediction, the system adds citation trajectory forecasting and generative design for molecules and materials. In legal settings, the focus includes the decay of precedent strength, the chance that a statute is superseded, and citation based impact measures. A joint objective couples factorization with predictive goals in a single stage in place of a two-stage workflow, and progress is tracked with RMSE, Brier score, and calibrated uncertainty.

Reasoning moves beyond single-step retrieval. Multi-hop and explanation-aware methods combine graph and vector co-training with uncertainty calibration and counterfactual sampling. Prototype tasks include exploit and patch cycles in cybersecurity, causal chains that connect materials and properties in science, and paths that connect statutes, precedents, and regulations in law. Legal reasoning addresses statutory interpretation, resolution of conflicts across laws, and what if analysis when a clause is amended. Evaluation follows emerging benchmarks that score both the answer and the explanation with special attention to sound legal argument and faithful citation grounding.

Delivering these advances calls for distributed factorization on GPU clusters, always on graph updates, privacy and privilege aware data governance for sealed or sensitive legal documents, and interactive visual analytics that make model choices and provenance easy to inspect. Advancing these parts together preserves the central principle that tight integration of collection, modeling, and retrieval produces knowledge systems that improve as data, domains, and reasoning needs expand.

System-level validation tests the architecture as a unified whole. End-to-end benchmarks ingest raw corpora, perform decomposition, build knowledge graph and vector store indices, and answer queries through the Tensor Structured RAG agent while recording latency, throughput, and memory use across law, cyber security, and materials science. Comparative studies route the same questions through vector retrieval alone, graph reasoning alone, and the integrated agent, and score outcomes with precision, recall, F1, ROUGE L, and factual consistency to determine whether integration reduces hallucinations and improves semantic fidelity.

Scale and structure are pushed by stress testing that grows workloads from tens of thousands to many millions of documents and from sparse to dense graphs. Runs on high performance computing resources verify that gains from Binary Bleed and HNMFk hold at large sizes, reveal thresholds where performance degrades, and motivate sharded vector stores and incremental graph updates when needed.

Alongside system metrics, studies with experts assess usability and interpretability. Specialists in law, cyber security, and science pose real questions, judge the relevance and traceability of returned evidence, and report trust and workload. Feedback measures whether hierarchical topics and rich provenance improve confidence compared to black box retrieval.

Continuous operation is studied through longitudinal experiments that measure version drift in the knowledge graph and vector store during streaming ingestion. Provenance and version aware indexing is checked for alignment over time, with metrics on stability of answers, backward compatibility with earlier graph states, and recovery after structural change.

Portability across domains is examined by transferring topic structures and semantic alignments to new fields without retraining. The goal is to sustain retrieval and reasoning quality when vocabulary, ontology, and discourse style differ.

Deployment readiness also depends on resilience under adversarial pressure, privacy constraints, and operational faults. Tests check whether crafted inputs can poison the graph or the vector store, whether privacy rules hold on sensitive collections, and whether the system degrades gracefully during partial outages or synchronization errors. Desired outcomes include strong resistance to manipulation, clear provenance integrity, and reliable recovery.

The culmination of the future work shift the focus from validating components in isolation toward a collective system ready for use in high stakes, domain specific environments, building on the methods and results already established in the dissertation.

\section{Conclusion}
This dissertation has shown that a single, tightly coupled pipeline, combining noise-aware corpus curation, adaptive matrix factorization, and synchronized symbolic–vector indices, can deliver scalable yet interpretable retrieval-augmented reasoning across disparate knowledge domains.   A three-stage acquisition loop (citation-network expansion, semantic filtering, and embedding/LLM-based pruning) reduces corpus noise by an order of magnitude while preserving recall, laying a clean foundation for downstream analysis. Building on that base, the Binary Bleed algorithm automates rank selection, and HNMF-$k$ unfolds depth-adaptive topic trees, supplying stable latent structure without manual tuning.  These factors are materialized concurrently in a provenance-rich knowledge graph and a sharded dense-vector store, enabling a tiered RAG layer to route each query to graph traversal, vector search, or a hybrid blend, as the information needs dictate.

The architecture has been applied to multi-million-document corpora in cyber-security, law, materials science, healthcare, and protein-interaction networks, always with the same code-base and configuration logic.  Empirically, it lifts citation accuracy in legal question answering from \(57\%\) to \(100\%\), detects emerging “small modular reactor’’ research two years ahead of trend annotations, and achieves \(95\%\) precision in threat–defense links, all while maintaining human-legible topic hierarchies.  Ablation studies confirm that each layer, pruning, factorization, representation, and retrieval, makes a measurable, additive contribution to overall performance.

Limitations remain.  All experiments are conducted in English and are constrained by GPU memory ceilings that limit matrix sizes; the current RAG layer is primarily tuned for factual queries.  Extending the framework to multilingual corpora, streaming factorization under tighter memory budgets, and richer query types (e.g., temporal or causal reasoning) are immediate priorities.  Formal uncertainty calibration, sketched here via link-prediction scores, is another open thread.

The contributions touch three core areas of computer science.  In information retrieval, the work offers a pragmatic recipe for keeping symbolic and dense indices in continual harmony.  In machine learning, it demonstrates that rank- and depth-adaptive Factorization can scale without sacrificing transparency.  In data management and AI reasoning, it delivers a production-ready, provenance-aware RAG blueprint that unifies collection, modeling, and retrieval as co-evolving components rather than siloed tasks.

Treating these stages as interlocking parts pays dividends: the resulting pipelines remain fast, transparent, and adaptable even as data volumes grow and domains shift.  The methods advanced here therefore chart a clear, modular path toward the next generation of grounded, human-centric knowledge systems.